\documentclass[12pt]{thesis}

\usepackage{titlesec}
   \titleformat{\chapter}
      {\normalfont\large}{Chapter \thechapter:}{1em}{}
\usepackage{tikz}
\usepackage{times}
\usepackage[english]{babel}
\usepackage{csquotes}   
\usepackage{subcaption}
\usepackage{graphicx}
\usepackage{lscape}
\usepackage{indentfirst}
\usepackage{latexsym}
\usepackage{multirow}
\usepackage{epstopdf}
\usepackage{tabls}
\usepackage{wrapfig}
\usepackage{slashbox}
\usepackage{float}
\usepackage{caption}
\usepackage{supertabular}
\usepackage[onehalfspacing]{setspace} 
\usepackage{enumitem}    

\usepackage[
    colorlinks=true,
    urlcolor=purple,
    linkcolor=blue,
    citecolor=blue,
]{hyperref}

\usepackage{amsmath,amssymb,mathtools}
\usepackage{bookmark}
\usepackage{newtxtext}       
\usepackage{caption}         
\usepackage{anyfontsize}     
\usepackage{dirtytalk}       
\usepackage{parskip}         
\usepackage{xcolor}          
\usepackage{pifont}          
\usepackage{lipsum}          
\usepackage{silence}         
\usepackage{textcomp}        
\usepackage{moresize}        
\usepackage{titlesec}        
\usepackage{bbding}          
\usepackage{siunitx}         
\usepackage{bm}              
\usepackage[counterclockwise,figuresleft]{rotating}

\usepackage{tabularray}      
\usepackage{booktabs}        
\usepackage{tabularx}        
\usepackage{array}           
\newcolumntype{C}{>{\centering\arraybackslash}X}

\RequirePackage{cleveref}
\crefname{figure}{Fig.}{Figs.}
\Crefname{figure}{Figure}{Figures}

\def\sayit#1{\say{\textit{{#1}}}}
\def\quote#1#2{\hspace{0.035\textwidth}\begin{tabular}{|p{0.82\textwidth}}\say{{#1}} --- {#2}\end{tabular}}

\DeclareMathOperator*{\argmin}{argmin_{\mathnormal{}}}
\DeclareMathOperator{\EE}{\mathbb{E}}
\DeclareMathOperator{\FF}{\mathcal{F}}

\DeclareMathOperator{\NN}{\mathcal{N}}
\DeclareMathOperator{\HH}{\mathcal{H}}

\newcommand{\norm}[1]{\left\lVert#1\right\rVert_2}
\newcommand{\KL}[2]{\mathcal{D}_{\mathrm{KL}}\Big{(}{#1}\,\big\|\,{#2}\Big{)}}
\newcommand{\expect}[2]{\EE_{#1} \Big{[} {#2} \Big{]}}
\newcommand{\EXPECT}[2]{\EE_{#1} \Bigg{[} {#2} \Bigg{]}}

\def\SP#1{\textsuperscript{{#1}}}
\def\SB#1{\textsubscript{{#1}}}
\def\SPSB#1#2{\rlap{\SP{#1}}\SB{#2}}
\newcommand{\CAS}{Ca\SPSB{2+}{\it slow} }
\newcommand{\CAF}{Ca\SPSB{2+}{\it fast} }
\newcommand{\CA}{Ca\texorpdfstring{\textsuperscript{2+}}{\texttwoinferior} }

\newcommand{\xmark}{{\Large\color{red}\ding{55}}}
\newcommand{\cmark}{{\Large\color{green}\ding{51}}}
\def\fixate#1{{{\ttfamily fixate-{#1}}}}
\def\obj#1{{{\ttfamily obj-{#1}}}}
\definecolor{MSBlue}{rgb}{.204,.353,.541}
\definecolor{MSLightBlue}{rgb}{.31,.506,.741}
\definecolor{cnvae_color}{HTML}{4878d0}
\definecolor{vae_color}{HTML}{ee854a}
\definecolor{cnae_color}{HTML}{6f6f6f}
\definecolor{ae_color}{HTML}{aeaeae}
\definecolor{enc_color}{HTML}{ED1C24}
\definecolor{dec_color}{HTML}{29ABE2}
\newcommand{\thenc}{{\color{enc_color}\bm{\theta}_{enc}}}
\newcommand{\thdec}{{\color{dec_color}\bm{\theta}_{dec}}}

\usepackage[
    backend=biber,
    sorting=none,
    style=numeric-comp,
    maxnames=2,
    minnames=1,
]{biblatex}
\AtBeginBibliography{\normalsize}

\usepackage{tocloft,calc}
\renewcommand{\cftlottitlefont}{\hspace*{\fill}\large}
\renewcommand{\cftafterlottitle}{\hspace*{\fill}}
\renewcommand{\cftloftitlefont}{\hspace*{\fill}\large}
\renewcommand{\cftafterloftitle}{\hspace*{\fill}}
\renewcommand{\cftchapaftersnum}{:\ }
\renewcommand{\cftchappresnum}{\chaptername\space}
\makeatletter
\g@addto@macro\appendix{%
  \addtocontents{toc}{%
    \protect%
  }%
}
\renewcommand{\baselinestretch}{2}
\begin{document}
\pagestyle{empty}

\hbox{\ } \vspace{.7in}
\renewcommand{\baselinestretch}{1}
\small \normalsize

\begin{center}
\large{{ABSTRACT}}

\vspace{3em}

\end{center}
\hspace{-.15in}
\begin{tabular}{ll}
Title of Dissertation: & {\large Unveiling Secrets of Brain Function with Generative Modeling:}\\
\ \\
& {\large  Motion Perception in Primates \&} \\
& {\large  Cortical Network Organization in Mice } \\
\ \\
& {\large  Hadi Vafaii} \\
& {\large Doctor of Philosophy, 2023} \\
\ \\
Dissertation Directed by: & {\large  Professor Luiz Pessoa} \\
& {\large  Department of Psychology} \\
\ \\
& {\large  Professor Daniel A.~Butts} \\
& {\large Department of Biology} \\

\end{tabular}

\vspace{3em}

\renewcommand{\baselinestretch}{2}
\large \normalsize

This Dissertation is comprised of two main projects, addressing questions in neuroscience through applications of generative modeling. \textbf{Project \#1} (Chapter 4) is concerned with how neurons in the brain \textit{encode}, or \textit{represent}, features of the external world.
A key challenge here is building artificial systems that represent the world similarly to biological neurons. In Chapter 4, I address this by combining Helmholtz's \say{Perception as Unconscious Inference}---paralleled by modern generative models like variational autoencoders (VAE)---with the hierarchical structure of the visual cortex. This combination results in the development of a hierarchical VAE model, which I subsequently test for its ability to mimic neurons from the primate visual cortex in response to motion stimuli. Results show that the hierarchical VAE perceives motion similar to the primate brain.
I also evaluate the model's capability to identify causal factors of retinal motion inputs, such as object- and self-motion. I find that hierarchical latent structure enhances the linear decodability of data generative factors and does so in a disentangled and sparse manner. A comparison with alternative models indicates the critical role of both hierarchy and probabilistic inference.
Collectively, these results suggest that hierarchical inference underlines the brain's understanding of the world, and hierarchical VAEs can effectively model this understanding.

\textbf{Project \#2} (Chapter 5) is about how spontaneous fluctuations in the brain are spatiotemporally structured and reflect brain states such as resting. The correlation structure of spontaneous brain activity has been used to identify large-scale functional brain networks, in both humans and rodents.
The majority of studies in this domain use functional MRI (fMRI), and assume a disjoint network structure, meaning that each brain region belongs to one and only one community. In Chapter 5, I apply a generative algorithm to a simultaneous fMRI and wide-field \CA imaging dataset and demonstrate that the mouse cortex can be decomposed into overlapping communities. Examining the overlap extent shows that around half of the mouse cortical regions belong to multiple communities.
Comparative analyses reveal that network structure derived from \CA signals reproduces many aspects of fMRI--derived network structure. Still, there are important differences as well, suggesting that the inferred network topologies are ultimately different across imaging modalities.
In conclusion, wide-field \CA imaging unveils overlapping functional organization in the mouse cortex, reflecting several but not all properties observed in fMRI signals.

The introduction (Chapter 1) is divided similarly to this abstract: sections \ref{sec:vis} to \ref{sec:motion} provide background information about \textbf{Project \#1}, and sections \ref{sec:orchestra} to \ref{sec:cafmri} are related to \textbf{Project \#2}. Chapter 2 includes historical background, Chapter 3 provides the necessary mathematical background, and finally, Chapter 6 contains concluding remarks and future directions.

\begin{figure}[ht!]
    \centering
    \includegraphics[width=1.0\linewidth]{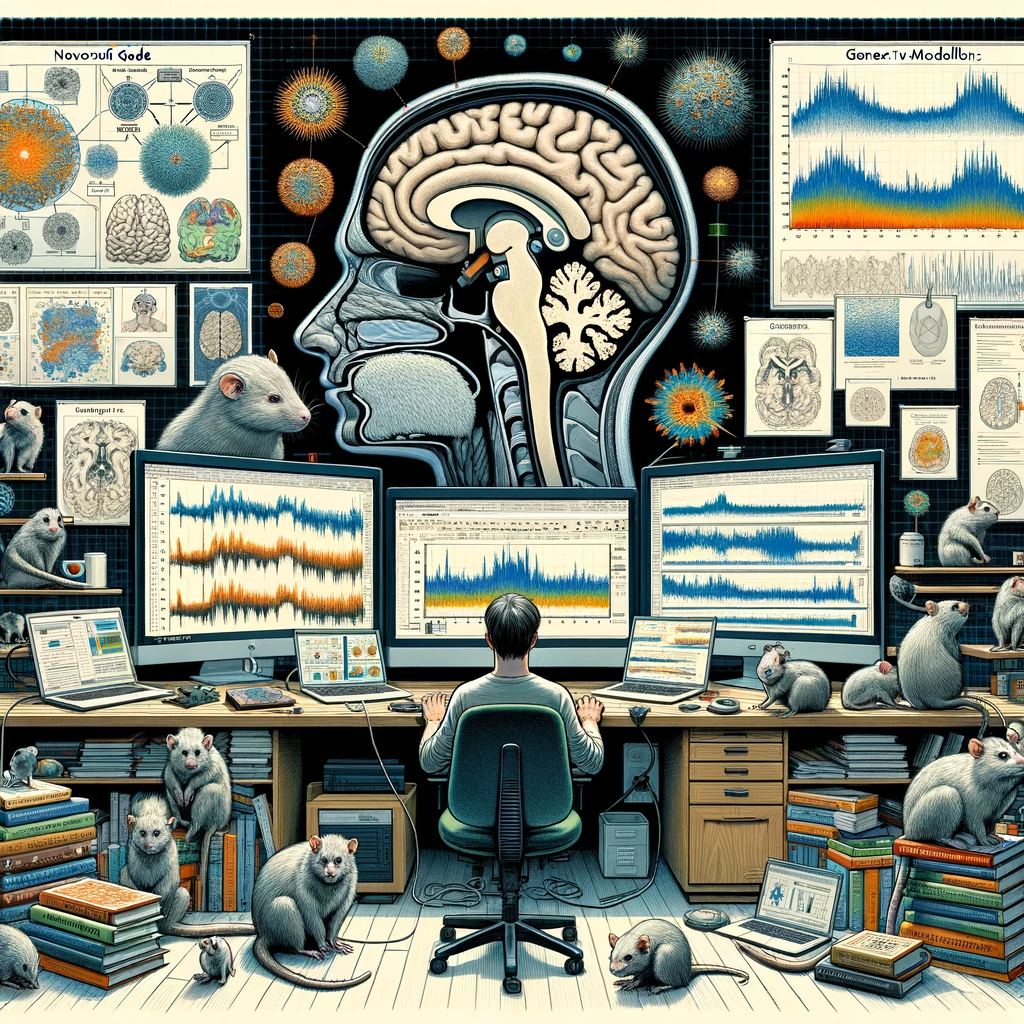}
    \caption[Graphical Abstract (AI Generated)]{
    This graphical abstract, created with the help of ChatGPT and DALL$\cdot$E, captures the essence of my PhD research. I used the October 2023 version, so if you're reading this in the future, you might laugh at how rudimentary it looks.
    To create it, I fed abstracts from my two main papers into ChatGPT, resulting in a representation with a mix of both accuracies and artistic liberties. The phrases ``neural code'' and ``generative modeling'' are subtly integrated into the background, aligning with my Dissertation's themes.
    The image features portrayals of a big mouse and what appears to be a sagittal slice from the primate brain, loosely reflecting the scope of my projects. Although, a macaque monkey brain would have been more precise than the depicted human brain.
    The artwork accurately reflects the need for three monitors and four laptops to do good work in science. However, contrary to the illustration, my office has never been home to such a sizable gathering of mice (as far as I'm aware).
    }
    \label{fig:my-phd}
\end{figure}


\thispagestyle{empty} \hbox{\ } \vspace{1.5in}
\renewcommand{\baselinestretch}{1}
\small\normalsize
\begin{center}

\large{{Unveiling Secrets of Brain Function with Generative Modeling:\\[1em]
Motion Perception in Primates \&\\
Cortical Network Organization in Mice}}\\
\ \\
\ \\
\large{by} \\
\ \\
\large{Hadi Vafaii}
\ \\
\ \\
\ \\
\ \\
\normalsize
Dissertation submitted to the Faculty of the Graduate School of the \\
University of Maryland, College Park in partial fulfillment \\
of the requirements for the degree of \\
Doctor of Philosophy \\
2023
\end{center}

\vspace{7.5em}

\noindent Advisory Committee: \\
\hbox{\ }\hspace{.5in}Professor Luiz Pessoa, Chair/Advisor \\
\hbox{\ }\hspace{.5in}Professor Daniel A.~Butts, Co-Chair/Advisor \\
\hbox{\ }\hspace{.5in}Professor Michelle Girvan \\
\hbox{\ }\hspace{.5in}Professor Evelyn MR.~Lake \\
\hbox{\ }\hspace{.5in}Professor Tom Goldstein, Dean’s Representative

\thispagestyle{empty}
\hbox{\ }

\vfill
\renewcommand{\baselinestretch}{1}
\small\normalsize

\vspace{.5in}

\begin{center}
\large{\copyright \hbox{ }Copyright by\\
Hadi Vafaii 
\\
2023}
\end{center}

\vfill

\newpage 

\pagestyle{plain} \pagenumbering{roman} \setcounter{page}{2}

\phantomsection 
\addcontentsline{toc}{chapter}{Preface}
\pagestyle{plain}\pagenumbering{roman} \setcounter{page}{2}
\renewcommand{\baselinestretch}{2}
\small\normalsize
\hbox{\ }

\vspace{.5in}

\begin{center}
\large{Preface}
\end{center}

I was admitted to the University of Maryland to work on condensed matter physics and quantum information theory. But here I am, writing a dissertation about neuroscience and artificial intelligence. In other words...~not physics. I feel fortunate that the University of Maryland generally provides a welcoming environment for interdisciplinary research, and I am grateful to the physics department for hosting me and allowing me to go after my interests. I never experienced any significant bureaucratic difficulties in this regard; however, I still feel compelled to write a few lines and justify my decision to switch the topic of my research.

I switched to neuroscience for the same fundamental reason that originally led me into physics: \textit{because I am interested in understanding the true nature of reality}.

Over time, I came to realize that an essential element is often overlooked in physics: the role of the observer. We tend to make theories about physical reality but rarely pause to think about how our limited access to the world and the limited computational capacities of the human brain might influence the construction of our theories about reality. Understanding how our imperfect perception and reasoning abilities influence our grasp of reality remains an unresolved challenge. This issue cannot be addressed by ignoring the role of the observer in physics. Instead, it necessitates an exploration of the human mind and intelligence, which constitute at least half of the equation.

The issues surrounding observers have fascinated some of the most interesting thinkers in physics. In particular, John Archibald Wheeler wrote an article in 1990, introducing his \sayit{it from bit} doctrine, which suggests that information is fundamental to the physics of the universe and all things physical are information-theoretic in origin \cite{zurek1990complexity}.

Information is a dyadic relationship between a sender and a receiver. It requires participation from both the \textit{observed}, and the \textit{observer}. Mainstream physics has been mostly preoccupied with the former, ignoring the latter \cite{fuchs2017on}. And maybe for good reason. Theorizing about some complex agent that is making observations about a system is much more challenging than making theories about reduced and simplified systems, as is typical in most of physics (e.g., the Standard Model). However, it turns out even a system comprised of a single quantum particle and an observer cannot be cleanly separated. Niels Bohr initially introduced the idea that there must be some nonzero overlap between the subject and the object, quantified by Plank's constant $\hbar$, which he referred to as \sayit{quantum of action}. In Bohr's own words \cite{bohr1934atomic}:\\[-5mm]

\quote{The finite magnitude of the quantum of action prevents altogether a sharp distinction being made between a phenomenon and the agency by which it is observed.}{Niels Bohr, \textit{Atomic Theory and the Description of Nature}, 1934}\\


\begin{figure}[ht!]
    \centering
    \includegraphics[width=0.45\linewidth]{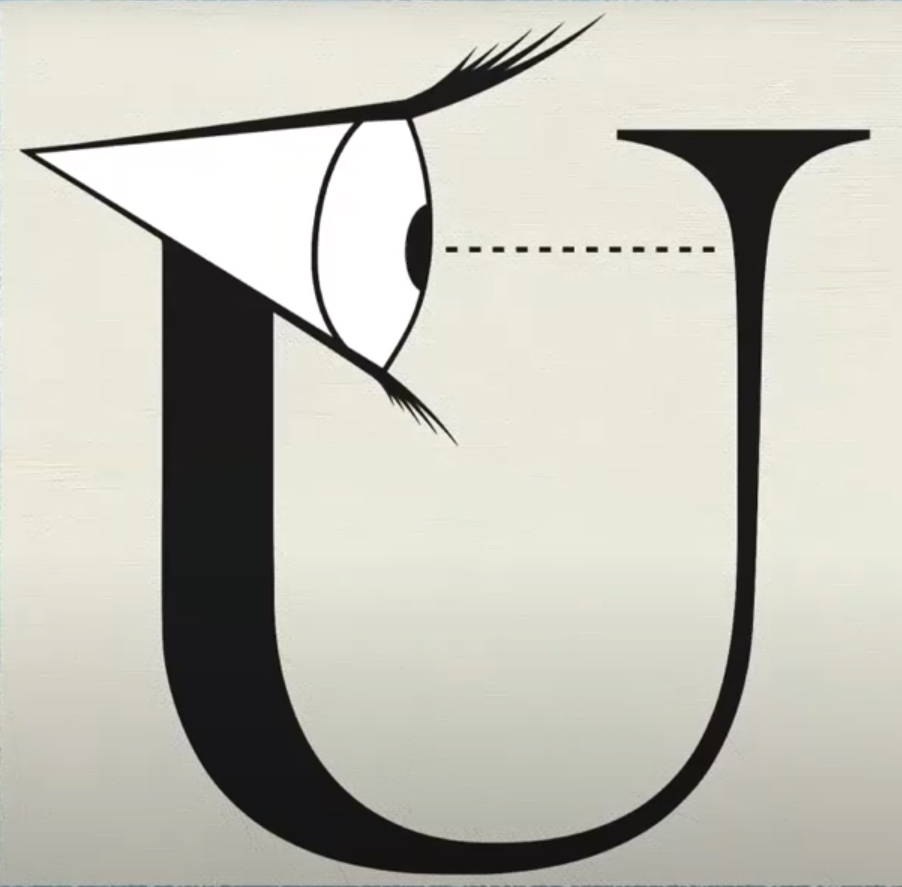}
    \caption[John Wheeler's Participatory Universe]{Wheeler's ``Participatory Universe" highlights the role of observers in physics. Image adapted from a lecture by \href{https://youtu.be/ftMULmgnLl8}{Amanda Gefter}, 2019.}
    \label{fig:wheeler}
\end{figure}

Bohr's ideas were the main source of inspiration for Wheeler when he introduced his notion of a \sayit{Participatory Universe}, stating that the observer and the observed cannot be cleanly separated. Instead, they both participate in creating reality (\cref{fig:wheeler}). Despite this, the majority of efforts in physics still neglect the observer \footnote{Notable exceptions include non-equilibrium statistical mechanics and thermodynamics, and in particular, the ongoing research on Maxwell's Demon \cite{knott1911life}; see refs.~\cite{landauer1961irreversibility, bennett2003notes, maruyama2009colloquium, mandal2012work, parrondo2015thermodynamics} for example.}.

I was convinced that calculating Feynman diagrams or diagonalizing Hamiltonians were not going to reveal any significantly new insights about the nature of reality. Instead, I became interested in studying the brain, as this seemed more conducive to my ultimate interests.


How does one study the brain? I did not know much about the brain or how research is performed in neuroscience. I did not know how to write code or work with complex biological data. All I knew was solving some equations in highly idealized cases. For these reasons, I still had my doubts, until I stumbled upon a quote from Ernst Mach, a physicist and philosopher who also made contributions to cognitive sciences and physiology:\\[-5mm]

\quote{The foundations of science as a whole, and of physics in particular, await their next greatest elucidations from the side of biology, and especially, from the analysis of the sensations.}{Ernst Mach, \textit{Analysis of Sensations}, 1890 \cite{mach1914analysis}}\\

At this point, I knew I had to go for it.  



\phantomsection 
\addcontentsline{toc}{chapter}{Acknowledgements}

\renewcommand{\baselinestretch}{2}
\small\normalsize
\hbox{\ }
 
\vspace{.5in}

\begin{center}
\large{Acknowledgments} 
\end{center} 

\vspace{1ex}

When I began my PhD, I set out on a path in physics. Little did I know that this path would lead me to the world of neuroscience and artificial intelligence. It was a highly nonlinear journey with many challenges, but one that has been incredibly rewarding.

I would like to thank my advisors, Dr.~Luiz Pessoa and Dr.~Daniel Butts, whose constant support and mentorship were crucial in guiding me through this journey. Luiz taught me how to always try to see the big picture first. He made sure I understood the broader implications of my work before getting caught up in minor details. This approach helped shape my growth as a scholar. When I initially switched to neuroscience, Dan supported my transition with patience, allowing me the freedom to explore and eventually find my interests. Both Luiz and Dan emphasized the importance of doing science the right way---carefully, thoroughly, and rigorously.

Collaboration played a big role in my research. I had the privilege to work with Dr.~Evelyn Lake, who is also an external member of my Dissertation committee. The insights and feedback I received during our weekly group meetings on Zoom enriched my work in many ways. There's much more to come, and I believe our best work lies ahead of us. I extend my gratitude to Eve and the entire team at Yale: Dr.~Francesca Mandino, Dr.~David O'Connor, Dr.~Marija Markicevic, Dr.~Xilin Shen, Dr.~Xinxin Ge, Dr.~Peter Herman, Dr.~Fahmeed Hyder, Dr.~Xenophon Papademetris, Dr.~Michael C.~Crair, and Dr.~R.~Todd Constable, for their contributions to this research. Dr.~Mallar Chakravarty's lab at McGill University, particularly Gabriel Desrosiers-Gr\'egoire and his preprocessing package (RABIES), provided indispensable contributions.

Similarly, I had the privilege of collaborating with Dr.~Jacob Yates on primate vision. I once heard someone describe Jake as a \say{scientific force}. This reputation is well earned, and my collaboration with Jake added a unique perspective to my research and broadened my understanding of vision science.

Being a part of the COMBINE (\underline{Co}mputation and \underline{M}athematics for \underline{Bi}ological \underline{Ne}tworks) program at the University of Maryland was another milestone in my PhD journey. As a COMBINE fellow, I was introduced to the importance of connecting ideas from different fields of research, and I was taught how to smoothly navigate such interdisciplinary landscapes. Along the way, learning about \say{network (relational) thinking} was one of the most crucial takeaways I had from being part of COMBINE. I am grateful to COMBINE director, Dr.~Michelle Girvan (a member of my Dissertation committee), along with Dr.~Daniel Serrano (aka the \textit{Dynamic Daniel}), and Glynis Smith for all their efforts in creating a thriving interdisciplinary environment.

I recall the time I was contemplating shifting from condensed matter theory to neuroscience. Dr.~Mohammad Hafezi, my then--physics advisor, gave me the advice that I needed. He emphasized that deep interest in the subject is the mainstay for research, advice that still resonates with me. Other mentors also left indelible marks on my academic journey. During my undergraduate years at Sharif University, Dr.~Hessamaddin Arfaei instilled in me the persistence to chase the truth. Dr.~Pouyan Ghaemi's guidance during my undergraduate research was instrumental in refining my interests.\\

I would like to thank my friends (in alphabetical order of surnames), Dr.~Zabdiel Alvarado, Dr.~Batoul Banihashemi, Debankur Bhattacharyya, Reza Ebadi, Dr.~Ali Fahimniya, Dr.~Ali Lavasani, Nick Mennona, Dr.~Anton Mitrokhin, Masoud Mohammadi, Alireza Parhizkar, Ali Izadi Rad, Dr.~Alireza Seif, Dr.~Anshuman Swain, and Sri Tata: I sincerely appreciate your friendship and our intellectually stimulating discussions that have enriched my PhD journey. I especially remember my trips to Popeyes with Anshuman and Zab, resulting in discussions about all sorts of things ranging from biology, physics, and philosophy to prog music and society, while consuming fried chicken. What a combination!

I also cherish memories with my musician friends who came to my jam sessions and invited me to their own: Sri Tata, Zabdiel Alvarado, Artur Perevalov, Mahour Manshaee, Parinaz Bina, MirSaleh Bahavarnia, Ali Ahmad Khostovan, Gelareh Hon, Ali Seyfi, James Baldassano, Jon Schenk, and Mathias Van Regemortel.

Let me also offer my coolest appreciation to Willis Carrier, the mastermind behind air conditioning and dehumidifiers. Without his ingenious inventions, I might have melted away during the oppressively muggy Maryland summers on this swampy terrain.

Finally, I extend my deepest gratitude to my family, whose support has been the cornerstone of all my pursuits and achievements.
    \cleardoublepage

\phantomsection 
\addcontentsline{toc}{chapter}{Table of Contents}
    \renewcommand{\contentsname}{Table of Contents}
\renewcommand{\baselinestretch}{1}
\small\normalsize
\tableofcontents 
\newpage

\phantomsection 
\addcontentsline{toc}{chapter}{List of Tables}
    \renewcommand{\contentsname}{List of Tables}
\listoftables 
\newpage

\phantomsection 
\addcontentsline{toc}{chapter}{List of Figures}
    \renewcommand{\contentsname}{List of Figures}
\listoffigures 
\newpage

\phantomsection 
\addcontentsline{toc}{chapter}{List of Abbreviations}

\renewcommand{\baselinestretch}{1}
\small\normalsize
\hbox{\ }

\vspace{.5in}

\begin{center}
\large{List of Abbreviations}
\end{center} 

\vspace{3pt}

\begin{supertabular}{ll}
cNVAE & compressed Nouveau VAE \\
OC & Overlapping Community \\
ROFL & Retinal Optic Flow Learning \\
VAE & Variational Autoencoder \\

\end{supertabular}

\newpage
\setlength{\parskip}{0em}
\renewcommand{\baselinestretch}{2}
\small\normalsize

\setcounter{page}{1}
\pagenumbering{arabic}

\renewcommand{\thechapter}{1}

\chapter{Introduction}

This Dissertation is composed of two projects. The first project is about the problem of neural coding, or how neurons in the brain \textit{encode} or \textit{represent} the external world. The second project is about how \textit{spontaneous activity} in the brain in the absence of any external stimulation is spatially organized. In the present Chapter, I will provide the necessary background information and motivation for these projects. Sections \ref{sec:vis} to \ref{sec:motion} correspond to the first project, and sections \ref{sec:orchestra} to \ref{sec:cafmri} are related to the second project. The rest of the Dissertation is organized as follows:\\

\begin{itemize}
    \item \textbf{Chapter 2:} Contains philosophical and historical background. When I switched from physics to neuroscience, I was told we have to \sayit{map these receptive fields}. But why? I was also told we need to \sayit{compute correlations of these brain regions}. But why? At the time, I could not fully appreciate why I was doing what I was doing. This was because of my limited understanding of different research traditions in neuroscience, and how they have come to be. The historical perspective provided in Chapter 2 illuminates these practices, tracing their historical evolution to the present day. If you are a physicist considering switching to neuroscience: Chapter 2 is for you.\\[-4mm]
    
    \item \textbf{Chapter 3:} Contains necessary technical background on \textit{variational inference} and generative models such as \textit{variational autoencoders} (VAE). I go over the basics of variational inference, derive the VAE loss function, and explore some other details related to VAEs. Variational inference is a statistical framework that has found application in a wide variety of problems, including both Chapters 4 and 5 in this Dissertation. In Chapter 4, I use a VAE as a perceptual model of motion processing and compare its representations to neurons from the motion processing pathway of primates. In Chapter 5, I employ an algorithm based on variational inference that infers a lower-dimensional description of the high-dimensional correlation structure of the mouse cortex.\\[-4mm]
    
    \item \textbf{Chapter 4:} Here I present the results for the first main project in this Dissertation, which is about neural coding of motion in primates. Chapter 4 is concerned with understanding the coding properties of biological and artificial neurons in the domain of motion perception. I demonstrate how drawing inspiration from Helmholtz's inferential approach to perception, combined with inspiration from the hierarchical architecture of the visual cortex, can be used to build effective computational models that resemble motion processing in the primate brain. Additionally, the model achieves \textit{optic flow parsing}, a difficult computation that has been debated by vision scientists since the 1950s.\\[-4mm]
    
    \item \textbf{Chapter 5:} Contains the results for the second main project in this Dissertation, which is about cortical network organization in mice. In Chapter 5, I show how a simultaneous functional MRI and wide-field calcium imaging dataset can be used to gain complementary insights into the functional organization of the mouse cortex during resting state.\\[-4mm]
    
    \item \textbf{Chapter 6:} I end with some concluding remarks and ideas for future work.\\
\end{itemize}

\section{Visual processing and representations in the brain}\label{sec:vis}

For millennia, humans have sought to understand the connection between the world we perceive and the reality that exists beyond our senses. Plato's Allegory of the Cave, one of the earliest recorded examples, proposed that our perceptions might be mere shadows of true existence.

This curiosity persisted into more modern work. Helmholtz, examining optical illusions and the limitations of human sensory organs, concluded that perception is not merely passive reception but involves an active process of inference about the causes of sensory inputs \cite{helmholtz1867handbuch}.

Recent advances in neuroscience have begun to identify how neurons in the brain represent specific aspects of the external world. The groundbreaking work of Hubel \& Wiesel uncovered that certain neurons in the visual cortex are activated by oriented edges \cite{Hubel1959ReceptiveFO, Hubel1968ReceptiveFA, kandel2009introduction}, revealing the concept of \textit{receptive field}---a specific area within the visual field that can be used to trigger a response in a neuron. See \textcite{alonso2009receptive} for a review.

\begin{figure}[ht!]
    \centering
    \includegraphics[width=1.0\linewidth]{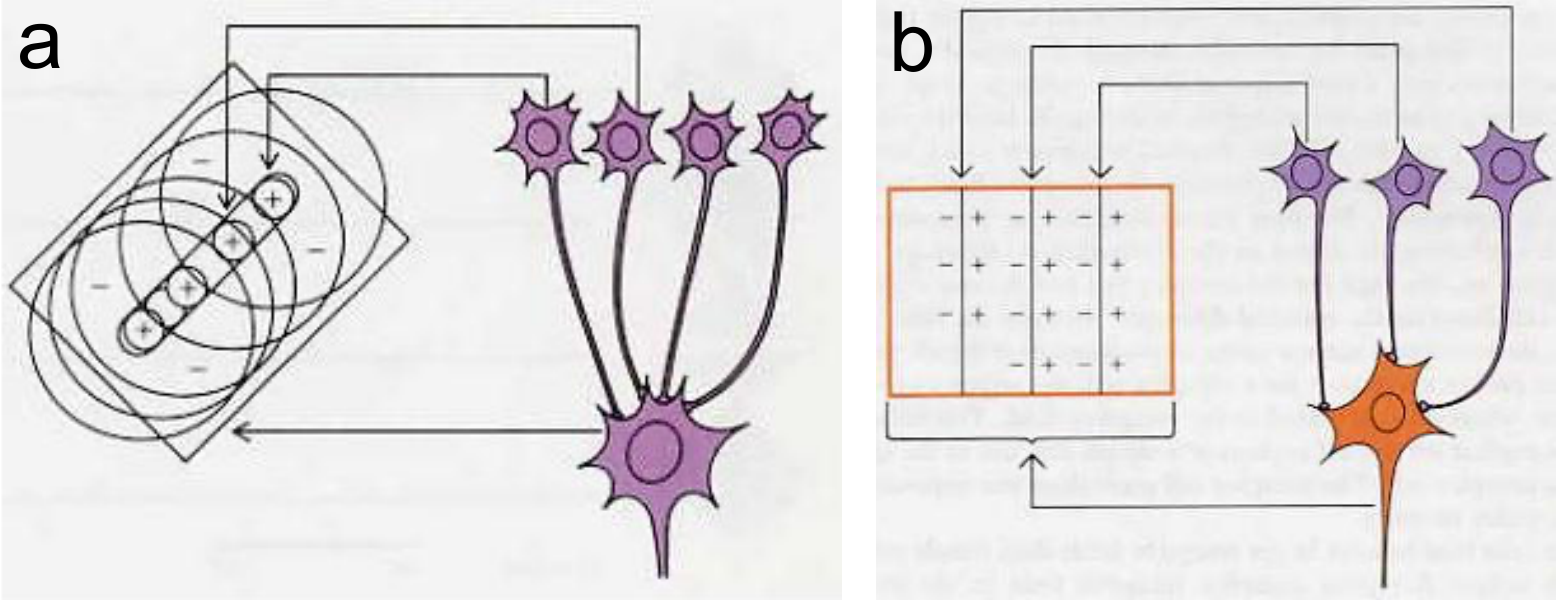}
    \caption[Receptive field integration in visual neurons]{Receptive field integration in visual neurons. \textbf{(a)} A mechanistic model depicting how orientation-selective simple V1 cell responses arise. The simple cell receives input from multiple center-surround cells that are excited by light presented in a small, circular central area (\say{$+$} signs) and inhibited by light in the surrounding area (\say{$-$} signs). \textbf{(b)} Similarly, complex V1 cells gain their selectivity by pooling from multiple simple cells that respond to edges of similar orientation. From David Hubel's book, \sayit{Eye, Brain, and Vision}, 1988 \cite{Hubel1988EyeBA}.}
    \label{fig:center-surround}
\end{figure}

The study of \sayit{what} neurons are selective to, and \sayit{how} they gain their selectivity, are considered two main pillars of systems neuroscience \cite{zeki1978functional}. For instance, the primary visual cortex contains edge-selective neurons that develop their specificity by integrating signals from thalamic neurons with center-surround receptive fields (\cref{fig:center-surround}).

To investigate the selectivity of neurons across various visual areas, it is crucial to examine the anatomical structure of the visual cortex, which is the subject of the next section.

\section{Hierarchical architecture of the visual cortex}

The term \sayit{hierarchy} has diverse interpretations across various contexts. Here, we address its anatomical or structural definition. Connections between cortical areas can be directionally characterized based on their layers of origination and termination \cite{rockland1979laminar}. Here, \say{layer} refers to the laminar organization of the primate cortex, another important organizational principle of cortical architecture. While a deep dive into this topic is beyond our current scope, interested readers can consult a review on the subject by \textcite{Douglas2004NeuronalCO}. See also \textcite{Mountcastle97Columnar}.

\begin{figure}[ht!]
    \centering
    \includegraphics[width=\linewidth]{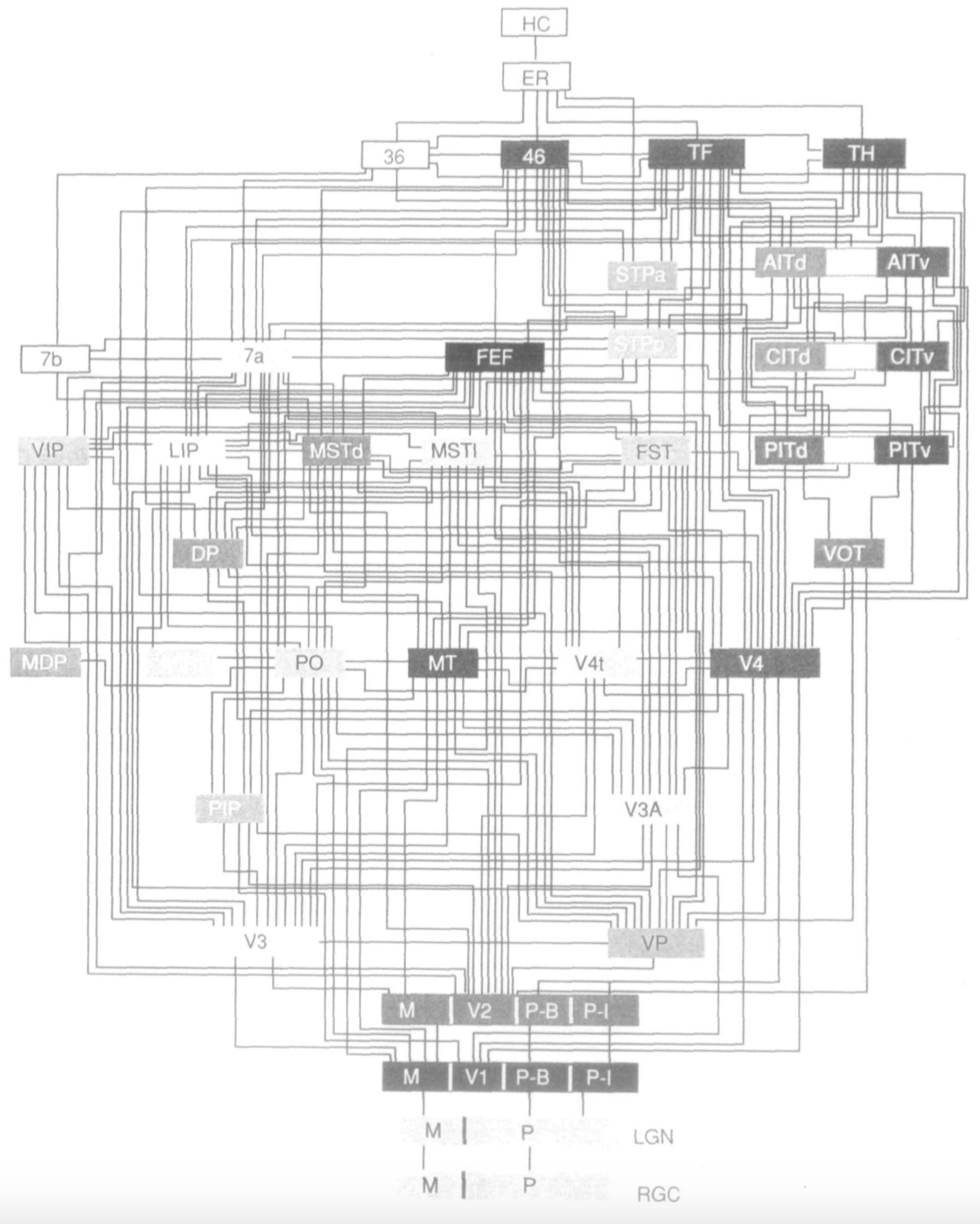}
    \caption[Hierarchy of the primate visual areas]{Hierarchy of the primate visual areas. From \textcite{felleman1991distributed}, 1991.}
    \label{fig:felleman-essen}
\end{figure}

Consider a pair of brain regions $A$ and $B$. When superficial layers of region $A$ project axons to layer 4 of region $B$, region $A$ is described as having \say{feedforward} connections to region $B$. This is because layer 4 is typically seen as the input layer of sensory cortical areas. Conversely, when axons from either superficial or deep layers of region $B$ connect to any layer (excluding layer 4) of region $A$, region $B$ is described as having \say{feedback} connections to region $A$. In a comprehensive review, \textcite{felleman1991distributed} employed these rules to label each member of any connected pair of cortical regions as \say{higher} or \say{lower}, allowing arrangement of the areas into a self-consistent hierarchy (\cref{fig:felleman-essen}).

While this anatomical configuration suggests a linear progression, the actual organization of the visual cortex deviates from a strict \say{tree-like} hierarchy. Other connectivity motifs like recurrent, lateral, and skip connections that bypass intermediate stages are also prevalent in the cortex. See \textcite{hegde2007reappraising, VezoliEvolvingHier} for a nuanced discussion around this topic. Nevertheless, for our purposes, considering this hierarchy as a first approximation is beneficial. With this structure in mind, let us now explore the features encoded by neurons in higher visual areas.


\section{The functional characteristics of higher visual areas}

In this section, we will examine the higher visual areas as delineated within the visual processing hierarchy, based on the framework proposed by \textcite{felleman1991distributed}. Our focus will be on the distinct functions and characteristics of these higher visual areas, exploring their contribution to complex visual perception and cognitive processes.

\subsection{Motion processing}

\begin{figure}[ht!]
    \centering
    \includegraphics[width=0.7\linewidth]{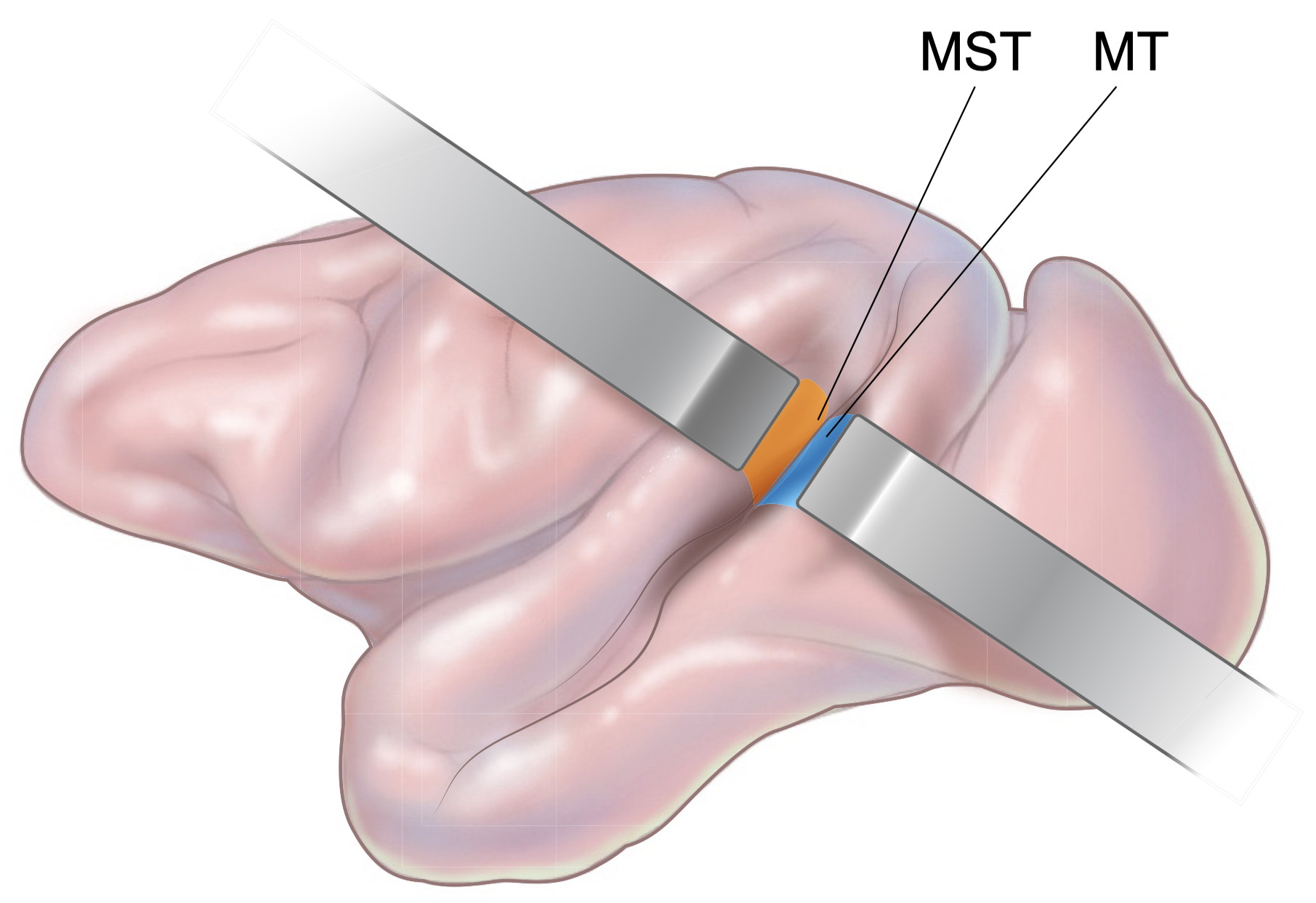}
    \caption[Anatomical location of area MT and MST]{Anatomical location of middle temporal (MT) and medial superior temporal (MST) areas in the macaque brain. MST receives most of its inputs from MT. Figure adapted from \textcite{Wild2021PrimateEC}.}
    \label{fig:mtmst}
\end{figure}

Examples of higher visual areas include regions like the middle temporal area (MT) and medial superior temporal area (MST), with MST primarily drawing inputs from MT. The anatomical positioning of MT and MST within the macaque monkey brain is shown in \cref{fig:mtmst}.

Studies performed on MT and MST have been central to our understanding of visual motion processing. A landmark paper by \textcite{Zeki1974FunctionalOO} was pivotal in highlighting the importance of area MT (also known as V5) for motion perception in primates, showing that it contains a high proportion of direction-selective cells. Following up, \textcite{Maunsell1983FunctionalPO} provided a detailed analysis of the receptive field properties of MT neurons, solidifying their role in processing specific aspects of visual motion such as velocity direction and magnitude (speed). Subsequent research began to explore the intricacies of motion integration. \textcite{Albright1984DirectionAO} demonstrated that MT neurons are sensitive not just to simple motion but also to more complex patterns like the direction of motion in three-dimensional space. \textcite{nishimoto2011three} developed a three-dimensional spatiotemporal receptive field model to successfully predict the responses of MT neurons to complex, naturalistic movie stimuli, highlighting the dynamic interplay between space and time in motion processing within the MT. Finally, \textcite{cui2013diverse} showed that MT neurons gain their complex motion selectivity through diverse surround-suppression influences.

Moving up the hierarchy, the adjacent MST region was shown by \textcite{Tanaka1986AnalysisOL} to contain neurons that respond to more complex motion patterns, such as radial, circular, and spiral motions, suggesting a role in analyzing more intricate motion dynamics and potentially aiding in self-motion perception. A seminal paper by \textcite{Duffy1991SensitivityOM} demonstrated that MST is crucial for the perception of self-motion (or optic flow) by identifying neurons within MST that respond to large-field motion consistent with an observer moving through the environment. Following this, \textcite{Graziano1994TuningOM} deﬁned a continuous circular spiral motion space composed of expansion, contraction, clockwise rotation, and counterclockwise rotation, and showed that MST neurons are tuned to such spiral motions. Further expanding on the diverse functions of MST, studies like that of \textcite{Page2003HeadingRI} explored its role in navigation, finding neurons in MST that seemed to respond to the animal's intended heading direction, hinting at a neural basis for navigation using visual cues.

Together, these foundational studies have established MT and MST as central players in the processing and integration of visual motion information, with MST exhibiting a blend of sensory and cognitive functions \cite{Wild2021PrimateEC}. For comprehensive reviews on MT and MST, see \textcite{Born2005StructureAF} and \textcite{Wild2021PrimateEC}, respectively.

\subsection{Object and face recognition}

\begin{figure}[ht!]
    \centering
    \includegraphics[width=0.80\linewidth]{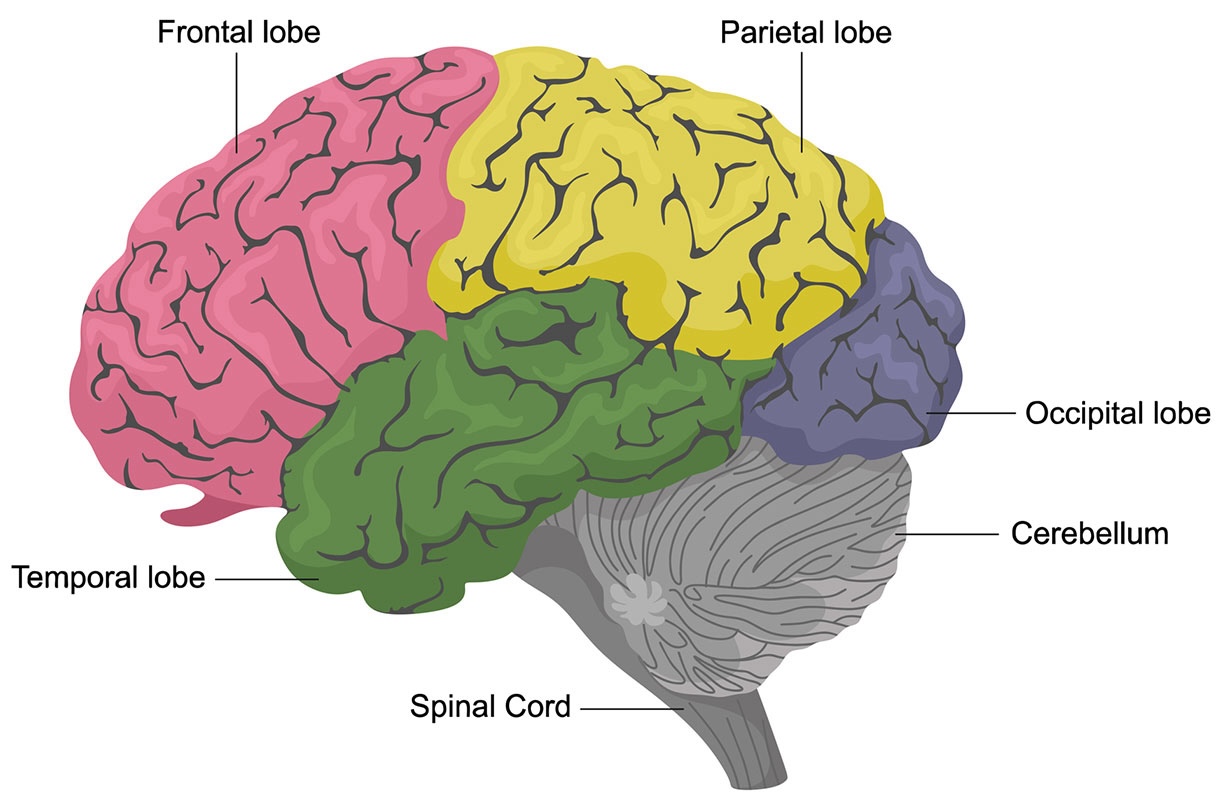}
    \caption[Major lobes of the human brain]{Major lobes of the human brain. Image adapted from \href{https://www.hopkinsmedicine.org/health/conditions-and-diseases/anatomy-of-the-brain}{hopkinsmedicine.org}.}
    \label{fig:lobes}
\end{figure}

In the 1980s, Charles Gross was inspired by stories about how lesions in the human temporal lobe (\cref{fig:lobes}, green) induced object \textit{agnosias} (inability to recognize objects), including \textit{prosopagnosia}, which is a condition where people struggle to recognize faces or can't interpret facial expressions and cues \cite{Squire2008TheHO}.

Intrigued by this, Gross and colleagues set out to explore the possibility that neurons in the temporal lobe are selective for objects and faces. They reported the discovery of neurons in the macaque inferior temporal area that responded selectively to various complex objects, such as trees, faces, and hands \cite{gross1972visual, bruce1981visual, desimone1984stimulus}.

More than a decade later, \textcite{kanwisher1997fusiform} used functional MRI (fMRI) to discover the fusiform face area, a region in the human brain that is largely selective to faces, as the name suggests. Inspired by this, Doris Tsao repeated a similar fMRI study in monkeys and discovered the macaque face patch system \cite{tsao2003faces}. Later, \textcite{tsao2006cortical} used electrophysiology to provide additional experimental support for the existence of the macaque face patch system.

Do neurons in macaque face patches encode individual faces? In 2017, \textcite{chang2017code} considered a brilliant alternative explanation for face selectivity: they asked what if single neurons in the face patch encode not individual faces, but rather individual \textit{axes of variation} in some abstract face space. They tested this hypothesis and found considerable evidence for it.

\subsection{The two-streams hypothesis}

\begin{figure}[ht!]
    \centering
    \includegraphics[width=0.8\linewidth]{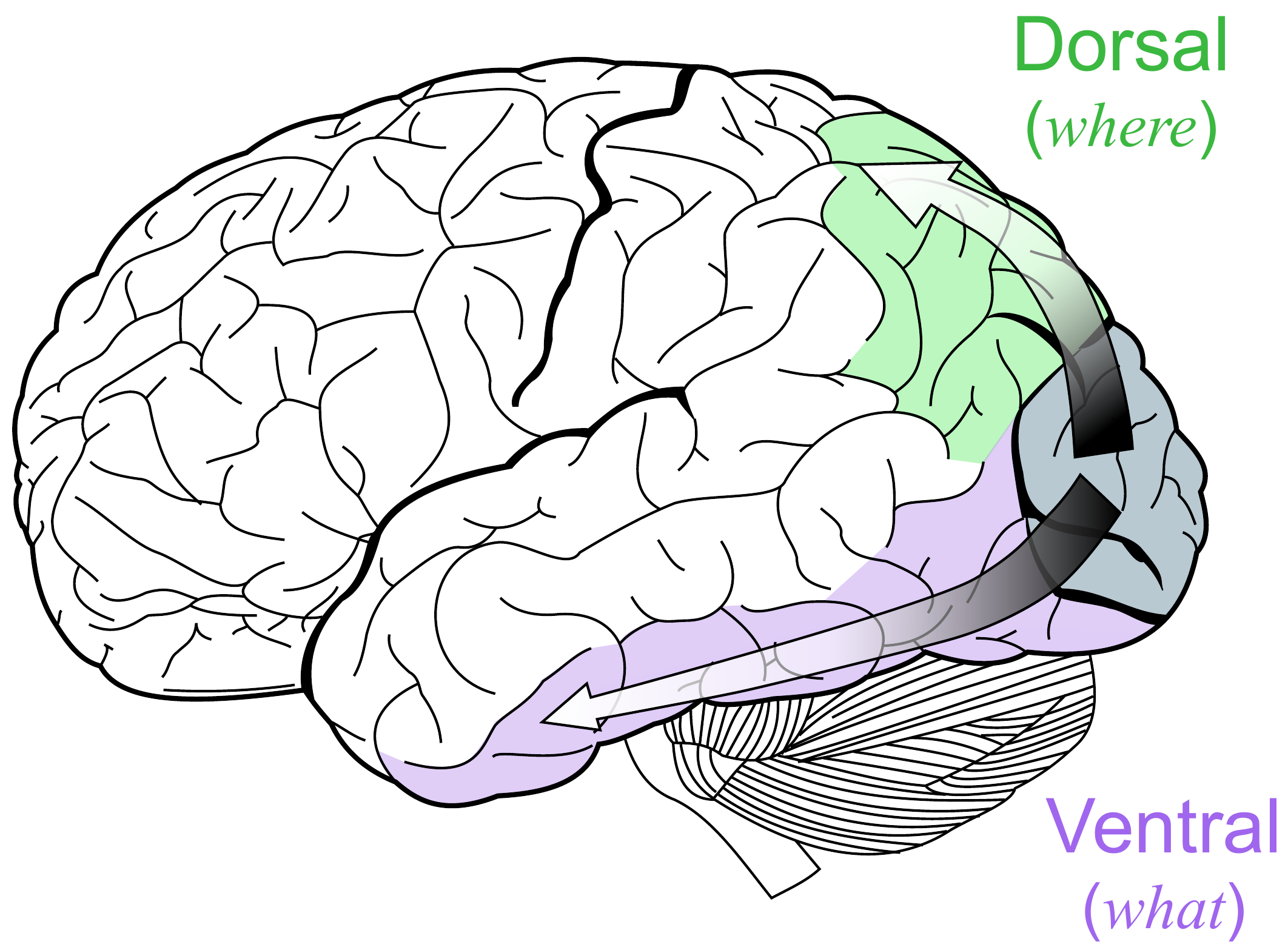}
    \caption[The two-streams hypothesis]{The two-streams hypothesis. Image adapted from \href{https://en.wikipedia.org/wiki/Two-streams_hypothesis}{Wikipedia}.}
    \label{fig:two-streams}
\end{figure}

Building on the understanding of higher visual areas, the \sayit{two-streams hypothesis} presents a fundamental framework in neuroscience that explains how visual information is processed along two distinct pathways in the brain. This hypothesis, first proposed by \textcite{ungerleider1982two} and later refined by \textcite{goodale1992separate}, posits that visual information diverges into two separate streams after initial processing in the primary visual cortex (V1). The ventral stream, often referred to as the \sayit{what} pathway, extends into the temporal lobe and is crucial for object recognition and form representation. It includes areas like the face patch system which is key for face recognition, as discussed above. On the other hand, the dorsal stream, or the \sayit{where} pathway, projects into the parietal lobe and plays a crucial role in spatial awareness, motion processing, and guiding actions in relation to objects (\cref{fig:two-streams}).

While the characterization of the ventral stream as \say{what} pathway has been relatively uncontroversial, we cannot say the same for the dorsal stream. The complex functional role of the dorsal pathway is discussed in a review by \textcite{kravitz2011new}. The authors identify three pathways emerging from the dorsal stream that consist of projections to the prefrontal and premotor cortices, and a major projection to the medial temporal lobe involving posterior cingulate and retrosplenial cortices. These three pathways support both conscious and non-conscious visuospatial processing, including spatial working memory, visually guided action, and navigation, respectively. 

Despite the complex nature of dorsal stream processing, the two-streams hypothesis aligns with and extends the insights gained from the study of MT and MST. Anatomically, areas MT and MST are considered as part of the dorsal stream. Functionally, they are known for processing motion and spatial orientation. However, research indicates that the function of areas along the dorsal pathway is deeply intertwined with the processing capabilities of the ventral stream. This is evidenced by studies showing that motion perception, a dorsal stream function, is enhanced by object recognition, typically a ventral stream function.

For instance, \textcite{ayzenberg2023temporal} showed that the dorsal pathway significantly contributes to object recognition, traditionally a ventral stream function. \textcite{kourtzi2002object} reported that regions like MT and MST, involved in motion processing, also engage in object shape analysis, suggesting a blending of motion and shape processing across the streams. Such dynamic interplay between the dorsal and ventral streams in visual processing challenges the notion of completely separable neural mechanisms for shape and motion processing.

More recently, \textcite{ayzenberg2022does} suggested that global shape information may be computed in the dorsal visual pathway and transmitted to the ventral visual pathway. In this context, \say{global} shape representations refer to a description of objects via the spatial arrangement of their local features, rather than by the appearance and details of the features themselves. Later, \textcite{goodale2023shape} challenged the arguments presented by \textcite{ayzenberg2022does}, citing evidence from patients with dorsal-stream lesions who were still able to perceive shapes \cite{goodale2023patients}. As is typical in science, the jury is still out regarding the function of dorsal stream.

Overall, it appears like a sharp and clean functional separation between dorsal and ventral streams is not fully supported by data. Nonetheless, recent neuroimaging and neuropsychological testing advancements have provided compelling evidence in partial support of the two-streams hypothesis. Patients with lesions in specific areas of either stream exhibit distinct types of visual deficits, further corroborating this model.

This division of labor in visual processing highlights the complexity and sophistication of the brain's approach to interpreting visual stimuli, suggesting that understanding each stream's unique contributions is crucial for a comprehensive view of visual cognition. Future research, particularly in the realm of integrative neuroscience, is likely to uncover even more about how these streams interact and function in concert, providing deeper insights into the nature of visual processing and perception. For more information, please see refs.~\cite{ungerleider2008what, sheth2016two, rossetti2017rise}. 

To conclude, our exploration in this section revealed that neurons in the brain can be selective to different aspects of the world, ranging from oriented edges to motion patterns, to more abstract concepts like faces. We also discussed how the separation of object identity and position/motion is to some extent paralleled in the brain by the ventral and dorsal streams, respectively. In the next section, we will explore how such complex selectivity may arise. We already briefly discussed how the selectivity of simple edge detectors in V1 is understood using mechanistic models (\cref{fig:center-surround}). What about higher visual areas? Could the selectivity in these higher areas be the result of a series of feedforward, nonlinear computations that progressively build in complexity?


\section{Visual processing as nonlinear feedforward computation}\label{sec:ff}

\begin{figure}[ht!]
    \centering
    \includegraphics[width=1.0\linewidth]{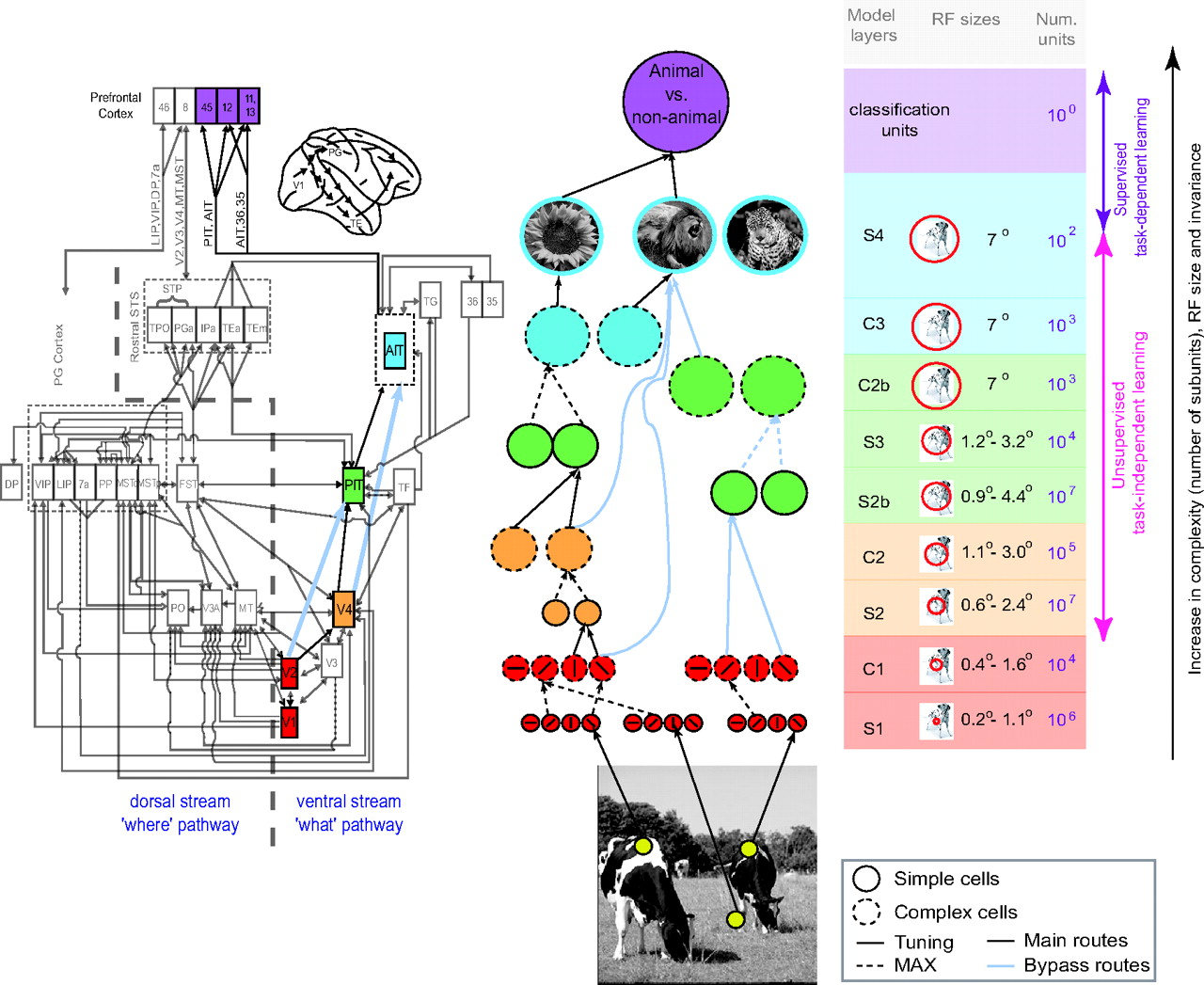}
    \caption[A feedforward model of visual processing by Serre, Oliva, and Poggio (2007)]{The sketch of the model by \textcite{serre2007feedforward}. A proposed correspondence between the ventral stream in the primate visual system (Left) and the basic building blocks of the predictive feedforward model (Right). Figure from \textcite{serre2007feedforward}, 2007.}
    \label{fig:serre-poggio}
\end{figure}

In 2007, \textcite{serre2007feedforward} published a seminal paper in which they presented a feedforward computational model of the ventral visual stream in primates that was able to achieve human-level performance on a rapid animal vs.~non-animal categorization task (\cref{fig:serre-poggio}). Their model was built on the idea that visual processing in the ventral stream progresses in a feedforward hierarchy from V1 to the inferotemporal cortex.

The model by \textcite{serre2007feedforward} quantitatively matched both neural response properties along the ventral stream and human behavioral performance on the rapid animal categorization task, providing support for the hypothesis that feedforward processing may be sufficient for \textit{core object recognition} abilities. Core object recognition refers to the rapid identification of object categories based on their basic shape and form.

Importantly, comparisons of model performance under different masking conditions showed that its accuracy drops off more quickly than human performance as masking increases. This observation suggests feedback signals may play a role in human recognition abilities when stimuli are heavily masked.

Overall, the work by \textcite{serre2007feedforward} provided a computational framework for testing models of feedforward visual processing in primate ventral stream against both neural and behavioral data. The results support the view that core object recognition may rely primarily on feedforward processing of visual inputs.

\begin{figure}[ht!]
    \centering
    \includegraphics[width=1.0\linewidth]{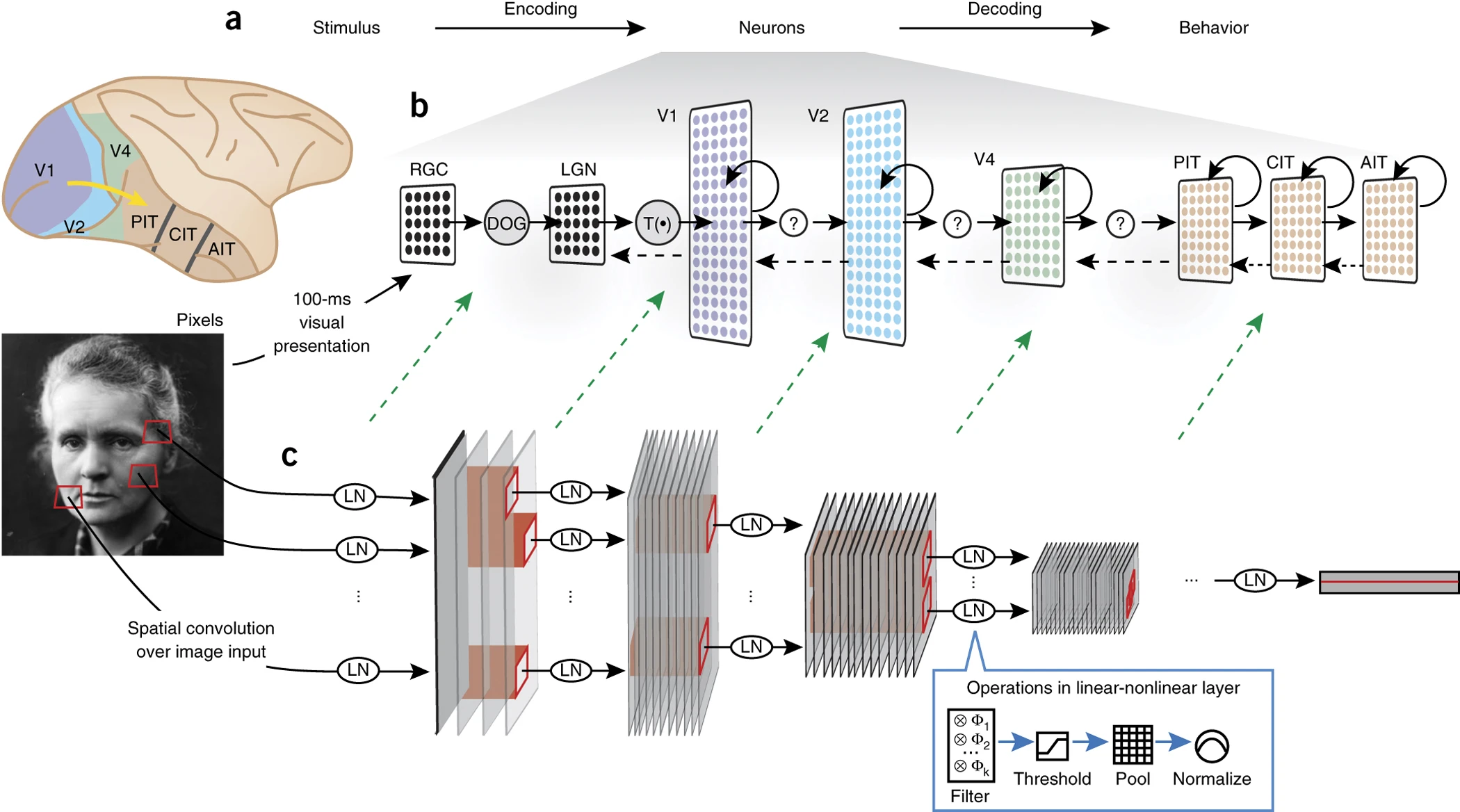}
    \caption[Hierarchical convolutional neural networks (CNN) as models of visual processing]{Hierarchical convolutional neural networks are used to model mapping of visual stimuli to neural responses as measured in the brain. Green arrows indicate that activations from artificial units within the CNN layers can predict the neuronal response patterns to large classes of stimuli. Figure from \textcite{yamins2016using}, 2016.}
    \label{fig:yamins-dicarlo}
\end{figure}

Building on this hierarchical processing concept, \textcite{yamins2014performance} showed in 2014 that deep convolutional neural networks (CNN) trained on object recognition tasks mimic visual processing along the ventral stream (\cref{fig:yamins-dicarlo}). They trained CNNs on image datasets and found the internal representations learned by the CNNs aligned closely with the primate inferotemporal (IT) cortex. As the CNNs were trained toward human-level performance, they developed unit-tuning properties and population coding similar to IT. Critically, the authors did not constrain the model to match neural data and the model's top output layer automatically developed IT-like representations. This work provided computational validation for the hypothesis that ventral stream representations are heavily shaped by exposure to real-world visual environments, and that feedforward processing may be sufficient for the emergence of such representations.

\textcite{yamins2014performance} also analyzed representational geometry and category clustering and found that IT and the trained nets both transitioned from low-level feature representations to behaviorally relevant categories. In sum, this work demonstrated how CNNs can provide insights into primate visual processing and offered a path toward artificial systems that better emulate biological vision.

Although both works highlighted in this section were on ventral stream object recognition, the idea of hierarchical feedforward processing has been successfully applied to the dorsal stream as well. For example, \textcite{mineault2012hierarchical} built a computational model demonstrating that MST neurons gain their selectivity by pooling from a population of MT-like direction-selective units, which varied in preferred direction, spatial position, and speed.

\section{Generative approaches to vision: beyond feedforward processing}\label{sec:generative}

The line of work starting from \textcite{serre2007feedforward} leading to \textcite{yamins2014performance} onward suggests that the responses of neurons in the IT cortex could be well modeled by linear regression onto responses of units in a deep CNN trained on the task of general object classification. The success of this approach has led to the dominant view that the only way to explain activity in high-level regions of the visual system is through black-box deep CNNs with millions of parameters. However, recent work has challenged this notion.

Instead, generative models that disentangle sensory inputs into interpretable latent factors show promise for an alternative approach to explaining neural representations. Some of the most interesting work in this domain has been on the macaque \textit{face patch system}, which refers to regions of the IT cortex in monkeys that exhibit highly selective responses to faces \cite{hesse2020macaque}.

For instance, \textcite{higgins2021unsupervised} compared single neuron responses in macaque face patches to units in a $\beta$-VAE model trained to disentangle generative factors of faces such as facial identity, gender, expression, etc. Intriguingly, some face patch neurons had direct counterparts in $\beta$-VAE, suggesting they may encode semantically meaningful generative factors.

Further evidence came from \textcite{yildirim2020efficient}, who used an inverse graphics model to predict face cell responses. This model separately represents intrinsic face properties like shape and extrinsic factors like illumination, enabling more accurate prediction of neural responses and behavior compared to feedforward classifiers. 

It is also known that models trained based on generative objectives better mimic human judgment of controversial stimuli \cite{golan2020controversial} and perceptual errors \cite{storrs2021unsupervised}. Overall, growing evidence suggests that generative models that decompose sensory inputs into causal factors align better with neural representations and behavior compared to black-box classifiers. This suggests visual areas in the brain may implement something like a generative model, encoding causal variables of raw sensory information.

Finally, a key distinction of generative approaches to vision is that these types of models usually assume a critical functional role for feedback pathways, alongside feedforward processing. Feedback connections are prevalent in the visual cortex \cite{markov2014anatomy}, and ignoring this architectural motif has been a fundamental limitation of the CNN/feedforward approaches mentioned above \cite{heeger2017theory}. We will explore this point more thoroughly in the next section.

\section{Normative approaches to vision: asking \textit{why} questions}\label{sec:normative}

Previously in this Chapter, we considered the \sayit{what} and \sayit{how} of neuronal selectivity in the visual system, examining both what neurons are selective for and how they acquire this selectivity. However, this discussion does not fully address the \sayit{why} question: Why do sensory neurons develop the specific patterns of selectivity we observe? What theoretical principles could explain the emergence of these empirical patterns?

Typically, answers to such \say{why} questions are framed in terms of optimization principles: Sensory neurons are thought to develop their selectivity as a means of optimizing specific objective functions. Theories rooted in this concept are termed \say{normative} theories. Within the realm of neuroscience, normative theories are not restricted to sensory processing and they generally present a framework in which the brain's structure and functions are interpreted through the lens of optimization. These theories suggest that both the architecture and operational mechanisms of the brain are fine-tuned to meet particular goals to efficiently address certain challenges.

In this section, we will explore some of the most prominent normative theories in neuroscience, detailing their success in elucidating various aspects of sensory processing and neural coding from first principles.

\subsection{Efficient Coding}

Efficient coding is a principle in neuroscience suggesting that sensory systems adapt to represent the most common features of their environment in the most efficient way possible. The idea, originating from \textcite{attneave1954some} in the mid-1950s and \textcite{barlow1961possible} in the 1960s, posits that our sensory neurons optimize the way they represent information to maximize information content while minimizing redundancy.

\begin{figure}[ht!]
    \centering
    \includegraphics[width=1.0\linewidth]{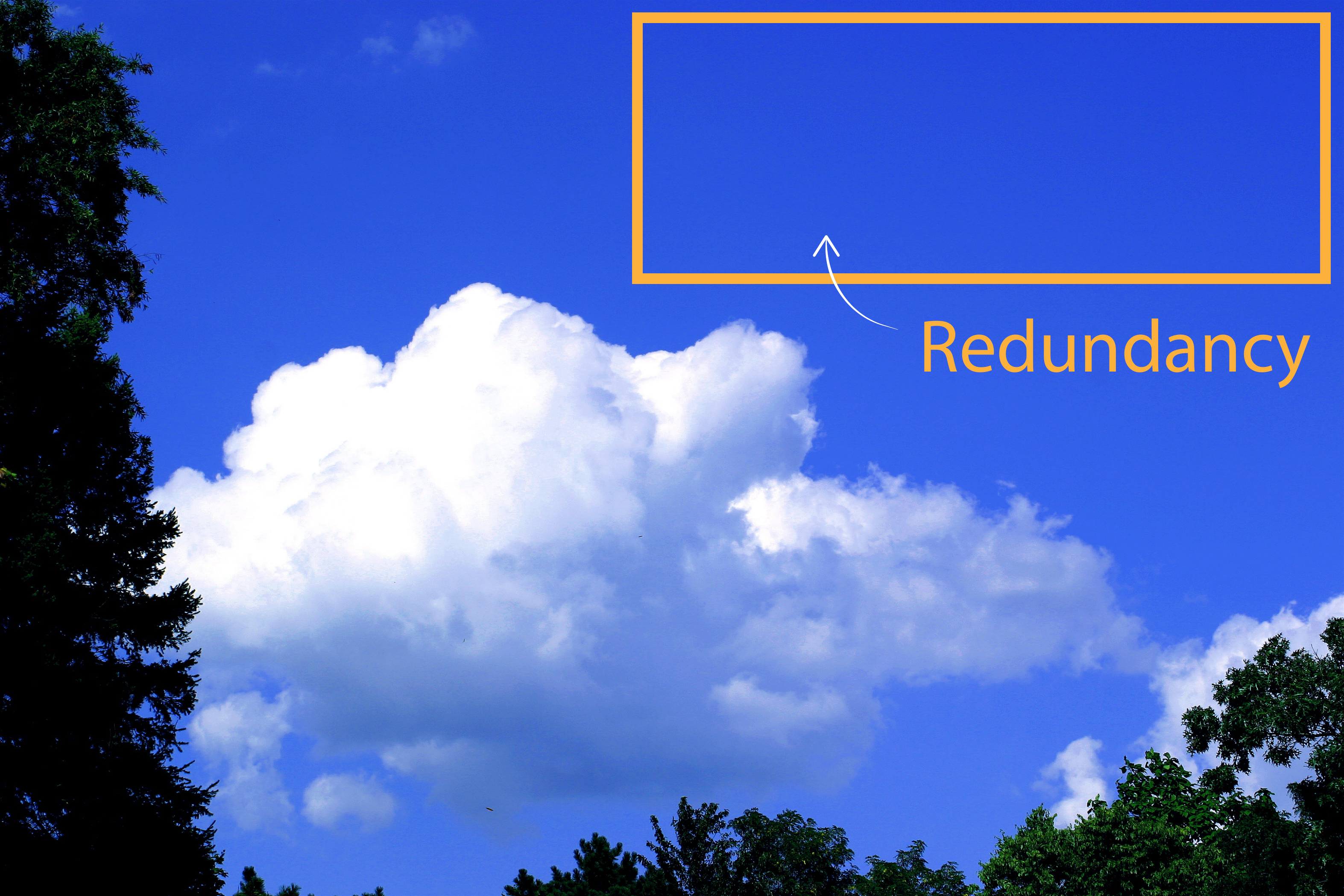}
    \caption[Efficient Coding: Local correlations and redundancy in natural images]{Local correlations and redundancy in natural images. The segment within the orange frame exhibits a high degree of perceptual uniformity, characteristic of the redundancy found in natural scenes. Such uniform areas can be encoded more efficiently by using fewer bits of information, allowing for image compression without significant loss of detail. This exemplifies the principle of ``efficient coding" \cite{attneave1954some, barlow1961possible, atick1992could}, where redundant information is minimized to optimize sensory data representation. Image adapted from \href{https://en.wikipedia.org/wiki/Sky_blue}{Wikipedia}.}
    \label{fig:sky}
\end{figure}

The efficient coding theory implies a form of data compression, where sensory information is encoded in as few neurons as possible, thus making the process more efficient (see \cref{fig:sky}). For instance, in vision, this can mean that certain neurons become specialized in detecting specific edges or orientations in the visual field, allowing for a more compact and efficient representation of the visual world.

The mathematics of efficient coding often involves concepts from information theory \cite{shannon1948mathematical}. The idea is to maximize the mutual information between the stimulus and the neural response while minimizing the number of active neurons.

Another key concept in efficient coding is \say{redundancy reduction}, which aims to minimize overlapping information among neurons, leading to the decorrelation of responses within a population of neurons. As a result, each neuron responds to specific, non-overlapping features of the stimulus which increases the efficiency of the neural code.

Efficient coding theory suggests that sensory systems should adapt their stimulus encoding strategies \cite{fairhall2001efficiency, lesica2007adaptation, gutnisky2008adaptive, ollerenshaw2014adaptive}, based on the statistical properties of sensory inputs they receive \cite{simoncelli2001natural, atick1992could, geisler2008visual}. For example, if certain stimuli become more common in an environment, the system should adapt to represent these stimuli more efficiently, a prediction that has been verified in various sensory and non-sensory systems \cite{lewicki2002efficient, ecker2010decorrelated, pitkow2012decorrelation, renart2010asynchronous, polania2019efficient, mlynarski2021efficient, bolanos2022efficient}.

A detailed exploration of the efficient coding hypothesis exceeds the scope of this Dissertation. Readers seeking a deeper understanding may consult Chapter 3 in \textcite{li2014understanding}, 2014. Finally, see Barlow's later work in 2001 \cite{barlow2001redundancy} for his updated perspective on efficient coding.

\subsection{Sparse Coding}

Sparse coding is closely related to efficient coding, emphasizing that neural representations of sensory information should involve the activation of only a small number of neurons at a time. This approach is based on the observation that at any given moment, the majority of neurons in sensory areas of the brain are relatively inactive, suggesting that information is represented by the strong activation of a relatively small subset of neurons \cite{wolfe2010sparse}.

\begin{figure}[ht!]
    \centering
    \includegraphics[width=1.0\linewidth]{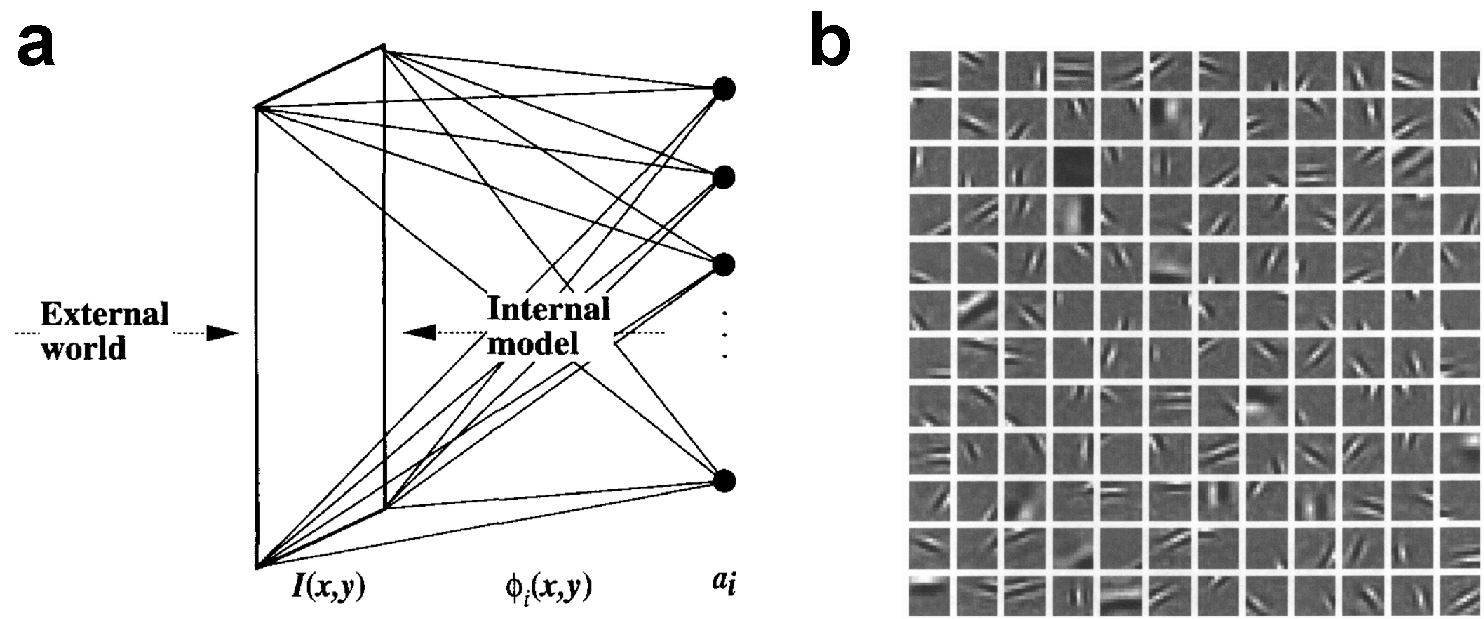}
    \caption[Sparse Coding (Olshausen \& Field)]{Sparse coding illustrated. \textbf{(a)} Schematic of a linear generative model where the activation of neurons represents sensory input from the external world. The brain contains an internal model of the sensory information. Each neuron is associated with a distinct basis vector, and the input image is reconstructed as a linear combination of these basis vectors, with the neuron's activation strength determining the contribution of each vector. \textbf{(b)} The basis vectors learned from a sparse coding algorithm optimizing \autoref{eq:sparse}. These vectors exhibit properties similar to the low-pass Gabor-like receptive fields found in the primary visual cortex (V1), demonstrating the algorithm's capacity to learn biologically plausible representations of visual information. Adapted from \textcite{olshausen1997sparse}, 1997.}
    \label{fig:sparse}
\end{figure}

Hubel \& Wiesel discovered the Gabor-like V1 receptive fields. They also explained how such receptive field properties may arise mechanistically from pooling operations (\cref{fig:center-surround}). Sparse coding, first proposed by \textcite{olshausen1996emergence}, is an attempt to answer \sayit{why} V1 receptive fields look the way they do: It is because the brain has to represent (synthesize) the external world in an efficient manner, using as few spikes as possible (\cref{fig:sparse}).

More formally, let us consider an image $\bm{I} \in \mathbb{R}_+^M$, where $M$ is the number of pixels, and a set of basis vectors $\{\phi_i: i = 1, \dots, N\}$, organized into a matrix $\bm{\Phi} \in \mathbb{R}^{M \times N}$. According to the linear generative model of Olshausen \& Field \cite{olshausen1996emergence, olshausen1997sparse}, the image can be represented as a linear combination of these basis vectors:

\begin{equation}
    \bm{I} = \bm{\Phi}\bm{a} + \bm{\epsilon},
\end{equation}

Here, $\bm{a} \in \mathbb{R}_+^N$ is a vector of neural activity, where each element $a_i$ corresponds to a basis vector $\phi_i$, and $\bm{\epsilon} \in \mathbb{R}^M$ is a Gaussian noise component. We consider features (columns) of $\bm{\Phi}$ to form an overcomplete dictionary, that is, $N \gg M$. The choice of basis vectors determines the coding strategy, i.e., the form and structure of $\bm{a}$ will change depending on $\bm{\Phi}$.

The goal of sparse coding is to learn an appropriate set of basis functions such that the corresponding neural activations contain as few non-zero elements as possible. In other words, the goal is to represent the world (i.e., image $\bm{I}$) in a sparse and efficient manner. This can be expressed using the following loss function:

\begin{equation}\label{eq:sparse}
    E(\bm{a}, \bm{\Phi})
    =
    \frac{1}{2}\norm{\bm{I} - \bm{\Phi}\bm{a}}^2
    +
    \lambda \cdot \sum_{i=1}^N c\left(a_i\right).
\end{equation}

The first term ensures robust reconstruction of the image, and the second term encourages sparsity of representations. There are different choices of regularizer $c(\cdot)$. A commonly used option is $\ell_1$ norm: $c\left(x\right) = |x|$. However, recent work has raised concerns with this choice \cite{rentzeperis2023beyond}, proposing the use of the $\ell_0$ pseudo-norm as a more effective alternative.

Sparse coding is efficient because it allows for a large number of possible stimuli to be represented by different combinations of a small number of active neurons. This leads to a robust and high-capacity coding scheme, as different stimuli can be represented distinctly without requiring a large number of neurons which would be metabolically costly \cite{laughlin1998metabolic, hasenstaub2010metabolic}.

Later developments in this domain explored how neural dynamics can be integrated into a sparse coding theory. In 2008, \textcite{rozell2008sparse} introduced a locally competitive algorithm (LCA) for sparse coding that uses thresholding functions and lateral inhibition between nodes in a neural population to calculate sparse codes that minimize an energy function, trading off reconstruction error and sparsity penalties. Simulations show LCAs achieve sparsity on par with widely used greedy methods like matching pursuit, and they demonstrate greater temporal smoothness beneficial for encoding video sequences.

Building on the neurally inspired LCA approach, a 2022 paper from \textcite{fang2022learning} proposed an efficient strategy for probabilistic inference and learning in sparse coding models using Langevin dynamics. By injecting noise into the LCA-style dynamics, continuous-time equations are obtained that naturally sample the posterior distribution over latent variables. This Langevin sparse coding (LSC) scheme enables dictionary learning while simultaneously adapting parameters like sparsity levels. Importantly, LSC provides an efficient way to sample, with explicit $\ell_0$ sparsity encouraging variables to be strictly set to zero. The proposed analog implementation maps neatly onto the LCA architecture with the additional requirement of harnessing intrinsic noise sources.

A growing body of experimental evidence supports the sparse coding hypothesis \cite{vinje2000sparse, young1992sparse, vinje2002natural, quiroga2008sparse, haider2010synaptic, froudarakis2014population, carlson2011sparse, willmore2011sparse, sachdev2012surround, beyeler2019neural, zaslaver2015hierarchical}. However, one key unsolved problem is developing a detailed biophysical understanding of how neural systems perform efficient probabilistic inference and learning in sensory coding.

It is not an exaggeration to state that the normative \say{way of thinking} embraced by Olshausen \& Field in their sparse coding work represents a paradigm shift. This generative approach to vision marks a departure from traditional feedforward receptive field mapping approaches that have dominated vision science for decades. For example, a V1 simple cell is typically thought of as a filter that processes images from the retina. In contrast, within the sparse coding framework, the concept of a receptive field expands beyond mere image processing; it is interpreted as an outward manifestation---or \say{shadow}---of the neuron's intrinsic role in the creation of the brain's sensory experiences. Hence, receptive fields, as defined and estimated by correlative methods, should be considered as secondary expressions, or \textit{epiphenomena}, of the neuron's underlying activity in the brain's generative sensory processing.

For further readings, please refer to Chapters 1--3 in \textcite{northoff2013unlocking}, 2013. See also a 2004 review by \textcite{olshausen2004sparse}, and a 2011 blog post by \textcite{mineault2011sparse}.

\subsection{Predictive Coding}

Predictive coding suggests the brain constantly makes predictions about incoming sensory information and uses differences between these predictions and actual sensory input to form perceptions \cite{clark2013whatever}. This theory posits that the brain minimizes the error between its predictions and actual sensory data, updating its internal model of the world based on this error \cite{rao1999predictive}. Predictive coding is efficient in a different sense; it reduces the amount of information the brain needs to process by focusing only on the unexpected elements of sensory input (\cref{fig:sky}). This theory has been influential in understanding various aspects of brain function, including perception \cite{rao1999predictive, srinivasan1982predictive, hohwy2008predictive, hohwy2016self}, motor control \cite{wolpert1995internal, miall1996forward, wolpert2003unifying, blakemore2000can}, and even the nature of hallucinations \cite{sterzer2018predictive, adams2013computational, seth2012interoceptive, powers2017pavlovian, corlett2019hallucinations}.

\begin{figure}[ht!]
    \centering
    \includegraphics[width=1.0\linewidth]{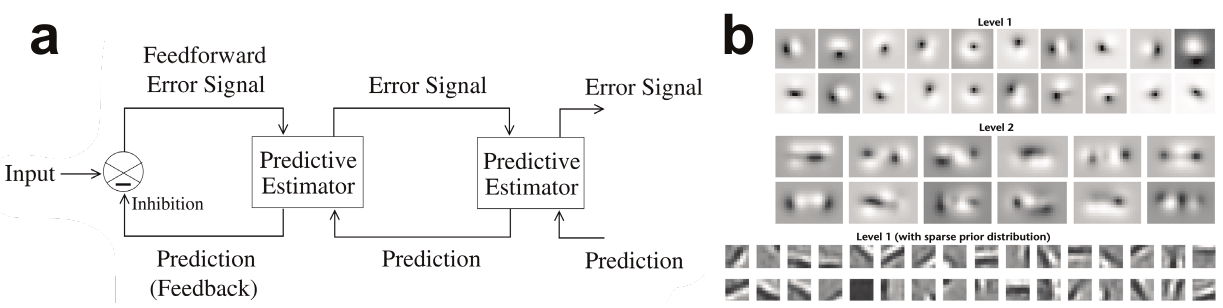}
    \caption[Predictive Coding (Rao \& Ballard)]{Predictive coding of \textcite{rao1999predictive}. \textbf{(a)} Model architecture, illustrating the computation of the error signal at each processing stage. The error signal is the result of the feedforward input minus the feedback prediction. In short, $error \,\, signal = input - prediction$. Only this residual error signal is propagated through the network, which becomes the input to the subsequent processing stage. \textbf{(b)} Features that emerge when the model is trained on natural images. Level 1 features (top row) resemble receptive fields found in thalamic relay nuclei, which decorrelate (or whiten) the inputs and reduce output redundancy, reminiscent of the efficient coding hypothesis. By incorporating a sparse kurtotic prior distribution for the network activities, the model's receptive fields take on characteristics similar to those in the primary visual cortex, resembling Gabor wavelets (bottom row). This is reminiscent of the sparse coding hypothesis. Image adapted from \textcite{rao1999predictive}, 1999.}
    \label{fig:pred}
\end{figure}

A specific implementation of predictive coding as proposed by \textcite{rao1999predictive} in their seminal 1999 paper involves a hierarchical model that aims to explain how the brain could naturally learn to represent the sensory input in an efficient way. In their model, each level of a sensory processing hierarchy is thought to generate predictions of lower-level sensory inputs and send these predictions down the hierarchy. The lower levels then send back up only the residual errors between the predictions and the actual sensory inputs, which is the information that was not predicted correctly (\cref{fig:pred}a).

At each level of the hierarchy, the predictions are generated by internal models that are continuously updated based on the error signals received. These internal models are essentially the brain's understanding or representation of the regularities of the world. This framework suggests that perception is largely a constructive process---it is about the brain's best guess of what is out there in the world, based on both the current sensory input and past experience.

Rao \& Ballard's model is particularly focused on explaining the receptive fields of neurons in the visual cortex. The model suggests that these receptive fields can be understood as kernels that are involved in the generation of predictions at each level of the hierarchy. The model was supported by simulations, showing that when the system was exposed to natural image sequences, the neurons in the hierarchy developed receptive fields similar to those found in the thalamus and the primary visual cortex (\cref{fig:pred}b).

Despite the elegance of these ideas, their experimental support is varied. A key challenge is the lack of specificity in predictive coding models, making them difficult to conclusively validate or refute \cite{solomon2021limited, kogo2015predictive, walsh2020evaluating}. Additionally, the term \say{predictive coding} is subject to varied interpretations. On one hand, it universally implies that the brain forms predictions, an idea with broad consensus. On the other, it specifically refers to the implementation as proposed by Rao \& Ballard, a concept some consider overly narrow.

The Rao \& Ballard version of predictive coding introduces two essential elements: error units, which calculate and transmit error signals, and representation units, responsible for making predictions. However, linking this theory to the characteristics of individual spiking neurons has been challenging \cite{kogo2015predictive, millidge2022predictive}. A recent and intriguing theory builds upon their work, suggesting that error computation and prediction occur within different dendritic compartments of the same pyramidal neuron \cite{mikulasch2023error}. Termed \say{dendritic hierarchical predictive coding (hPC)}, this approach suggests that basal dendritic compartments represent bottom-up errors, while apical compartments, receiving higher-level cortical feedback \cite{harris2015neocortical}, capture the top-down prediction error for the neuron's activity. A further advancement in dendritic hPC theory is the notion that both bottom-up and top-down signals serve as predictive mechanisms \cite{spratling2008predictive, mikulasch2023error}, with prediction neurons directly competing through lateral inhibition on basal dendrites.

Another limitation of predictive coding is its assumption of a predominantly feedforward error correction process. This perspective might not fully capture the brain's complex feedback loops and parallel processing. For instance, the interplay between various sensory modalities and the role of prior knowledge and memory in perception are challenging to explain within the confines of a strict predictive coding framework.

A more encompassing approach is hierarchical Bayesian inference. This model allows for a wider range of interactions between feedforward and feedback signals, extending beyond simple subtraction. Hierarchical Bayesian inference accommodates more complex relationships between top-down and bottom-up signals, such as gating or modulation, which are absent in Rao \& Ballard's predictive coding model.

\subsection{Hierarchical Bayesian Inference}\label{sec:hier-bayes-inf}

Hierarchical Bayesian inference is a fundamental framework for understanding cognitive processes, particularly in how the brain interprets and integrates sensory information. This framework extends beyond the scope of predictive coding by positing that the brain operates not just on immediate sensory input, but also on a wealth of accumulated knowledge and expectations. In this model, perception is viewed as a probabilistic process, where the brain continually refines its understanding of the world by integrating new sensory data with existing beliefs and predictions. This integration occurs at multiple hierarchical levels, each refining the interpretation of sensory input in light of higher-order cognitive factors. This perspective of perception as an inferential process has certain elements in common with principles of predictive coding, yet it encompasses a broader, more intricate network of interactions.

\begin{figure}[ht!]
    \centering
    \includegraphics[width=1.0\linewidth]{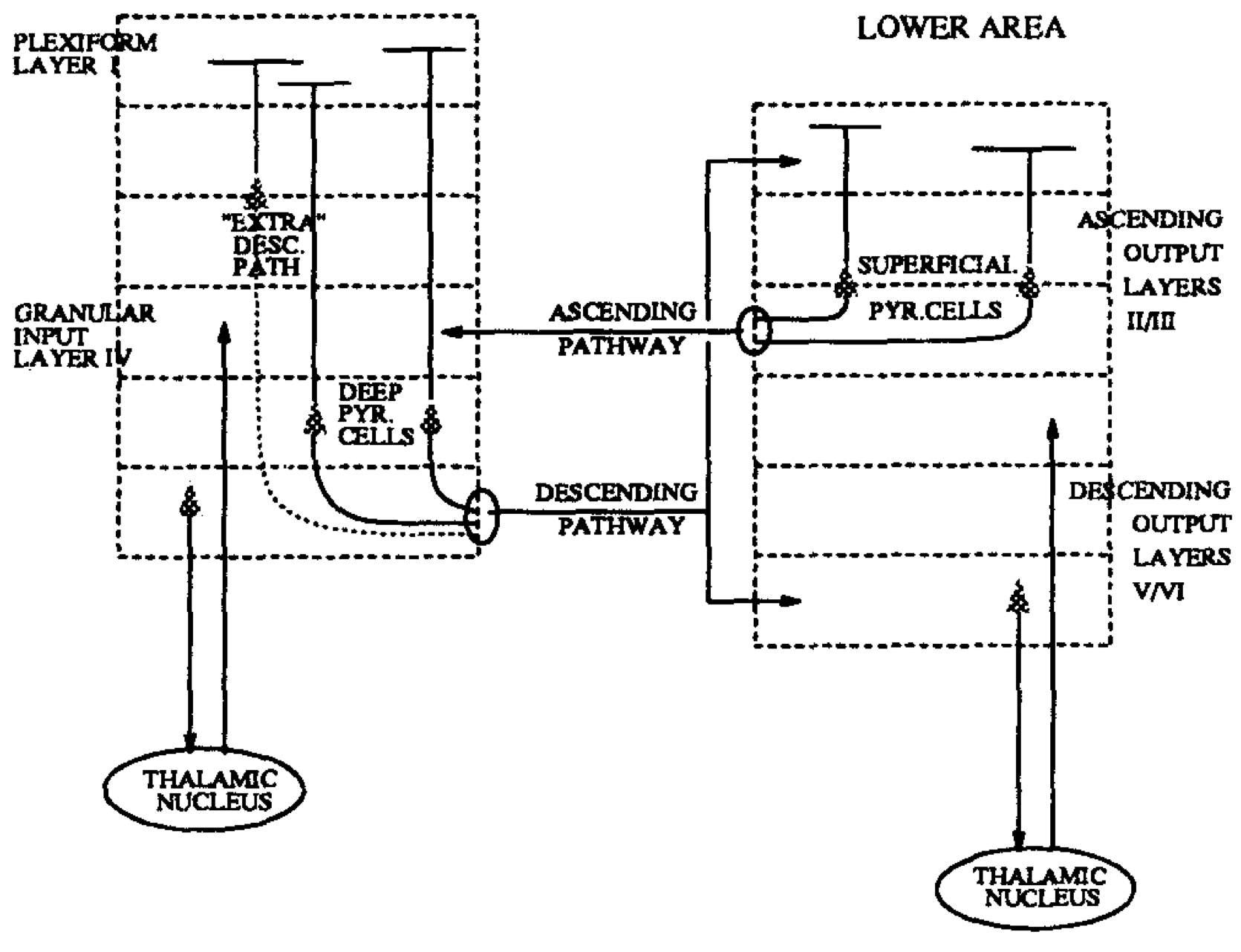}
    \caption[Mumford's conjecture of visual perception as hierarchical Bayesian inference]{Cortico-cortical pathways by layer. Foodforward (ascending) and feedback (descending) connections in the visual cortex are hypothesized to carry different types of information (see text for details). Figure from \textcite{mumford1992computational}, 1992.}
    \label{fig:mumford92}
\end{figure}

The genesis of this theory dates back to the early 1990s. During this period, Mumford, influenced by the seminal review by \textcite{felleman1991distributed}, pondered the functional roles of different cortical areas and the nature of the connections between them. A key observation he made was the predominance of feedback connections over feedforward connections in the cortex (\cref{fig:mumford92}, also see \textcite{markov2014anatomy} for a more recent review). Integrating this observation with Helmholtz's concept of \say{perception as inference} \cite{helmholtz1867handbuch}, Mumford formulated a bold conjecture: The brain engages in hierarchical Bayesian inference to make sense of the world, with feedback connections carrying signals about the brain's prior expectations about incoming data \cite{mumford1992computational, lee2003hierarchical}.

The essence of Mumford's conjecture is that, at each level of the visual hierarchy, prior information descending from feedback connections is combined with the ascending feedforward sensory information, a process that constructs our perceptions \cite{lee2003hierarchical}.

\begin{figure}[ht!]
    \centering
    \includegraphics[width=0.95\linewidth]{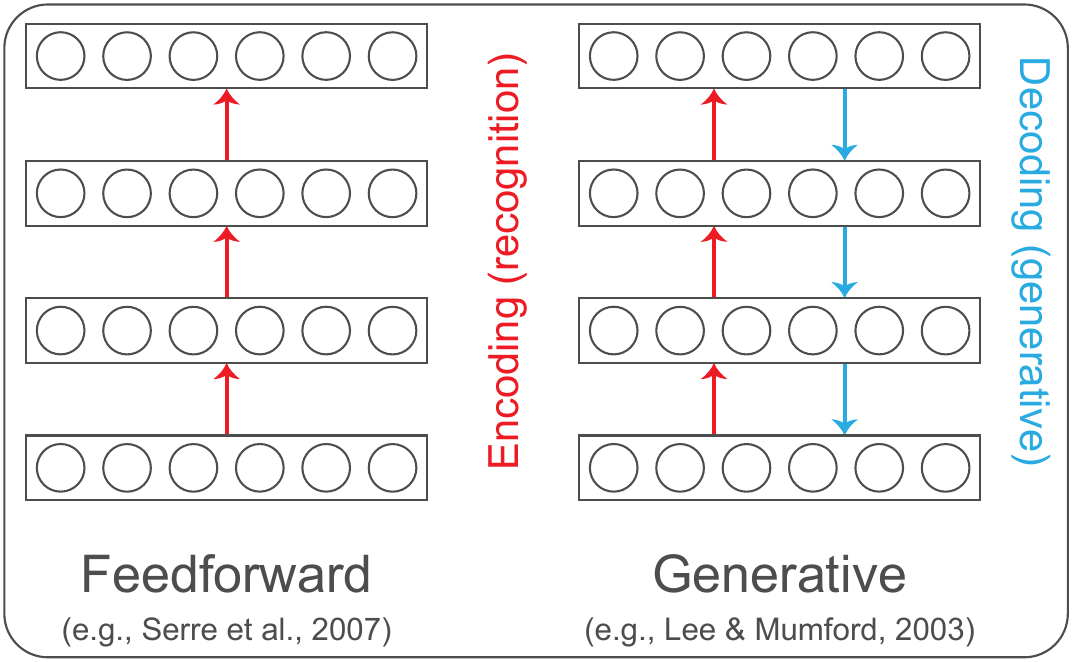}
    \caption[Feedforward versus Generative approaches to vision]{Feedforward (\cref{sec:ff}; e.g., \textcite{serre2007feedforward}) vs.~Generative (\cref{sec:generative,sec:normative}; e.g., \textcite{lee2003hierarchical}) approaches to vision. In this Dissertation, we are team generative.}
    \label{fig:ff-vs-fb}
\end{figure}

This concludes our exploration of feedforward (\cref{sec:ff}), generative (\cref{sec:generative}), and normative (\cref{sec:normative}) approaches to vision. See \cref{fig:ff-vs-fb} for a visual summary. In Chapter 4, I will adopt a Helmholtzian generative approach, similar to the conceptual frameworks employed in the study of the macaque face patch system by \textcite{higgins2021unsupervised} and \textcite{yildirim2020efficient}. I will apply this framework to elucidate neuronal responses in the motion processing area MT. Drawing inspiration from Mumford, I will develop a hierarchical Variational Autoencoder (VAE) model featuring both feedforward and feedback connections. This model demonstrates an ability to process motion in a manner that resembles MT neurons.\\[-3mm]

\section{Neuronal encoding and systems identification}

Building encoding models of visual neurons is a key challenge in vision science \cite{dayan2001theoretical}. This is the type of question that papers like \textcite{serre2007feedforward, mineault2012hierarchical, yamins2014performance} were trying to address, as we discussed above. In this section, I will provide a more detailed exposition of the encoding problem.

Suppose in an experiment we show stimuli $S$ to monkeys and we record from neurons in its visual cortex. Let us use $R$ to denote the responses of these neurons. Can we build a computational model $f$ that can approximate these visual responses? In other words, can we find $f$ such that $\hat{R} = f(S)$, where $\hat{R}$ is the model predicted responses? The function $f$ is an encoding model that accepts arbitrary stimuli and produces outputs that match the recorded neurons from the brain. $f$ can be learned from data by requiring that $\hat{R}$ should be close to $R$. This process of estimating a mapping between some inputs to some output is technically known as \say{systems identification} \cite{wu2006complete}.

\subsection{Linear-nonlinear (LN) encoding models}
\begin{figure}[ht!]
    \centering
    \includegraphics[width=0.77\linewidth]{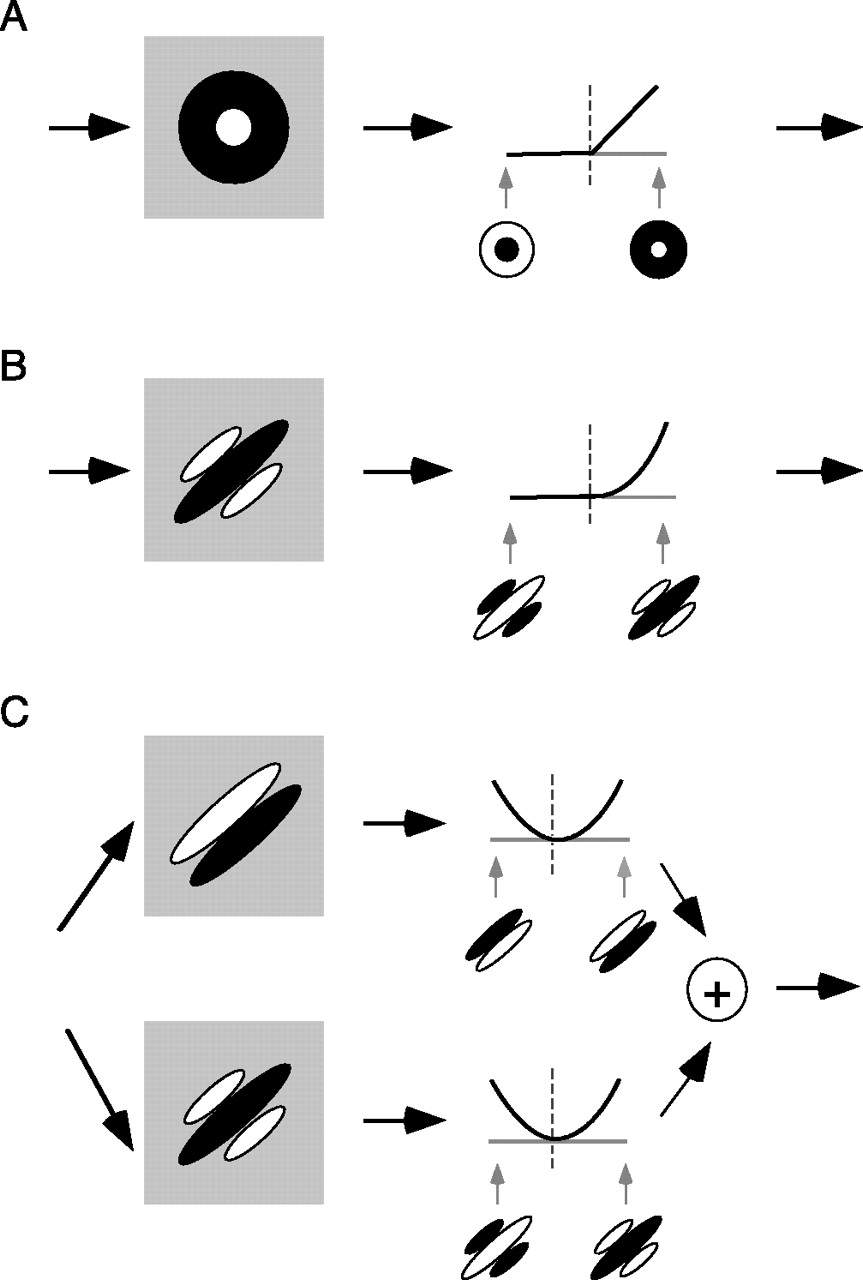}
    \caption[Linear-nonlinear (LN) encoding models]{Basic models of neurons involved in early visual processing. In all models, the response of a neuron is described by passing an image through one or more linear filters (by taking the dot product or projection of an image and a filter). The outputs of the linear filters are passed through a pointwise nonlinear function. Panels show examples of LN encoding models. \textbf{(a)} Simple model of a retinal ganglion cell or of a thalamic relay neuron. \textbf{(b)} Model of a V1 simple cell. \textbf{(c)} Energy model of a V1 complex cell. Figure from \textcite{carandini2005we}, 2005.}
    \label{fig:ln}
\end{figure}

There are different choices for $f$. A simple but effective approach is based on the ubiquitous linear-nonlinear (LN) computational motif \cite{carandini2005we}. LN operation includes two steps: (1) linear filtering that takes the dot product of local patches in the input stimulus with a set of templates, followed by (2) a pointwise nonlinearity such as a rectified linear threshold or a sigmoid. LN encoding models are highly interpretable but they may not be flexible. The linear filter applied in step (1) is considered the receptive field of the neuron. See \cref{fig:ln}.

\subsection{The nonlinear input model (NIM)}

In 2013, \textcite{mcfarland2013inferring} presented the Nonlinear Input Model (NIM), a framework for characterizing how sensory neurons compute responses to stimuli. The key idea in an NIM is that complex nonlinear computations can be approximated as a summation of rectified inputs, motivated by the physiology of neurons receiving excitatory and inhibitory synaptic inputs. The NIM expands on previous approaches like LN models by adding a layer of upstream nonlinearities to account for the rectification of inputs. The authors show that this model structure allows the NIM to capture a range of nonlinear neural computations in a biologically interpretable way.

Key advantages are the ability to infer separate excitatory and inhibitory receptive fields, model complex computations with many nonlinear subunits, and fit high-dimensional spatiotemporal stimuli. Overall, the NIM provides a relatively more flexible yet physiologically grounded framework for understanding nonlinear sensory processing building on earlier LN models. The modeling approach leads to realistic components that match known biology while improving the characterization of diverse neural computations. The NIM was employed to analyze the receptive field properties of motion-processing MT neurons by \textcite{cui2013diverse}, a topic we will return to later.

\subsection{Artificial neural networks as models of sensory processing}

Another way of obtaining an encoding model for biological neurons is by making use of artificial neural networks (ANN), an approach that has gained a lot of traction in recent years. As discussed above, goal-directed convolutional neural networks have been used for this purpose to great effect \cite{serre2007feedforward, yamins2014performance, yamins2016using}. In general, when using ANNs as candidates for encoding models of sensory neurons, we need to make four major choices:

\begin{enumerate}
    \item \textbf{Model architecture:} What is the network topology? Is it feedforward, recurrent, or something in between? Making the network architecture biologically plausible is shown to be an effective way of increasing its representational alignment with the brain \cite{kubilius18cornet, bakhtiari2021functional}.
    \item \textbf{Loss function:} What is the objective used to train the network? For instance, data-driven supervised models can be trained to directly map stimuli to neuronal responses \cite{McIntosh16Deep, Klindt17neural, sinz18stimulus, ecker2018rotation, butts2019data, wang23towards}. Alternatively, one can utilize transfer learning in combination with goal-directed networks \cite{yamins2014performance, yamins2016using, cadena2019deep}. There's also the option of self-supervised approaches \cite{zhuang2021unsupervised, nayebi2023mouse}, or using generative objectives focused on inference, such as the approach by \textcite{higgins2021unsupervised}.
    \item \textbf{Training dataset:} What is the visual diet we train the network on? It is crucial to choose a training set that contains rich, ecologically relevant stimuli \cite{david2004natural, mehrer2021ecologically, konkle2022self}.
    \item \textbf{Learning rule:} How do we adjust the parameters of the network to reduce the loss? Stochastic gradient descent with backpropagation is the most common choice here. However, the biological plausibility of backpropagation is actively debated \cite{grossberg1987competitive, lillicrap2020backpropagation}.\\[-4mm]
\end{enumerate}

\begin{figure}[ht!]
    \centering
    \includegraphics[width=1.0\linewidth]{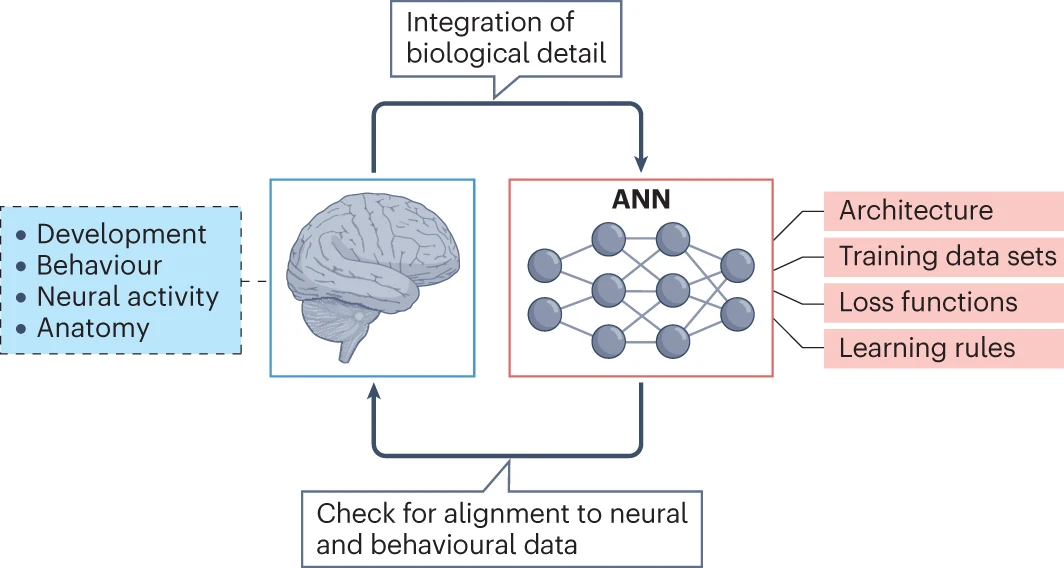}
    \caption[The neuroconnectionist research programme]{The \textit{neuroconnectionist} research programme. This iterative research approach merges detailed biological insights from neural and behavioral studies across levels to inform the development of new artificial neural network (ANN) models with different components. The ANNs are then tested for alignment with neural and/or behavioral data (left), leading to further cycles of model creation and model testing. Model creation involves four central ingredients: architectures, objective functions, training data sets, and learning rules. Model evaluation involves hypothesis testing via tools such as representational similarity analysis, encoding models, and in silico experimentation. Figure from \textcite{doerig2023neuroconnectionist}, 2023.}
    \label{fig:doerig23}
\end{figure}

These ideas are discussed in depth in a recent review article by \textcite{doerig2023neuroconnectionist} (\cref{fig:doerig23}), with similar arguments previously made by \textcite{richards2019deep}.

In Chapter 4, I address the first 3 out of these 4 challenges in building an ANN that perceives motion. First, I design my models to be hierarchical like the visual cortex (\cref{fig:felleman-essen}; \cref{fig:mumford92}). Second, I use Helmholtzian inference as my choice of loss function  \cite{helmholtz1867handbuch}. The combination of these two ideas leads to a hierarchical VAE model that performs hierarchical Bayesian inference on its sensory inputs. As for the third point (training set), I introduce a novel synthetic data framework that combines ecologically relevant causes of motion (more details below). Lastly, for the fourth point (learning rule), I use backpropagation with stochastic gradient descent.

\section{Motion perception in primates}\label{sec:motion}
In this section, I explore the concept of motion perception and highlight it as a foundational problem in vision. I start with a brief review of past efforts in developing encoding models for neurons in area MT, which play a crucial role in cortical motion processing. I will then discuss the key challenges involved in motion perception that the brain has to overcome. The section concludes with an explanation of the motivation behind my research on cortical motion processing and provides a succinct overview of the findings presented in Chapter 4.

\subsection{Previous efforts in building encoding models of MT neurons}

Here I will briefly review two key papers that have looked into modeling MT neurons. The first one is by \textcite{nishimoto2011three} who developed a successful mechanistic model that explained MT responses to naturalistic stimuli. The second one is by \textcite{cui2013diverse} who used NIMs to show how MT neuron responses are shaped by diverse suppressive surround effects.

\subsubsection{Mechanistic modeling of MT neurons}

\begin{figure}[ht!]
    \centering
    \includegraphics[width=1.0\linewidth]{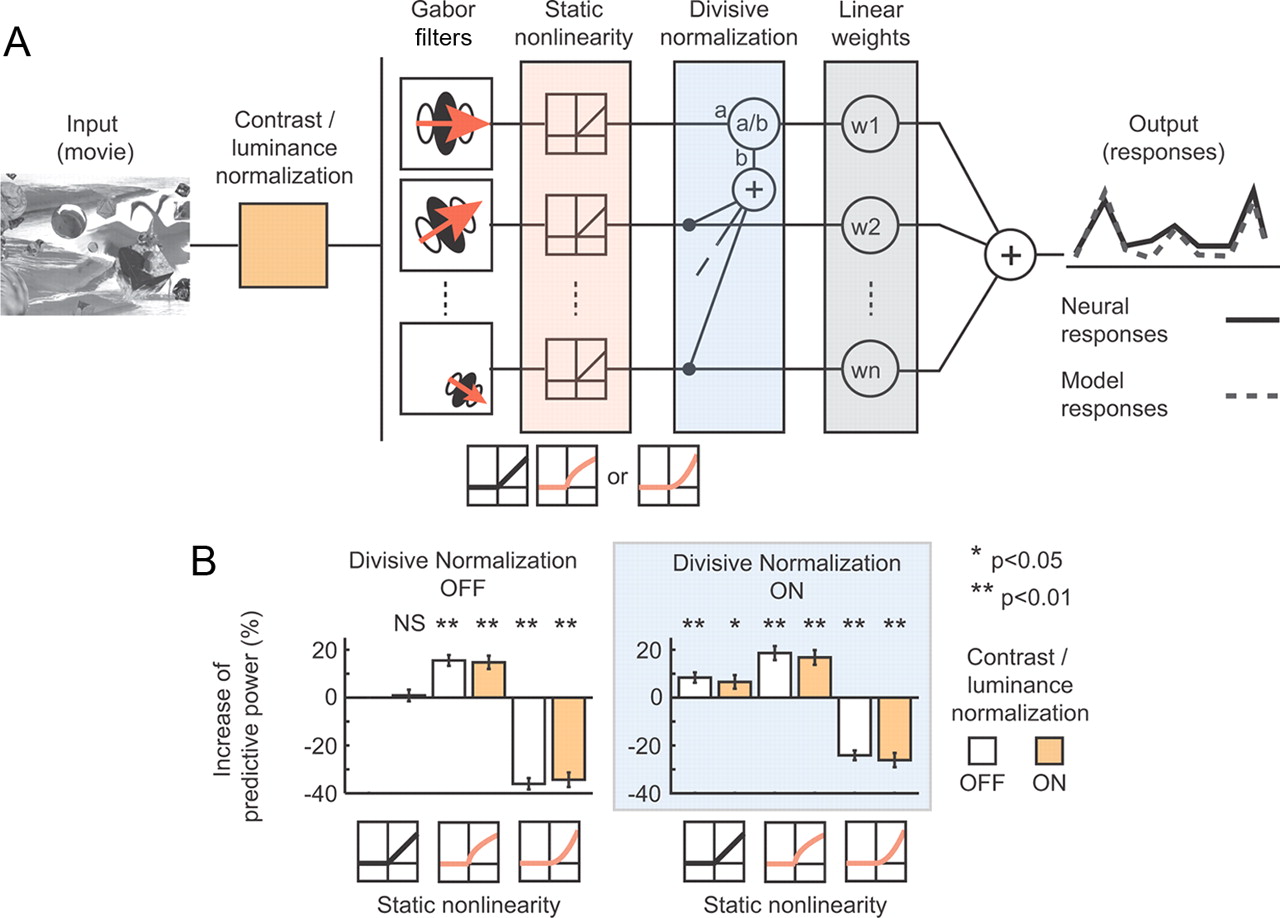}
    \caption[A mechanistic model of MT neurons by \textcite{nishimoto2011three}]{Analysis of MT neurons using the switched model of \textcite{nishimoto2011three}. \textbf{(a)} Schematic of the switched model framework for MT neurons, featuring stages like nonlinear contrast/luminance normalization (orange), simple and complex V1 filters, static nonlinearities with rectification (pink), divisive normalization (blue), and summation by weighted linear regression (gray). \textbf{(b)} Evaluation of model performance by varying nonlinear stages, shown in 12 configurations across 52 MT neurons. Bars represent predictive power increase over the simplest model, differentiated by the presence (filled bars) or absence (open bars) of luminance and contrast normalization (refer to legend). Left and right panels show results without and with divisive normalization, respectively. Figure from \textcite{nishimoto2011three}, 2011.}
    \label{fig:nishimoto-gallant}
\end{figure}

In 2011, \textcite{nishimoto2011three} investigated how neurons in visual area MT represent motion information under natural viewing conditions. The authors recorded MT neuron responses in macaques viewing natural movies with added foreground motion. Using a mechanistic model, they found MT responses could be explained by a model with V1-like spatiotemporal filters followed by compressive and divisive nonlinearities and weighted linear pooling (\cref{fig:nishimoto-gallant}).

Analysis of the model filters showed MT neurons have planar tuning in 3D frequency space, with most forming a partial ring avoiding low temporal frequencies. This matches predictions that MT neurons should be tuned to velocity planes while avoiding sensitivity to ambiguous static patterns. The suppressive filters tended to lie off this plane, sharpening velocity tuning. This shows MT motion computations characterized by simple stimuli generally extend to natural vision. The methods quantify 3D spectral tuning directly and show most MT neurons represent motion in a way that avoids responding to static stimuli. Overall, the paper by \textcite{nishimoto2011three} illustrates how mechanistic modeling can provide new insights even for well-studied visual areas.

\subsubsection{Diverse suppressive influences in area MT}

In 2013, \textcite{cui2013diverse} utilized the then--recently developed Nonlinear Input Models (NIM; \cite{mcfarland2013inferring}) to investigate the processing of MT neurons of macaques viewing naturalistic optic flow stimuli. The authors found that MT neurons have distinct excitatory and suppressive receptive field components. Excitation matched the neuron's preferred direction tuning whereas suppression has multiple roles. One suppressive component acted like an untuned surround, consistent with normalization. However, another component showed diverse direction tuning, often offset from excitation spatially and temporally. This direction-selective suppression contributes to MT selectivity for complex motion features. Simulations showed that the inclusion of such heterogeneous suppression (compared to standard motion opponency models) improves the decoding of motion information. Overall, the results provide new insight into the functional roles of suppression in MT by linking its diversity to improved encoding of natural motion stimuli. The findings illustrate how suppression interacts with excitation to enable selectivity to higher-order stimulus features.

\begin{figure}[ht!]
    \centering
    \includegraphics[width=0.85\linewidth]{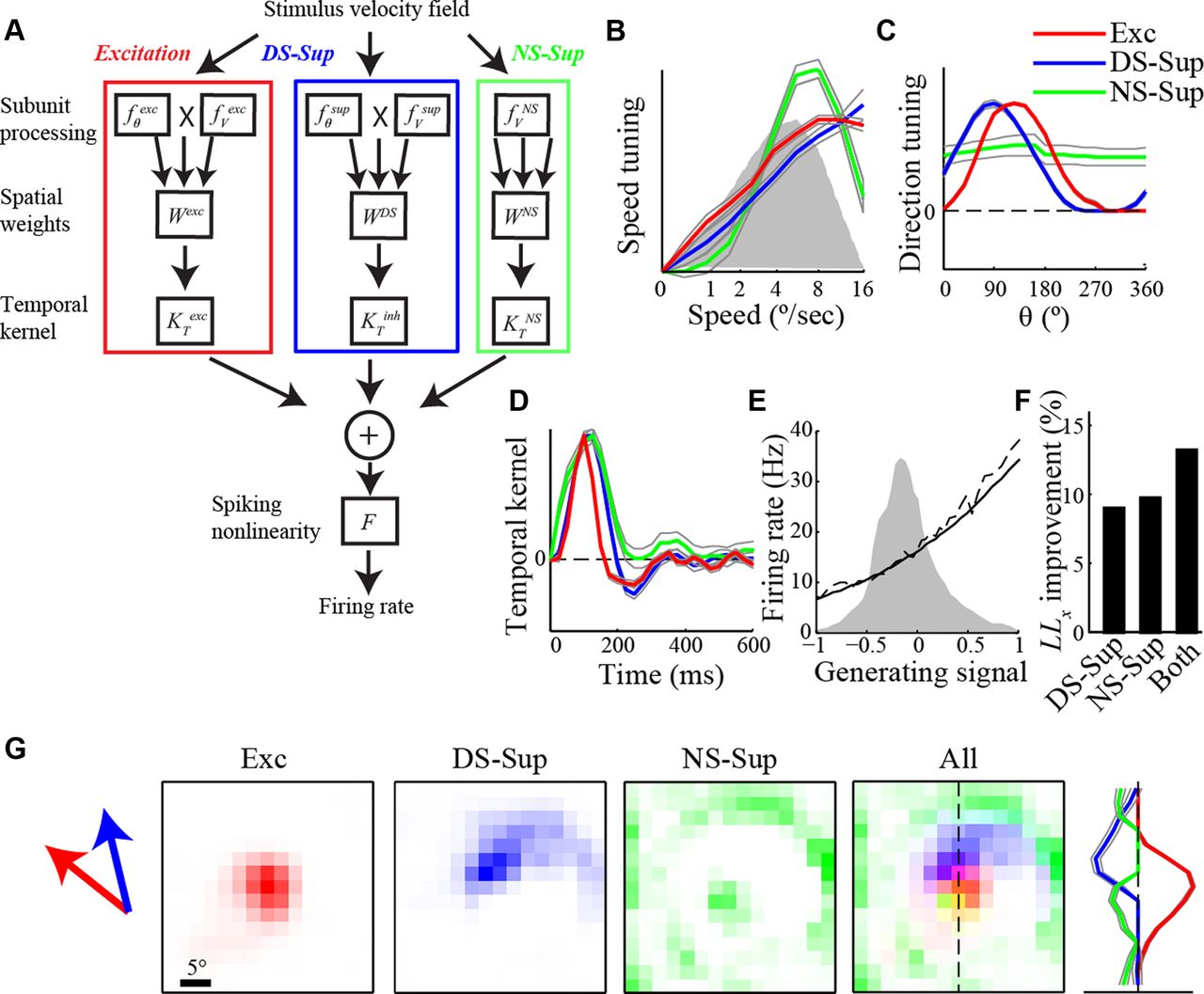}
    \caption[Diverse suppressive components contribute to MT processing]{Incorporating suppression into models of MT processing. \textbf{(a)} Shows different NIM subunits. Excitation (Exc) in red, direction-selective suppression (DS-Sup) in blue, and non-direction-selective suppression (NS-Sup) in green. \textbf{(b)} The model components have similar speed tuning. \textbf{(c)} Different direction tuning, with Exc and DS-Sup antagonistic and NS-Sup untuned. \textbf{(d)} Temporal kernels show suppression is delayed relative to excitation. \textbf{(f)} Adding suppression improves model performance over excitation alone. \textbf{(g)} Spatial kernels show distinct profiles for Exc, DS-Sup, and NS-Sup. Figure from \textcite{cui2013diverse}, 2013.}
    \label{fig:cui}
\end{figure}

\subsection{Unsupervised models of MT?}

The journey to unravel the intricacies of cortical motion processing in area MT has been significantly advanced by the mechanistic models of \textcite{nishimoto2011three} and the NIM approach by \textcite{cui2013diverse}. Both of these papers share something in common: they use stimulus/response pairs to train models using supervision. However, my Dissertation explores an emerging, potentially transformative approach: the use of unsupervised ANNs to build encoding models of neurons. 

This paradigm shift from supervised to unsupervised learning presents both remarkable opportunities and formidable challenges. As evidenced in \cref{fig:doerig23} and discussed by many before \cite{richards2019deep, doerig2023neuroconnectionist, sinz2019engineering}, the success of this approach hinges on making precise choices to avoid the pitfalls of learning representations that are irrelevant to biological computation. Chapter 4 of this Dissertation explores the unsupervised approach, addressing the challenges head-on and uncovering new ways in the understanding of cortical motion processing.

In the following section, we will broaden our perspective to examine the fundamental challenges inherent in motion perception. This exploration is important in understanding how our brains interpret and respond to the dynamic visual environment. This discussion not only contextualizes the specific focus of my research but also sheds light on the overarching questions that drive advancements in our understanding of visual perception.

\subsection{Motion perception: A foundational problem in vision}

Almost all animals locomote. Moving around is essential for staying alive (running away from predators, finding food) and passing on the genes (finding a mate). But it has a side effect on vision: locomotion dynamically changes the patterns of light landing on the retina, leading to the perception of motion even when the environment is static. As a result, a substantial portion of the motion signals received by the brain arise due to self-motion.

\begin{figure}[ht!]
    \centering
    \includegraphics[width=\linewidth]{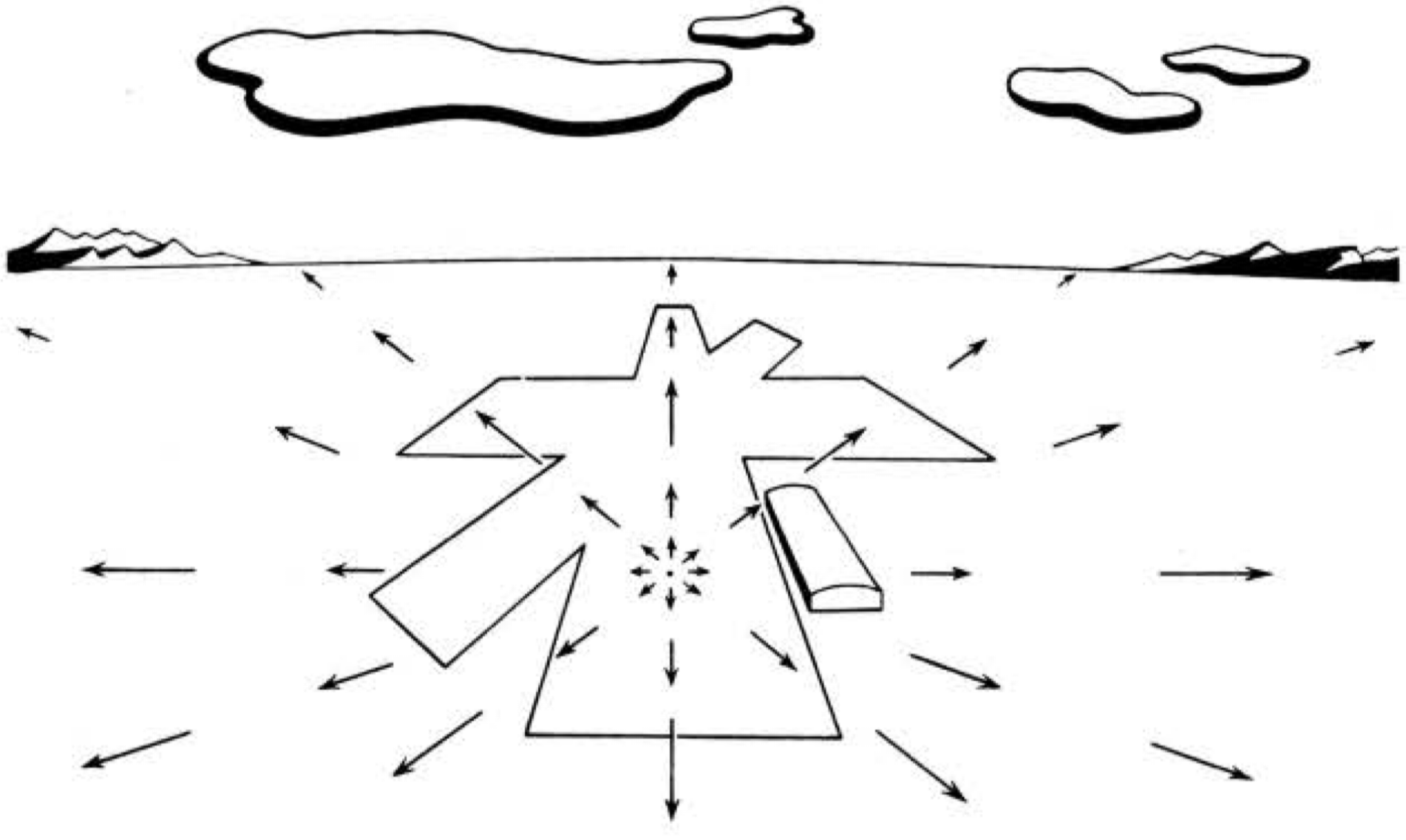}
    \caption[Gibson drawings of optical flow]{One of Gibson's earliest drawings of optical flow. The arrows depict flow patterns that arise during the landing of an aircraft. Image adapted from \textcite{gibson1947motion}, 1947.}
    \label{fig:opticflow}
\end{figure}

There is a special structure to motion patterns that arise from self-motion. For instance, the patterns cover the entire scene in a non-uniform manner. American psychologist, James Gibson was the first to recognize how an observer's movement through the environment results in certain radial expansion patterns, which he called optic flow (\cite{gibson1947motion, gibson1950perception}; \cref{fig:opticflow}).

Gibson also contemplated the intricate interplay between translational and rotational self-motion components \cite{gibson1950perception, warren1988direction}, and their impact on resulting optic flows. He then proposed that the translational and rotational components of the optic flow can be decomposed visually, although he did not elaborate on how this could be accomplished \cite{gibson1950perception, gibson1954visual}. Other theories have proposed that, to estimate self-motion, the brain relies on a combination of visual and non-visual cues \cite{DeAngelisAngelaki2012}, such as efference copies of motor commands \cite{von1954relations}, vestibular inputs \cite{macneilage2012vestibular}, and proprioception \cite{cullen2021proprioception}.

Further complications arise when independently moving objects are added to the mix \cite{gibson1954visual, royden1996human, warren1995perceiving}. Even when the objects are not moving, the depth structure in the scene causes different optic flows in different locations: self-motion causes their image to move differently than the background. Given all these complex relationships, how can the brain understand the motion of an object when self-motion is non-trivially affecting the entire flow field \cite{horrocks2023walking}? Teasing apart the contributions from different sources of motion is a fundamental problem that brains need to solve \cite{lee1980optic}, manifested in diverse ecological scenarios such as prey-predator dynamics, crossing the street, or playing soccer.

\begin{figure}[ht!]
    \centering
    \includegraphics[width=1.0\linewidth]{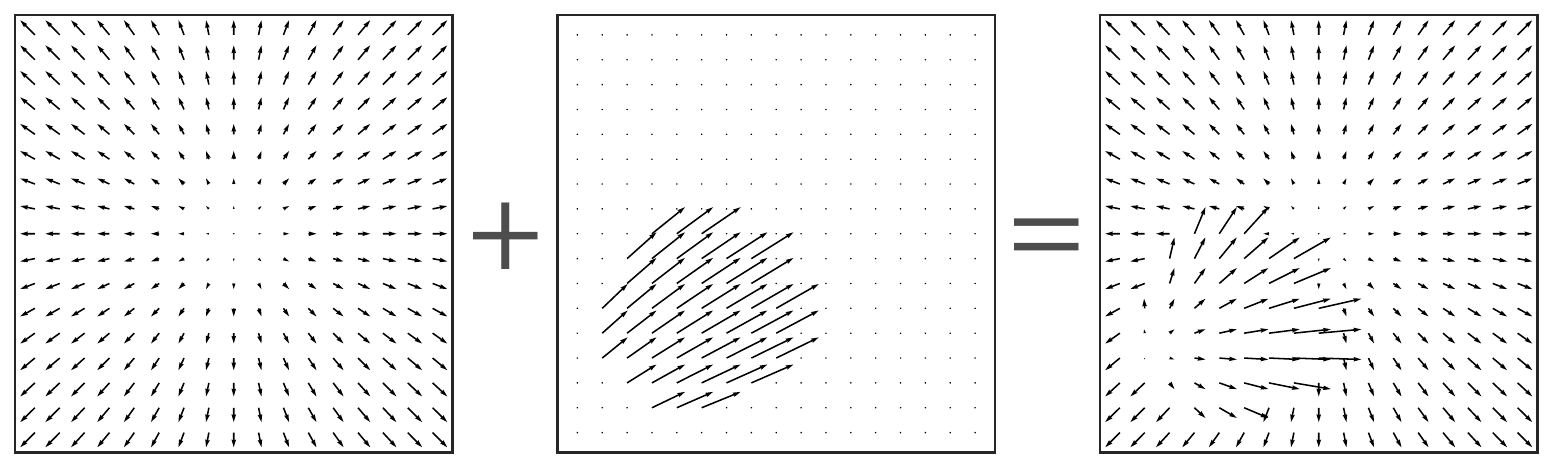}
    \caption[Optic flow parsing hypothesis]{The interactions between self-motion (left) and object motion (middle) result in a distorted optic flow pattern (right). Optic flow parsing hypothesis refers to the idea that such distorted flow patterns can, in principle, be decomposed into their constituent components.}
    \label{fig:self+obj}
\end{figure}

\Cref{fig:self+obj} illustrates this problem. Here, an observer is moving forward, which causes the radial expansion patterns (\cref{fig:self+obj}, left). The scene also contains an object that is moving to the right (\cref{fig:self+obj}, middle). The combination of both self and object motion components results in a distorted flow pattern (\cref{fig:self+obj}, right). We see that accurate estimation of object velocity in the presence of self-motion is not an easy task, and several requirements need to be met. First, the brain needs to have access to the direction and magnitude of its own velocity. Second, the brain needs to understand the underlying compositional rules that govern optic flow generation. Only then can the brain subtract self-motion from the entire flow patterns to infer object motion.

This separation of different components of optic flow is a difficult problem in vision. Are retinal visual-only cues sufficient to extract information about different self-motion components, as well as object motion? Vision scientists have debated this for decades, and still, there is no clear answer (see \cite{lappe1999perception} for a review). A solution to this problem was proposed by Rushton and Warren who introduced the \sayit{flow-parsing hypothesis}: that retinal information alone, specifically optic flow, can be used to dissociate object motion from observer's self-motion \cite{rushton2005moving, warren2009optic, warren2007perception}.

\subsection{Summary of results presented in Chapter 4}

In Chapter 4, I create a simulated world in which an observer is moving around while maintaining fixation on a background point. Thus, the simulated optic flow patterns include both translational and rotational components of self-motion. The world also includes an independently moving object. The inclusion of the object results in all sorts of interactions between flow patterns arising from self-motion and object velocity, thus effectively capturing the complications that are present in motion processing in the real world. Finally, the object has a fixed size but appears at different depths which further complicates the inference of its 3D velocity.

I train unsupervised neural network models on this simulated data that aim to understand the underlying structure and composition of the optic flow. I evaluate the performance of the models on their ability to parse optic flow and extract self-motion and object motion components. The premise is that if an unsupervised model has achieved a good understanding of its training dataset, then its internal representations should contain information about the hidden causes of their sensory inputs. I consider the following two core ideas when building the model:

\begin{itemize}
    \item \textit{Brains comprehend the world by inferring the underlying causes of their sensory inputs.}
    \item \textit{The visual cortex is organized hierarchically because it has to make sense of a multi-scale visual world.}
\end{itemize}

I demonstrate that optic flow parsing is possible if we train a neural network model that (a) is structured hierarchically, and (b) is trained using an objective function based on Helmholtzian inference. Our results provide a proof of concept that, in principle, optic flow parsing is possible without relying on any extra-retinal information.

Finally, I use the models as encoding models of neurons from macaque area MT and evaluate the models in their ability to predict MT neuronal activation in response to motion stimuli. The hierarchical VAE model trained using an inference-based loss function surpasses the previous state-of-the-art model by a gain of over 2x in predictive power. Collectively, these results suggest that hierarchical Bayesian inference underlies the brain's interpretation of the visual world, and this understanding can be modeled using hierarchical VAEs.

This concludes our introductory background for the first main project included in this Dissertation. The other project is about the intrinsic \sayit{organization} of the mouse cortex. But what does that mean? In the following section, I will use an illustrative example to motivate the concept of organization.

\section{The intrinsic functional organization of a symphony orchestra}\label{sec:orchestra}

For a moment, let's shift our attention away from the brain and towards something more tangible: a symphony orchestra. An orchestra consists of a collection of musicians, each wielding instruments like violins, cellos, flutes, and so forth. Typically, within each instrument group, several musicians play the same note, as dictated by the composer's score for that ensemble.

Imagine we're entirely unfamiliar with the inherent organization of a symphony orchestra. Let's pretend we're unaware that it consists of grouped musicians playing identical notes. Our goal is to infer this organizational principle merely by observing various orchestras performing diverse compositions.

Unraveling the correct organization of an orchestra begins with a fundamental question: what is its core purpose? At its heart, an orchestra's mission is to create music, a realization that shifts our attention away from peripheral details such as the identities of the musicians. The question of whether they are humans or futuristic alien robots fades into the background, as does the significance of their attire. Although it is intriguing to consider how physical attributes might subtly shape the quality of the music, the essence of an orchestra lies in the outcome---the notes played by the musicians. By embracing such a functionalist, systems-oriented view, we recognize that these other elements, intriguing as they may be, ultimately play a secondary role.


Bearing this in mind, we document the specific notes each musician produces. After extensive observations across a variety of orchestras and pieces, patterns emerge: certain musicians consistently produce identical notes. Consequently, we group them together based on these similarities. Recognizing these redundancies allows for more concise descriptions of an orchestra, because knowing the notes of one musician reveals the notes of others within that group.

Upon completing this functional grouping, we can now label these clusters. Violins, flutes, and so on. Through this process, we have uncovered the \sayit{lower-dimensional intrinsic functional organization} of a symphony orchestra:\\[-4mm]

\begin{itemize}
    \item \textbf{Lower-dimensional}: because the number of instrument groups is substantially smaller than the total count of individual musicians.\\[-5mm]
    \item \textbf{Intrinsic}: because we did not provide any external stimulation. As far as we know, the orchestra was playing spontaneously (for the sake of argument, let’s consider the conductor as an internal member, invisible to us).\\[-5mm]
    \item \textbf{Functional}: because the function of the orchestra, playing music, was our main consideration. We did not care about the musicians and their individual properties. Rather, we were concerned primarily with the notes they played.\\[-5mm]
    \item \textbf{Organization}: because we now know how individual musicians within an orchestra are functionally organized, or related to each other: some are violinists performing the function of playing the violin, others are oboists, and so on.
\end{itemize}

\section{Network science and community detection}

In the orchestra example above, there were several possible ways we could go about clustering the musicians into groups. We could either apply a simple clustering algorithm based on some notion of similarity between the notes being played. For instance, we could apply KMeans clustering to the time series data of the notes played by the musicians.

Alternatively, we could achieve a similar outcome by generating a network of interactions derived from our observation of the notes being played. We would start by computing the pairwise correlations, or the covariance matrix of the musicians. This is also known as \textit{functional connectivity}. We would then apply a threshold to the covariance matrix and keep only the highest correlation values while zeroing out the rest. This would result in a binary interaction graph, where nodes are musicians and links indicate the existence of a relationship between them. In this case, a link between a pair of musicians means they play the same notes over an extended duration of time. Such binarization ensures that the spurious correlations are filtered out and only the strongest ones survive. Once we have the interaction graph, we can employ a community detection algorithm from network science to detect communities or clusters of nodes.

In sum, the idea is to represent a complex system as a network of interconnected nodes. Then, you can use various community detection algorithms to decompose the graph into communities, or groups of functionally \textit{similar} nodes.

\subsection{The Girvan-Newman algorithm}
The Girvan-Newman algorithm \cite{girvan2002community}, published in 2002, is the first paper on community detection that arguably kick-started this whole line of research. The algorithm is a hierarchical and iterative method, progressively knocking out edges with high \textit{betweenness} centrality. In network science, betweenness is a measure of centrality based on shortest paths. The betweenness centrality of a node $i$ is given by the expression:

\begin{equation}
    g(i) = \sum_{s \neq i \neq t} \frac{\sigma_{st}(i)}{\sigma_{st}},
\end{equation}
where $\sigma_{st}$ is the total number of shortest paths from node $s$ to node $t$, and $\sigma_{st}(i)$ is the number of those paths that pass through $i$.

The intuition behind the Girvan-Newman algorithm is that the edges connecting communities will have high edge betweenness. By removing these edges, the algorithm separates groups from one another, revealing the underlying community structure of the network::

\begin{enumerate}[leftmargin=25mm]
    \item Calculate the betweenness centrality for all edges in the network.
    \item Remove the edge with the highest betweenness.
    \item Recalculate betweennesses for all edges affected by the removal.
    \item Repeat from step 2 until no edges remain.
\end{enumerate}

After \textcite{girvan2002community}, there was an explosion of research looking into developing new community detection algorithms. Over time, three major classes of community detection algorithms have emerged. We will briefly explore each below.

\subsection{Algorithms based on optimization}
In this class of algorithms, we assign a score to a decomposition of a network based on some measure of goodness. Then we look for a decomposition that maximizes the score. Based on the domain of application, different scoring approaches are used.

The most commonly used score is \textit{modularity}, which explicitly favors divisions that result in many within-group edges and fewer cross-group edges. In the context of network science, modularity is defined as follows \cite{newman2006modularity, newman2018networks, sporns2016modular}:

\begin{equation}
    Q = \frac{1}{2m}\sum_{ij} \Bigg[ A_{ij} - \frac{k_ik_j}{2m} \Bigg] \delta_{g_ig_j}.
\end{equation}

$A$ is the \textit{adjacency matrix}: $A_{ij} = 1$ of there is a link between nodes $i$ and $j$, and zero otherwise. $m$ is the total number of links in the network. $g_i$ is the community assignment of $i$: it is an integer $\in \left[1, 2, \dots, N\right]$, with $N$ being the total number of communities. Intuitively, the modularity score measures the density of connections within a module or community. $Q$ will be large when there are many links connecting nodes within communities and few links across communities. 

The term $\frac{k_ik_j}{2m}$ can be interpreted as the probability of the existence of edge $(i, j)$ according to a null model known as the \textit{configuration model} \cite{fosdick2018configuring, vavsa2022null}. The configuration model of a graph preserves its degree distribution but is random otherwise \cite{newman2018networks}.

Examples in this class include a simple node-moving algorithm proposed (\textcite{newman2006modularity}, 2006), the Louvain algorithm (\textcite{blondel2008fast}, 2008), and a more recent extension called the Leiden algorithm (\textcite{traag2019louvain}, 2019). Over the years, the Louvain algorithm has become the first go-to algorithm that can be easily applied to different types of networks.

Despite the popularity of modularity maximization approaches, some authors have raised issues with it. For example, \textcite{good2010performance} explored the degeneracy of community decompositions that result from modularity maximization. The authors observed that many different network decompositions have the same modularity score. This degeneracy makes it difficult to choose between alternative decompositions. Based on these observations, the authors recommend that the output of any modularity maximization procedure should be interpreted cautiously in scientific contexts \cite{good2010performance}. 

\textcite{peixoto2023descriptive} even goes as far as calling modularity maximization \say{harmful} because of its descriptive nature, and recommends we instead use methods based on statistical inference.

\subsection{Algorithms based on simulating random walks}
These classes of algorithms rely on links between community structure and simulating random walks taking place on networks. The intuition behind these types of algorithms is that nodes within communities are typically tightly connected, with some sparse connectivity linking communities together. Therefore, a random walker on a network would typically linger a lot longer within a community, than traversing across communities. Hence, one can use the probability distribution of random walks for community detection.

The most popular method in this class is known as the InfoMap algorithm (\textcite{rosvall2008maps}, 2008), which considers the sequence of communities visited by a random walker, and a partition is considered good if the sequence can be described using a minimum amount information, in the sense of Shannon entropy. In 2011, \textcite{Power2011FunctionalNO} applied this algorithm to uncover the community structure of human fMRI networks.

\subsection{Algorithms based on statistical inference}
Recall Axiom 1 from above. A core assumption in network science is that the global topology of the network alone encodes information about its function. Network topology, on the other hand, depends on how links are placed locally. In this sense, the existence or non-existence of a link between pairs of nodes is informative and merits an explanation.

The third class of community detection algorithms, those based on statistical inference, aim to answer this question using a latent variable modeling approach. Namely, the algorithms assume there is a hidden generative process that explains the existence of links. Given this assumption, methods from statistical inference can be used to infer the generative process from observing the graph and its links alone. As a result, in this approach communities are not merely a descriptive \textit{feature} of the network structure but a primary \textit{driver} of it.

The placement of links is most commonly modeled using the stochastic block model (SBM) \cite{holland1983stochastic}, a probabilistic modeling approach in which the probability that two nodes are connected depends solely on the communities they belong to. For instance, in the symphony orchestra example above, there is a high likelihood that two violin players will be connected with a link because they belong to the same (latent) group.

In Chapter 5, I use a stochastic mixed membership \cite{airoldi2008mixed} community detection algorithm that assumes pairs of nodes are more likely to be connected if they belong to one or more \textit{overlapping} communities 
together (\cite{Gopalan2013EfficientDO}; see \cref{fig:disj-ovp}). A concrete generative model is constructed based on this idea, where observed links in the network are explained by latent (or hidden) community membership values. The algorithm uses variational inference and stochastic gradient ascent \cite{Blei2016VariationalIA, Hoffman2013StochasticVI} to infer latent community memberships from observed networks.

In conclusion, network science offers a wide variety of community detection algorithms that have found applications in various domains. See \textcite{fortunato2010community} or a comprehensive survey, and \textcite{fortunato202220} for a more recent, concise overview.

\section{Communities can be disjoint or overlapping}
\begin{figure}[ht!]
    \centering
    \includegraphics[width=0.8\linewidth]{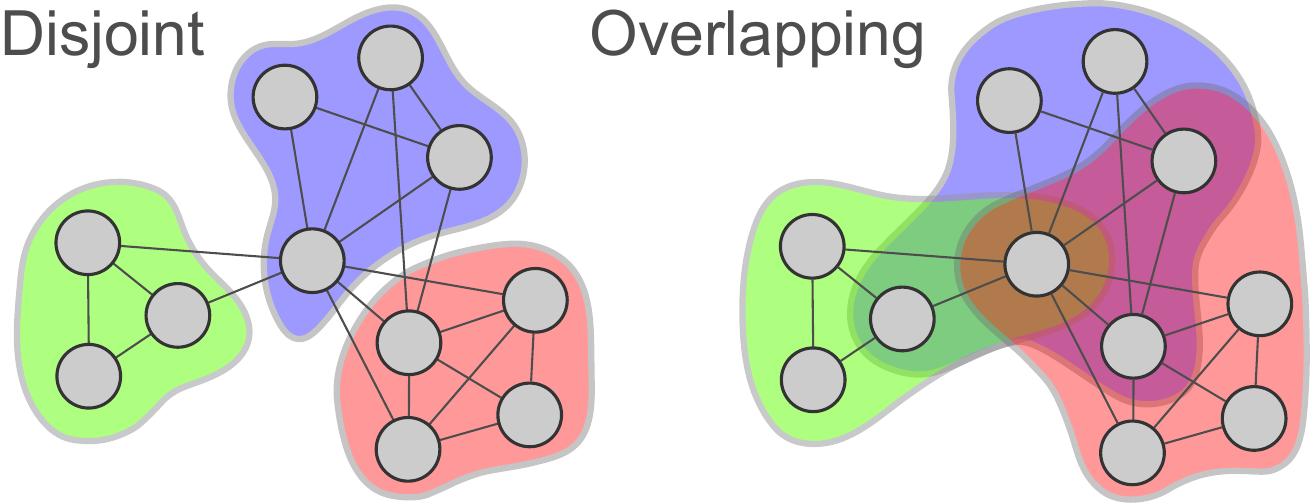}
    \caption[Disjoint and overlapping community structure]{Two types of community structure. In the disjoint case, every node belongs to a single community. In the overlapping case, nodes are allowed to belong to multiple communities.}
    \label{fig:disj-ovp}
\end{figure}

In the preceding section, we discussed different approaches to community detection in terms of the three main classes of algorithms. Here, we will consider this question from another angle: community detection algorithms can be classified as either \textit{disjoint}, or \textit{overlapping}, depending on whether they allow nodes to belong to a single community or multiple communities (\cref{fig:disj-ovp}).

Disjoint community detection algorithms assign each node to a single community. The membership variable of a node is therefore a binary all-or-none type of belonging: given a community, a node either belongs or does not belong to it.

Overlapping community detection algorithms relax this constraint by allowing nodes to belong to multiple communities. In some cases, the algorithms produce a \say{fuzzy} membership strength for nodes, which quantifies the extent to which the node belongs to a given community.

To further elucidate this concept, let us use $\pi_i$ to denote the \textit{membership} vector of node $i$. If there are $K$ communities, $\pi_i$ will be a vector of length $K$, where each entry indicates the belonging of $i$ to a given community. In other words, $\pi_{ik}$ indicates the membership strength of node $i$ in the community $k$. In disjoint communities, $\pi_{ik} = 0 \text{ or } 1$; whereas, in overlapping or fuzzy communities, $\pi_{ik} \in [0, 1]$.

Some overlapping community detection algorithms begin by grouping network links into distinct categories (\textit{link communities}; \cite{ahn2010link}), similar to assigning a color to each link. This categorization of links defines overlapping communities at the node level: nodes connected by links of the same color belong to the same community. Nodes with links of multiple colors are considered overlapping, while those with links of a single color are disjoint \cite{Faskowitz2020EdgecentricFN, betzel2023living, faskowitz2022edges, jo2021diversity}.

\subsection{Harmony in a symphony orchestra and the brain}

In the orchestra example from above, we observed the notes played by musicians, computed their functional connectivity via correlations, and used this information to infer the underlying organization. Here, we will apply a similar idea to the brain. The musicians are replaced by neurons or brain regions, and the notes played are replaced by spontaneous neural activity. Given this setup, it is interesting to determine which kind of community structure (disjoint or overlapping; \cref{fig:disj-ovp}) best captures the underlying organization of these systems.

Let us consider the case of the symphony orchestra first. In an orchestra, the composer never specifies notes for individual musicians. As such, the musicians within the same group \textit{always} play the same note by design. Therefore, it is reasonable to assume that a disjoint organization is an appropriate choice for the orchestra.

But what about the brain? Do different neurons always play the same tune within their group? Over the past three decades, considerable evidence has accrued suggesting that the answer is no. Let us go over some examples below.

\subsection{The stomatogastric nervous system in crabs and lobsters}

The stomatogastric nervous system has been used as a primary system for analyzing how circuit dynamics arise from the properties of its neurons and their connections \cite{marder2007understanding}. The system is composed of the stomatogastric ganglion (STG), among other components. A ganglion refers to a collection of neurons. The STG consists of $\sim\!30$ neurons and contains the motor neurons responsible for generating a variety of rhythmic movements in the crab stomach, such as a slow \textit{gastric mill rhythm} or a faster \textit{pyloric rhythm} \cite{marder2007understanding}.

The simplicity of STG circuits has allowed neuroscientists to gain extremely valuable insights into circuit function in nervous systems. One of these insights is the importance and prevalence of \textit{neuromodulation} in neuronal circuits \cite{marder2012neuromodulation, lee2012neuromodulation, mccormick2020neuromodulation, shine2019neuromodulatory, shine2023neuromodulatory}. Neuromodulation is a physiological process by which neural activity is regulated by way of controlling the concentration levels of different types of neurotransmitters, often affecting various neural functions and potentially altering mood, cognition, and behavior. A consequence of neuromodulation is dynamic \textit{reconfiguration} of the effective connectivity patterns in neural circuits.

\begin{figure}[ht!]
    \centering
    \includegraphics[width=\linewidth]{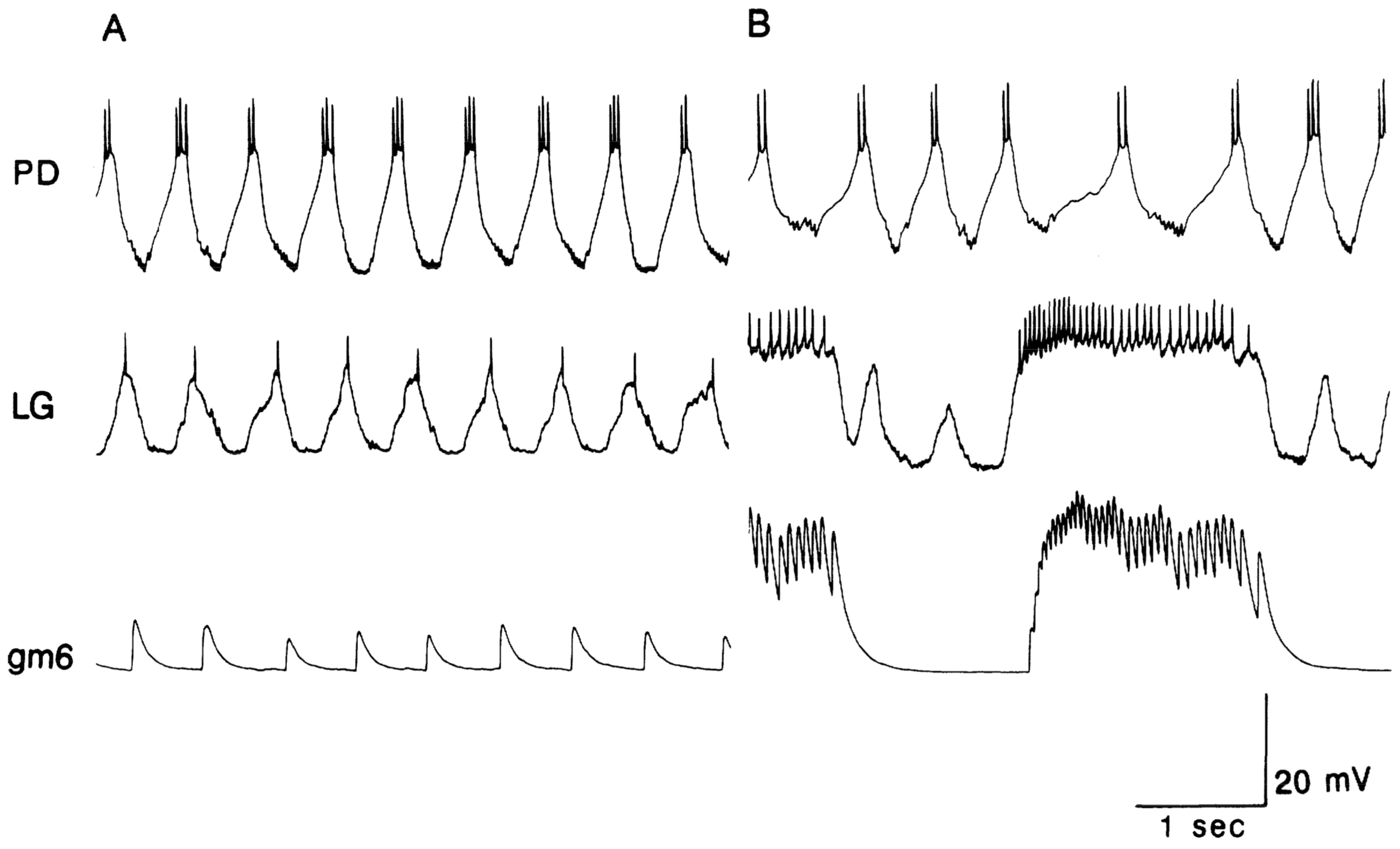}
    \caption[Circuit switching in crab STG]{One of the first experimental demonstrations of circuit switching in crab stomatogastric ganglion (STG). Neurons switch from being part of the pyloric or gastric circuits. \textbf{(a)} In the absence of rhythmic gastric mill activity, the LG neuron fired in time with the pyloric rhythm (illustrated by the PD neuron). \textbf{(b)} Recordings from the same cells as shown in part A. The spontaneous sequence of rhythmic gastric activity is associated with long bursts of high-frequency action potentials in the LG neuron. Figure adapted from \textcite{weimann1991neurons}, 1991.}
    \label{fig:stg-reconfig}
\end{figure}

One of the earliest experimental demonstrations of this idea came out in 1991, where \textcite{weimann1991neurons} reported dynamic circuit switching between certain neurons in crab STG (\cref{fig:stg-reconfig}). In \cref{fig:stg-reconfig}a, the LG neuron is firing in time with the PD neuron, while the gm6 muscle fiber is silent, an indication of the \textit{pyloric rhythm} being active. In \cref{fig:stg-reconfig}b, the \textit{gastric rhythm} becomes active as evidenced by the activity of gm6, recruiting the LG neuron along the way. This experiment shows how LG neurons can switch back and forth between two different types of rhythms, thus effectively coupling themselves to different pattern-generating circuits.

\subsection{Dynamic circuit reconfiguration implies overlapping communities}

The study by \textcite{weimann1991neurons} provides proof of existence for the idea that functional connectivity patterns between neurons can reconfigure in short time scales. Something like this would never happen in an orchestra: imagine a situation in which a violinist suddenly decides to ignore the notes written for the violin group, and starts playing in tune with the oboes. By design, an orchestra does not involve such switching, but this happens to be a common theme governing brain dynamics across spatiotemporal scales.

Circuit switching is not limited to STG. There is substantial evidence that this organizational principle is present across spatiotemporal scales, from circuits of $\sim\!30$ neurons, all the way up to brain regions comprising millions of neurons. See \textcite{shine2018principles} for a review.

\begin{figure}[ht!]
    \centering
    \includegraphics[width=0.7\linewidth]{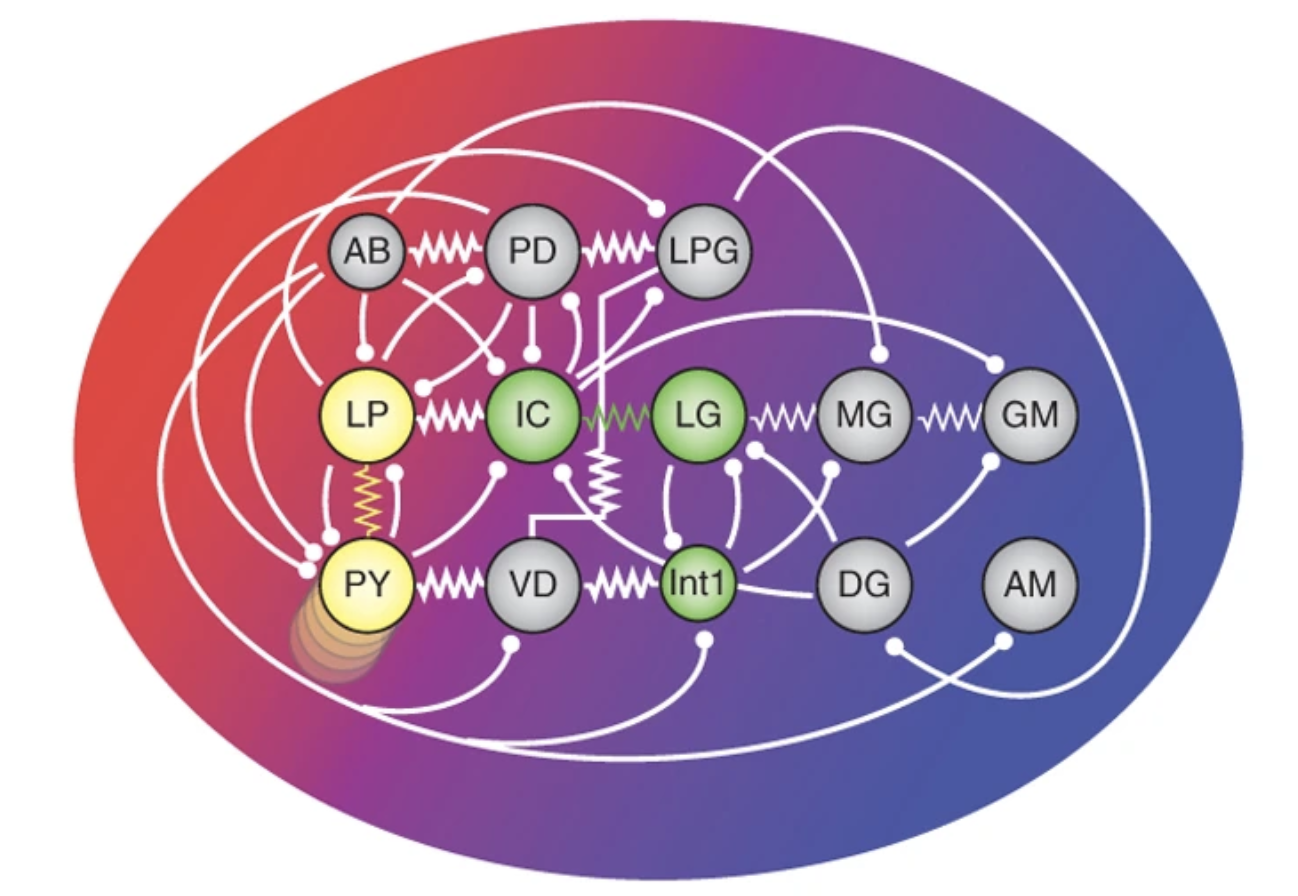}
    \caption[Crab STG as an overlapping circuit]{Connectivity diagram of the crab STG based on electrophysiological recordings. Red and blue background shading indicates neurons that are primarily part of the \textit{pyloric} and \textit{gastric} circuits, respectively. Purple shading indicates that some neurons switch between firing in pyloric and gastric time and that there is no fixed boundary between the pyloric and gastric circuits. Figure adapted from \textcite{bargmann2013connectome}, 2013.}
    \label{fig:stg-overlap}
\end{figure}

A consequence of such dynamic reconfiguration is that brain regions cannot be clustered into sharply delineated communities. A node (neuron, brain region) that is functionally connected (correlated) with two distinct groups, will have its links distributed within both groups. As a result, it becomes difficult to choose a single community for such a node. Overlapping communities offer a natural solution to this conundrum. \Cref{fig:stg-overlap} showcases the crab STG as an overlapping circuit of neurons.

In Chapter 5, I ask whether the mouse cortical networks are also overlapping, and find confirming evidence for this. Quantifying the overlap extent suggests it is substantial and non-negligible.

\section{The mystery of spontaneous brain activity}

Spontaneous activity refers to ongoing brain activity in the absence of any explicit, external stimulation. What is this intrinsic activity? In the symphony orchestra example from above, it is the conductor who elicits the music playing. As a result, it is easy to interpret the ongoing activity in an orchestra. In a simple nervous system such as the STG, the purpose of the generated rhythms is also relatively unambiguous: they are used for food movement and digestion. But what about the spontaneous activity in the mammalian brain? What is the tune being played there?

Studying spontaneous brain activity is interesting from both theoretical and empirical perspectives. In a short but thought-provoking article from 2006, \textcite{raichle2006dark} approaches this question from the perspective of energy consumption. He observes that the adult human brain represents about $2\%$ of the body weight, yet accounts for about $20\%$ of the body’s total energy consumption \cite{raichle2002appraising}---10 times that predicted by its weight alone. \textcite{raichle2006dark} further notes that momentary task demands from the environment may require as little as $0.5\text{ to }1.0\%$ of the total energy budget \cite{raichle2006brain}. This implies that intrinsic activity may be far more significant than evoked activity in terms of overall brain function. In sum, the brain consumes a substantial proportion of the body's total energy and uses most of its energy for functions unaccounted for. For these reasons, \textcite{raichle2006dark} refers to spontaneous brain activity as \sayit{The Brain’s Dark Energy}.

Another conceptually interesting question is whether spontaneous brain activity is structured or random. Imagine the brain has $N$ independent degrees of freedom that exhibit ongoing, spontaneous activity. The very nature of these degrees of freedom is not essential to the argument I will be making here. They could be neurons, voxels, or brain regions. For simplicity, let's assume the brain has N neurons, and at a given point in time, each neuron is either active or inactive. Let us now collect all possible brain states and put them in a set called $\mathcal{H}$ (this alludes to the notion of a \textit{Hilbert} space, but is not central to our argument here). In this setup, the space of all brain states, $\mathcal{H}$, contains $2^N$ independent states. Each brain state at a given time point is a vector comprised of zeros and ones, indicating whether or not a neuron was active at that time.

\begin{figure}[t!]
    \centering
    \includegraphics[width=1.0\linewidth]{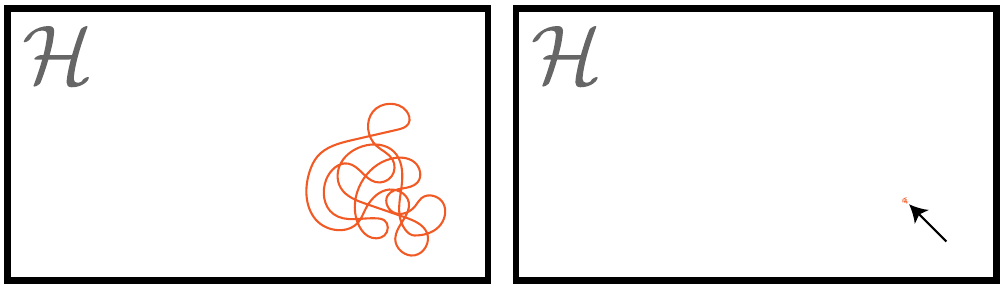}
    \caption[The massive space of possible brain states]{The massive space of possible brain states, denoted as $\mathcal{H}$ here. \textbf{(Left)} Experimental evidence has shown that spontaneous brain activity, represented as orange trajectory, is highly structured and it dynamically (re)visits only certain corners or subspaces of $\mathcal{H}$. \textbf{(Right)} A more realistic illustration indicates that the dimensionality of biologically plausible brain states is vanishingly small compared to the number of all possible states, which can be considered infinitely large (see text for details).}
    \label{fig:hilbert-space}
\end{figure}

The human brain contains $N \approx 100,\!000,\!000,\!000$ neurons. Therefore, for all practical purposes, the cardinality of $\mathcal{H}$ tends to infinity. In other words, $\vert \mathcal{H} \vert  \approx 2^{100,000,000,000} \rightarrow \infty$. Given such a massive space of possibilities, it is interesting to ask whether spontaneous brain dynamics explores all corners of $\mathcal{H}$ with equal probability. In other words, is spontaneous activity simply composed of infinite-dimensional random fluctuations? Alternatively, the dynamics could be restricted to specific corners of $\mathcal{H}$, exhibiting a non-trivial structure (\cref{fig:hilbert-space}). Data across a variety of model organisms and experimental techniques suggest that the latter is the case. I will review some of the evidence below.

From an empirical perspective, it is known that spontaneous brain activity is not random and is spatiotemporally structured. This was first reported by \textcite{biswal1995functional} in 1995, who employed a task involving right-hand movement and observed fMRI activations not only in the expected motor regions in the left hemisphere but also in the contralateral hemisphere, indicating a correlated activity between the two hemispheres.

\begin{figure}[ht!]
    \centering
    \includegraphics[width=\linewidth]{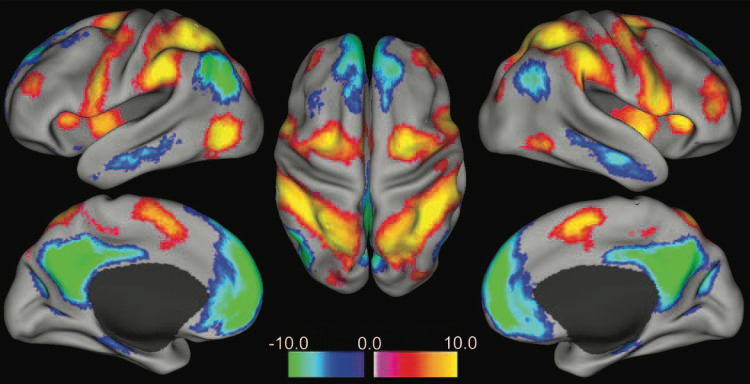}
    \caption[Intrinsically deﬁned anti-correlated networks in the human brain]{Intrinsically deﬁned anti-correlated networks in the human brain. Positive nodes (warm color) are signiﬁcantly correlated with brain regions that are activated during task demands. Negative nodes (cool color) are signiﬁcantly correlated with regions that are deactivated during task demands. From \textcite{fox2005human}, 2005.}
    \label{fig:fox05}
\end{figure}

Furthermore, \textcite{fox2005human} found that the intrinsic functional connectivity structure of the human brain can be understood in terms of two anti-correlated networks of brain regions. One comprises regions that typically show task-related activations and the other, regions that exhibit task-related deactivations (\cref{fig:fox05}). This nuanced view further emphasized the non-random nature of spontaneous brain activity and its structured organization.

Finally, in a recent paper, \textcite{bolt2022parsimonious} employed a complex-valued extension of PCA to show that spontaneous fMRI fluctuations in the human brain can be described parsimoniously using only three spatiotemporal patterns \cite{bolt2022parsimonious, gonzalez2022traveling}. Collectively, these results establish the low-dimensional and highly structured nature of spontaneous brain fluctuations.

\subsection{Spontaneous and evoked responses}
An intriguing perspective emerges if we consider evoked brain responses alongside spontaneous activity and their potential interactions. A wealth of evidence across experimental techniques suggests that evoked neural responses can show pronounced variability even when the presented stimulus is identical across trials. Neural variability, driven by stochastic mechanisms and complex dynamics within the central nervous system, is inherent in sensory processing \cite{shadlen1998variable, tolhurst1983statistical, werner1963variability, arieli1996dynamics, tomko1974neuronal} and motor actions \cite{faisal2008noise}. Such variability is evident not just at the neuronal level, but also across larger brain regions as reported in fMRI studies \cite{zhang2022brain, he2013spontaneous}.

Spontaneous brain activity may contribute to trial-to-trial variability. In an influential study, \textcite{fox2006coherent} explored this hypothesis using a basic motor or visual stimulus task. They found that subtracting spontaneous activity from the contralateral hemisphere enhanced the signal-to-noise ratio in the evoked responses. At first sight, this result implies that the relationship between spontaneous and evoked activity can be explained as a simple superposition of independent mechanisms at play. Over the past two decades, intensive research has aimed to elucidate this relationship. However, emerging evidence indicates that the interaction between spontaneous and evoked responses is more intricate than a linear superposition.

\begin{figure}[t!]
    \centering
    \includegraphics[width=1.0\linewidth]{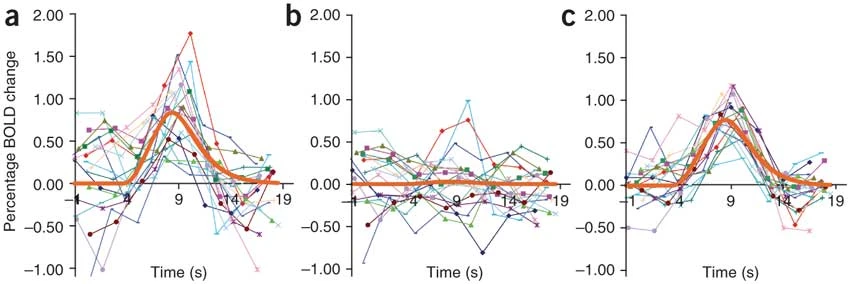}
    \caption[Variability in evoked brain responses explained by spontaneous activity]{Variability in evoked brain responses explained by spontaneous activity. \textbf{(a)} Raw responses in the left somatomotor cortex during right-handed button presses. \textbf{(b)} The corresponding activity in the right hemisphere for each button press. \textbf{(c)} Task-related responses in the left somatomotor cortex after subtraction of spontaneous fluctuations measured in the right somatomotor cortex. From \textcite{fox2006coherent}, 2006.}
    \label{fig:fox06}
\end{figure}

In a now-classic paper published in 2010, \textcite{churchland2010stimulus} combined data from an impressively diverse set of experiments across visual and motor tasks to study neural variability before and after stimulus presentation. Results were compiled both from sensory areas during a stimulus presentation, including visual areas V1, V4, and area MT; and from motor-related areas during the execution of a stereotyped task, including the lateral intraparietal area, dorsal premotor cortex, and orbitofrontal cortex. A common computational principle emerged in each of these cortical regions: stimulus or motor onset reduces, or \sayit{quenches}, the level of neural variability. Through careful analyses, the authors showed that such neural variability quenching occurs in a correlated way across large populations of neurons; therefore, it likely is a network-level phenomenon. This work has been explored in commentaries by \textcite{nogueira2018neuronal} and \textcite{fairhall2019whither}, summarizing the results, and discussing the paper's long-lasting impact.

Over the years, more evidence has accumulated, establishing this \say{decrease of variability with respect to the spontaneous level} as one of the most robust hallmarks of task or stimulus-driven brain dynamics, also observed in human fMRI experiments \cite{he2013spontaneous, ponce2015task, huang2015there, zhang2022brain}.

The functional significance of such interactions between spontaneous fluctuations and evoked responses may be profound. Spontaneous activity within the visual cortex not only reflects the orientation preference maps seen with visual stimulation \cite{kenet2003spontaneously, ringach2003states}, but also evolves in complexity throughout development \cite{berkes2011spontaneous}. Moreover, it can recapitulate aspects of recently encountered visual stimuli \cite{carrillo2015endogenous}. Such activity may be indicative of the brain's gradual internalization of the statistics of sensory inputs. It has been postulated that spontaneous activity might represent a stochastic sampling from a distribution of experiences ingrained within brain networks \cite{hoyer02interpreting, fiser2010statistically, berkes2011spontaneous, orban2016neural, pitkow2016probability, echeveste2020cortical, shrinivasan2023taking, lange2023bayesian}. These insights provide additional support to the significance of spontaneous brain activity, and the need to study and characterize its structure and dynamics.

Finally, clinical research has underscored the potential of spontaneous brain activity as a diagnostic and prognostic tool for various brain diseases \cite{fox2010clinical, broyd2009default, lau2019resting, northoff2016abnormalities, watanabe2017brain, perovnik2023functional, sorg2007selective, zhu2021altered}. For instance, in Alzheimer's disease, significant alterations in spontaneous brain activity have been observed even before the apparent cognitive decline, suggesting its utility in early diagnosis \cite{sorg2007selective, zhu2021altered, perovnik2023functional}.

In conclusion, studying spontaneous brain activity has emerged as a cornerstone of understanding the inner workings of the mammalian brain \cite{fox2007spontaneous, Power2014StudyingBO, northoff2016self, Northoff2018TheSB, liu2022decoding, Grandjean2020CommonFN, grandjean2023consensus}. Far from being mere \say{background noise} \cite{uddin2020bring}, this activity holds implications for understanding how intrinsic brain states interact with external perturbations to sculpt our perceptions \cite{raichle2006dark, cole2014intrinsic, orban2016neural, Northoff2018TheSB}.

\section{Cortical network organization in mice}\label{sec:cafmri}

The purpose of this section is to discuss the motivations behind my work on estimating mouse cortical network structure, which is the content of Chapter 5. I will start with some background information building up toward study motivations, and end with a concise overview of results reported in Chapter 5.

\subsection{The study of large-scale brain network organization}

Since the mid-1990s, the discovery that spontaneous brain fluctuations are highly structured \cite{biswal1995functional}, has inspired many subsequent studies on the intrinsic organization of brain activity during resting state. These efforts have established an understanding of the organization of large-scale brain networks as a central problem in neuroscience. These types of studies have been mostly in humans, using fMRI \cite{Raichle2001ADM, fox2005human, damoiseaux2006consistent, Yeo2011TheOO, Power2011FunctionalNO, smith2013resting, Faskowitz2020EdgecentricFN}. In recent years, resting-state fMRI has gained a lot of traction in animal models \cite{pagani2023mapping, mandino2020animal}, particularly in rodents \cite{hutchison2010functional, white2011imaging, jonckers2011functional, becerra2011robust, nasiriavanaki2014high, stafford2014large, Sforazzini2014DistributedBA, Zerbi2015MappingTM, sierakowiak2015default, Liska2015FunctionalCH, Gozzi2016LargescaleFC, Coletta2020NetworkSO, Grandjean2020CommonFN, bergmann2020individual, Markicevic2021EmergingIM, mandino2022triple, Rocchi2022IncreasedFC, GutierrezBarragan2019InfraslowSF, GutierrezBarragan2022UniqueSF, ragone2023modular, mandino2023where, grandjean2023consensus}.

Within the context of studying the large-scale organization of the brain, animal models are particularly interesting because they allow invasive measurements and perturbations that are impossible in humans. For example, my collaborators \textcite{Lake2020SimultaneousCF} developed an imaging setup that combines brain-wide fMRI with cortex-wide calcium (\CA\!) imaging in mice. This concurrent imaging technique yields valuable data that contribute to our understanding of the cellular foundations of the BOLD signal. Furthermore, the fMRI component serves to connect mouse research with human studies, offering promising avenues for clinical applications.

In Chapter 5, I use this dataset to interrogate fMRI--derived network structure and test whether its properties are reproduced by \CA\!\!--derived network structure.

\subsection{Resting-state fMRI networks: disjoint or overlapping?}

Most research in resting-state network organization in humans and animals assumes \textit{disjoint} functional organization, such that a brain region belongs to one and only one network (\cref{fig:disj-ovp}). However, many complex networks are inherently overlapping, including biological, technological, and social ones \cite{palla2005uncovering, barabasi2011network, menche2015uncovering}. In the brain, regions that are multi-functional are likely to participate in multiple networks \cite{cole2013multi, fedorenko2013broad, yeo2015functional}. This leaves the question of whether a disjoint approach would provide an accurate representation of the underlying organization.

Work in humans sought to address this question using a variety of methods and approaches. \textcite{Yeo2014EstimatesOS} (2014) used latent Dirichlet allocation, \textcite{Najafi2016OverlappingCR} (2016) used a mixed-membership stochastic blockmodel, and \textcite{Faskowitz2020EdgecentricFN} (2020) edge-centric functional connectivity to determine overlap in human resting-state fMRI networks. Despite the diverse methodology used, all these studies converged to the same answer: that human brain networks are substantially overlapping. However, it is still unclear whether the putative overlapping organization is driven, at least in part, by the nature of BOLD signals.

Interpreting BOLD is difficult because it is a noisy measure. This includes physiological noise such as respiratory and cardiac effects, as well as noise due to technological reasons such as issues with the scanners. For instance, magnetic field inhomogeneities could cause spatial distortions and signal loss, especially near air-tissue boundaries. This can particularly affect regions near sinuses and ear canals, leading to artifacts that complicate data interpretation.

Another example is thermal noise inherent to the MRI system, which is influenced by the scanner's electronics and environment. This random fluctuation in the MR signal can obscure the subtle BOLD signal changes associated with brain activity, making it challenging to distinguish between true neural events and technical artifacts. Subject-related effects such as head motion add to the difficulty of data preprocessing.

\begin{figure}[t!]
    \centering
    \includegraphics[width=\linewidth]{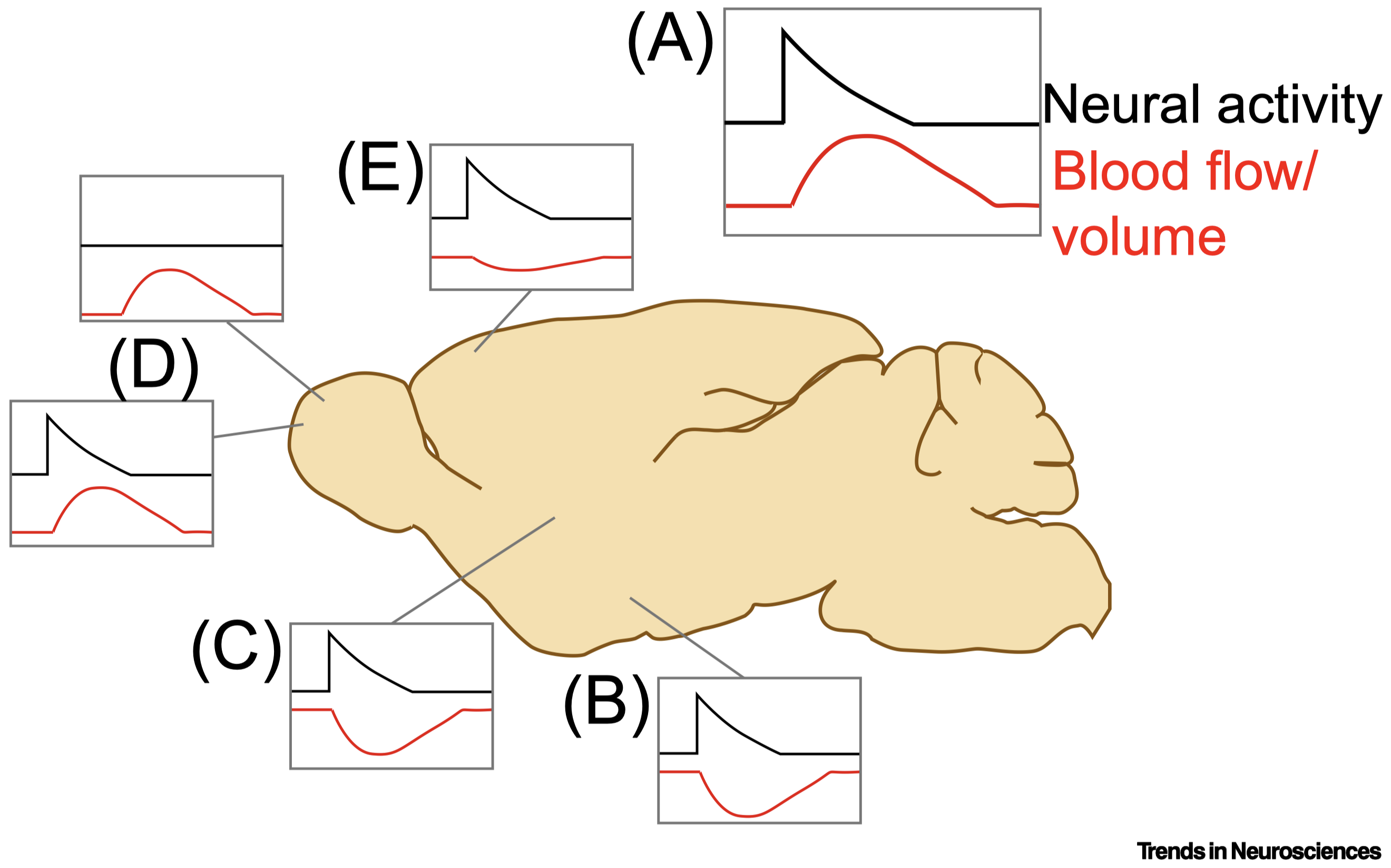}
    \caption[Neurovascular coupling is variable across brain regions]{Neurovascular coupling is variable across brain regions. The increase in blood flow and oxygenation elicited by neural activity provides the contrast mechanism for fMRI. However, as shown here, the relationship between neural activity and blood flow can be irregular or even inverted in some brain areas. This variability in neurovascular coupling across regions complicates the interpretation of fMRI-BOLD signals as a proxy for neural activity. From \textcite{drew2022neurovascular}, 2022.}
    \label{fig:motive-unkown}
\end{figure}

Finally, BOLD is difficult to interpret due to a lack of neuronal specificity \cite{logothetis2001neurophysiological, logothetis2004interpreting, logothetis2008we, hillman2014coupling}, and confounding vascular effects \cite{drew2022neurovascular, drew2020ultra, drew2019vascular}. \Cref{fig:motive-unkown} nicely summarizes this difficulty.

Given these challenges with fMRI-BOLD, the simultaneous \CA\!/ fMRI dataset provides an excellent opportunity to test the extent to which the overlap observed in human studies is driven by the nature of BOLD contrast. Alternatively, it is possible that network overlap is a generalizable property of brain organization, in which case we should observe substantial overlap in network structures inferred from both \CA and fMRI signals.

\subsection{Summary of results presented in Chapter 5}

The results reported in Chapter 5 indicate that both fMRI--derived and \CA\!\!--derived networks are substantially overlapping. I quantify these using a variety of methods and metrics, and find converging evidence in support of the overlapping hypothesis. For both imaging modalities, around $50\%$ of the entire cortex is overlapping, meaning that nearly half of the brain regions belong to multiple networks (see \cref{fig:disj-ovp}). 

Comparative analyses reveal that certain aspects of fMRI--derived network structure are reproduced by \CA\!\!--derived network structure. For instance, several communities exhibit similar spatial organization across modalities, and the principal functional gradient is nearly identical for fMRI-BOLD and \CA signals. Despite the similarities, important differences also emerged in certain communities and centrality measures such as functional connectivity strength and diversity (degree and entropy, respectively). These robust differences demonstrate that \CA and fMRI-BOLD signals capture complementary features of brain organization.

To wrap up Chapter 1, let's consolidate the key details regarding the data modalities and model organisms used in this Dissertation, which I will summarize in the next section.

\section{Data modalities \& model organisms}

\quote{When you can measure what you are speaking about, and express it in numbers, you know something about it; but when you cannot measure it, when you cannot express it in numbers, your knowledge is of a meagre and unsatisfactory kind.}{Lord Kelvin, \textit{Electrical Units of Measurement}, lecture given in 1883}\\

\begin{figure}[ht!]
    \centering
    \includegraphics[width=1.0\linewidth]{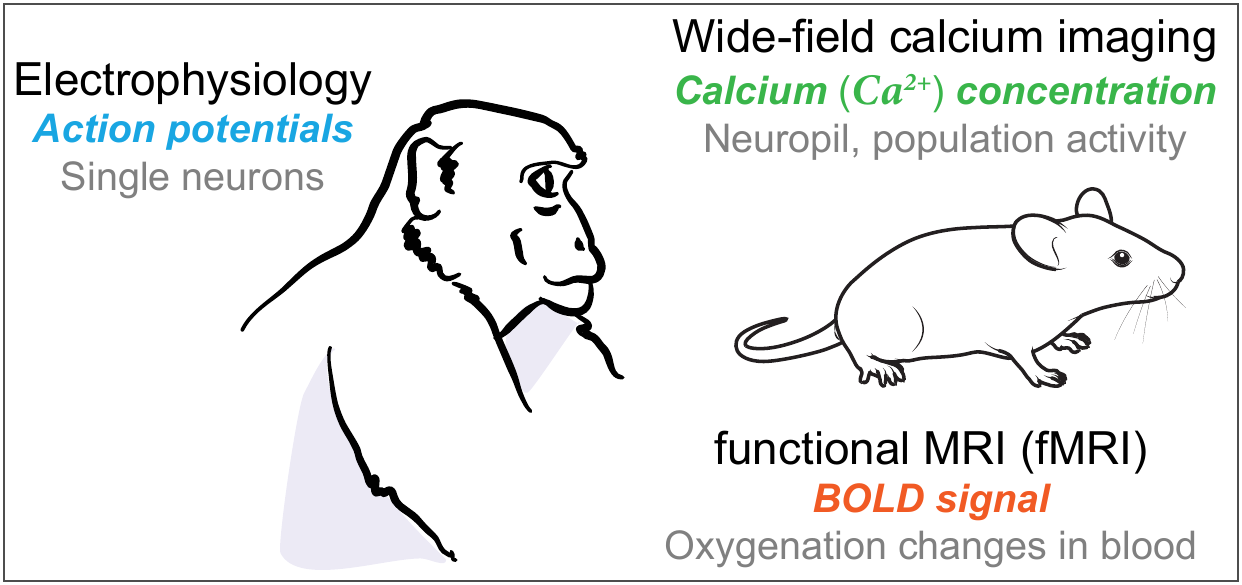}
    \caption[Model organisms and data modalities used in this Dissertation]{Model organisms and data modalities used in this Dissertation. \textbf{(Left)} Electrophysiology in macaque monkey. Electrophysiology measures action potentials, or spikes, from individual neurons \cite{Kandel2021}. \textbf{(Right)} Simultaneous fMRI and calcium imaging in mice \cite{Lake2020SimultaneousCF}. Wide-field calcium imaging detects the concentration of \CA ions in brain tissue, a proxy for neural activity \cite{grienberger2012imaging}. The main source of the variance detected by calcium imaging comes from neuropil in shallow cortical layers \cite{waters2020sources}. Finally, fMRI measures blood-oxygen-level-dependent signal or BOLD \cite{logothetis2001neurophysiological, logothetis2004interpreting, logothetis2008we, hillman2014coupling}. The BOLD signal stems from a variety of neuronal \cite{arthurs2002well, drew2019vascular, moon2021contribution, howarth2021more, matsui2019neuronal} and non-neuronal sources, including astrocytes \cite{takata2018optogenetic} and neurovascular effects \cite{hillman2014coupling, o2016neural, drew2020ultra, drew2022neurovascular}.}
    \label{fig:model-organisms}
\end{figure}

To understand brain function, we must turn it into numbers. There are many different ways of measuring brain activity. This Dissertation covers three of the most commonly used approaches. Namely, electrophysiology measures action potentials or spike trains from individual neurons; functional magnetic resonance imaging (fMRI) measures blood-oxygen-level-dependent signal (BOLD), which is related to underlying neuronal activity through a process termed neurovascular coupling; and wide-field calcium imaging (\CA) measures fluctuations in the concentration of \CA ions in brain tissues, which correlates with neuronal activity (\cref{fig:model-organisms}).\\[2mm]

\renewcommand{\thechapter}{2}

\chapter{Philosophical and historical background}

\section{The relationship between perception and reality}
Does the human mind have access to the world \textit{out there}, in its entirety? This question has occupied philosophers for millennia. Like any other interesting topic, the Greeks had something important to contribute here as well. In our case, it is Plato and his \sayit{Allegory of the Cave}.

\subsection{Plato and the Allegory of the Cave}
Consider some prisoners who have been held captive in a cave since birth. They face a stone wall at the end of the cave, with another wall separating the prisoners from the outside world (\cref{fig:allegory-cave}). As a result, all they have access to are some reflections from the objects in the real world. These shadows constitute their entire reality. How could these prisoners know about the \textit{true} nature of the objects that cause those shadows if they never had a chance to leave the cave? They will still develop a limited understanding of the external world based on the preserved \textit{relationships} between objects visible in the shadows. But this understanding will never be complete. Over 2400 years ago, Plato told this story in his book \textit{Republic} to argue that life is like being chained in a cave, forced to watch shadows in a stone wall: we only observe incomplete portions of the world.

\begin{figure}[t!]
    \centering
    \includegraphics[width=\linewidth]{figs/intro/allegory-cave}
    \caption[Plato's Allegory of the Cave]{Plato's ``Allegory of the Cave" explores the idea that human perception may not unveil the full scope of external reality. Image from \href{https://en.wikipedia.org/wiki/Allegory_of_the_cave}{Wikipedia}.}
    \label{fig:allegory-cave}
\end{figure}

\subsection{Ibn Al-Haytham and subjectivity of perception}\label{sec:alhazen}
Fast-forwarding some 1400 years, we encounter Ibn Al-Haytham, also known as Alhazen, who conducted groundbreaking research on vision and arrived at conclusions echoing those of Plato. In his landmark work, \textit{Book of Optics (Kitab al-Manazir)} published in 1011--1021 AD \cite{alhazen}, Alhazen was the first to argue that vision occurs in the brain rather than the eyes. He was also the first to point out that personal experiences affect what people see and how they see, and that perception is ultimately a subjective experience. In other words, each observer's experience-dependent prior knowledge about the world shapes their perceptions, even when viewing the same fixed object. These conjectures were later confirmed by modern psychologists \cite{baker2012oxford}.

In another contribution, Alhazen challenged the prevailing \textit{extramission} theories of vision---the idea that visual perception involves eye beams emitted by the eyes, a belief held by ancient Greek philosophers like Euclid and Ptolemy. Instead, Alhazen (and later, Avicenna \cite{baker2012oxford, adamson2016philosophy}) argued that vision is not solely the result of light emanating from the eyes, but rather, it involves something representative of the external objects that enter the eyes. Thus, the \textit{intromission} theory of vision was born. Thanks to modern physics and physiology, we now know centuries later that the things representative of the object are photons that are detected by photoreceptors in the retina.

Some scholars consider Alhazen to be the \say{first scientist} who practiced what we now call the modern scientific method \cite{steffens2007ibn}. Here is a quote that best encapsulates Alhazen's approach to doing science:\\[-3mm]

\quote{The seeker after the truth is not one who studies the writings of the ancients and, following his[/her] natural disposition, puts his trust in them, but rather the one who suspends his faith in them and questions what he gathers from them, the one who submits to argument and demonstration, and not to the sayings of a human being whose nature is fraught with all kinds of imperfection and deficiency. Thus the duty of the man who investigates the writings of scientists, if learning the truth is his goal, is to make himself an enemy of all that he reads, and, applying his mind to the core and margins of its content, attack it from every side. He should also suspect himself as he performs his critical examination of it, so that he may avoid falling into either prejudice or leniency.}{Ibn al-Haytham (\textit{aka} Alhazen), circa 1000 AD}\\[3mm]

Alhazen conceived of ideas that were hundreds of years ahead of his time. Much later, the Latin translation of his books dominated Western optical thought \cite{lindberg1967alhazen} and inspired thinkers like Roger Bacon, Johannes Kepler \cite{adamson2016philosophy}, and possibly Isaac Newton and John Locke, among others.

\subsection{John Locke and how perceptions are changed by judgment}

John Locke, a prominent philosopher of the Enlightenment era, made profound contributions to the understanding of human perception. His monumental book, \textit{An Essay Concerning Human Understanding} (1689, \cite{locke1690an}), laid the foundations of modern empiricism and is concerned with the limits of human understanding in a variety of topics \cite{sep22locke}. Of particular interest to us is Chapter IX of the book, entitled \textit{Of Perception}. Let us consider the following excerpt:\\

\quote{...the ideas we receive by sensation are often, in grown people, altered by the judgment, without our taking notice of it.}{\textcite{locke1690an}, Chapter IX, part 8}\\

What does he mean by \textit{ideas} and \textit{judgment}? To elaborate, Locke provides a concrete example of seeing a \say{round globe of any uniform colour} and uses it to make a case for how our previous experience might alter our interpretation of what we see. Locke argues that when we look at the globe, the initial impression \say{imprinted on our mind} is of a flat, colored circle. However, he suggests that our eventual \textit{judgment}, influenced by past experiences, can deviate from what is apparent from these sensory first impressions. For instance, we may associate the appearance of the circle (sensation) with the idea of a three-dimensional, convex shape (perception) because we have seen many more three-dimensional globes rather than flat circles. Therefore, it is more likely that the thing causing the sensory \textit{ideas} is actually a round globe rather than a flat circle, even though both circle and globe would result in the same sensory impressions.

Locke's example touches upon what we now call the \textit{degeneracy} problem in vision. The sensory data sent to the brain are limited in many ways, as they are two-dimensional projections of a three-dimensional world. These limitations, the argument goes, make it difficult for the brain to decide between competing explanations of a visual scene based on raw sensory data alone. Here, the knowledge we have accumulated by living in the world helps us by shaping what we expect to see. Our expectations, in turn, break the degeneracy by assigning larger likelihoods to interpretations that are more consistent with our past experiences. Thus, both what we see (sensory data) and what we expect to see (prior knowledge) contribute to what we eventually perceive (scene interpretation).

Locke's notion of \textit{judgment} almost sounds like the concept of posterior probability distribution in a Bayesian sense. Similarly, his usage of the word \textit{idea} resembles the notion of a likelihood function. Therefore, in \say{grown people}, as Locke puts it, the eventual perception of a visual phenomenon (= judgment), depends not only on the sensory evidence or likelihood (= ideas), but also on the observer's prior experiences (for which he uses the phrase being \say{accustomed}). In this sense, Locke's theory of perception is conceptually very similar to posterior inference in Bayesian statistics. But at the time, these terminologies did not exist, because Thomas Bayes was two years old in 1704 when Locke died.

Our brains are so good at vision that we often forget how remarkable it is that we can see at all. Even more remarkable is the fast time scale at which this operation occurs. Later in the chapter, Locke also contemplates this instantaneous aspect of perception:\\

\quote{How, by habit, ideas of sensation are \textit{unconsciously} changed into ideas of judgment. Nor need we wonder that this is done with \textit{so little notice}, if we consider how quick the actions of the mind are performed.}{\textcite{locke1690an}, Chapter IX, part 10 (\textit{italics} mine)}\\

\noindent Finally, this seamless and habitual nature of perception leads Locke to conclude that:\\

\quote{And therefore it is not so strange, that our mind should often change the idea of its sensation into that of its judgment, and make one serve only to excite the other, without our taking notice of it.}{\textcite{locke1690an}, Chapter IX, part 10}

\subsection{Hermann von Helmholtz and perception as unconscious inference}

Helmholtz is renowned in the realm of physics for introducing the concept of free energy in 1882, and a free energy equation that he did not write but is named after him: $F = U - TS$. However, this contribution was not his first significant scientific achievement. In 1847, Helmholtz published a treatise on the conservation of energy, marking a crucial milestone in his career when he was 26 years old. His fascination with energy conservation stemmed from an unconventional source: the study of muscle metabolism. Helmholtz sought to demonstrate that no energy is lost during muscle movement, an idea that would later contribute to a rigorous formulation of the concept of energy and inspire Rudolf Clausius's first law of thermodynamics in 1850.

What may be less widely known among physicists is that Hermann von Helmholtz was not only a dominant force in 19th-century physics but also one of the most influential physiologists and vision researchers of his era, and some would argue, of all time \cite{warren1984helmholtz}.

Helmholtz carefully examined the optical characteristics of the human eye and found many serious flaws \cite{warren1984helmholtz}, including chromatic aberration, spherical aberration, and turbidity caused by colloidal particles in the intraocular fluids, to name a few (see \textcite{navarro2009optical} for a comprehensive review). He concluded that the shape of the human eye and the centering of the imaging surfaces (cornea and lens) are even worse than a cheap camera \cite{ditzinger2021illusions}. Because of these observations, Helmholtz once remarked: \say{I would throw out an optician who would bring me an instrument like the human eye.}

In addition to his optical investigations, Helmholtz explored acoustical defects in the human ear and found that nonlinear distortions could introduce spectral components not present in the stimulus \cite{helmholtz2009sensations}. The design flaws in human sensory hardware led Helmholtz to ponder: how can perception be accurate when the stimuli are so badly distorted? These thoughts, along with his experiments on optical illusions, culminated in one of his most groundbreaking discoveries: that perception must involve a process of unconscious inference.

This notion of \say{perception as unconscious inference}, published in his three-volume \textit{Treatise on Physiological Optics} (1860s, \cite{helmholtz1867handbuch}), was largely neglected or criticized at the time \cite{boring1942sensation}. However, modern research has revived the idea and has positioned unconscious inference as our current best theory of perception \cite{olshausen2014}.

Helmholtz went further to discuss the existence of \textit{invariant relations} between neural responses and environmental conditions. He was also aware of Riemann's non-Euclidean geometry (1868) and examined how a person in such a non-Euclidean world would adapt perceptually based on the world's geometrical axioms \cite{warren1984helmholtz}. His experiments with distorting spectacles demonstrated the connection between perceptual inferences and \say{real-world} geometry \cite{warren1968helmholtz}.

Almost 150 years later, conceptually similar ideas are currently being pursued in the field of \say{Geometric Deep Learning} \cite{bronstein2021geometric}, a subset of modern machine learning research that aims to derive different architectures based on geometric and group theoretical considerations. Briefly, the premise is that stimuli generated by nature stem from complex underlying geometries, and designing our deep networks based on these geometries will facilitate learning. In other words, a deep network designed based on the same geometry of its world (= training data set) will better understand (perceive) the world. Similar insights drove the development of convolutional neural networks by Fukushima in 1980 \cite{fukushima1980neocognitron}, and by LeCun and colleagues in the 90s \cite{lecun1995convolutional, lecun1998gradient}. 

Helmholtz was a textbook definition of a polymath who made a diverse set of contributions ranging from thermodynamics to neurophysiology, neuroanatomy, optics, acoustics, and visual and auditory perception: an accomplishment that was unusual even for his time \cite{cahan2018helmholtz}. His journey from physics to medicine and physiology and back to physics, where he secured Germany's top physics chair at the time \cite{cropper2001great}, remains an extraordinary achievement. Can you imagine someone in our time obtaining an MD/PhD, receiving a Nobel prize in medicine, making important strides in physics, and then winning another Nobel in theoretical physics?

\section{Modern neuroscience and the study of perception}

We have covered a lot of ground so far. We began with philosophical propositions from Plato all the way to Locke and ended with an exploration of Helmholtz's discoveries in the study of perception. These investigations collectively point in the same direction: that perception likely involves a process of inference. But can we do better than that? Can we open up the black box that is the brain, and search for concrete evidence that may confirm this hypothesis? Beyond philosophical musings, can we build a framework based on the concept of inference that would allow us to study neuronal operations underlying perception? 

Before presenting my own efforts in addressing these questions in Chapter 4, I will first provide a concise overview of modern neuroscience in the remainder of this chapter to ensure proper contextual grounding.

\subsection{The Neuron Doctrine}
\begin{figure}[ht!]
    \centering
    \includegraphics[width=0.80\linewidth]{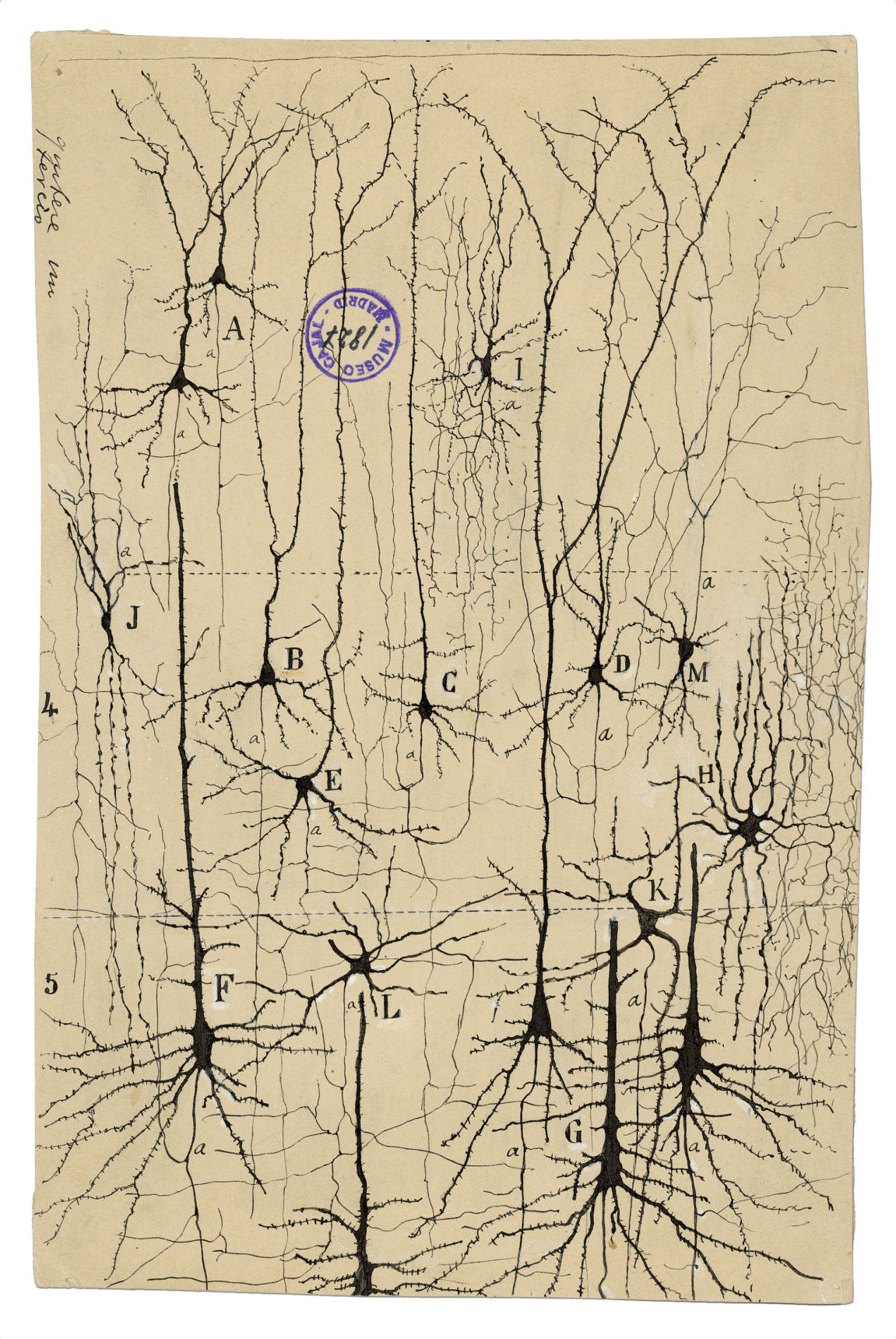}
    \caption[Cajal's drawing of neurons]{Cajal's ink-on-paper drawing of neurons shows them as separate, individual cells. Apart from being a great neuroscientist and generally being recognized as the field's founder, Cajal was also a great artist. Image courtesy of \href{https://nobelprizemuseum.se/en/art-and-science-meet-when-the-nobel-prize-museum-displays-cajals-nerve-cell-drawings/}{Nobel Prize Museum}.}
    \label{fig:cajal}
\end{figure}

\begin{figure}[ht!]
    \centering
    \includegraphics[width=0.9\linewidth]{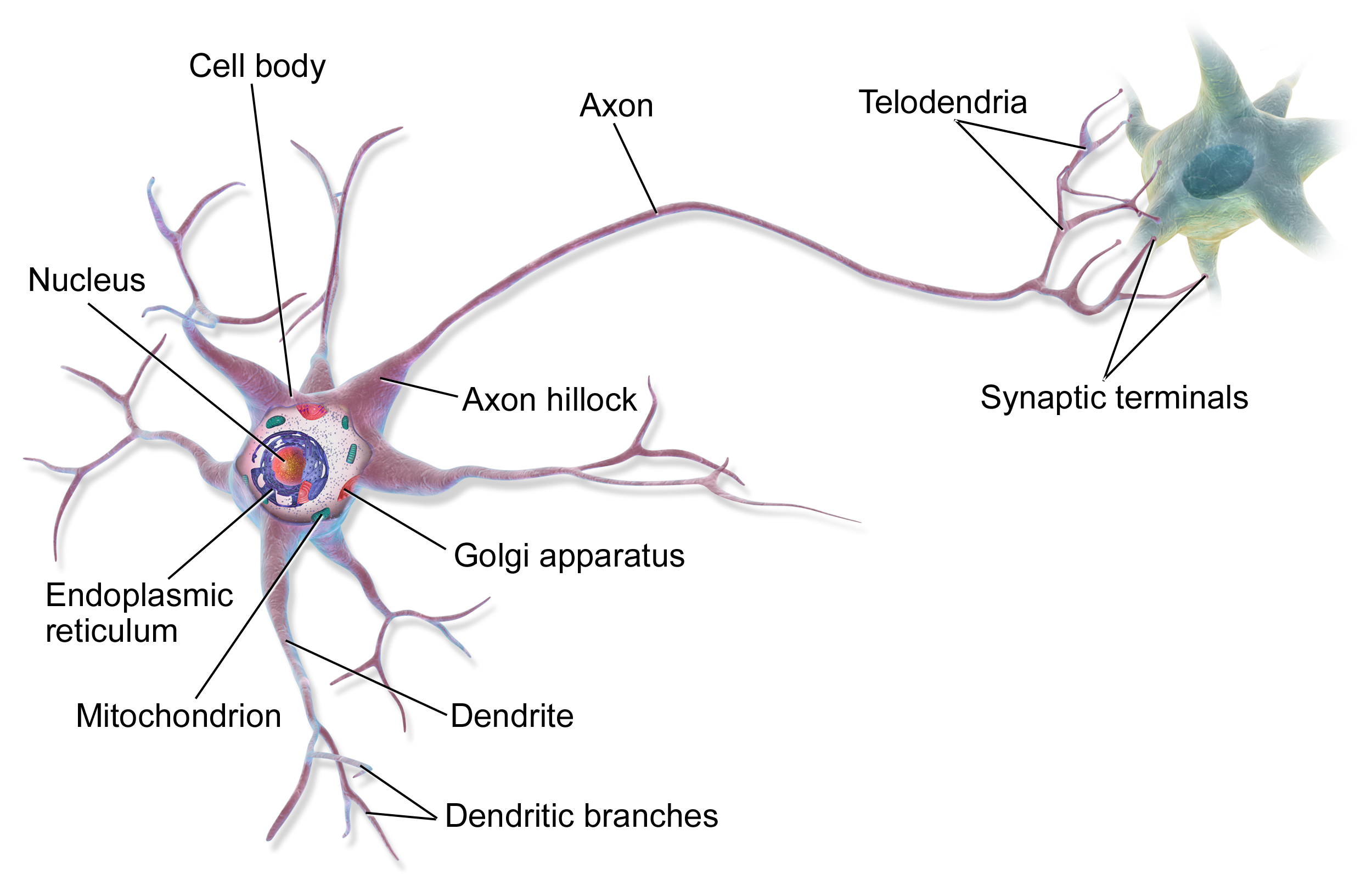}
    \caption[Anatomy of a neuron]{Anatomy of a neuron. Image from \href{https://en.wikipedia.org/wiki/Neuron}{Wikipedia}.}
    \label{fig:neuron}
\end{figure}

What is the basic building block of the nervous system? This question kept neuroscientists busy during the late 19th century and early 20th century. At the time, the prevailing view was that everything in the brain is a single continuous network, devoid of any granular structure. This idea, known as \textit{reticular} theory, was introduced by German anatomist Joseph von Gerlach in 1871 and later popularized by Italian physician Camillo Golgi. At the time, studying the brain was difficult because of technological limitations. A major breakthrough came when Golgi invented a cell staining technique that allowed neuroscientists to examine the structure of the nervous system in unprecedented detail. Spanish neuroscientist, Ram\'on y Cajal, used Golgi's method to collect structural data from various parts of the nervous system. These experiments allowed Cajal to clearly see that the nervous tissue was comprised of individual cells that were separate from one another, rather than forming a continuous network (\cref{fig:cajal}). Eventually, these observations led Cajal to introduce the \textit{Neuron Doctrine}, the idea that nervous tissue, like other tissues in the body, is made of discrete cells called \textit{neurons} (\cref{fig:neuron}).

Interestingly, Cajal was not the only scientist who came up with the idea that neurons are independent of one another. This observation was made previously by Norwegian Fridtjof Nansen in 1887, and even earlier by Englishman Edward Sch\"afer in 1877. But Cajal collected the most convincing data, and he also made sure to travel extensively, sharing his findings with the world. As a result, today we recognize the Neuron Doctrine almost synonymously with Cajal's name. See \textcite{bock2013cajal} for a historical review. In the end, both Golgi and Cajal shared the Nobel Prize in Physiology or Medicine in 1906 \say{in recognition of their work on the structure of the nervous system}, with Cajal being the clear winner of the reticular versus neuron debate.

\subsection{The Neuron Doctrine as a plan of attack}

A century ago, science was primarily a reductionist enterprise \footnote{And it still is to a good extent.}. Before Cajal's work, the basic \textit{unit} of the nervous system was disputed, which hindered progress. There was no clear path forward. Cajal and his doctrine provided a clear plan of attack: If you want to understand the nervous system, you study the neuron. If you want to understand the brain, you study the neuron. This sentiment is best encapsulated in a current re-statement of the doctrine:\\

\quote{\textbf{Neuron Doctrine}: The neuron is the structural unit, the embryological unit, the functional unit and the trophic unit of the nervous system.}{From \textcite{jones1994neuron}}\\

Cajal is sometimes credited as the founder of the field of neuroscience. This is because of his numerous contributions and discoveries, but also because of his strong influence on neuroscientists that came after him. In what followed, many neuroscientists started investigating various properties of single neurons, including how they interact with one another and propagate and transmit signals. British neurophysiologist, Sir Charles Scott Sherrington, is particularly noteworthy here.

\subsection{Charles Sherrington and the synapse}
Sherrington's work was critical in settling the debate between reticular theory and neuron doctrine \cite{Burke2006SirCS}. By the early 20th century, it was established that neurons are independent entities, but it wasn't clear how they interacted. Neuroscientists knew that neurons were tightly connected to each other, but it was debated whether the connections were continuous or separated by small gaps.

\begin{figure}[ht!]
    \centering
    \includegraphics[width=0.9\linewidth]{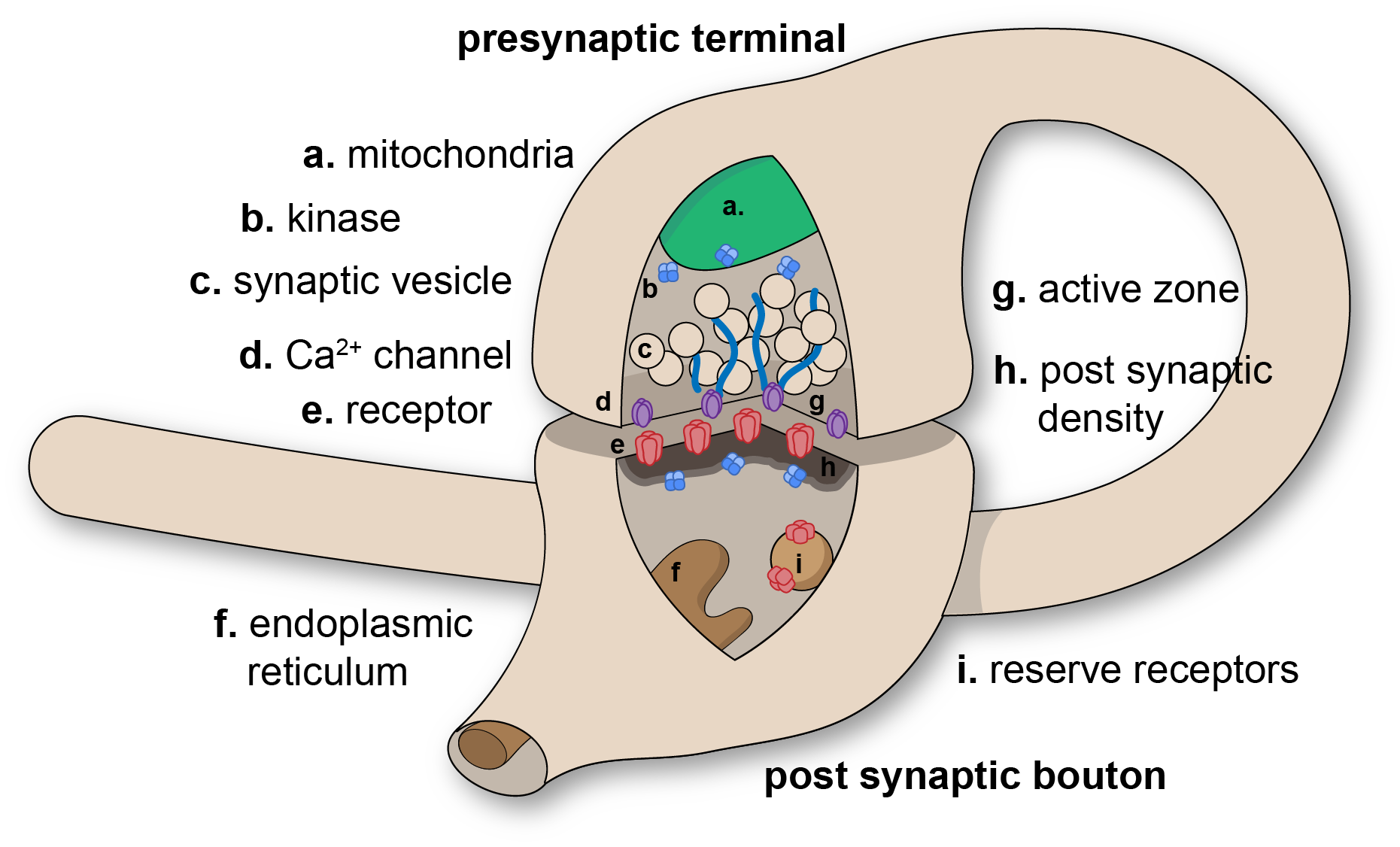}
    \caption[A synapse]{A chemical synapse between two neurons. Information is transmitted from the presynaptic neuron to the postsynaptic neuron. Image from \href{https://en.wikipedia.org/wiki/Synapse}{Wikipedia}.}
    \label{fig:synapse}
\end{figure}

One of Sherrington's important contributions was solidifying the idea of discrete communications between neurons, which meant that there was indeed a gap or a junction. He coined the term \sayit{synapse} (from Greek \textit{synapsis} meaning \textit{conjunction}) to describe this junction (see \textcite{shepherd1997centenary} for a historical review). Sherrington's experiments on reflex arcs provided evidence for this synaptic communication. Later, electron microscopy (developed after Sherrington's time) would provide visual confirmation of the synaptic cleft and the intricate structures involved in synaptic transmission (\cref{fig:synapse}). 

\begin{figure}[ht!]
    \centering
    \includegraphics[width=0.55\linewidth]{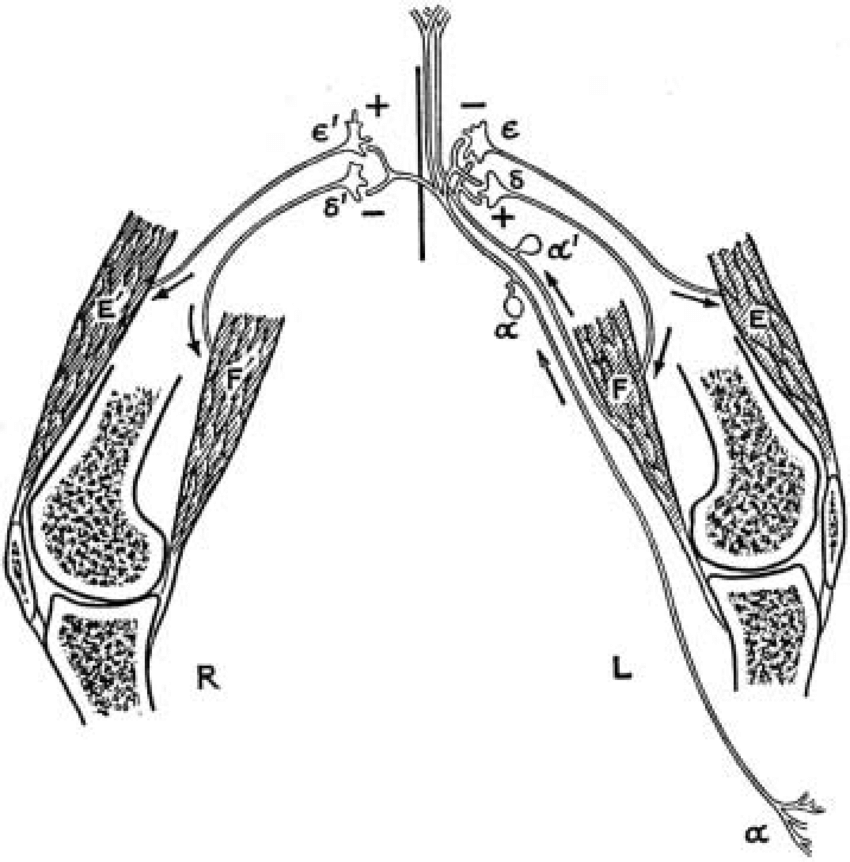}
    \caption[Circuit diagram from Sherrington’s \textit{Integrative Action}]{A circuit diagram from Sherrington’s \textit{The Integrative Action of the Nervous System}, illustrating the fundamental interactions and pathways of two primary afferent neurons ($\alpha$) and their reflex influence. The plus and minus signs indicate excitatory and inhibitory interactions. Image adapted from \textcite{Grant2006The1A} (originally published in \textcite{sherrington1906integrative}, 1906).}
    \label{fig:reflex-circuit}
\end{figure}

Sherrington's magnum opus, \textit{The Integrative Action of the Nervous System} (1906, \cite{sherrington1906integrative}), is derived from over two decades of his experiments primarily on reflex actions. The book profoundly influenced how subsequent generations of neuroscientists viewed the organization of the central nervous system \cite{Burke2006SirCS}. Years later, in a 1947 review, \textit{The Integrative Action} was compared to Newton's \textit{Principia} \cite{Swazey1968SherringtonsCO}, citing as a reason that it is not merely a compilation of empirical observations, but rather, it is a \say{product of sustained thought} on the function of nervous system \cite{Swazey1968SherringtonsCO}. Interestingly, the book showcased circuit diagrams involving multiple neurons, encapsulating the true essence of the word \sayit{integrative} (\cref{fig:reflex-circuit}).

It is worth noting how Sherrington's \textit{integrative} perspective was visionary and ahead of its time. It foreshadowed ideas that would surface over a century later, challenging the traditional neuron doctrine that emphasized the neuron as the sole functional unit \cite{bullock2005neuron, Yuste2015FromTN}. These contemporary perspectives propose a shift towards understanding brain function at the level of populations, networks, ensembles, or circuits (you name it) of interconnected neurons. We will revisit this point later in the chapter.

Building on his integrative perspective, Sherrington considered the cellular diversity of the brain, discovering that different cell types served distinct functional roles. Specifically, he noted that certain neurons, through their \say{nerve impulses}---what we now understand as action potentials---induced excitation in the subsequent muscle units they influenced. In contrast, other neurons led to an inhibitory response (note \say{$+$} and \say{$-$} signs in \cref{fig:reflex-circuit}). Thanks to advances in our understanding of neurotransmitters and neuronal morphologies, now we identify these as \textit{excitatory} and \textit{inhibitory} neurons, but Sherrington's era lacked such detailed knowledge.

Sherrington's pioneering work laid down foundational principles of brain function that the next generation of neuroscientists would further elucidate. While he recognized the significance of \textit{nerve impulses} in neural function and communication, the true nature of these impulses as electrical phenomena can be traced back even further. In the late 18th century, Luigi Galvani provided the initial glimpses of electrical activity in the nervous system. Then, nearly a century later, Julius Bernstein illustrated this electric pulse, laying the groundwork for the modern understanding of the \say{action potential} with his 1868 publication (see \cite{Schuetze1983TheDO}). However, the bridge between the electrical nature of nerve impulses and their functional implications in response to the outside world was first demonstrated by Edgar Adrian \cite{Finger1999MindsBT}, which we will explore next.

\subsection{Edgar Adrian and the first single neuron recording}

By the 1920s, it was known that the brain consisted of discrete neurons interconnected via synapses, transmitting information through electrical impulses known as action potentials. Yet, a deeper question remained---what is the \textit{meaning} of an action potential? Put another way, the basic vocabulary of the brain's communication was identified, but the language in which the brain spoke remained elusive.

\begin{figure}[ht!]
    \centering
    \includegraphics[width=0.8\linewidth]{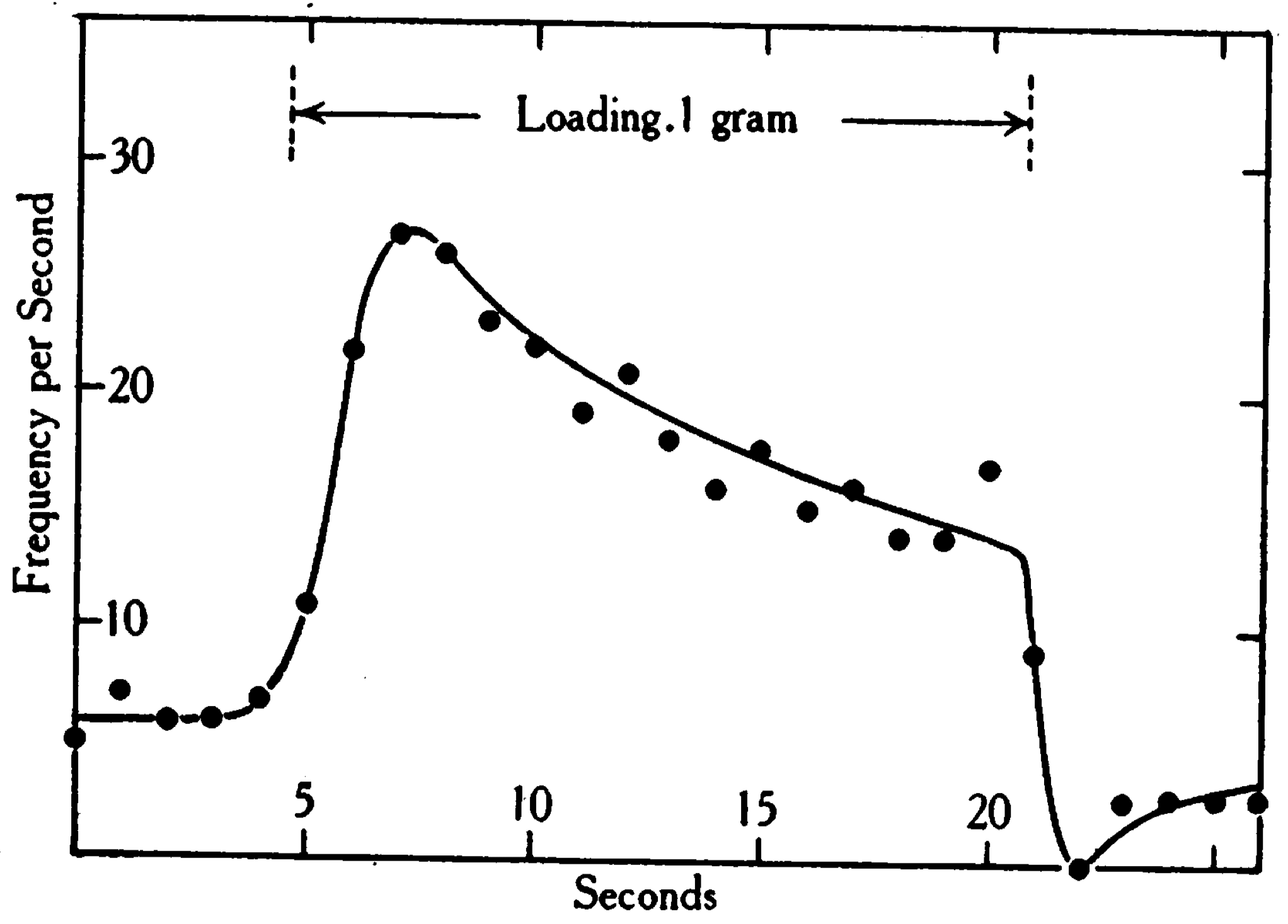}
    \caption[The \textit{rate code}, Adrian \& Zotterman 1926]{The first experimental demonstration of the \textit{rate coding hypothesis}, which states that the frequency of action potentials rather than their shape, amplitude, or other properties, encodes information. Image adapted from \textcite{Adrian1926TheIP}, 1926.}
    \label{fig:adrian}
\end{figure}

A significant advancement came from the pioneering work of Adrian \& Zotterman in 1926 \cite{Adrian1926TheIP}. Adrian had earlier developed a device that amplified the faint electrical activity of neurons \cite{Finger1999MindsBT}. Using this invention, Adrian \& Zotterman achieved the unprecedented feat of recording from a single neuron in response to an external stimulus. Their experimental setup involved a neuron linked to a frog's muscle, and through careful observation, they deduced that this neuron encoded the extent to which the muscle was stretched. The main finding of \textcite{Adrian1926TheIP} was that the frequency (or rate) of action potentials in a sensory neuron was related to the intensity of the stimulus applied. In other words, they had discovered the \sayit{rate code} (\cref{fig:adrian}).

\begin{figure}[ht!]
    \centering
    \includegraphics[width=0.75\linewidth]{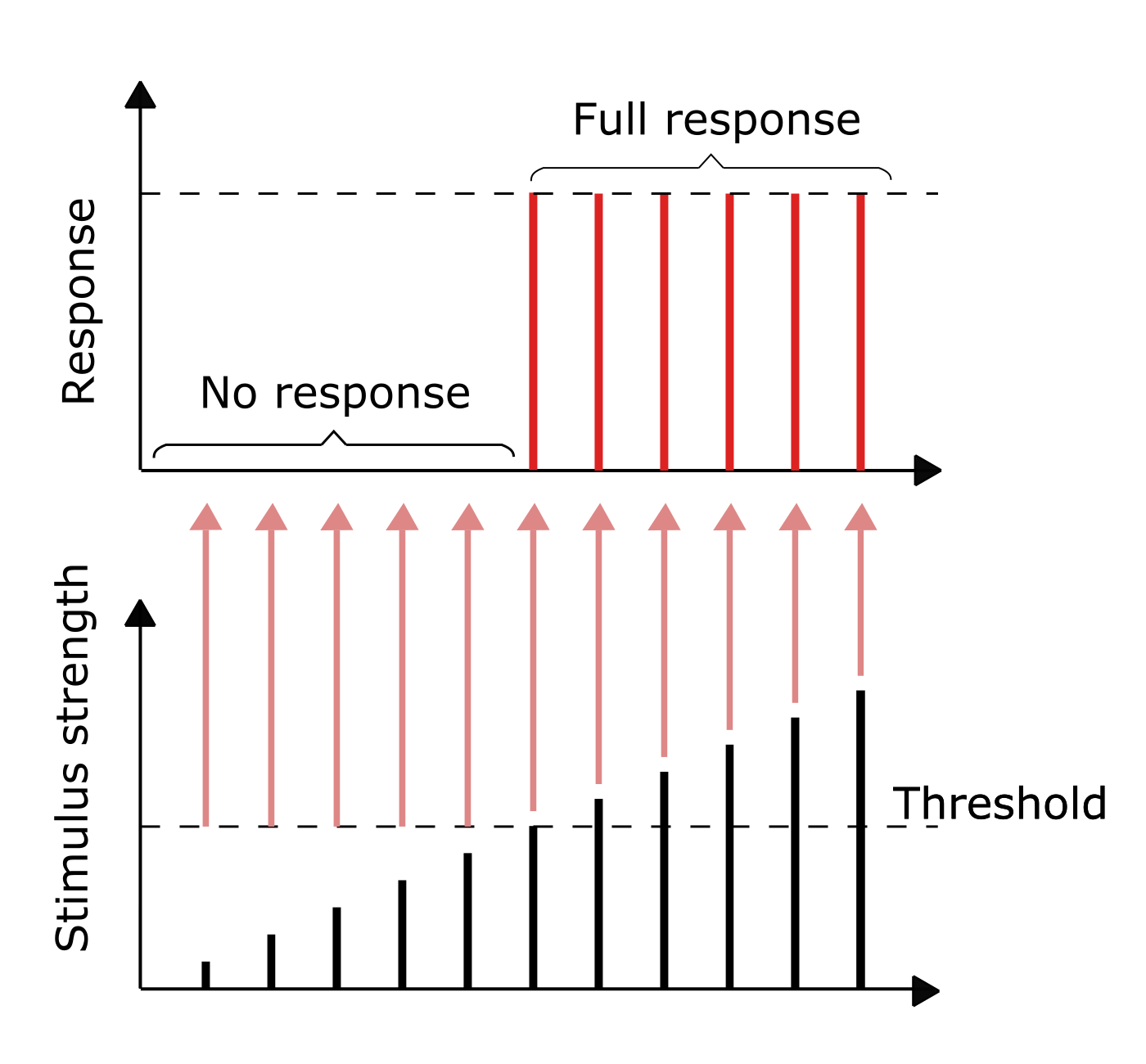}
    \caption[The ``all-or-none" law of nerve impulses]{The ``all-or-none" law of nerve impulses was first described by Henry Pickering Bowditch in 1871 in the context of cardiac muscle contractions. He noted that a minimum threshold of stimulus was needed to elicit a contraction, and once that threshold was crossed, the muscle would contract fully. Later, in the 1920s, Edgar Adrian provided experimental evidence that solidified the understanding of the ``all-or-none" law for action potentials in neurons. Image from \href{https://en.wikipedia.org/wiki/All-or-none_law}{Wikipedia}.}
    \label{fig:all-or-none}
\end{figure}

Adrian's subsequent work largely revolved around the exploration and elucidation of the \sayit{all-or-none} law of nerve impulses (\cref{fig:all-or-none}). Astonished by the simplicity of the rate code, he once remarked that the sensory messages were \say{scarcely more complex than a succession of dots in the Morse Code} \cite{Finger1999MindsBT}. Adrian expanded his studies to encompass various types of sensory receptors, exploring how they encode different forms of stimuli, like touch, sound, and vision, into neuronal signals. Eventually, he would share the 1932 Nobel Prize in Physiology or Medicine with Sherrington for \say{their discoveries regarding the functions of neurons}.

In retrospect, the 1926 study by Adrian \& Zotterman stands as a watershed moment in neuroscience. It offered a mechanistic explanation for how sensory stimuli could be translated and encoded into a neuronal language that the central nervous system could interpret. By linking stimulus intensity to the firing rate of sensory neurons, they offered a framework that would go on to influence much of the subsequent research in sensory neuroscience.

\subsection{A triumph of the neuron doctrine: the story of Hubel \& Wiesel}

By the mid-20th century, foundational concepts were well-established in neuroscience. The brain was understood to consist of discrete neurons communicating through discrete junctions, or synapses. Further, it was understood that these neurons transmit discrete all-or-none signals known as action potentials, and the frequency of these signals could relate to external phenomena, such as the intensity of stimuli on a muscle. But what about other aspects of the external world? How did the brain interpret them?

Recall the 1926 discovery made by \textcite{Adrian1926TheIP} of a frog neuron specifically tasked with signaling muscle stretch intensity through its firing rate. Naturally, this led to curiosity about neurons connected to other sensory organs, especially the eyes. After all, eyes enable vision, much like wings enable flight \cite{barlow1961possible}. Could there be neurons connected to the eyes, encoding visual information similarly to how muscle stretch was encoded? A 1959 study by \textcite{lettvin1959frog} entitled \say{What the Frog's Eye Tells the Frog's Brain}, reported the discovery of \say{bug detector} neurons in frogs. This raised the question: are there neurons in the mammalian visual cortex that encode specific features of the visual environment? If so, what kinds of features?

\begin{figure}[ht!]
    \centering
    \includegraphics[width=\linewidth]{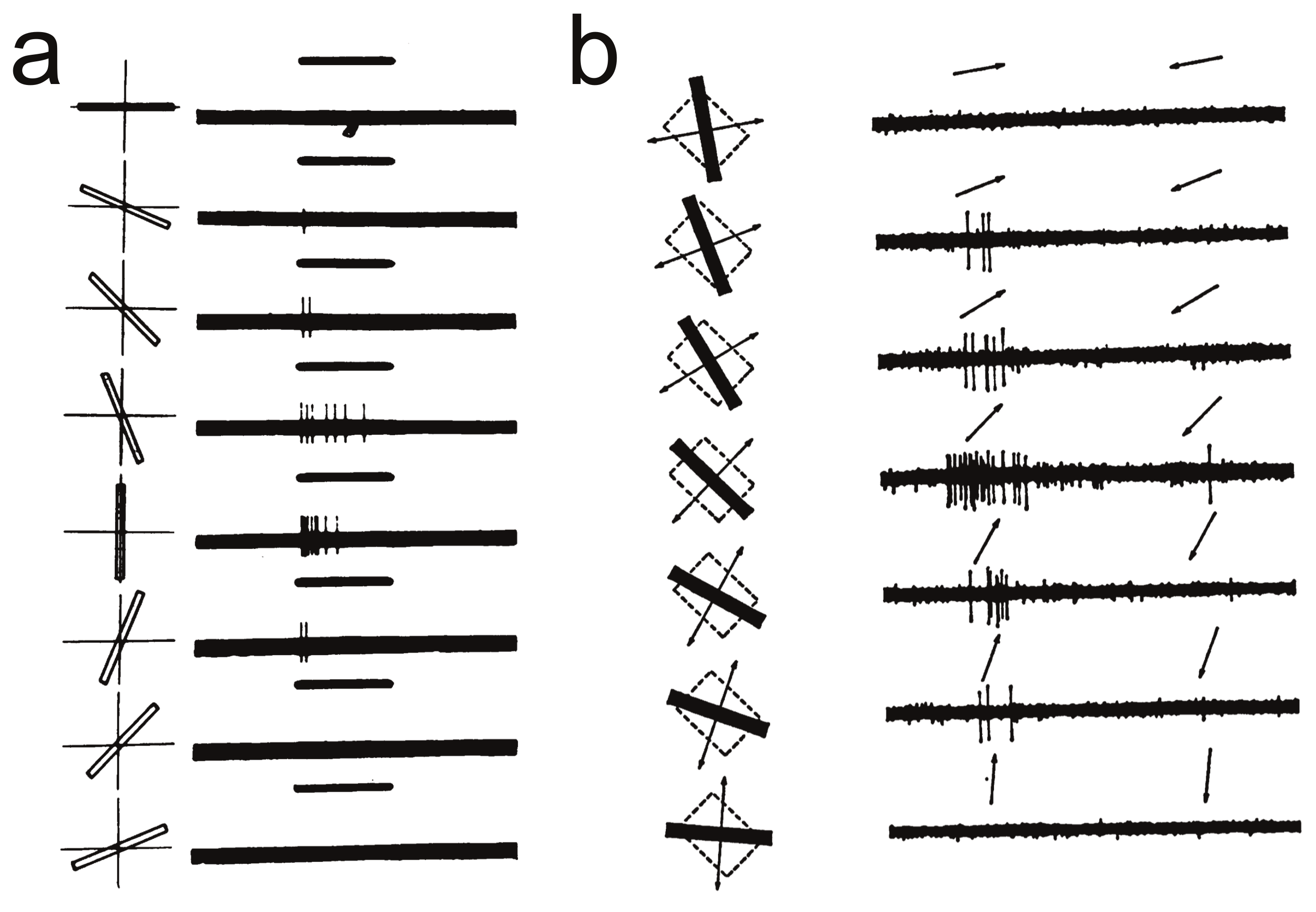}
    \caption[Orientation and direction selectivity in the primary visual cortex]{Orientation and direction selectivity in the primary visual cortex. \textbf{(a)} A neuron from the cat primary visual cortex (V1) responds strongly to a bar of light but only when the bar is oriented vertically. This is an example of a ``\textit{simple cell}" that is tuned to specific orientations. Black horizontal lines indicate stimulus duration. \textbf{(b)} A ``\textit{complex cell}" from macaque V1 signals the presence of certain orientations, but only when the edge is moving in a specific direction. \textbf{(a)} Adapted from \cite{Hubel1959ReceptiveFO}, 1959. \textbf{(b)} Adapted from \cite{Hubel1968ReceptiveFA}, 1968.}
    \label{fig:simple-complex-cell}
\end{figure}

Driven by these very questions, David Hubel and Thorsten Wiesel undertook a series of experiments to unveil the functioning of the visual cortex. They used tungsten electrodes \cite{hubel1957tungsten} to probe the brain while presenting various visual stimuli on a screen, aiming to elicit responses from neurons in the cat visual cortex. Initially, traditional stimuli like dots of light yielded little reaction. A turning point came when, during a routine slide change, they noticed a neuron reacting vigorously to the edge of the slide. This serendipitous observation led them to the discovery of cells specialized in detecting oriented edges---a basic feature of the visual world. Their 1959 paper \cite{Hubel1959ReceptiveFO} documented neurons that were \say{tuned} to specific orientations of bars of light (\cref{fig:simple-complex-cell}a). Building on this, by 1968 \cite{Hubel1968ReceptiveFA}, Hubel \& Wiesel had also identified neurons responsive to the movement direction of oriented edges (\cref{fig:simple-complex-cell}b).

Hubel \& Wiesel's research provided a major advance in our understanding of the relationship between external stimuli and their internal neural representations. From Plato's allegory of the cave to Alhazen's \textit{intromission} theory of vision, and Helmholtz's \textit{inferential} approach to perception, there has always been a persistent drive throughout history to decipher this connection. Hubel \& Wiesel's experiments provided the first concrete evidence of this relationship, revealing that specific neurons in the mammalian brain respond to distinct elements, or \sayit{features}, of the visual environment.

In 1981, Hubel \& Wiesel were awarded the Nobel Prize in Physiology or Medicine in recognition of their \say{discoveries concerning information processing in the visual system}, an honor they shared with Roger W.~Sperry. Their seminal findings not only advanced our understanding of the visual system but also paved the way for subsequent research probing the neural representation of various features of the external world.

This period marked a transformative shift in neuroscience, transitioning from a primary focus on the basic biophysical properties of single neurons to a broader \say{system-level} understanding. Within this context, Hubel \& Wiesel's work played a pivotal role. It ushered in a new era of neuroscience, wherein researchers probed diverse stimuli to uncover the encoding properties of different neurons. Thus, systems neuroscience just embraced a new objective: to determine how various aspects of the environment---be it stimuli, places, and values---are represented at the neuronal level.

\subsection{Neural representation of space in the hippocampal complex}

In the latter half of the 20th century, systems neuroscience took on the challenge of identifying neural \textit{correlates} or \textit{representations} of various aspects of the external world. How do specific neurons in the brain \textit{encode} diverse elements of physical reality? The discoveries of Hubel \& Wiesel revealed how the visual cortex encodes specific visual features of the environment. But around the same time, researchers were exploring more abstract and complex aspects of neural representation. One of the most intriguing questions was: how does the brain represent space? If there are neurons dedicated to encoding edges and bars in the visual field, might there also be neurons that capture position within an environment? If yes, where in the brain do we even begin to look for them?

\begin{figure}[t!]
    \centering
    \includegraphics[width=\linewidth]{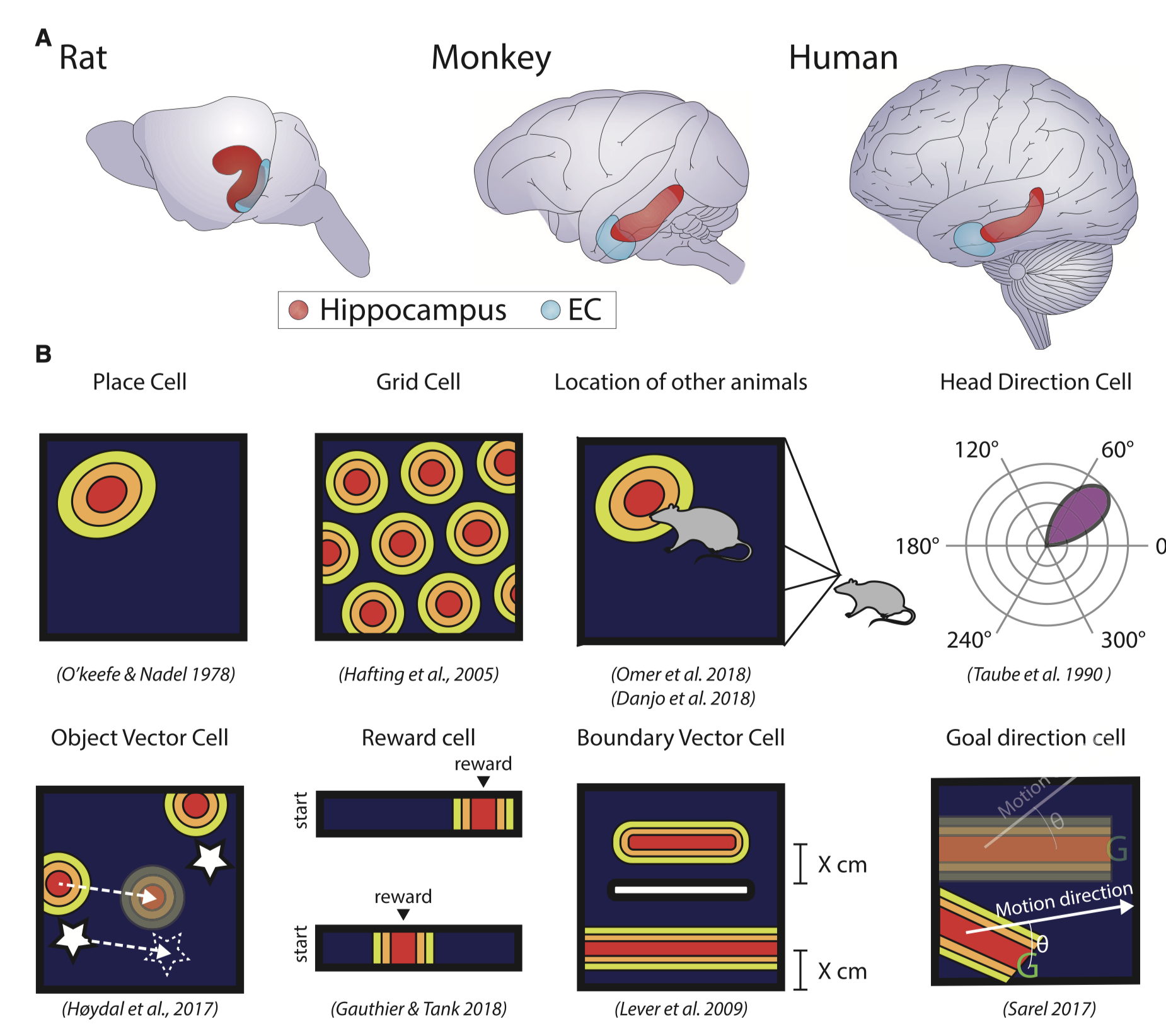}
    \caption[The Hippocampal Zoo]{The Hippocampal Zoo. \textbf{(A)} Anatomical location of the hippocampus and entorhinal cortex (EC) in different species (originally from \textcite{Strange2014FunctionalOO}). \textbf{(B)} Various cells in the hippocampal formation encode distinct spatial variables. See the original publication by \textcite{Behrens2018WhatIA} for a description of different panels. Color gradients indicate the firing rate of neurons (hot colors mean a larger firing rate and so on). Figure from \textcite{Behrens2018WhatIA}, 2018.}
    \label{fig:hippo-ec}
\end{figure}

While visual stimuli travel from the eyes through the thalamus to the visual cortex, spatial representation is more abstract, making it less evident which brain regions might handle such encoding. A significant insight came from William Scoville and Brenda Milner in 1957 \cite{Scoville1957LOSSOR}. They studied patients with hippocampal lesions and documented the critical role of the hippocampus in declarative memories about people, objects, events, and crucially, \textit{places}. This discovery hinted at the possibility of the hippocampus housing neurons specific to place encoding (\cref{fig:hippo-ec}A).

John O'Keefe, in the 1970s, took up this thread. Recording from the hippocampi of rats navigating mazes, O'Keefe found neurons that fired specifically when the rat occupied particular spatial regions. He coined the term \say{place cells} for these neurons, suggesting they form a \textit{cognitive map} of space (\textcite{OKeefe1978TheHA}; see \textcite{passingham79the} for an interesting book review). For a rat in a maze, different place cells fire in various sections, hinting at a neural representation of the spatial layout (\cref{fig:hippo-ec}B, top left panel).

Fast forward three decades to the mid-2000s: Edvard and May-Britt Moser, studying the entorhinal cortex (EC; \cref{fig:hippo-ec}A), a region feeding into the hippocampus \footnote{Entorhinal cortex is the primary cortical input source to the hippocampus \cite{Garcia2020AnatomyAF}.}, stumbled upon \say{grid cells} \cite{Hafting2005MicrostructureOA}. These neurons fired when animals occupied specific locations arranged in a triangular grid pattern (\cref{fig:hippo-ec}B). This discovery demonstrated the brain's intricate spatial encoding system, with grid cells varying in their spatial frequencies while exhibiting beautiful symmetries. Since then, researchers have identified other cell types encoding spatial and even social spatial information, such as \say{social place cells} (\textcite{Danjo2018SpatialRO}; \textcite{Omer2018SocialPI}), \say{head-direction cells} (in rats: \textcite{Taube1990HeaddirectionCR}; and even insects: \textcite{Seelig2015NeuralDF}), \say{object-vector cells} (\textcite{Hydal2019ObjectvectorCI}), \say{reward cells} (\textcite{Gauthier2018ADP}), \say{boundary vector cells} (\textcite{Lever2009BoundaryVC}), and \say{goal direction cells} (\textcite{Sarel2017VectorialRO}). See \cref{fig:hippo-ec}B and \textcite{Behrens2018WhatIA} for a review.

These discoveries highlight that brain activity reflects bits and pieces of the external environment, encompassing visual, spatial, and even cognitive aspects like goals and rewards. These components are encoded in the firing patterns of neurons across various regions, aiding the brain in predicting and interacting with the world. In recognition of their pioneering work, O'Keefe and the Mosers were awarded the 2014 Nobel Prize \say{for their discoveries of cells that constitute a positioning system in the brain} \cite{Kandel2014APA}.

\subsection{Higher visual areas represent more complex visual features}

Let's return to vision. As discussed in Chapter 1, following the pioneering work of Hubel \& Wiesel \cite{kandel2009introduction}, subsequent generations of vision scientists expanded their horizons beyond the primary visual cortex (V1), leading to explorations of other visual areas. Notably, the middle temporal (MT) area became recognized for its critical role in motion perception, with neurons finely tuned to specific movement directions \cite{newsome1988selective}. Adjacent to MT, the medial superior temporal (MST) area processes more intricate motion patterns, contributing to our understanding of self-movement and spatial orientation \cite{Duffy1991SensitivityOM}. 
Meanwhile, the monkey inferotemporal (IT) cortex has been identified as a key player in object recognition, particularly for complex entities like faces. In humans, this function is carried out by the fusiform face area (FFA), which exhibits specialized face processing capabilities \cite{kanwisher1997fusiform}. The IT cortex also contains neurons sensitive to color, contributing to the rich and detailed perception of our surroundings \cite{conway2007specialized, conway2009color}.

These brain regions exhibit distinct functional roles, like focusing on movement or recognizing faces and colors, but there's more to the story, as we will explore below.

\section{Beyond the Neuron Doctrine}

In the first half of this chapter, we explored how the \textit{research programme} laid out by Cajal’s discovery of the Neuron Doctrine and Adrian’s discovery of the rate code transformed 20th-century neuroscience. This line of work led to a myriad of subsequent discoveries, revealing how individual neurons represent various aspects of the outside world. But this is not all there is to neuroscience. Another line of research was being developed in parallel: that of neural \say{oscillations}.

\subsection{Hans Berger's discovery of spontaneous neural oscillations}

\begin{figure}[ht!]
    \centering
    \includegraphics[width=0.8\linewidth]{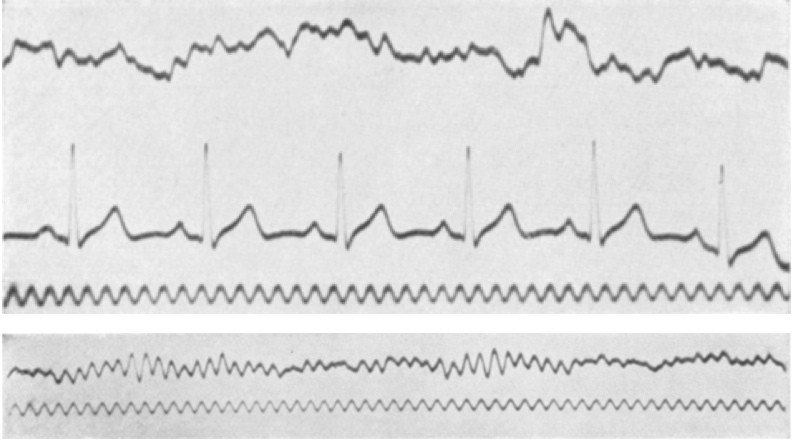}
    \caption[The invention of EEG, Berger 1929]{The first experimental demonstration of spontaneous neural oscillations in the human brain. Both samples represent EEG recordings from Berger's 16-year-old son, Klaus. Bottom, ``alpha rhythm" ($10\text{--}12~Hz$); top, ``beta rhythm" ($10\text{--}30~Hz$). The lowest tracing in both graphs is a generated $10~Hz$ sine wave, and the middle tracing from the top figure is the ECG. Image from \href{https://brainclinics.com/history-of-the-eeg-and-qeeg/}{The Brainclinics Foundation}, originally published in \textcite{berger1929elektroenkephalogramm}, 1929.}
    \label{fig:berger}
\end{figure}

Three years after the landmark study of \textcite{Adrian1926TheIP}, another incredible paper came out. In 1929, German psychiatrist Hans Berger wrote a paper announcing his invention of electroencephalography (EEG), which he used to measure neural \say{oscillations} or \say{brainwaves} \cite{berger1929elektroenkephalogramm}. This was the first experimental demonstration that the brain is spontaneously active and generates its own rhythms (\cref{fig:berger}).

Neural oscillations arise from the collective activity of many neurons in the brain \cite{buzsaki2004neuronal, wang2010neurophysiological}, and critically, they can occur spontaneously \cite{raichle2006dark}, in the absence of any external sensory drive. For these reasons, the concept of neural oscillations appears to be at odds with the neuron doctrine, and the Adrian tradition of looking for neural correlates of experimentally controlled variables such as stimuli \cite{llinas2002vortex}.

This raises the question, what is the significance of such oscillations if the neuron is the functional unit of the nervous system? This question has been revisited by contemporary neuroscientists, and generally, the consensus is that the neuron doctrine requires an upgrade, or possibly a replacement.

In an article entitled \sayit{The Neuron Doctrine, Redux}, \textcite{bullock2005neuron} proposed that although neurons are anatomically discrete units, they may not be the sole functional units in the sense envisioned by early proponents of the Neuron Doctrine. In support of this sentiment, the authors discuss evidence including the effects of neuromodulation \cite{marder2012neuromodulation}, dendritic computations \cite{london2005dendritic}, and the functional role of glial cells contributing to information processing in the brain \cite{mu2019glia}. The authors conclude that it is unlikely that studying single neurons in isolation will provide all the insights needed to understand brain function:\\[-3mm]

\quote{...the complexity of the human brain and likely other regions of the nervous system derive from some organizational features that make use of the permutations of scores of \textit{integrative} variables and thousands or millions of connectivity variables, and perhaps \textit{integrative emergents} yet to be discovered.}{\textcite{bullock2005neuron} (\textit{italics} mine)}\\

Notably, \textcite{bullock2005neuron} invoke the word \say{integrative} twice in their concluding remark. A word that was thoroughly emphasized by Sherrington over 100 years before this paper was published \footnote{It even made it to his book title: \textit{The integrative action of the nervous system} \cite{sherrington1906integrative}.}. This makes you wonder, what are the concepts highlighted by today's visionaries that will take the rest of us another century to fully appreciate and embrace?

\subsection{Possible ways of expanding upon the neuron doctrine}

\begin{figure}[ht!]
    \centering
    \includegraphics[width=1.0\linewidth]{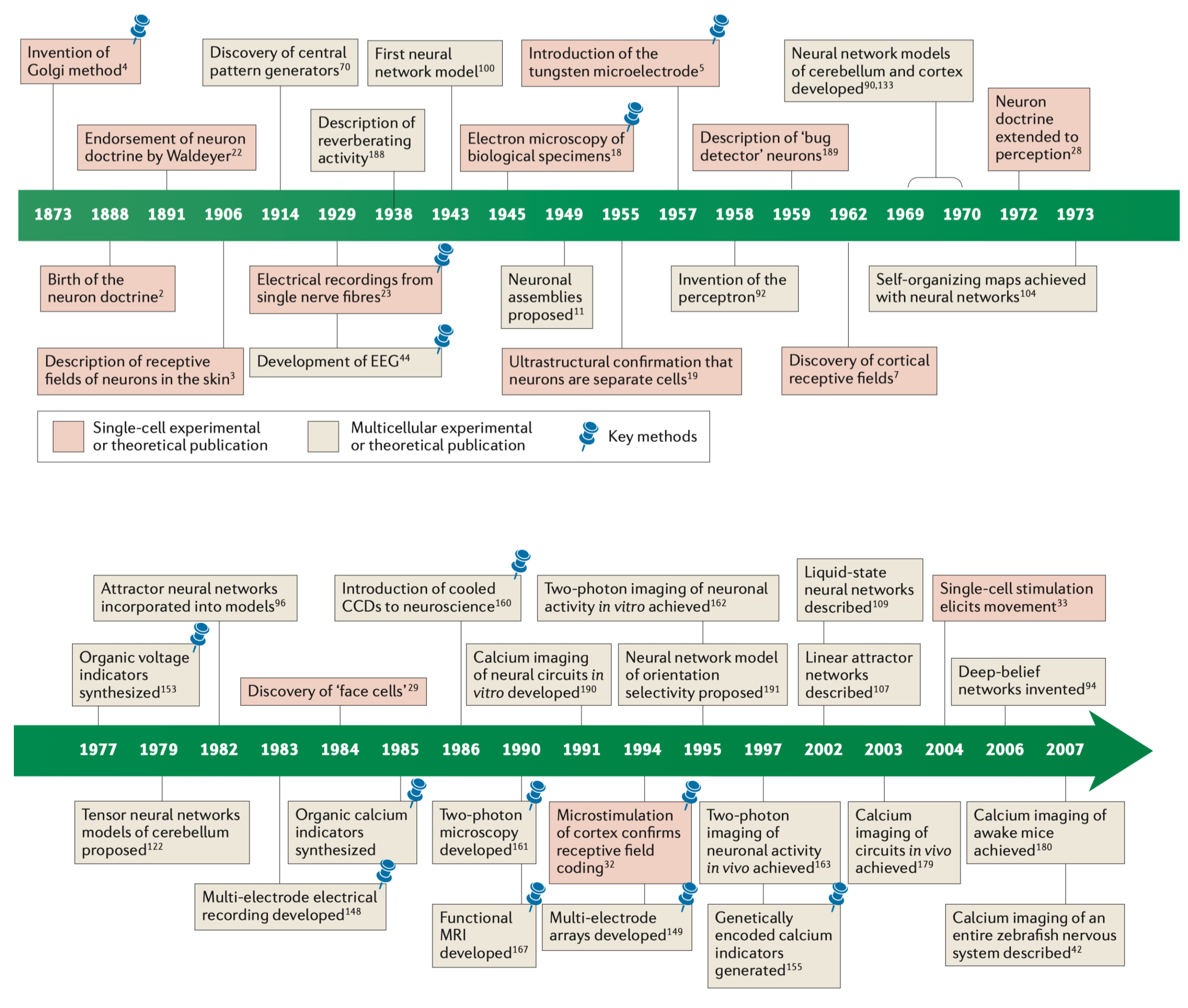}
    \caption[A condensed history of neuroscience]{A condensed history of neuroscience. From \textcite{Yuste2015FromTN}, 2015.}
    \label{fig:yuste}
\end{figure}

The neuron doctrine, foundational as it was, has now been expanded upon. To illustrate this, let's look at two recent examples. In a perspective piece from 2015, \textcite{Yuste2015FromTN} started by exploring major events that shaped the evolution of modern neuroscience, much like we did in this Chapter (\cref{fig:yuste}). He then suggested a shift from the traditional neuron doctrine, proposing neural \say{networks}, \say{ensembles}, or \say{circuits} as potential functional units of the brain. In 2019, another perspective by \textcite{saxena2019towards} posited that future research should focus on neuronal populations as the brain's functional unit. They then formalized this notion by suggesting the term \say{neural population doctrine} \cite{saxena2019towards}, which was later embraced by cognitive neuroscientists \cite{ebitz2021population}.

At their core, these proposals share a common emphasis: the vital \sayit{interactions} within groups of neurons. Without factoring in these interactions, the rationale behind focusing on a population of neurons would seem incomplete. Consequently, it is fair to say that these recent ideas, in general, underscore the importance of considering the \sayit{relationships} between neurons.

\section{A relational perspective on brain function}

In our previous discussion, we highlighted perspective pieces from the 2010s published by neuroscientists emphasizing the importance of considering neuronal interactions in understanding brain function. However, this concept isn't entirely new. Similar ideas were already being explored much earlier, in lesser-known realms of biology.

During the 1970s, American theoretical biologist and scholar, Robert Rosen, was considering an alternative way of understanding biological complexity. Rather than focusing solely on the individual properties of biological entities, he proposed that a richer understanding could be gained by examining the \textit{relationships} between them. Rosen was influenced by his mentor, Nicolas Rashevsky---a notable character in the history of mathematical biology \cite{cull2007mathematical}---to undertake a relational approach to biology.

Central to Rosen's thought process is the idea of \sayit{organization}. He discusses these ideas extensively in his work on relational biology and, notably, in his 1991 book \say{Life Itself} \cite{rosen1991life}. Rosen's concept of organization differentiates itself from the mere structural composition of a system. Simply put, while the tangible components or \textit{stuff} of a system define its structure or mechanisms, the unique interplay or \textit{patterns of interrelations} between these components form its organization.

Building upon this perspective, Rosen then went on to develop a specific definition of complexity based on category theoretical models of autonomous living organisms. A comprehensive exploration of this concept exceeds the scope of this Dissertation. Instead, let us consider a few simple examples below to build some intuitive insights.

\subsection{The relational approach may be unavoidable}

The human body, comprised of cells, constantly undergoes cellular replacement \cite{sender2021distribution}. Yet, we retain our unique identities, memories, personalities, and thoughts. This raises the question: if cells are regularly being replaced, then what remains invariant in the human body, keeping our essence intact? Answering this question requires identifying organizational principle that remains unaffected by these cellular changes.

Furthermore, it is an oversimplification to label the human body merely as a \say{collection of replaceable cells}. In actuality, we are more like a \say{walking ecosystem}, carrying around with us gut bacteria that interact with our cells and even influence our mood and thoughts \cite{chevalier2020effect}. It is essential to incorporate all these interactions that give rise to the continuity and stability of our existence, including elements like gut-brain interactions \cite{mayer2022gut}.


A counterargument could suggest that our identity stems from our neurons, which are not frequently replaced, thereby preserving our sense of self. While it is true that the majority of neurons do not regenerate, their internal components, such as proteins, are in a constant state of flux. This turnover occurs at various time scales, from minutes to weeks, indicating a dynamic process within the neurons themselves \cite{marder2006variability}.

Adding to the complexity, neurons exhibit notable variability in their activity, even in short time scales of minutes and seconds. It is now a well-established fact that across experimental trials, neuronal activity can widely fluctuate in response to repeated presentations of the exact same stimulus \cite{shadlen1998variable, arieli1996dynamics, stein2005neuronal, faisal2008noise}. Moreover, recent evidence suggests that sensory neurons undergo representational drift---a phenomenon where a single neuron represents different features of the environment at different times \cite{schoonover2021representational, deitch2021representational, rule2019causes, driscoll2022representational}.

Irrespective of flux in the individual cellular components, trial-to-trial variability, and representational drift in single neurons, we still maintain invariance in our perceptions. We still see the same way today as we did yesterday, and will likely continue to do so tomorrow. If the individual neuron exhibits all sorts of structural and functional variability as discussed above, then where does this invariance and stability in our perception originate from?

Both empirical evidence and theoretical discourses, spanning from Rosen to Yuste above, indicate that this invariance is likely rooted in the collective attributes of neurons rather than their individual characteristics. To understand a complex system such as the brain, we must view it holistically. Once we make this leap, a relational approach, which considers the interactions between components, becomes indispensable. To ignore these relationships is to reduce a system to a mere mathematical set---a conglomerate of elements devoid of any interactions. Over a century ago, Poincar\'e echoed a similar sentiment:\\[-5mm]

\quote{The aim of science is not things themselves, as the dogmatists in their simplicity imagine, but the relations between things; outside those relations there is no reality knowable.}{Henri Poincar\'e, \textit{Science and Hypothesis}, 1902 \cite{Poincar1906ScienceAH}}\\

Taken literally, even the Adrian tradition of deciphering the neural code aligns with this perspective: the neural code represents a (cor)relation between a stimulus and a neuron's firing rate. Incorporating more neurons simply expands this network of relations.

\subsection{Network science and relational thinking}

Relational thinking is at the core of \textit{network science}, a relatively new branch of physics and applied math \cite{barabasi2016network, newman2018networks}. As the name suggests, network science focuses on the study of networks (or graphs), which are constituted of two main components: (1) nodes (or vertices) and (2) links (or edges) connecting these nodes. A link, denoted as $e_{ij}$, signifies the existence of a relationship or interaction between nodes $i$ and $j$. For example, within a social network, nodes might represent individuals, and links could represent friendships.

\begin{figure}[ht!]
    \centering
    \includegraphics[width=\linewidth]{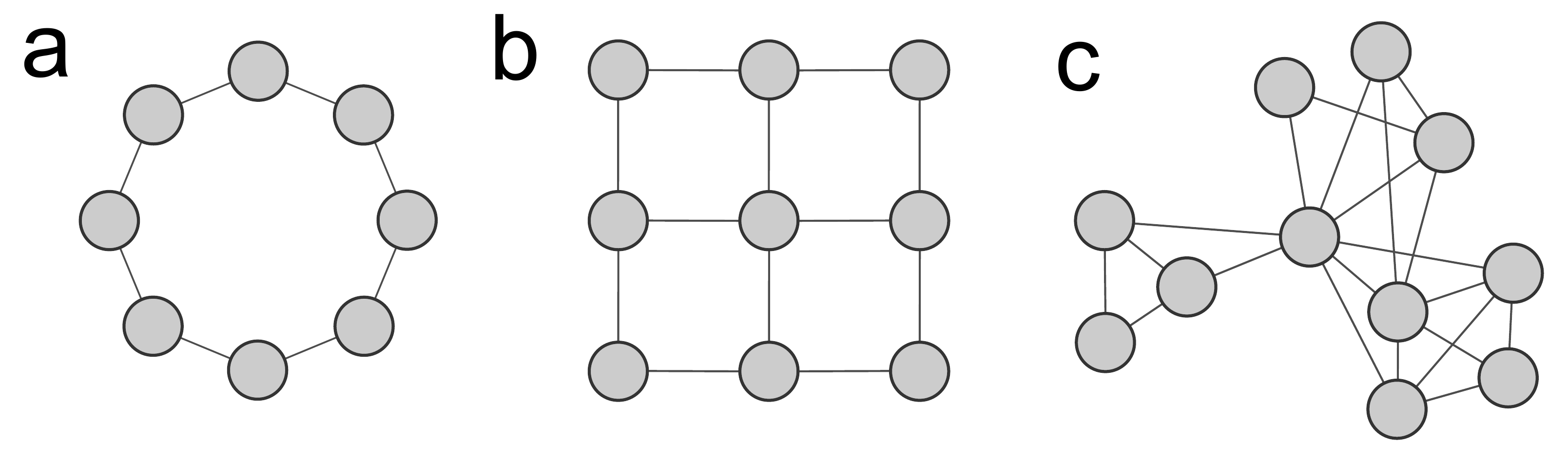}
    \caption[Different network topologies]{Different network topologies. \textbf{(a)} Ring. \textbf{(b)} Lattice. \textbf{(c)} A more complex topology.}
    \label{fig:topo}
\end{figure}

In the context of network science, the intrinsic properties of the nodes are secondary. The primary focus lies in how these nodes are connected to each other. For illustration, the nodes might form a circle (\cref{fig:topo}a), adopt a lattice pattern (\cref{fig:topo}b), or exhibit more intricate patterns like the one in \cref{fig:topo}c. Such patterns or arrangements of links within networks are referred to as \say{topologies}. Using this terminology, one might say the networks demonstrate different topologies such as a ring topology as shown in \cref{fig:topo}a, and so forth. This brings us to a fundamental axiom in network science:\\[1mm]

\hspace{0.08\textwidth}\begin{tabular}{|p{0.70\textwidth}}\textbf{Axiom 1}: \sayit{Network topology, or the specific manner in which nodes are interconnected, defines the most essential property of a network.}\end{tabular}\\[5mm]

\begin{figure}[ht!]
    \centering
    \includegraphics[width=0.85\linewidth]{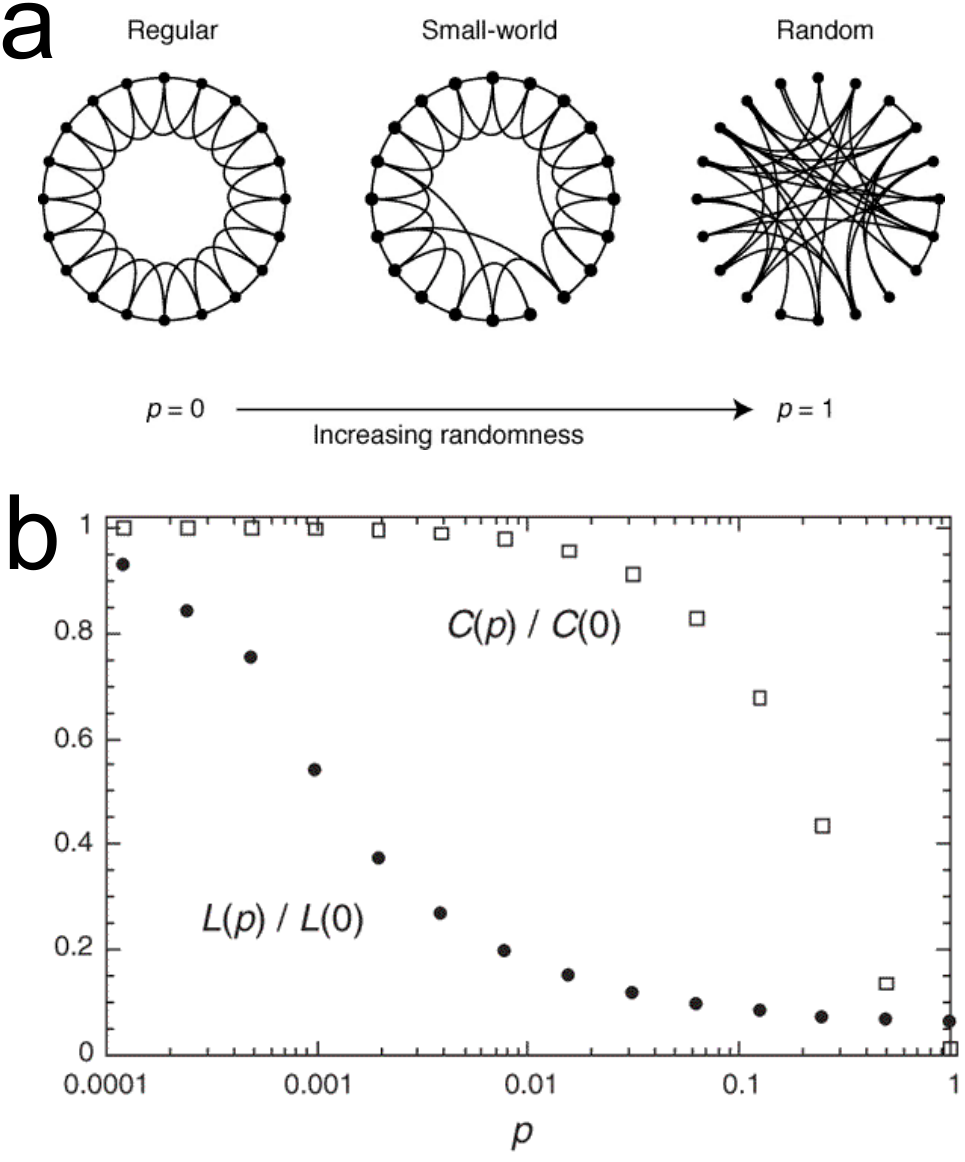}
    \caption[Small-world networks]{Small-world networks. \textbf{(a)} Random rewiring procedure of \textcite{watts1998collective} allows interpolating between a regular ring lattice and a random network. With probability $p$, they reconnect one side of an edge to randomly chosen nodes over the entire ring. \textbf{(b)} At around $p=0.01$, the network simultaneously exhibits (1) a \textit{large clustering coefficient} ($C(p)$, which measures the degree to which nodes in a network tend to cluster together, often quantified by the number of closed triplets or triangles); and (2) a small \textit{characteristic path length} ($L(p)$, which measures the average number of nodes that need to be traversed to link any one node to all others in the network). These are two defining properties of a small-world network. Thus, a small-world topology reconciles measures of integration and segregation. Figure adapted from \textcite{watts1998collective}, 1998.}
    \label{fig:small-world}
\end{figure}

Adopting this network-centric perspective has unveiled certain topologies with unique and significant properties. For instance, in 1998, Watts \& Strogatz identified what's known as the \textit{small-world} network topology \cite{watts1998collective}. In such a network, multiple clusters of closely interconnected nodes exist, indicative of a measure of \textit{segregation}. However, there are also a few scattered long-range links facilitating the flow of information to distant areas of the network, indicative of \textit{integration}. Therefore, a small-world topology reconciles these two fundamental organizational principles of integration and segregation (\cref{fig:small-world}). This network topology provides an explanation for Milgram's earlier empirical observation, the \say{6 degrees of separation} concept, suggesting that any two individuals are, on average, separated by just six connections \cite{milgram1967small}. Finally, it has been suggested that the structural networks in the brain are also consistent with a small-world topology \cite{bassett2006small, bassett2017small, humphries2006brainstem}.

Network science has provided an extensive array of tools and concepts, which have proven to be beneficial in various scientific fields, including neuroscience \cite{Bassett2017NetworkN, sporns2010networks, fornito2016fundamentals, pessoa2014understanding, bassett2018nature, betzel2023community, barabasi2023neuroscience}. As we move forward, the contribution of network science to neuroscience is expected to become increasingly significant, helping to unravel the complexities of brain structure and functionality.

\subsection{The study of large-scale brain network structure in humans}

In the early 1990s, the invention of functional magnetic resonance imaging (fMRI) revolutionized our capacity to map brain activity \cite{ogawa1990oxygenation, kwong1992dynamic}. fMRI measures the blood-oxygen-level-dependent (BOLD) signal \cite{ogawa1990oxygenation}, which captures changes in blood flow, serving as an indirect indicator of neural activity \cite{logothetis2001neurophysiological, hillman2014coupling}.

By the mid-1990s, researchers began to uncover the mysteries of spontaneous fluctuations in the brain, hinting at the presence of \textit{resting-state} networks \cite{biswal1995functional}. The resting state refers to an experimental condition where there are no specified tasks or controlled stimulation.

As the 2000s rolled in, advances in data analysis techniques, epitomized by studies like \textcite{beckmann2005investigations}, deepened our understanding of these networks. \textcite{damoiseaux2006consistent} provided more evidence for the consistent existence of resting-state networks across individuals. This period also witnessed the discovery of default mode network (DMN) \cite{Raichle2001ADM}, a network of brain regions more active during rest than during task engagement, and has been linked to mind wandering \cite{mittner2016neural}. \textcite{greicius2007resting} expanded on the significance and function of the DMN. Finally, \textcite{vincent2007intrinsic} demonstrated evolutionary conservation of resting-state activity patterns in anesthetized monkey brain.

As the understanding of intrinsic brain networks matured, there emerged a need for a comprehensive map that could systematically delineate the intrinsic functional architecture of the human brain.

\begin{figure}[ht!]
    \centering
    \includegraphics[width=0.8\linewidth]{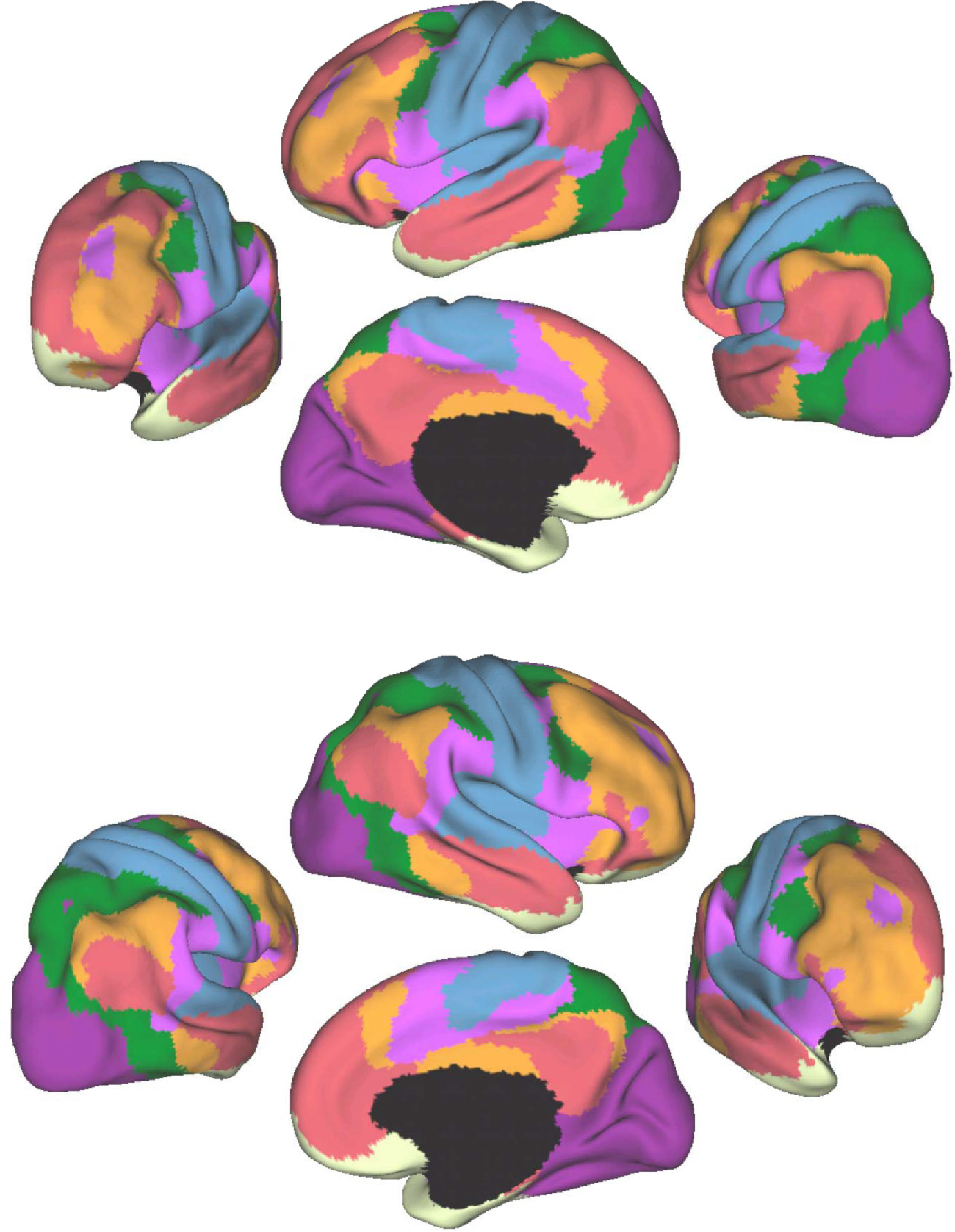}
    \caption[The Yeo 7-network parcellation of the human cortex]{The 7-network parcellation of the human cerebral cortex. Purple (visual), Blue (Somatomotor), Green (Dorsal Attention), Violet (Ventral Attention), Cream (Limbic), Orange (Frontoparietal), and Red (Default). Figure adapted from \textcite{Yeo2011TheOO}, 2011.}
    \label{fig:yeo11}
\end{figure}

In 2011, \textcite{Yeo2011TheOO} applied a clustering algorithm on a large resting-state fMRI dataset to parcellate the human cerebral cortex into multiple distinct networks (\cref{fig:yeo11}). Their findings presented a more nuanced and detailed partition than previously recognized. The significance of \textcite{Yeo2011TheOO} lies not just in its refined parcellation, but also in how it underscored the complexity and rich organization of the brain's intrinsic functional architecture. This study became a standard reference point for subsequent work exploring the interplay and functionalities of these identified networks.

\section{Multimodal imaging as a next frontier in neuroscience}

The progress of neuroscience research has consistently been shaped by technological advancements. Every once in a while, novel methodologies arise, granting researchers the tools to pose previously unimaginable questions about the brain. However, while these inquiries often reveal new insights, they remain inherently constrained by the limitations of the technologies that made them possible. In essence, each technological tool defines its own scope or \say{box} of inquiry. By integrating different methodologies, we can fuse these individual boxes together to form a bigger one.

Multimodal imaging has emerged as a promising frontier in neuroscience, integrating information from various modalities, and allowing researchers to capitalize on the strengths of each method while compensating for their individual weaknesses \cite{uludaug2014general, lake2022building}. For instance, fMRI excels in providing full-brain maps of activity, capturing the hemodynamic response associated with neuronal activity. However, its spatial and temporal resolution is limited. Conversely, electroencephalography (EEG) provides high temporal resolution \cite{ritter2006simultaneous}, suitable for tracking the dynamics of brain activity over milliseconds, but lacks the spatial coverage of fMRI. Magnetoencephalography (MEG) offers a balance, with better spatial resolution than EEG and superior temporal resolution compared to fMRI, making it particularly useful for exploring the electrophysiological activity of the brain \cite{hall2014relationship}.

The integration of these techniques, such as simultaneous EEG-fMRI, allows researchers to leverage the temporal accuracy of EEG and the spatial resolution of fMRI in a single study \cite{cichy2020eeg, laufs2012personalized, he2008multimodal}. Such multimodal approaches not only enhance our understanding of the brain's structure and function but also hold great potential in clinical applications for diagnosing and monitoring neurological disorders \cite{liu2015multimodal, tulay2019multimodal}.

In animal models, moderately invasive imaging techniques like wide-field calcium (\CA) imaging have shown great potential to improve our understanding of large-scale brain dynamics \cite{grienberger2012imaging, Cardin2020MesoscopicIS, ren2021characterizing, Engel2021TheDA}. This technique utilizes genetically encoded \CA indicators like GCaMP \cite{tian2009imaging} to capture cortex-wide neural activity with a high signal-to-noise ratio, making it possible to measure neural activity with high fidelity in behaving animals.

The adoption of dual-color imaging within the wide-field \CA imaging technique has further enhanced its capability. For example, simultaneous neural and hemodynamic imaging can reveal neuronal underpinnings of hemodynamic signals \cite{Ma2016RestingstateHA, Matsui2016TransientNC}, and dual-color imaging of neuronal activity alongside neurotransmitter dynamics, like acetylcholine, can uncover interactions between neurotransmitter release and neural activity patterns \cite{lohani2022spatiotemporally}. These advancements offer insights into complex cognitive processes from sensorimotor integration to decision-making and pave the way for understanding the large-scale circuit dynamics that underlie behavior and cognition. See \textcite{Cardin2020MesoscopicIS} and \textcite{ren2021characterizing} for a review.

\begin{figure}[ht!]
    \centering
    \includegraphics[width=0.9\linewidth]{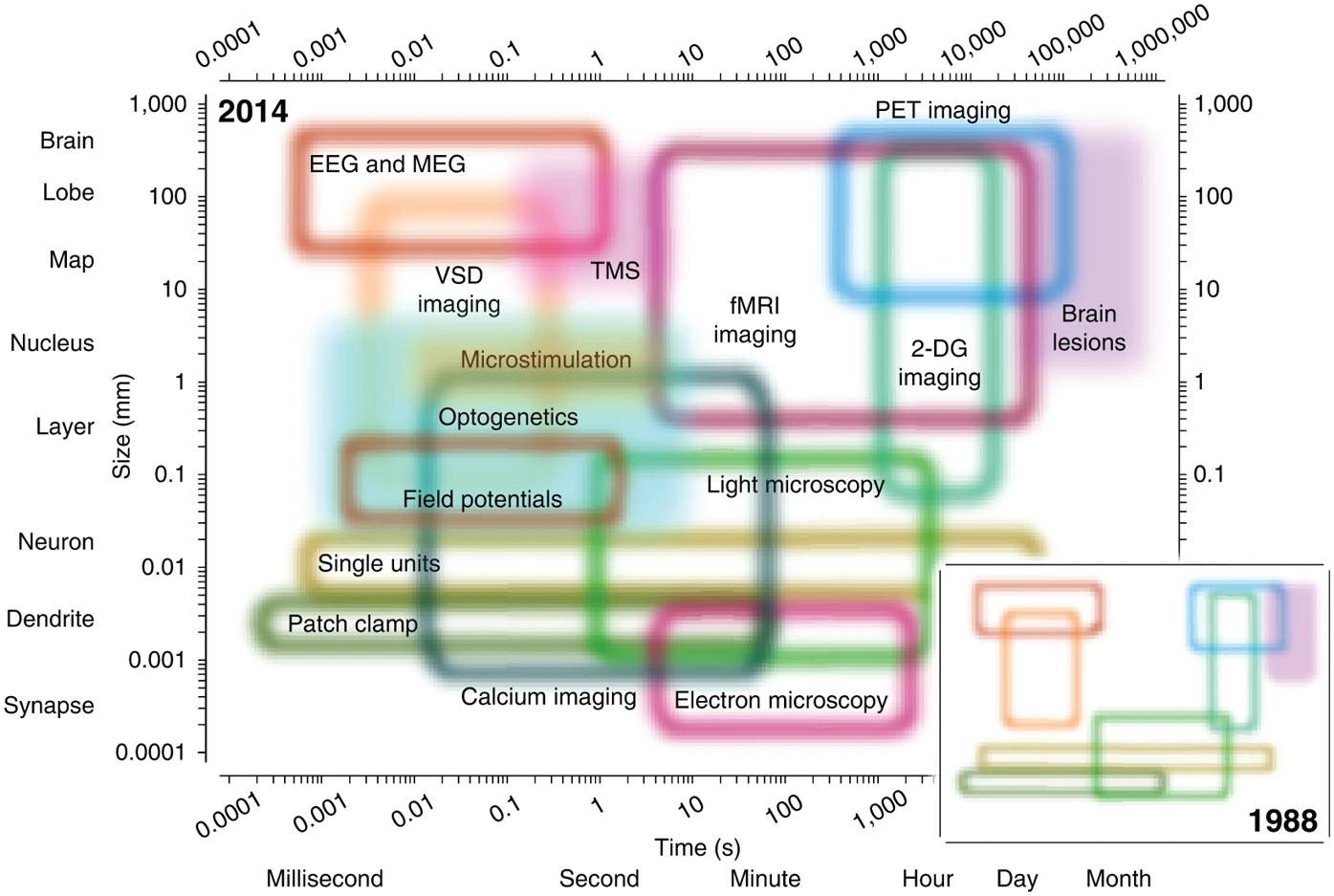}
    \caption[The spatiotemporal domain of neuroscience methodologies]{The spatiotemporal domain of neuroscience methodologies. Each colored region depicts the optimal spatial and temporal resolution range for a specific method. Note the progress from 1988 to 2014, where significant gaps in our methodological coverage were addressed. Image adapted from \textcite{sejnowski2014putting}, 2014.}
    \label{fig:data-scales}
\end{figure}

The combination of fMRI and wide-field \CA imaging is particularly interesting. While fMRI excels in providing whole-brain information, its low spatial and temporal resolution and lack of neuronal specificity can be limiting. In contrast, wide-field \CA imaging offers higher spatiotemporal resolution and relatively higher neuronal specificity, but it has limited coverage (cortical surface only). Furthermore, the non-invasive nature of fMRI has led to its wide usage in humans, while \CA imaging has been only possible in animal models. By fusing fMRI and \CA data, it becomes possible to capture a more comprehensive representation of brain activity, combining across spatiotemporal scales (\cref{fig:data-scales}). See \textcite{lake2022building} for a review.

\begin{figure}[ht!]
    \centering
    \includegraphics[width=0.97\linewidth]{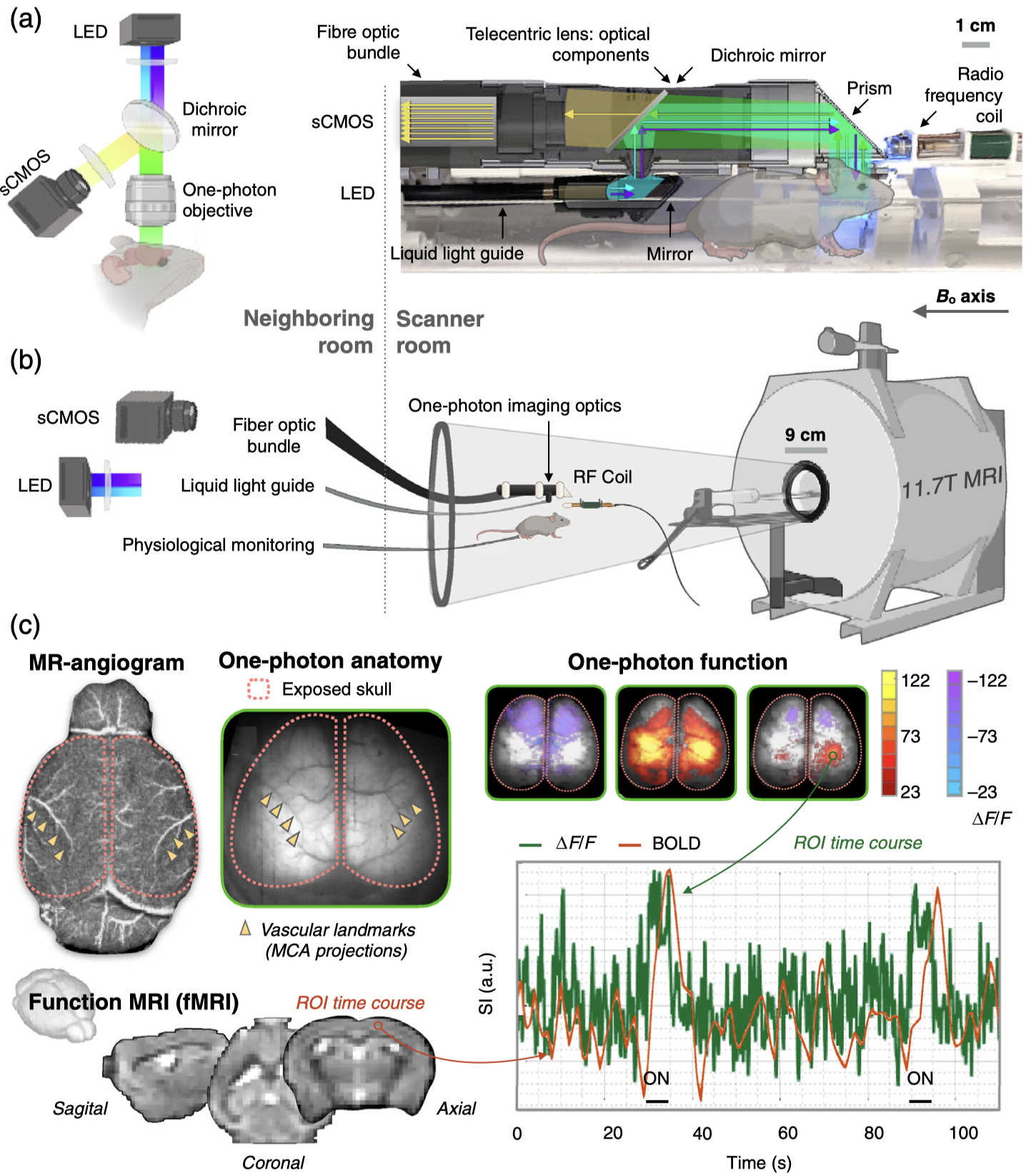}
    \caption[Simultaneous fMRI and wide-field calcium imaging in mice]{Simultaneous cortex-wide \CA imaging and brain-wide fMRI in mice, introduced by \textcite{Lake2020SimultaneousCF} in 2020. Image adapted from \textcite{lake2022building}, 2022.}
    \label{fig:multimodal}
\end{figure}

Motivated by this perspective, \textcite{Lake2020SimultaneousCF} developed an imaging setup that combines brain-wide fMRI with cortex-wide \CA imaging in mice (\cref{fig:multimodal}). This concurrent imaging approach provides data that contribute to our understanding of the cellular underpinnings of the BOLD signal. Additionally, the fMRI component bridges the gap between mouse and human studies, with potential clinical applications. Finally, note how \CA imaging and fMRI, when combined, cover large and complementary portions of the spatiotemporal domain of neuroscience methodologies shown in \cref{fig:data-scales}.

\section{Concluding remarks}

Throughout history, humanity has been deeply intrigued by the enigmatic bridge between the inner workings of our minds and the vast expanse of external reality we see and live in. Long ago, thinkers like Plato wrote about it, and it's possible that even the earliest humans, living in caves, wondered about this too. Now, thanks to modern science and technology, we can measure brain's activity and see how it links to things around us.

For a long time, neuroscientists mainly studied individual neurons to see how they work. Such endeavors taught us extremely valuable insights about brain function at the level of individual neurons, but now we're starting to look at the bigger picture. With new brain imaging tools, we can measure large populations of neurons at the same time, or maybe even the whole brain. These tools give us the chance to think about how groups of neurons, and importantly, the way they interact, might be important for how the brain works as a whole.

Looking ahead, with improved brain imaging and data collection methods, paired with interpretive theoretical frameworks, we foresee an exhilarating voyage of discovery that constantly pushes the boundaries of our understanding. This Chapter is a testament to our journey so far, and a prologue to the novel and exciting advancements that lie ahead. We are steering towards a future rich with potential, ready to unravel the deeper secrets of the brain with every innovative tool and technique at our disposal.


\renewcommand{\thechapter}{3}

\chapter{Variational Inference \& Variational Autoencoders (VAE)}

In this Dissertation, we assume a subjective Bayesian interpretation of probabilities. As Bruno de Finetti put it, \sayit{probability does not exist} \cite{de2017theory, nau2001finetti, caves2002quantum}. We interpret probabilities as subjective degrees of belief or personal judgments about uncertainty, rather than intrinsic characteristics of the world or events. More formally, we assume $P(A)$ reflects the subjective state of belief of an agent about the probability of event $A$. Likewise, $P(A \vert B)$ is assumed to reflect the probability of $A$ given the agent's knowledge that $B$ happened. Finally, the agent's belief is updated according to Bayes' theorem:

\begin{equation}\label{eq:belief}
P(A \vert B) = \frac{P(B \vert A) P(A)}{P(B)}.
\end{equation}

\section{Introduction}
The idea that brains try to infer the causes of their sensory inputs dates back to Helmholtz \cite{helmholtz1867handbuch}, as discussed in Chapter 2. More recent formulations of this idea in terms of hierarchical Bayesian inference \cite{mumford1992computational} and Predictive Coding \cite{rao1999predictive} have a lot in common with Variational Autoencoders (VAE, \cite{kingma2014auto}) in machine learning. Interestingly, while these ideas share many similarities, they have been developed almost independently in theoretical neuroscience and machine learning.

Most notably, both lines of research aim to model the data using latent variables and variational inference, where the ultimate goal is to learn a generative model of the observations. The deep connections between VAEs and ideas in theoretical neuroscience are explored more thoroughly in an excellent review by \textcite{marino2022predictive}.

In the present Chapter, our goal is to introduce the mathematical formalism powering VAEs. We will begin by introducing variational inference from a statistician's point of view. Then, we will show how the VAE naturally arises from the marriage of variational inference with the autoencoder architecture. Finally, we will end with how these ideas can be used to model data and answer scientific questions. For further readings, see \textcite{kingma2019introduction} which is a great review on VAEs written by the researchers who first introduced the VAE. See also \textcite{doersch2016tutorial} for a tutorial.

\section{Latent variable modeling}
In latent variable models, the goal is to relate a set of observable variables to a set of latent (or hidden) variables. This is a general framework that can be applied to many different classes of problems. Experiments on real complex systems often capture an incomplete picture of the general system properties. This is because usually, many unobserved degrees of freedom influence the dynamics. But crucially, these variables are absent from the recorded data.  We call these unobserved degrees of freedom \textit{latent variables}.

Formally, let us denote the observed data points by $\bm{x} \in \mathbb{R}^D$ that are sampled from a distribution $\bm{x} \sim P(X)$. We can write the generative process as:

\begin{equation}\label{eq:lvm}
P(X) = \int P(X, Z) \, dZ = \int P(X \vert Z) P(Z) \, dZ.
\end{equation}

Here, $\bm{z} \in \mathbb{R}^d$ are latent variables, and $P(X, Z)$ is the joint distribution of $X$ and $Z$. Usually, we prefer $d \ll D$, which achieves dimensionality reduction, or compression of the input data. \autoref{eq:lvm} can be thought of as a mixture model, because, for every possible value of $Z$, we add another conditional distribution $P(X \vert Z)$ to $P(X)$, weighted by its prior probability $P(Z)$.

As an example, imagine a dataset that consists of $N \times N$ color pictures of natural images. For $N = 100$ pixels, the dimensionality of the raw data would be $D = 3N^2 = 30,000$; however, a considerably smaller number of latent dimensions is sufficient to model this dataset since the samples lie on a low-dimensional manifold (i.e., the \sayit{manifold hypothesis}, \textcite{bengio2013representation}).

\subsection{Bayesian inference problem}

Given a formulation such as in \autoref{eq:lvm}, it is interesting to ask what latent variables $\bm{z}$ are likely, given an observation $\bm{x}$. This way, we can gain insights into the latent structure of the data. To achieve this, we have to estimate the \textit{posterior} distribution $P(Z \vert X)$. This is known as (approximate) Bayesian inference which is usually a difficult task. To see why, let us consider how we can calculate the posterior using Bayes' rule:

\begin{equation}\label{eq:bayes}
P(Z \vert X) = \frac{P(X \vert Z)P(Z)}{P(X)} = \frac{P(X \vert Z) P(Z)}{\int P(X \vert Z) P(Z) \, dZ},
\end{equation}

where on the right-hand side we used the integral from \autoref{eq:lvm} to replace $P(X)$. Computing this integral is difficult and is usually intractable. We need to sum over all configurations of the latent variables which is computationally expensive.

Approximating such difficult-to-compute distributions is one of the core problems of modern statistics \cite{blei2017variational}. There are various approaches to accomplish this, such as Markov Chain Monte Carlo (MCMC) or Variational Inference (VI). MCMC is a \say{shallow} method, where the idea is to construct a Markov chain that asymptotically produces exact samples from the target density \cite{hastings1970monte, gelfand1990sampling, Jones2022markov}. These methods can be more computationally expensive but come with some theoretical guarantees of convergence, usually good for small but valuable datasets. VI, on the other hand, is preferred when we have large amounts of (possibly noisy) data.

\section{Variational inference}

This section closely follows \textcite{blei2017variational}. The goal of VI is to approximate the target density of latent variables given observed variables. We define a family of variational densities $\mathcal{Q}$ and find a member $q \in \mathcal{Q}$ such that it sufficiently approximates the true posterior. The closeness is measured by the Kullback-Leibler (KL) divergence.

\begin{equation}
\text{VI:} \quad \text{choose a } q \in \mathcal{Q} \text{ such that } q(Z) \text{ is a good approximation of } p(Z \vert X).
\end{equation}

We need to consider two factors when choosing the variational family $\mathcal{Q}$.  First, we want the family to be expressive enough so that we can approximate complicated densities. Second, we want to constrain the family to prevent unnecessary computational costs. Given this well-defined family, all that remains is to find the desired density $q^*$ that minimizes the KL divergence term. Crucially, this approach turns the problem of posterior estimation into an optimization problem. We can then conveniently use for example stochastic gradient methods and neural nets as universal function approximators to achieve this goal.

We will provide some definitions below. Then, in the next section, we will go over variational autoencoders (VAE) and study some of their properties.

\paragraph{\textbf{Expectations.}} Suppose $X$ is a random variable that is distributed according to $P(X)$. We will use the following notation for expectations:

$$\expect{\bm{x} \sim P(X)}{F(X)} = \sum_{\bm{x} \in X} P(\bm{x})F(\bm{x}).$$

\paragraph{\textbf{Kullback–Leibler (KL) divergence.}} For distributions $P$ and $Q$ defined over $X$, the KL divergence is defined as follows:

$$\KL{Q(X)}{P(X)} = \EXPECT{\bm{x} \sim Q(X)}{\log\frac{Q(X)}{P(X)}} = \sum_{\bm{x} \in X} Q(\bm{x}) \log\frac{Q(\bm{x})}{P(\bm{x})}.$$

It can be shown that $\mathcal{D}_{\mathrm{KL}} \geq 0$ and equality happens when $Q = P$. Hence, the KL divergence is a measure of how \textit{close} $Q$ is to $P$. Although, unlike proper distance metrics, KL divergence is not symmetric, i.e., $\KL{Q(X)}{P(X)} \neq \KL{P(X)}{Q(X)}$.

\section{Why KL divergence?}

In VI, the goal is to find a $q$ that is not too different from the true posterior. KL divergence is the most commonly used distance measure, but it's certainly not the only one in existence. Various other distance measures can be used to quantify this discrepancy, including the Jensen-Shannon (JS) divergence which is a symmetrized KL divergence, and the Wasserstein-1 distance. They each have interesting properties that could make them a more suitable choice depending on the domain of application. Here, we will stick with the standard approach of using KL divergence.

\begin{figure}[ht!]
    \centering
    \includegraphics[width=0.9\textwidth]{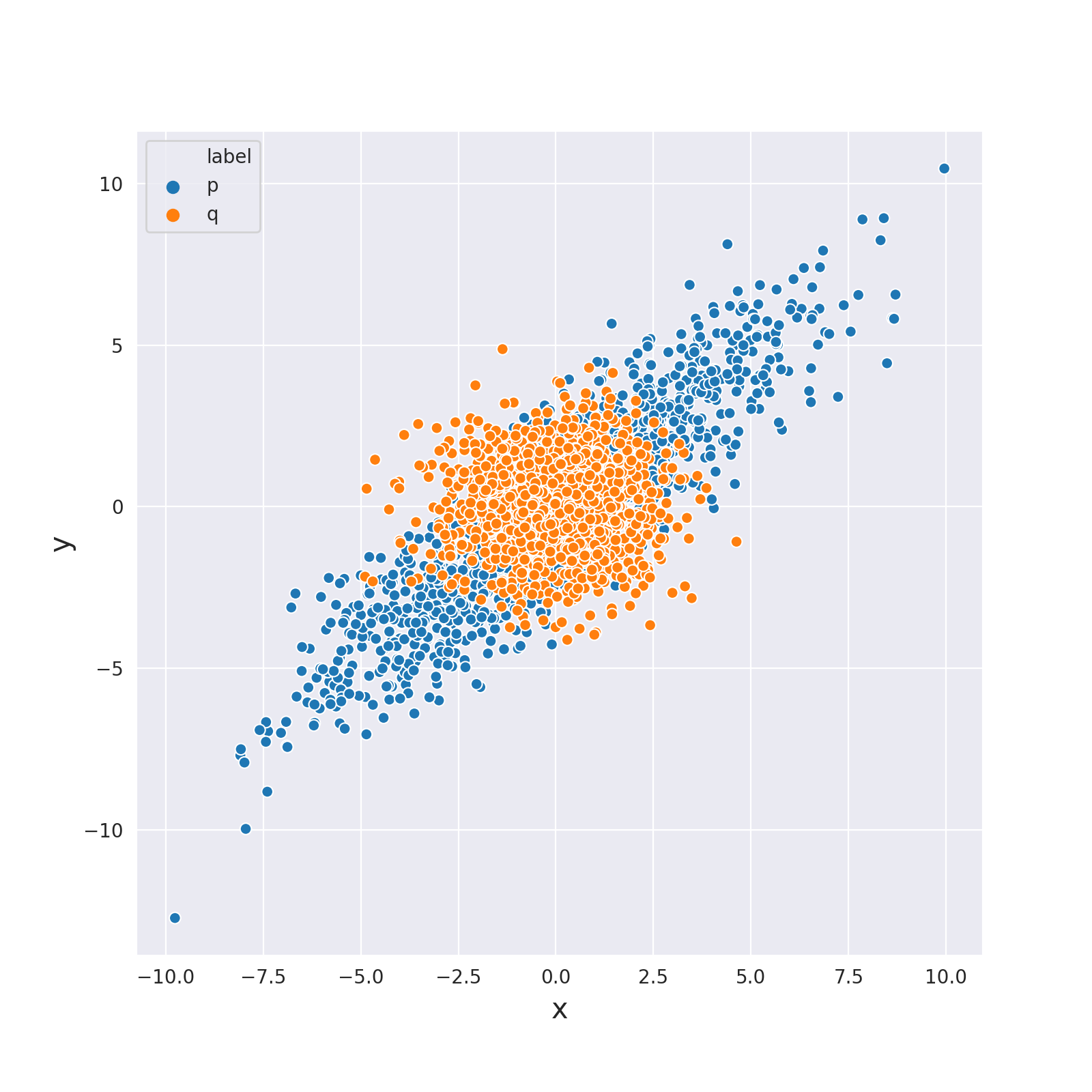}
    \caption[Kullback-Leibler divergence intuition]{Blue dots represent $p = \NN(0, \Sigma)$ where $ \Sigma = \begin{bmatrix} 10 & 9\\ 9 & 10 \end{bmatrix}$, and orange dots represent $q = \NN(0, 1.9)$.  In this example $\KL{q}{p} \approx 0.83$ nats. Conversely, we have $\KL{p}{q} \approx 7.5$ nats. Remember that we don't fully know what $p$ is. In regions where we know it, $q$ does a good job of covering its probability mass. Therefore, $q$ is a decent approximation of $p$, and a small value of $\KL{q}{p}$ is a reasonable quantification of the difference compared to the other way around.}
    \label{fig:kl-eg}
\end{figure}

Recall the definition of KL divergence, $\KL{q}{p} = \expect{q}{\log \big(\frac{q}{p} \big)}$.  From this expression, we see that the distance between $q$ and $p$ will be small as long as $q \approx p$ within the support of $q$ (i.e., regions where $q$ has nonzero weight). In the rest of the space where $q$ is zero, it doesn't matter whether it is similar to $p$ or not. This corresponds to a situation where we know there are \say{things that we don't even know we don't know}. As long as $q$ sufficiently covers what we already know about $p$, it is good enough. Later, we might encounter a previously unseen event from regions outside the support of $q$, where $p$ has nonzero weight. When this happens, we will be surprised at first, then we will reduce the surprisal by updating our $q$ to incorporate the new information. If we live long enough in that environment we will eventually learn everything about it. See \cref{fig:kl-eg} below for an illustrative example.

\section{Variational Autoencoders}

Let us briefly recap. So far we have shown that within the context of latent variable modeling, estimating the posterior $P(Z \vert X)$ is an interesting goal. This can be accomplished by finding a variational density $q \in \mathcal{Q}$ that minimizes the KL divergence. Formally, this is expressed as an optimization problem:

\begin{equation}\label{eq:optim-def}
q^* = \argmin_{q \, \in \mathcal{Q}} \KL{q(Z)}{p(Z \vert X)}.
\end{equation}

Let me emphasize again that the choice of variational family $\mathcal{Q}$ is entirely problem-dependent. It can be a mixture of Gaussians or a neural network parametrized by some weights $\theta$. Furthermore, the variational densities $q(Z)$ may or may not be conditioned on $X$. This will become clear when we discuss an explicit example below.

VAEs leverage the expressive power of neural networks to address this problem. Given large enough data and the flexibility of deep neural networks, any complex distribution can be sufficiently approximated in principle. We will now derive the objective for VAEs.

\subsection{VAE loss}

To compute the VAE loss, let us start from \autoref{eq:optim-def} and expand the KL divergence. Rearranging some terms, we have:

\begin{equation}
    \begin{aligned}
    \KL{q(Z)}{p(Z \vert X)} &=
    \EXPECT{\bm{z} \sim q(Z)}{\log\frac{q(Z)}{p(Z \vert X)}}
    \\
    &= \EXPECT{\bm{z} \sim q(Z)}{\log\frac{q(Z)p(X)}{p(Z \vert X)p(X)}}
    \\
    &= \EXPECT{\bm{z} \sim q(Z)}{\log\frac{q(Z)p(X)}{p(X, Z)}}
    \\
    &= \EXPECT{\bm{z} \sim q(Z)}{\log\frac{q(Z)}{p(X, Z)}} + \log p(X).
    \\
    \end{aligned}
\end{equation}

The last term comes out of the expectation since $p(X)$ does not depend on $\bm{z}$. Next, we rearrange the terms to get:

\begin{equation}\label{eq:vaeloss}
    \begin{aligned}
    \log p(X) &= \KL{q(Z)}{p(Z \vert X)} - \FF(q),
    \\
    \FF(q) &= \EXPECT{\bm{z} \sim q(Z)}{\log\frac{q(Z)}{p(X, Z)}}.
    \end{aligned}
\end{equation}

Here, $\FF$ is known as the variational free energy and it is a non-negative quantity. We see that $\log p(X) \geq - \FF$ because $\mathcal{D}_{\mathrm{KL}} \geq 0$. In other words, the model evidence $\log p(X)$ is bounded from below by negative free energy. In machine learning literature $- \FF$ is referred to as the evidence lower bound, or ELBO:

\begin{equation}
\begin{aligned}
    \text{\it model evidence} \,\, &:= \,\, \log p(X) \,\geq\, - \FF(q) \,\, = \,\, \text{ELBO},
    \\
    \text{\it surprisal} \,\, &:= \,\, - \log p(X) \,\leq\, \FF(q).
    \end{aligned}
\end{equation}

Let us look back at \autoref{eq:optim-def} once again. This is an optimization problem where our goal is to find the best $q$ that minimizes the KL term. If we compare this to \autoref{eq:vaeloss}, we can see that the same goal can be accomplished by minimizing free energy $\FF(q)$. This is because the left-hand side in \autoref{eq:vaeloss} is constant with respect to $q$; therefore, by trying to minimize $\FF(q)$ we would force the KL term to also shrink so that the left-hand side remains constant. Formally, the free energy is a functional (i.e. function of a function) of $q$. Differentiating both sides of \autoref{eq:vaeloss} with respect to $q$ gives:

$$
0 = \frac{\delta [ \log p(X) ]}{\delta q} = \frac{\delta \big[ \KL{q(Z)}{p(Z \vert X)} \big] }{\delta q} - \frac{\delta [ \FF(q) ]}{\delta q}, \quad \Rightarrow \quad \frac{\delta \big[ \mathcal{D}_{\mathrm{KL}}(q) \big] }{\delta q} = \frac{\delta \big[ \FF(q) \big]}{\delta q}.
$$

As a result, the objective can be equivalently expressed in terms of the free energy:

\begin{equation}\label{eq:optim-free-energy}
q^* = \argmin_{q \in \mathcal{Q}} \FF(q) = \argmin_{q \in \mathcal{Q}} \EXPECT{\bm{z} \sim q(Z)}{\log\frac{q(Z)}{p(X, Z)}}.
\end{equation}

Let us now focus on the free energy term and examine some of its properties.

\subsection{Variational free energy}

We already showed that the variational free energy is an upper bound on \textit{surprisal}. Minimizing it will indirectly maximize model evidence. It is called model evidence because the larger its value, the more apt the model is in explaining the data.  In physics, (Helmholtz) free energy is defined as $F = U - TS$, where $U$ is the internal energy, $T$ is the temperature, and $S$ is the entropy of the system. If we expand the variational free energy we see that:

\begin{equation}\label{eq:free-energy}
    \begin{aligned}
    \FF(q) &= \EXPECT{\bm{z} \sim q(Z)}{\log\frac{q(Z)}{p(X, Z)}}
    \\
    &= - \expect{\bm{z} \sim q(Z)}{\log p(X, Z)} + \expect{\bm{z} \sim q(Z)}{\log q(Z)}
    \\
    &= - \expect{\bm{z} \sim q(Z)}{\log p(X, Z)} - \HH(q).
    \end{aligned}
\end{equation}

Here, the first term can be interpreted as some sort of energy related to the joint density because for any distribution $p$, some energy function $E$ can be defined such that $p \propto e^{-E}$. Furthermore, $\HH(q)$ is just the information-theoretic entropy of $q$. This resemblance is why $\FF(q)$ is referred to as free energy.

\subsection{Why is it called an autoencoder?}

So far, we have only covered the probability theory perspective on VAEs. It is now time to discuss the neural net perspective. To achieve this, we first start from the definition of the free energy and rewrite it using Bayes' rule:

\begin{equation}
    \begin{aligned}
    \FF(q) &= \EXPECT{\bm{z} \sim q(Z)}{\log\frac{q(Z)}{p(X, Z)}}
    \\
    &= \EXPECT{\bm{z} \sim q(Z)}{\log\frac{q(Z)}{p(X \vert Z)p(Z)}}
    \\
    &= - \expect{\bm{z} \sim q(Z)}{\log p(X \vert Z)} + \EXPECT{\bm{z} \sim q(Z)}{\log\frac{q(Z)}{p(Z)}}
    \\
    &= - \expect{\bm{z} \sim q(Z)}{\log p(X \vert Z)} + \KL{q(Z)}{p(Z)}.
    \end{aligned}
\end{equation}

Let us consider the first term that appeared on the right-hand side. It involves an expectation value over $p(X \vert Z)$. This conditional distribution can be thought of as a \textit{decoder} since it quantifies which observations $X$ are likely given a latent code $Z$. In other words, a mapping from latents $Z$ to observed data $X$ (i.e., data generation). Along this direction, we can also condition $q(Z)$ on $X$ and interpret it as an \textit{encoder}. This is equivalent to making an additional assumption on our original variational family $\mathcal{Q}$. We are allowed to choose variational densities that depend on the data, i.e.~$q(Z) \rightarrow q(Z \vert X)$. Remember that thus far we kept things pretty general, but it is totally fine to assume such dependencies in the variational family $\mathcal{Q}$. With this new assumption, we get:

\begin{equation}\label{eq:free-energy-enc-dec}
\FF(q) = - \expect{\bm{z} \sim q(Z \vert X)}{\log p(X \vert Z)} + \KL{q(Z \vert X)}{p(Z)}.
\end{equation}

From a coding theory perspective, $q(Z \vert X)$ resembles an encoder that takes data in and encodes it in the latent space. Similarly, $p(X \vert Z)$ is a decoder that maps from the latent space to observations $X$. This is why it's called variational \textit{autoencoder}. VAEs are not actually autoencoders, but rather, they are generative models that aim to learn $p(X)$ using encoder-decoder bottleneck architectures.

With this new light shed on the VAE loss, we can now express the objective function using neural net encoders and decoders. When the encoder is parametrized as a neural net, we can think of it as a function that maps each observation $\bm{x}$, to a distribution $q(Z \vert X\!=\!\bm{x})$. Technically, this is known as \textit{amortized inference}, but that's just a detail. Let us use $\nu$ and $\theta$ to parameterize the encoder and the decoder neural networks, respectively. We have:

\begin{equation}\label{eq:vaeloss-nn}
    \begin{aligned}
    \{\theta^*; \nu^*\} &= \argmin_{\{\theta; \nu\}} \FF(\theta; \nu),
    \\
    \FF(\theta; \nu) &= - \expect{\bm{z} \sim q(Z \vert X; \nu)}{\log p(X \vert Z; \theta)} + \KL{q(Z \vert X; \nu)}{p(Z)}.
    \end{aligned}
\end{equation}

\subsection{Interpreting the loss}

We derived a neat expression for the loss in \autoref{eq:vaeloss-nn}, but still, there are details that we need to figure out. The VAE loss involves two terms (the right-hand side of \autoref{eq:vaeloss-nn}). We will refer to them as \textit{reconstruction loss} and the \textit{KL term}, respectively. Below, we will elaborate on the meaning of each of these terms.

\paragraph{\textbf{Reconstruction Loss.}} This is an expectation over $\log p(X \vert Z; \theta)$, which we would like to maximize. In some ways, this measures the expressive power of the model. We would like to learn a model that can recreate all of the samples $\bm{x}$ in the dataset. For example, if we have a regression problem, $p(X \vert Z; \theta)$ can assume a Gaussian form. In this case, $\log p(X \vert Z; \theta)$ will correspond to the mean-squared error (MSE) between the actual data points and the corresponding model-reconstructed counterparts.

\paragraph{\textbf{KL term.}} This term measures the closeness of the variational density $q(Z \vert X; \nu)$ to a prior on latent variables $p(Z)$. What is this prior and how do we learn it? In the majority of VAE literature, including the first paper that introduced VAEs, it is assumed that the prior on $Z$ is a simple Gaussian with zero mean and unit variance. Is this assumption justified? It depends. For some problems this assumption is sufficient and it works, and for some, it doesn't. I believe the priors should be learned somehow, along with the rest of the model. Or there should be some structure on it that can be added as an inductive bias to the model, reflecting our intuitions about the latent space. Recent variants of hierarchical VAEs address this issue with some success \cite{vahdat2020nvae}. In Chapter 4, we will introduce a new hierarchical VAE that also contains a learned prior.

\subsection{Vanilla VAE}

In this section, we will define the simplest possible VAE that is trained using the objective in \autoref{eq:vaeloss-nn}. Before that, we still need to make some further assumptions and choices.

\paragraph{\textbf{i. Decoder $\bm{p(X \vert Z)}$.}} The choice of decoder distribution is problem-dependent. As mentioned earlier, a Gaussian distribution would be suitable for regression problems. One can use Bernoulli distribution when the data is binary, and Poisson loss for spiking data.

\paragraph{\textbf{ii. Encoder $\bm{q(Z \vert X)}$.}} A Gaussian distribution with mean $\mu(\bm{x})$ and covariance $\Sigma(\bm{x})$ is commonly used for the encoder. That is, $q(Z \vert X\!=\!\bm{x}) = \NN\big(\mu(\bm{x}), \Sigma(\bm{x})\big)$. Mean and covariance $\mu(\bm{x})$ and $\Sigma(\bm{x})$ can be implemented using neural networks and the sampling can be handled using the so-called \textit{reparameterization} trick (described below). Given a $d$-dimensional latent space, we will have $\mu(\bm{x}) \in \mathbb{R}^d$ and $\Sigma(\bm{x}) \in \mathbb{R}^{d \times d}$.

\paragraph{\textbf{iii. Prior $\bm{p(Z)}$.}} As discussed above, it is usually assumed that $p(Z) = \NN\big(0, I \big)$ which sometimes works for simpler problems but can fail for real-world data.

\paragraph{\textbf{iv. Mean-field approximation.}} This assumes latent variables don't interact with one another which allows us to factorize the variational density to $d$ independent distributions.  That is:

$$q(Z) = \prod_{i=1}^d \, q_i(Z_i).$$

Mean-field approximation adds further simplification to the problem and is commonly used with decent results. Under this approximation and using the Gaussian assumption for the encoder, the covariance $\Sigma(\bm{x})$ reduces to a diagonal matrix.

Now we are ready to write down the final form of the vanilla VAE objective by combining these assumptions with \autoref{eq:vaeloss-nn}, which leads us to \footnote{KL divergence of two Gaussians can be computed analytically. This is where the second term comes from.}:

\begin{equation}\label{eq:vanilla-vae-loss-nn}
    \begin{aligned}
    \{\theta^*; \nu^*\} &= \argmin_{\{\theta; \nu\}} \FF(\theta; \nu),
    \\
    \FF(\theta; \nu) &= - \expect{\bm{z} \sim q(Z \vert X; \nu)}{\log p(X \vert Z; \theta)}
    \\
    & + \frac{1}{2} \Big( \text{tr} \, \Sigma + \mu^T\mu - \log \text{det} \, \Sigma - d \Big).
    \end{aligned}
\end{equation}

For regression problems with Gaussian decoder distribution, the reconstruction loss will be reduced to:

\begin{equation}
    - \expect{\bm{z} \sim q(Z \vert X; \nu)}{\log p(X \vert Z; \theta)} \propto \sum_{\bm{x} \in X}\sum_{\bm{z} \sim q} \norm{\bm{x} - f_\theta(\bm{z})}^2.
\end{equation}

Here, $f_\theta$ is the neural net implementation of the decoder. With these simplifying assumptions, we now have concluded the derivation of the loss in closed form.

There is one last thing that we need to take care of: the loss in \autoref{eq:vanilla-vae-loss-nn} involves a sampling step. During model training, how can we backpropagate through a stochastic node? The answer is something known as the \textit{reparameterization} trick, which was introduced as a key idea in the original VAE paper \cite{kingma2014auto}.

\subsection{The reparameterization trick}

The idea here is to take the sampling step outside of the optimization loop by sampling from a normal Gaussian with zero mean and unit variance instead of $q(Z \vert X) = \NN\big(\mu(X), \Sigma(X)\big)$.  This is possible because given $\mu(X)$ and $\Sigma(X)$, instead of sampling $\bm{z} \sim \NN\big(\mu(X), \Sigma(X)\big)$ we can sample $\bm{\epsilon} \sim \NN\big(0, I \big)$ and compute $\bm{z} = \mu(x) + \Sigma^{\frac{1}{2}}(x) * \bm{\epsilon}$. Thus, we actually take the gradient of the following equation:

\begin{equation}\label{eq:reparam-grad}
    \begin{aligned}
    - \sum_{\bm{x} \in X} \,\,\, & \EXPECT{\bm{\epsilon} \sim \NN\big(0, I \big)}{\log p\big{(}X \vert Z\!=\!\mu(\bm{x}) + \Sigma^{\frac{1}{2}}(\bm{x}) * \bm{\epsilon}; \theta\big{)}} \\
     &+ \frac{1}{2} \Big{(} \text{tr} \, \Sigma(\bm{x}) + \mu(\bm{x})^T\mu(\bm{x}) - \log \text{det} \, \Sigma(\bm{x}) - d \Big{)}.
    \end{aligned}
\end{equation}

\begin{figure}[ht!]
    \centering
    \includegraphics[width=\textwidth]{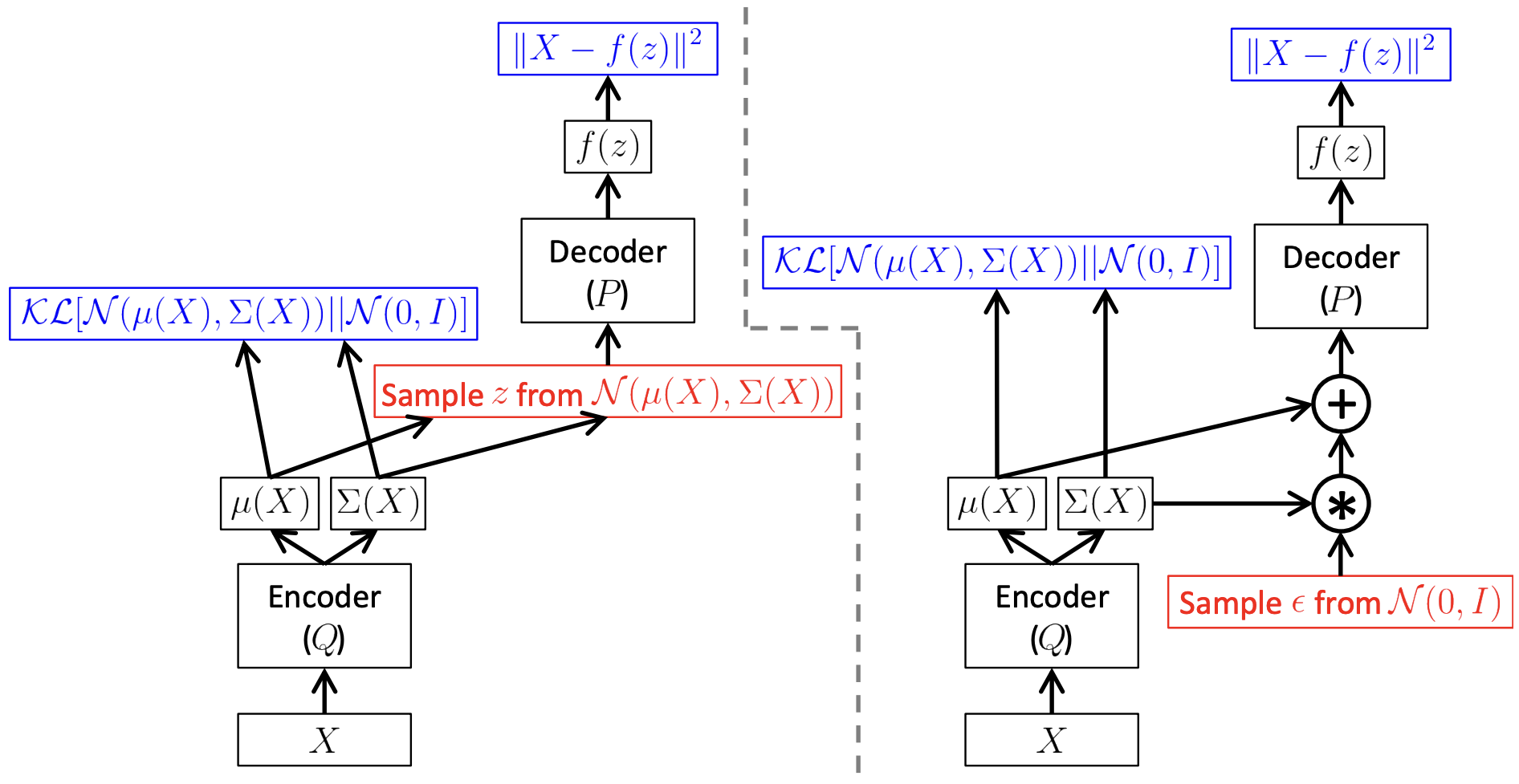}
    \caption[The reparameterization trick]{The reparameterization trick. Left shows the computation graph w/o the ``reparameterization trick", and right, with it. Red indicates sampling operations that are non-differentiable. Blue indicates loss layers. The feedforward behavior of these networks is identical, but backpropagation can be applied only to the network on the right. Figure adapted from \textcite{doersch2016tutorial}.}
    \label{fig:reparam}
\end{figure}

\Cref{fig:reparam} shows an illustration of the idea. During training, the expectation on $\bm{\epsilon}$ is replaced by a single sample drawn from $\NN\big(0, I \big)$ to get a $\bm{z}$ which can be used to compute the reconstruction loss. The expectation on $\bm{x}$ is replaced with a batch of data. During test time an $\bm{\epsilon}$ is simply drawn from $\NN\big(0, I \big)$ to generate new instances of the data (e.g., fake faces) using the learned model.\\

This concludes our introduction to VAE. In Chapter 4, we will expand upon this by endowing a VAE with a hierarchical latent structure. We will see that this single inductive bias enhances the learned VAE representations in many ways.

\section{A simple example with code}

In this section, I will present the results of training a simple VAE on the MNIST dataset\footnote{Code is available here: \href{https://github.com/hadivafaii/ConvVAE}{https://github.com/hadivafaii/ConvVAE}}. The data consists of $28 \times 28$ digits. The latent space is 2 dimensional which is perfect for visualization and is sufficient for simple datasets such as MNIST. The encoder is a stack of convolutional layers, and the decoder is made of transposed convolutions that reconstruct digits from the $2$-dimensional latent vectors. The choice of encoder and decoder architecture is irrelevant here. Similar results could be achieved using simple feedforward networks. Results are summarized in the figures below.

\subsection{Evaluating the results}

\begin{figure}[ht!]
    \centering
    \subfloat[Reconstruction performance. 1st row, real data samples; 2nd row, reconstructed versions.]{\includegraphics[width=0.47\textwidth]{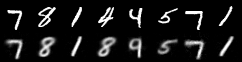}}
    \hfill
    \subfloat[Sample form prior $\NN(0, 1)$ to generate new instances of data.]{\includegraphics[width=0.47\textwidth]{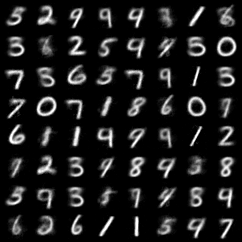}}
    \caption[VAE example: reconstruction and synthesis]{The model has achieved a decent \textit{understanding} of the data.}
    \label{fig:vae-recon}
\end{figure}

When evaluating VAEs, the ability to reconstruct existing samples from the dataset is a primary measure of performance. This reconstruction ability indicates how well the VAE has learned to encode and subsequently decode the input data back to a form that closely matches the original. The quality of these reconstructions is often assessed visually and quantitatively, for instance, through the use of error metrics such as MSE between the original and reconstructed images. High-fidelity reconstructions suggest that the VAE has effectively captured the essential features in the data. The reconstruction results shown in \cref{fig:vae-recon}a indicate that our toy VAE with a simple 2-dimensional latent space can achieve a decent performance on the MNIST dataset.

Another crucial aspect of VAE evaluation is the model's capacity to generate new, synthetic samples that are plausible within the context of the dataset. This generative capability is demonstrated by sampling from the latent space and passing these samples through the decoder to produce new outputs. The diversity and realism of these generated digits are indicative of the model's understanding of the data distribution. The generative performance can be qualitatively evaluated by human observers or quantitatively using metrics like the Fr\'echet Inception Distance (FID), which compares the feature distribution of generated images to real ones \cite{heusel2017gans}. The synthetic data generation results are shown in \cref{fig:vae-recon}b.

\begin{figure}[ht!]
    \centering
    \includegraphics[width=0.99\textwidth]{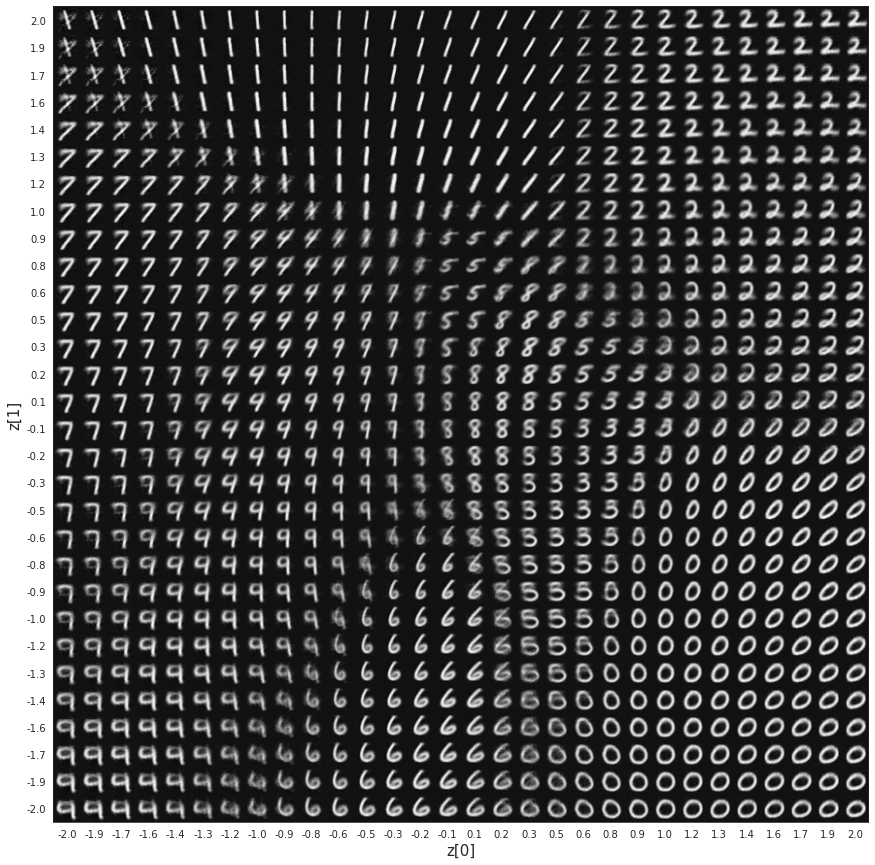}
    \caption[VAE example: latent space visualized]{Latent space visualized. This grid map shows data points generated by feeding values of the latent vector $\bm{z} \in [-2.0, 2.0]^2$ the decoder.}
    \label{fig:vae-latent}
\end{figure}

\begin{figure}[t!]
    \centering
    \includegraphics[width=\textwidth]{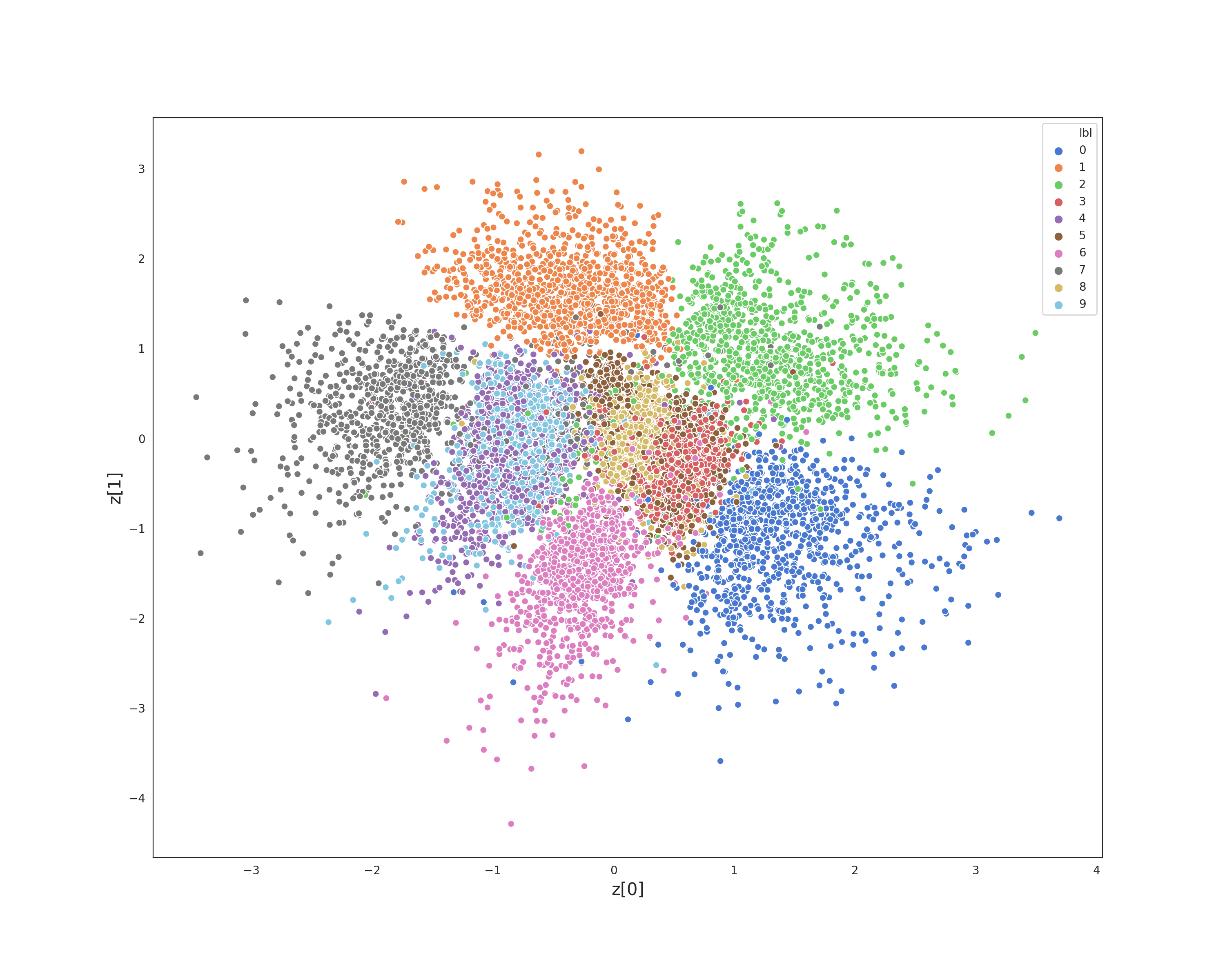}
    \caption[VAE example: clustering MNIST digits]{Clustering the digits.  Generated by pushing data points in the validation set into the encoder to get their corresponding $\bm{z}$. The emergence of such structures in the latent space is a sign of successful training in VAEs.}
    \label{fig:vae-cluster}
\end{figure}

Moreover, the organization of the latent space is a significant indicator of a well-trained VAE. A well-structured latent space is one in which similar digits are clustered together and different digits are separated. This can be visualized by plotting the latent representations of the dataset and observing the emergent structure. For a 2D latent space, one can use scatter plots to represent different classes (digits) and assess whether the VAE has learned a meaningful manifold. A good latent space should allow for smooth transitions and interpolations between points, which in turn should result in smooth variations in the generated images. In \cref{fig:vae-latent}, we show the result of interpolating along the two latent dimensions while plotting the generated outputs from the decoder. The digits are clearly clustered in this latent space, which we confirm in \cref{fig:vae-cluster}.

In summary, a VAE's performance is multifaceted and is evaluated based on its reconstruction accuracy, generative quality, and the interpretability of its latent space. By considering these aspects together, one gains a comprehensive understanding of the model's capabilities and limitations.

\section{Mixed-membership stochastic blockmodels}

In Chapter 5, we will use an algorithm to decompose the correlation structure of mouse cortical networks into overlapping communities of brain regions. The algorithm is called SVINET \cite{Gopalan2013EfficientDO}, and it belongs to a family of community detection algorithms known as \say{mixed-membership stochastic blockmodels} (MMSB, \cite{airoldi2008mixed}). In MMSB approaches, each community in a network corresponds to a \textit{latent functional role}, and these latent roles are typically inferred using techniques from variational inference. See \textcite{airoldi2015handbook} for an extensive review.

In the context of MMSB, variational inference plays a critical role. The model's objective is to infer the latent community memberships of each node, which involves estimating complex posterior distributions. Given the high-dimensional nature of network data, direct computation of these distributions is often impractical. Variational inference addresses this challenge by providing an approximate solution that is both efficient and scalable.

This approximation is achieved through the introduction of a simpler, tractable family of distributions. The variational inference algorithm iteratively adjusts the parameters of these distributions to closely match the true posterior distributions of community memberships. This approach not only enhances computational efficiency but also retains a high degree of accuracy in the membership estimations.

Specifically, SVINET \cite{Gopalan2013EfficientDO}, the algorithm we employ in Chapter 5, constructs and optimizes a generative model of binary graphs that aims to explain the existence (or non-existence) of a link between a pair of nodes based on their latent community memberships. Intuitively, pairs of nodes are more likely to be connected via a link if they belong to several latent communities together. More formally, SVINET has the following generative model:


\begin{equation}
\begin{aligned}
    q(\bm{\beta}, \bm{\theta}, \bm{z})
    \quad = \quad
    & \prod_{k=1}^K q\left(\beta_k \mid \lambda_k\right) \,\, \prod_{n=1}^N q\left(\theta_n \mid \gamma_n\right) \\
    & \prod_{i<j} q\left(z_{i \rightarrow j} \mid \phi_{i \rightarrow j}\right) q\left(z_{i \leftarrow j} \mid \phi_{i \leftarrow j}\right)
\end{aligned}
\end{equation}

Here, $N$ is the number of nodes and $K$ is the number of communities. The posterior over community indicators $\bm{z}$ is parameterized by the interaction parameters $\bm{\Phi}$, the posterior over $\bm{\theta}$ is parameterized by the community memberships $\bm{\gamma}$, and the posterior over $\bm{\beta}$ is parameterized by the community strengths $\bm{\lambda}$. Finally, the ELBO loss is used to optimize free model parameters.

The big picture summary of this algorithm is that it accepts binary graphs as observations, and infers an underlying latent community structure that would be the most likely \textit{explanation} or \textit{cause} of the observed graph and its particular patterns of links. For more information, please refer to the original publication by \textcite{Gopalan2013EfficientDO}.

In Chapter 5, we will apply this idea to functional graphs from the mouse cortex, where each link indicates the existence of a large correlation between pairs of brain regions. We will then infer an underlying lower-dimensional explanation of the observed correlation structure, which will reveal the \textit{functional organization} of the mouse cortex. 


\renewcommand{\thechapter}{4}

\chapter{Motion perception in primates}

The content of this Chapter has been published as a conference paper entitled: \sayit{Hierarchical VAEs provide a normative account of motion processing in the primate brain}, to be presented at the $37^{th}$ Annual Conference on Neural Information Processing Systems (NeurIPS 2023) \cite{vafaii2023hierarchical}. This was joint work between myself, Jacob Yates, and my advisor Dan Butts.

\section{Introduction}

Intelligent interactions with the world require representation of its underlying composition. This inferential process has long been postulated to underlie human perception \cite{helmholtz1867handbuch, alhazen, mumford1992computational, lee2003hierarchical, rao1999predictive, knill2004bayesian, yuille2006vision, friston2005theory, clark2013whatever}, and is paralleled in modern machine learning by generative models \cite{dayan1995helmholtz, kingma2014auto, rezende2014stochastic, schott2018towards, yildirim2020efficient, storrs2021unsupervised, higgins2021unsupervised, marino2022predictive}. Generative models aim to learn the underlying processes that cause their sensory observations. To achieve this goal, the models need to learn a \say{good} representation of the data.

The question of what constitutes a \say{good} representation has no clear answer \cite{higgins2018towards, locatello2019challenging}, but several desirable features have been proposed. In the field of neuroscience, studies focused on object recognition have suggested that effective representations \sayit{untangle} the various factors of variation in the input, rendering them linearly decodable \cite{dicarlo2007untangling, dicarlo2012does}. This intuitive notion has independently emerged in the machine learning community under different names such as \sayit{informativeness} \cite{eastwood2018framework} or \sayit{explicitness} \cite{ridgeway2018learning}. Additionally, it has been suggested that \sayit{disentangled} representations are desirable, wherein distinct, informative factors of variations in the data are separated \cite{bengio2013representation, lake2017building, peters2017elements, lecun2015deep, schmidhuber1992learning, tschannen2018recent}.

Artificial neural networks (ANN) are also increasingly evaluated based on their alignment with biological neural processing \cite{schrimpf2018brain, yamins2014performance, khaligh2014deep, yamins2016using, mineault2021your, elmoznino2022high, conwell2023can, sexton2022reassessing, tuckute2022many}, because of the shared goals of ANNs and the brain's sensory processing \cite{lake2017building, richards2022application, zador2023catalyzing}. Such alignment also provides the possibility of gaining insights into the brain by understanding the operations within an ANN \cite{doerig2023neuroconnectionist, kanwisher2023using, cao2021explanatory, richards2019deep, barrett2019analyzing, serre2019deep, kriegeskorte2015deep}.

In this Chapter, we investigate how the combination of (i) model architecture, (ii) loss function, and (iii) training set, affects learned representations, and whether this is related to the brain-alignment of the ANN \cite{richards2019deep}. We focus specifically on understanding the representation of motion because processing motion constitutes a fundamental visual task \cite{lee1980optic, simoncelli1993distributed}, with large sections of the visual cortex devoted to this task \cite{mineault2021your}. Furthermore, the causes of retinal motion (moving objects and self-motion \cite{gibson1954visual}) can be manipulated systematically.

Crucially, motion in an image can be described irrespective of the identity and specific visual features that are moving, just as the identity of objects is invariant to how they are moving. This separation of motion and object processing mirrors the division of primate visual processing into dorsal (motion) and ventral (object) streams \cite{ungerleider1982two, goodale1992separate, ungerleider2008what}.

We designed a \textit{naturalistic} motion simulation based on distributions of ground truth variables corresponding to the location and depth of objects, motion of these objects, motion of the observer, and their direction of gaze (i.e., the fixation point; \cref{fig:1_rofl}). Using our simulated retinal flow, we then trained and evaluated an ensemble of autoencoder-based models. These models were trained in an unsupervised way, without any access to ground truth factors during training.

We based our model evaluation on (1) whether the models untangle and disentangle the ground truth variables in our simulation; and (2) the degree to which their latent spaces could be directly related to neural data recorded in the dorsal stream of primates (area MT).

We introduce a new hierarchical variational autoencoder (cNVAE) based on compressing NVAE \cite{vahdat2020nvae}. The cNVAE exhibited superior performance compared to other models in all of our evaluation metrics. First, it discovered latent factors that accurately captured the ground-truth variables in the simulation in a more disentangled manner than other models. Second, it achieved significant improvements in predicting neural responses compared to the previous state-of-the-art model \cite{mineault2021your}, with a performance gain of over 2x and sparse mapping from its latent space to neuronal responses. 

Taken together, these observations demonstrate the power of the synthetic data framework and show that a single inductive bias---hierarchical latent structure---leads to many desirable features of representations, including brain alignment.

\section{Background \& Related Work}
\subsection{Neuroscience and VAEs}
It has long been argued that perception and inference are tightly linked. This concept is commonly referred to as \say{perception as unconscious inference}, initially proposed by Helmholtz in the 19th century \cite{helmholtz1867handbuch}. Similar ideas can be traced back more than a millennium to Alhazen \cite{alhazen}, and possibly even earlier.

More recently, the hierarchical architecture of the cortex, paired with the larger volume of feedback connections relative to feedforward connections, inspired \textcite{mumford1992computational} to conjecture that brains engage in hierarchical Bayesian inference to comprehend the world \cite{mumford1992computational, lee2003hierarchical}. We explored this concept more thoroughly in Chapter 1, specifically see \cref{sec:hier-bayes-inf} and \cref{fig:mumford92}.

There has been much influential work following different aspects of how this idea connects to neuroscience, including Predictive Coding \cite{srinivasan1982predictive, rao1999predictive, clark2013whatever, bastos2012canonical, dong1995temporal, hohwy2008predictive, hohwy2016self, lotter2017deep, singer2021recurrent, mikulasch2023error, millidge2022predictive}, Bayesian Brain Hypothesis \cite{knill2004bayesian, knill1996perception, weiss2002motion, geisler2002illusions, vilares2011bayesian, gregory1980perceptions, lochmann2011neural, shivkumar2018probabilistic}, Analysis-by-Synthesis \cite{yuille2006vision}, and Free Energy Principle \cite{friston2005theory}.

Notably, these ideas are closely paralleled by VAEs whose loss function can be derived from the requirement of effective posterior inference---a normative requirement that resonates well with Helmholtz's original conjecture. See Chapter 3 for more details.

The deep connections between VAEs and neuroscience is explored in a review by \textcite{marino2022predictive}, but work looking at the synergy between neuroscience and VAEs has thus far been limited to the following recent papers.

\textcite{higgins2021unsupervised} trained $\beta$-VAE \cite{higgins2017beta} models on face images and evaluated the models on predicting responses of individual inferotemporal neurons from Macaque face patch. They found that $\beta$-VAE discovers individual latents that are aligned with IT neurons and that this alignment linearly increases with increased beta values.

\textcite{storrs2021unsupervised} trained PixelVAE \cite{gulrajani2017pixelvae} models on synthesized images of glossy surfaces and found that the model spontaneously learned to cluster images according to underlying physical factors like material reflectance and illumination, despite receiving no explicit supervision. Most notably, PixelVAE mimicked human perceptual judgment of gloss in that it made errors that were similar to those made by humans.

\textcite{csikor2022topdown} investigated the properties of representations and inference in a biologically inspired hierarchical VAE called Topdown VAE that is capable of reproducing some key properties of representations emerging in V1 and V2 of the visual cortex. The authors showed that architectures that feature top-down influences in their recognition model give rise to a richer representation, such that specific information that is not present in mean activations becomes linearly decodable from posterior correlations.

Finally, \textcite{miliotou23gen} learned mapping from the latent space of a hierarchical VAE to fMRI voxel space. This allowed the use of the trained decoder to reconstruct images directly from brain responses. Importantly, they perform ablation experiments and find hierarchy is an essential component of their models.

\subsection{Hierarchical VAEs}
An early influential work that introduced hierarchy to the latent space of VAEs was Ladder VAE (LVAE, \textcite{sonderby2016ladder}), which improved upon standard VAEs by sharing information between the inference and generative networks, allowing LVAEs to learn deeper and more distributed latent representations compared to regular VAEs. Building on LVAE, \textcite{maaloe2019biva} introduced Bidirectional-Inference Variational Autoencoder (BIVA), a model that uses skip-connections to further enhance the flow of information among latent variables and better avoids inactive units. Both the LVAE and BIVA made valuable contributions towards enabling VAEs to effectively leverage deep stochastic hierarchies, through innovations in the inference model design and its relationship to the generative network. Finally, \textcite{havtorn19hierarchical} showed that hierarchical VAEs learn low-level features that are shared across datasets, enabling out-of-distribution generalization.

More recent hierarchical VAE variants further improved performance. The Nouveau VAE (NVAE, \textcite{vahdat2020nvae}), is a VAE equipped with a hierarchical architecture that achieved state of the art in several benchmarks and generated high-quality faces. Very deep VAE (vdvae, \textcite{child2021vdvae}) achieved similarly impressive performance on complex image benchmarks. Additionally, the vdvae paper demonstrated that hierarchical VAEs perform well because they are deep in a stochastic sense (i.e., several stages of sampling vs.~just once for the vanilla VAE). This key conceptual advance provided an explanation of why non-hierarchical VAEs previously struggled to compete against autoregressive models \cite{child2021vdvae}.

In sum, both NVAE and vdvae achieved impressive performance in difficult benchmarks, while a central motivation of NVAE was to improve the quality of generated samples. However, neither work focused on evaluating how the hierarchical latent structure alters and potentially enhances the quality of learned representations in downstream applications. 

Despite their success, NVAE and vdvae have an undesirable property: the convolutional nature of their latents results in an overly redundant latent space that is several orders of magnitude larger than the number of variables in the input space. This is at odds with the principle of \say{parsimony} (promoting compact and efficient representations) and defeats one of the main purposes of the autoencoder family. \textcite{hazami2022efficientvdvae} showed that for vdvae, only a tiny subset $\left(\sim 3\%\right)$ of its latent space is necessary, and the rest is there for marginal improvements on log-likelihood performance.

Here, we demonstrated that it is possible to compress hierarchical VAEs and focus on investigating the functionality of individual latents. To the best of our knowledge, this work is the first study focused on the evaluation of representations in hierarchical VAEs with applications to neuroscience data.

\subsection{Evaluating ANNs on predicting biological neurons}
Several studies have focused on evaluating ANNs on their performance in predicting brain responses, but almost entirely on describing static (\say{ventral stream}) image processing \cite{schrimpf2018brain, yamins2016using, conwell2023can}. In contrast, motion processing (corresponding to the dorsal stream) has only been considered thus far in \textcite{mineault2021your}, who used a 3D ResNet (DorsalNet) to extract ground truth variables about self-motion from drone footage (\say{AirSim}, \cite{shah2018airsim}) in a supervised manner. DorsalNet learned representations with receptive fields that matched known features of the primate dorsal stream and achieved state-of-the-art on predicting neural responses on the dataset that we consider here.

There are many differences between our approach and that of \textcite{mineault2021your}, including model architecture and training set. However, the most fundamental difference is that \textcite{mineault2021your} train their models using direct supervision, and we do not. As such, their models have access to the ground truth variables at all times. Here, we demonstrate that it is possible to obtain ground-truth variables \say{for free}, in a completely unsupervised manner, while achieving much better performance in predicting responses of biological neurons.

\subsection{Using synthetic data to train ANNs}

A core component of a reductionist approach to studying the brain is to characterize neurons based on their selectivity to a particular subset of pre-identified visual \say{features}, usually by presenting sets of \say{feature-isolating} stimuli \cite{rust2005praise}. In the extreme, stimuli are designed that remove all other features except the one under investigation \cite{julesz1971foundations}. While these approaches can inform how pre-selected feature sets are represented by neural networks, it is often difficult to generalize this understanding to more natural stimuli, which are not necessarily well-described by any one feature set. As a result, here we generate synthetic data representing a \textit{naturalistic} distribution of natural motion stimuli. Such synthetic datasets allow us to manipulate the causal structure of the world, in order to make hypotheses about what aspects of the world matter for the representations learned by brains and ANNs \cite{golan2020controversial}.

Like previous work on synthesized textures \cite{storrs2021unsupervised}, here we specifically manipulate the data generative structure to contain factors of variation due to known ground truth variables.

\section{Retinal Optic Flow Learning (ROFL)}

\begin{figure}[ht!]
    \centering
    \includegraphics[width=0.56\linewidth]{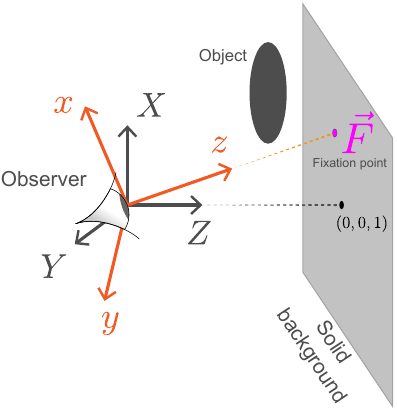}
    \caption[Introducing Retinal Optic Flow Learning (ROFL)]{Introducing Retinal Optic Flow Learning (ROFL), a simulation platform for synthesizing naturalistic optic flow patterns. The general setup includes a moving or stationary observer and a solid background, with optional moving object(s) in the scene (more details in \cref{sec:rofl}).}
    \label{fig:1_rofl}
\end{figure}

Our synthetic dataset framework, ROFL, generates the resulting optic flow from different world structures, self-motion trajectories, and object motion (\cref{fig:1_rofl}).

ROFL can be used to generate \textit{naturalistic} flow fields that share key elements with those experienced in navigation through 3-D environments. Specifically, each frame contains global patterns that are due to self-motion, including rotation that can arise due to eye or head movement \cite{gibson1950perception, warren1988direction}. In addition, local motion patterns can be present due to objects that move independently of the observer \cite{gibson1954visual}. The overall flow pattern is also affected by the observer's direction of gaze (fixation point \cite{thomas1994spherical}, \cref{fig:1_rofl}).

ROFL generates flow vectors that are instantaneous in time, representing the velocity across the visual field resulting from the spatial configuration of the scene and motion vectors of self and object. Ignoring the time-evolution of a given scene (which can arguably be considered separably \cite{matthis2022retinal}) dramatically reduces the input space from $\left[3 \times H \times W \times T\right]$ to $\left[2 \times H \times W\right]$, and allows a broader sampling of configurations without introducing changes in luminance and texture. As a result of this simplifying choice, we can explore the role of different causal structures in representation learning in ANNs.

From an observer's perspective, the retinal flow patterns generated by a moving object depend on the observer's self-motion and the rotation of their eyes as they maintain fixation in the world. One such interaction is visualized in \cref{fig:1_rofl_trav}, top row. In this example, the observer is moving forward, and the object is moving to the right. As a result of interactions between self and object motion, an object on the left side will have its flow patterns distorted, while an object on the right will have its flow patterns largely unaffected because its flow vectors are parallel with that of the self-motion.

In summary, ROFL allows us to simulate retinal optic flow with a known ground truth structure driven by object and self-motion.

\begin{table}[t]
    \caption[ROFL categories used in this Dissertation]{ROFL categories used in this Dissertation. Ground truth variables include fixation point $(+2)$; velocity of the observer when self-motion is present $(+3)$; and, object position \& velocity $(+6)$. \Cref{fig:1_rofl-eg} showcases a few example frames for each category. The stimuli can be rendered at any given spatial scale $N$, yielding an input shape of $2 \times N \times N$. Here we work with $N=17$.}
    \label{tab:rofl}
    \centering
    \begin{tabular}{lcr}
        \toprule
        Category & Description & Dimensionality \\
        \midrule
        \fixate{1} & \begin{tabular}[x]{@{}c@{}}A moving observer maintains fixation on a background point.\\In addition, the scene contains one independently moving object.\end{tabular} & $11 = 2 + 3 + 6$\\
        \midrule
        \fixate{0} & Same as \fixate{1} but without the object. & $5 = 2 + 3$\\
        \midrule
        \obj{1} & A single moving object, stationary observer. & $8 = 2 + 6$\\
        \bottomrule
    \end{tabular}
\end{table}

\begin{figure}[ht!]
    \centering
    \includegraphics[width=\linewidth]{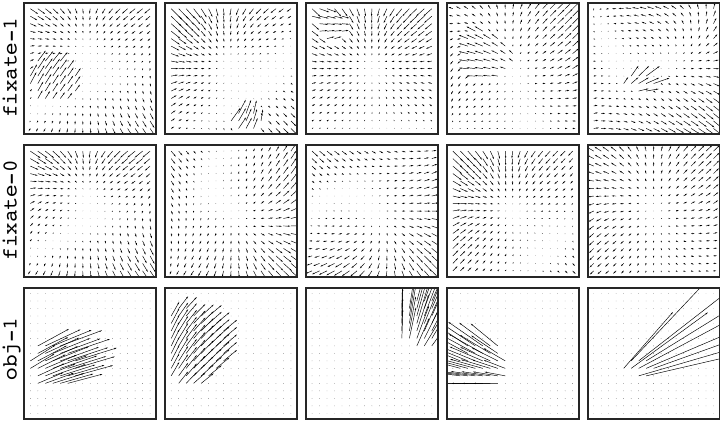}
    \caption[ROFL categories]{Example frames showcasing different categories. See Table~\ref{tab:rofl} for definitions.}
    \label{fig:1_rofl-eg}
\end{figure}

\begin{figure}[ht!]
    \centering
    \includegraphics[width=\linewidth]{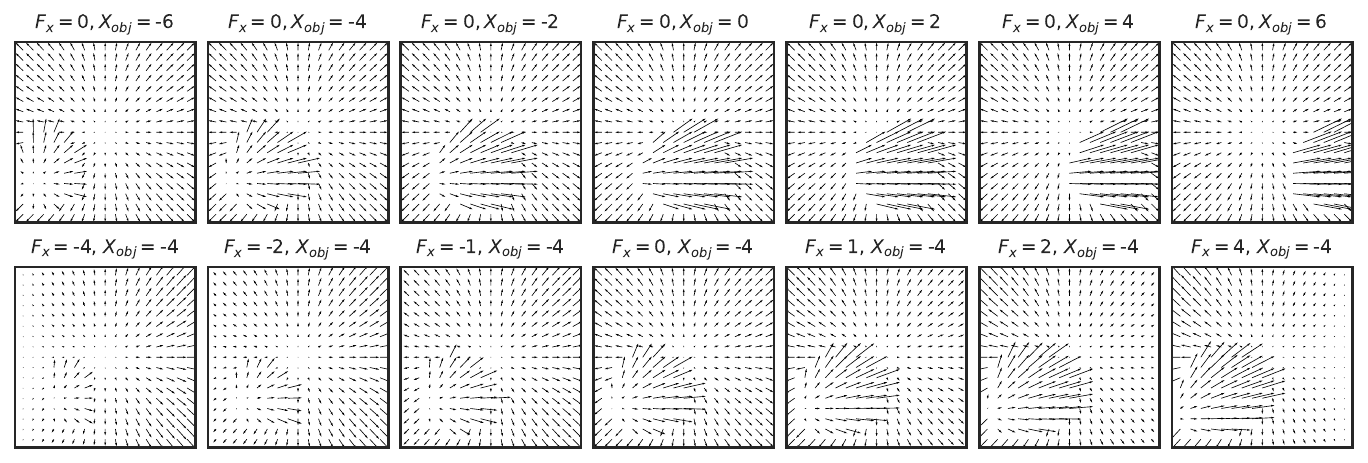}
    \caption[Interpolating ROFL ground truth factors]{Demonstrating the causal effects of varying a single ground truth variable while keeping all others fixed. Top: $X_{obj}$, the $x$ component of object position (measured in retinal coordinates, orange). Bottom: $F_x$, the $X$ component of the fixation point (measured in fixed coordinates, gray).}
    \label{fig:1_rofl_trav}
\end{figure}

\section{The compressed NVAE (cNVAE)}

\begin{figure}[ht!]
    \centering
    \caption[Model architecture comparison]{Architecture comparison. Left, compressed NVAE (cNVAE); right, non-hierarchical VAE. We modified the NVAE \emph{sampler} layer (grey trapezoids) and introduced a deconvolution \emph{expand} layer (yellow trapezoids). Numbers like $2 \times 2$ and so on indicate the spatial scale of latent groups. The encoder (inference) and decoder (generation) pathways are depicted in red and blue, respectively. $x$, input data (i.e., optic flow fields); $\hat{x}$, model reconstruction of the input; $z_i$, hierarchical latent groups; $z$, non-hierarchical latent variables; $r$, residual block; $h$, trainable parameter; $+$, feature combination, such as addition or concatenation.}
    \label{fig:2_model}
    \includegraphics[width=0.95\linewidth]{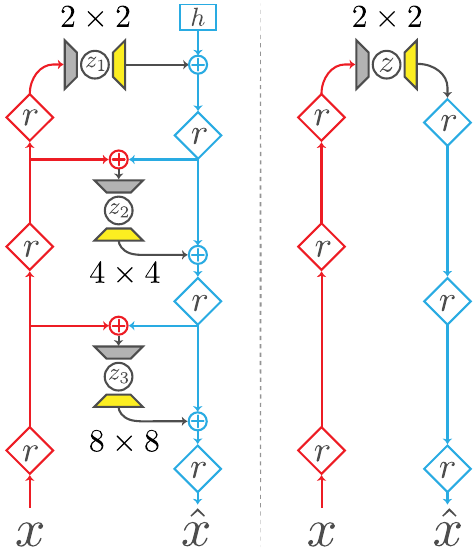}
\end{figure}

Our model architecture is shown in \cref{fig:2_model}. The cNVAE closely follows the NVAE \cite{vahdat2020nvae} with one important difference: the original NVAE latent space is convolutional, and ours is not. We modified the \textit{sampler} layers (grey trapezoids, \cref{fig:2_model}) such that their receptive field sizes match the spatial scale they operate on, which effectively integrates over spatial information before sampling from the approximate posterior. To project latents back to convolution space, we introduce \textit{expand} modules (yellow trapezoids, \cref{fig:2_model}), which are a simple deconvolution step.

Our modification of the NVAE serves two purposes. First, it decouples spatial information from the functionality of latent groups, allowing them to capture abstract features that are invariant to particular spatial locations. Second, it has the effect of compressing the input space into a lower-dimensional latent code.

A convolutional NVAE latent group operating at spatial scale $s$ has $d \times s \times s$ latent dimensions, where $d$ is the number of latents per group. In contrast, each latent group in our model has just $d$ dimensions. Thus, a cNVAE with $21$ latent groups, each with $20$ latent dimensions, has a $420$ dimensional latent space (see below), which is smaller than the input dimensionality of $578$ ($= 2 \times 17 \times 17$). In contrast, the NVAE would have a $17,520$-dimensional latent space using otherwise similar settings as above. In the next section, we will provide more details regarding the processing of the sampler and expand layers and model dimensionality.

\subsection{Key architectural differences between cNVAE and NVAE}\label{sec:arch-details}

Our main contribution to architecture design lies in the modification of the NVAE \sayit{sampler} layer and the introduction of the \sayit{expand} layer, which are simple deconvolutions. In the NVAE, the purpose of a sampler layer is to map encoder features to mean and variance vectors $\bm{\mu}(\bm{x})$ and $\bm{\sigma}^2(\bm{x})$, which are then used to construct approximate posterior distributions $q\left(\bm{z}\vert\bm{x}\right) = \mathcal{N}(\bm{\mu}(\bm{x}), \bm{\sigma}^2(\bm{x}))$. Note that during inference, downsampling operations will progressively reduce the spatial scale of representations along the encoder pathway, such that processed stimuli will have different spatial dimensions at different stages of inference. In the NVAE, all sampler layers are convolutional with a fixed kernel size of $3 \times 3$ and padding of 1. Thus, the application of a sampler layer does not alter the spatial dimension of hidden encoder features, resulting in a convolutional latent space.

For example, at level $\ell$ of the hierarchy, $\bm{\mu}_\ell\left(\bm{x}\right)$ and $\bm{\sigma}_\ell^2\left(\bm{x}\right)$ will have shapes like $d \times s_\ell \times s_\ell$, where $d$ is number of latent variables per latent group, and $s_\ell$ is the spatial scale of the processed encoder features at level $\ell$. As a result, NVAE latents themselves end up being convolutional in nature, with shapes like $d \times s_\ell \times s_\ell$. This design choice leads to the massive expansion of NVAE latent dimensionality, which is often orders of magnitude larger than the input dimensionality.

To address this, we modified NVAE sampler layers in a manner that would integrate out spatial information before sampling, achieving a compressed, non-convolutional, and more abstract latent space. Specifically, we designed the cNVAE sampler layers such that their kernel sizes are selected adaptively to match the spatial scale of their input feature. For example, if latent group $\ell$ operates at scale $s_\ell$, then the sampler layer will have a kernel size of $s_\ell \times s_\ell$, effectively integrating over space before the sampling step. This results in latents with shape $d \times 1 \times 1$.

Finally, combining the latents with decoder features requires projecting them back into convolutional space. We achieve this by introducing \sayit{expand} modules which are simple deconvolution layers with a kernel size of $s_\ell \times s_\ell$. Thus, processing a $d \times 1 \times 1$ latent by an expand layer results in a $d \times s_\ell \times s_\ell$ output, allowing them to be concatenated with the decoder features (blue \say{+} in \cref{fig:2_model}).

\subsection{Structure of cNVAE latent space}

The latent space of cNVAE contains a hierarchy of latent groups sampled sequentially: \say{top} latents are sampled first, all the way down to \say{bottom} latents that are closest to the stimulus (\cref{fig:2_model}). \textcite{child2021vdvae} demonstrated that model depth, in this statistical sense, was responsible for the success of hierarchical VAEs. Here, we increased the depth of cNVAE architecture while ensuring that training remained stable. Our resulting model had the following structure: 3 latent groups operating at the scale of $2 \times 2$; 6 groups at the scale of $4 \times 4$; and 12 groups at the scale of $8 \times 8$ (\cref{fig:2_model}, \Cref{tab:latent-dims}). Therefore, the model had $3 + 6 + 12 = 21$ hierarchical latent groups in total. Each latent group had $20$ latents, which resulted in an overall latent dimensionality of $21 \times 20 = 420$. More details are provided below in \cref{sec:latent-dims}. See also \Cref{tab:latent-dims}.

\begin{table}[t]
    \caption[Model and loss function details]{Model details. Here, \emph{hierarchical} means that there are parallel pathways for information to flow from the encoder to the decoder (\cref{fig:2_model}), which is slightly different from the conventional notion. For variational models, this implies hierarchical dependencies between latents in a statistical sense \cite{child2021vdvae}. This hierarchical dependence is reflected in the KL term for the cNVAE, where $L$ is the number of hierarchical latent groups. See \cref{sec:latent-dims} for more details and \cref{sec:hier-derive} for a derivation. All models have an equal \# of latent dimensions ($420$, see \Cref{tab:latent-dims}), approximately the same \# of convolutional layers, and \# of parameters $(\sim24~M)$. EPE, endpoint error.}
    \label{tab:model}
    \centering
    \begin{tabular}{lllc}
        \toprule
        Model & Architecture & Loss & Kullback–Leibler term (KL) \\
        \midrule
        {\color{cnvae_color}cNVAE} & Hierarchical & EPE $+ \beta * \mathrm{KL}$ & \begin{tabular}[x]{@{}c@{}c@{}}$\mathrm{KL} = \sum_{\ell = 1}^L \mathbb{E}_{q(\bm{z}_{<\ell} \vert \bm{x})}\big[\mathrm{KL}_\ell\big],$ where\\[0.8ex] $\mathrm{KL}_\ell \coloneqq \mathcal{D}_{\mathrm{KL}}\big[q(\bm{z}_\ell \vert \bm{x}, \bm{z}_{<\ell}) \,\|\, p(\bm{z}_\ell \vert \bm{z}_{<\ell})\big]$\end{tabular}\\
        \midrule
        {\color{vae_color}VAE} & Non-hierarchical & EPE $+ \beta * \mathrm{KL}$ & $\mathrm{KL} = \mathcal{D}_{\mathrm{KL}}\big[q(\bm{z} \vert \bm{x}) \,\|\, p(\bm{z})\big]$\\
        \midrule
        {\color{cnae_color}cNAE} & Hierarchical & EPE & - \\
        \midrule
        {\color{ae_color}AE} & Non-hierarchical & EPE & - \\
        \bottomrule
    \end{tabular}
\end{table}

\section{Approach}

\subsection{Alternative models}

We evaluated a range of unsupervised models alongside cNVAE, including standard (non-hierarchical) VAEs \cite{kingma2014auto, rezende2014stochastic}, a hierarchical autoencoder with identical architecture as the cNVAE but trained only with reconstruction loss (cNAE), and an autoencoder (AE) counterpart for the VAE. All models had the same latent dimensionality, and approximately the same number of parameters and number of convolutional layers. See \Cref{tab:model} for a summary.

We used endpoint error as our measure of reconstruction loss, which is the Euclidean norm of the difference between actual and reconstructed flow vectors. This metric is known to work well with optical flow data \cite{ilg2017flownet}.

\subsection{Model representations}

We define a model's internal representation to be either the mean of each Gaussian for variational models (i.e., samples drawn from $q\left(\bm{z}|\bm{x}\right)$ at zero temperature), or the bottleneck activations for autoencoders. For hierarchical models (cNVAE, cNAE), we concatenate representations across all levels (\Cref{tab:latent-dims}).

\subsection{Training details}

Models were trained for $160~k$ steps at an input scale of $17 \times 17$, requiring slightly over a day on Quadro RTX 5000 GPUs. Please refer to the supplementary \cref{sec:train-details} for additional details.

\subsection{Disentanglement and \texorpdfstring{$\beta$}{beta}-VAEs}

A critical decision when optimizing VAEs involves determining the weight assigned to the KL term in the loss function compared to the reconstruction loss. Prior research has demonstrated that modifying a single parameter, denoted as $\beta$, which scales the KL term, can lead to the emergence of disentangled representations \cite{higgins2017beta, burgess2018understanding}.

Most studies employing VAEs for image reconstruction typically optimize the standard evidence lower bound (ELBO) loss, where $\beta$ is fixed at a value of 1 \cite{kingma2014auto, vahdat2020nvae, child2021vdvae}. However, it should be noted that due to the dependence of the reconstruction loss on the input size, any changes in the dimensionality of the input will inevitably alter the relative contribution of the KL term, and thus the \say{effective} $\beta$ \cite{higgins2017beta}.

Furthermore, \textcite{higgins2021unsupervised} recently established a strong correspondence between the generative factors discovered by $\beta$-VAEs and the factors encoded by inferotemporal (IT) neurons in the primate ventral stream. The alignment between these factors and IT neurons exhibited a linear relationship with the value of $\beta$. In light of these findings, we explicitly manipulate the parameter $\beta$ within a range spanning from $0.01$ to $10$ to investigate the extent to which our results depend on its value.

\section{Results}

Our approach is based on the premise that the visual world contains a hierarchical structure. We use a simulation containing a hierarchical structure (ROFL, described above) and a hierarchical VAE (the cNVAE, above) to investigate how this architectural choice affects latent representations. While we are using a relatively simple simulation generated from a small number of ground-truth variables, $\bm{g}$, we do not specify how $\bm{g}$ should be represented in our model or include $\bm{g}$ in the loss. Rather, we allow the model to develop its own latent representation in a purely unsupervised manner. More details about our approach are provided in \cref{sec:approach}.

\subsection{Robust reconstruction performance across all models}
\begin{figure}[ht!]
    \centering
    \includegraphics[width=1.0\linewidth]{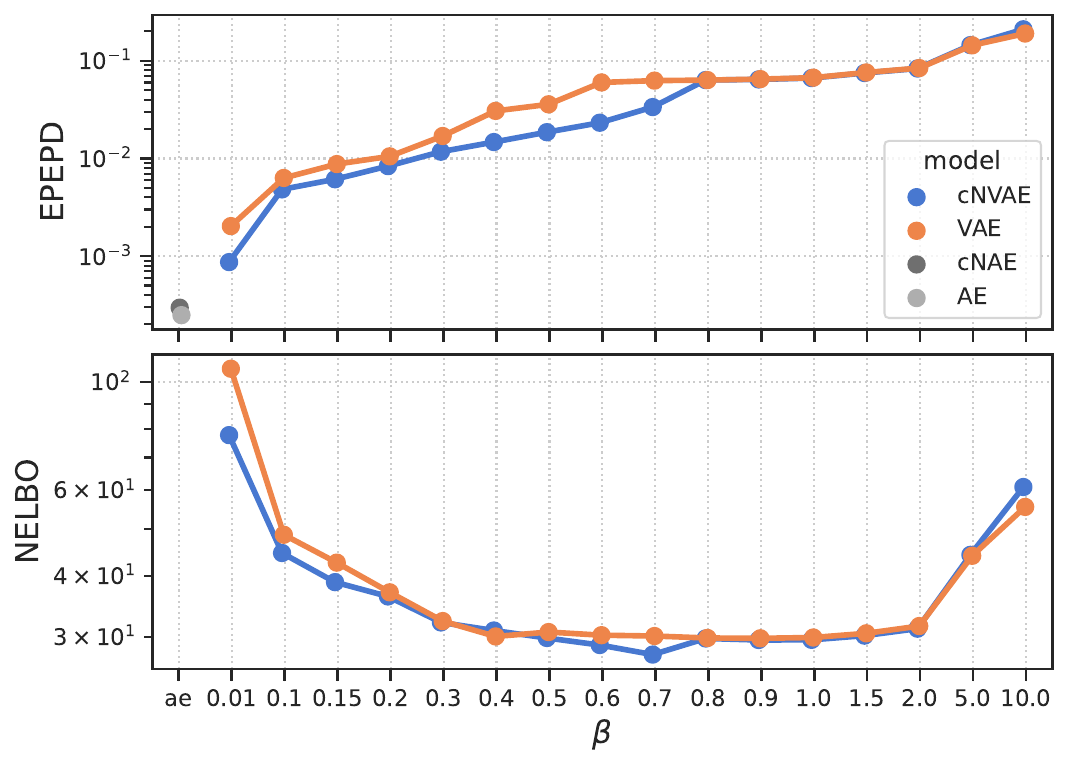}
    \caption[Reconstruction performance of cNVAE vs.~VAE]{Both models demonstrate robust reconstruction performance, with cNVAE exhibiting a slight improvement. EPEPD, endpoint error per dim (lower is better). NELBO, negative evidence lower bound (lower is better).}
    \label{fig:13_losses_vs_beta}
\end{figure}

The first step in evaluating VAEs is to test their reconstruction performance. \Cref{fig:13_losses_vs_beta} illustrates how the reconstruction loss (EPEPD, endpoint error per dim) and negative evidence lower bound (NELBO) depend on different values of $\beta$ and model types. Both hierarchical and non-hierarchical VAE models perform well, with cNVAE exhibiting slightly superior performance for $\beta < 0.8$.

\subsection{Functional specialization emerges in the cNVAE} 

For our first experiment in representation learning, we trained and tested hierarchical and non-hierarchical VAEs on \fixate{1} (see \Cref{tab:rofl}; throughout this work, \fixate{1} is used unless stated otherwise). We then extracted latent representations from each model and estimated the mutual information (MI) between the representations and ground truth variables such as self-motion, etc. For \fixate{1}, each data sample is uniquely determined using 11 ground truth variables (\Cref{tab:rofl}), and the models have latent dimensionality of $420$ (\Cref{tab:latent-dims}). Thus, the resulting MI matrix has shape $11 \times 420$, where each entry shows how much information is contained in that latent variable about a given ground truth variable.

\begin{figure}[ht!]
    \centering
    \includegraphics[width=1.0\linewidth]{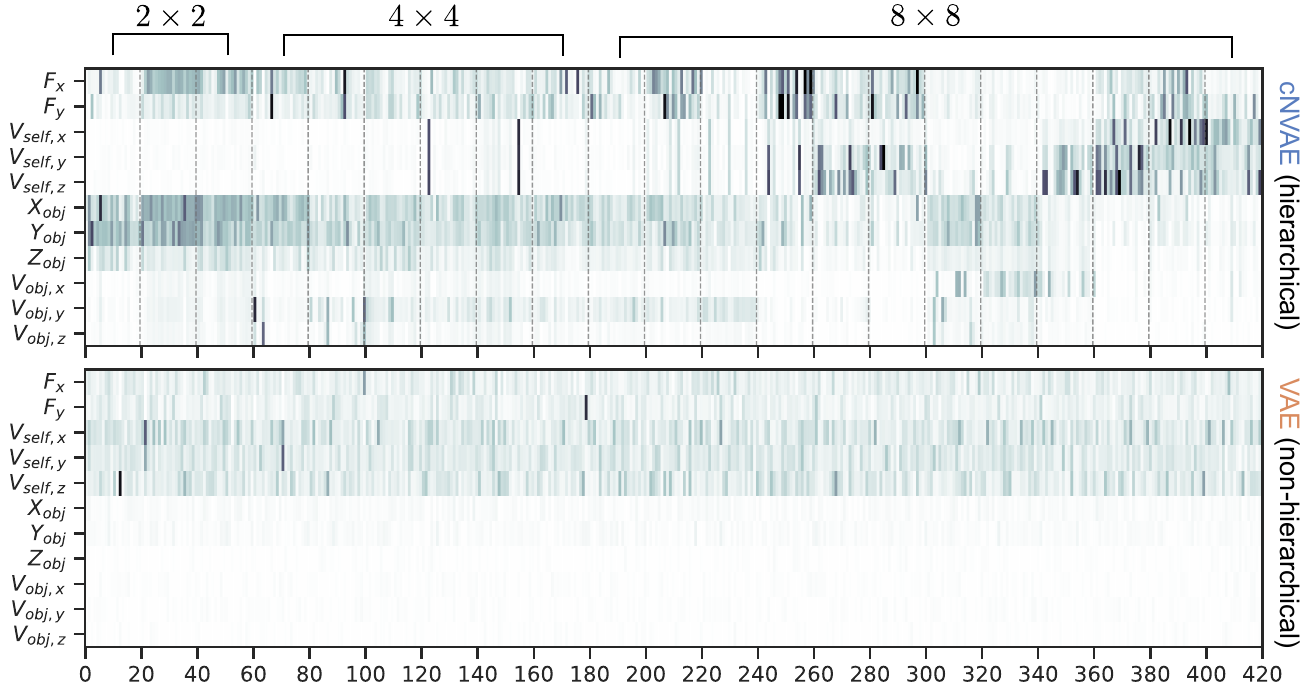}
    \caption[Mutual information between latent variables and ground truth factors]{Mutual information between latent variables (x-axis) and ground truth variables (y-axis) is shown for cNVAE (top) and VAE (bottom). Dashed lines indicate $21$ hierarchical latent groups of $20$ latents each, comprising a $420$-dimensional latent space. These groups operate at three different spatial scales, as indicated. In contrast, the VAE latent space lacks such grouping and operates solely at the spatial scale of $2 \times 2$ (see \cref{fig:2_model} and \Cref{tab:latent-dims} for details on model latent configurations).}
    \label{fig:3_mi}
\end{figure}

\Cref{fig:3_mi} shows the MI matrix for the latent space of cNVAE (top) and VAE (bottom). While both models achieved a good reconstruction of validation data as shown above (\cref{fig:13_losses_vs_beta}), the MI matrix for cNVAE exhibits clusters corresponding to distinct ground-truth variables at different levels of the hierarchy. Specifically, object-related factors of variation are largely captured at the top $2 \times 2$ scale, while information about fixation point can be found across the hierarchy, and self-motion is largely captured by $8 \times 8$ latent groups.

In contrast, non-hierarchical VAE has no such structure, suggesting that the inductive bias of hierarchy enhances the quality of latent spaces, which we quantify next.

To quantitatively measure the clear relationship between ground truth variables and latent representations discovered by the cNVAE visible in \cref{fig:3_mi}, we apply metrics referred to as \say{untangling}, \say{disentengling}. Additionally, in a separate set of experiments, we also evaluate model representations by relating them to MT neuron responses, which we call \say{brain-alignment}. We discuss each of these in detail in the following sections.

\subsection{Untangling: the cNVAE untangles factors of variation}

One desirable feature of a latent representation is whether it makes information about ground truth factors easily (linearly) decodable \cite{kriegeskorte2019peeling, dicarlo2007untangling, dicarlo2012does}. This concept has been introduced in the context of core object recognition as \sayit{untangling}. Information about object identity that is \say{tangled} in the retinal input is untangled through successive nonlinear transforms, thus making it easier for higher brain regions to extract \cite{dicarlo2007untangling}. This concept is closely related to \sayit{informativeness} \cite{eastwood2018framework} and \sayit{explicitness} \cite{ridgeway2018learning} metrics.

\begin{figure}[ht!]
    \centering
    \includegraphics[width=1.0\linewidth]{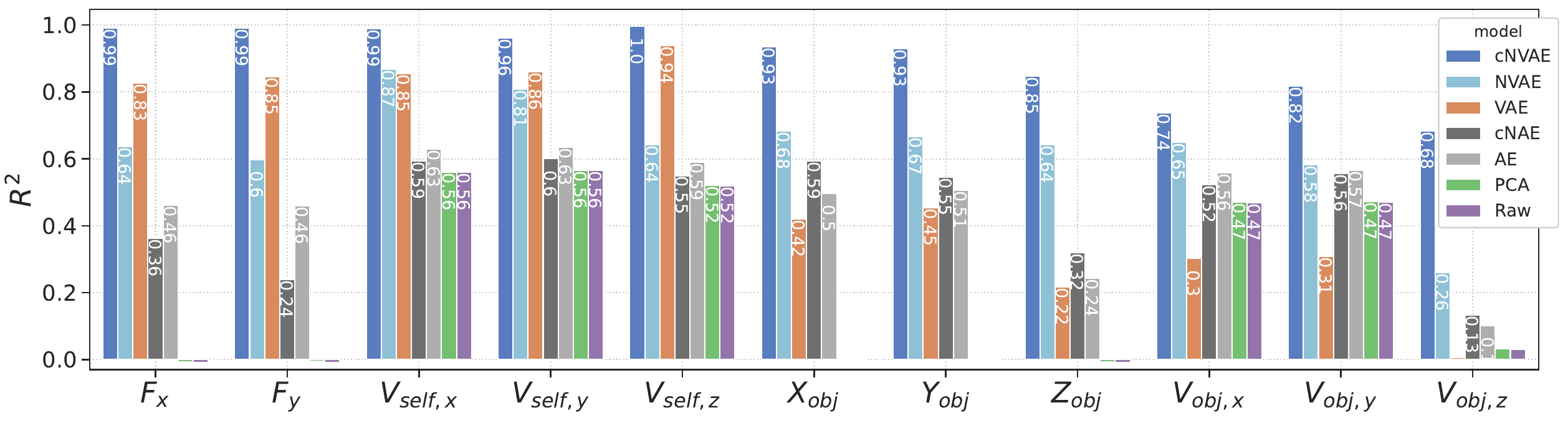}
    \caption[Hierarchical VAE untangles underlying factors of variation in data]{Hierarchical VAE untangles underlying factors of variation in data. The linear decodability of ground truth factors (x-axis) from different latent codes is shown. Untangling scores averaged across all ground truth variables are cNVAE $ = 0.898$, NVAE $ = 0.639$, VAE $ = 0.548$, cNAE $ = 0.456$, AE $ = 0.477$, PCA $ = 0.236$, and Raw $ = 0.235$. For variational models, the best performing $\beta$ values were selected: cNVAE, $\beta=0.15$; VAE, $\beta=1.5$. (see \cref{sec:choosing-beta} for more details). The full range of $\beta$ values is explored in \cref{fig:5_dci}.}
    \label{fig:4_untangle}
\end{figure}

\paragraph{\textbf{The cNVAE achieves optic flow parsing.}} To assess the performance of our models, we evaluated the linear decodability of the ground truth variables, $\bm{g}$, from model latent codes, $\bm{z}$. Based on the $R^2$ scores obtained by predicting $\bm{g}$ from $\bm{z}$ using linear regression (\cref{fig:4_untangle}), the cNVAE greatly outperforms competing models, faithfully capturing all ground-truth variables. In contrast, the non-hierarchical VAE fails to capture object-related variables. Notably, the cNVAE can recover the fixation point location ($F_X$, $F_Y$) in physical space almost perfectly. The fixation location has a highly nontrivial effect on the flow patterns, and varying it causes both global and local changes in the flow patterns (\cref{fig:1_rofl_trav}, bottom).

Furthermore, cNVAE is the only model that reliably captures object position and velocity: especially note $V_{obj,z}$ (last column in \cref{fig:4_untangle}). Inferring object motion from complex optic flow patterns involves two key components. First, the model must extract self-motion from flow patterns. Second, the model must understand how self-motion influences flow patterns globally. Only then can the model subtract self-motion from global flow vectors to obtain object motion. In vision science, this is known as the \sayit{flow-parsing hypothesis} \cite{rushton2005moving, warren2009optic, warren2007perception, peltier2020optic}. Such flow-parsing is achieved by the cNVAE but none of the other models. See \cref{sec:supp-discussion} for further discussion of this result and its implications.

\subsection{The cNVAE produces more disentangled representations}

The pursuit of disentanglement in neural representations has garnered considerable attention \cite{higgins2017beta, kumar2018variational, kim2018disentangling, chen2018isolating, ridgeway2018learning, rolinek2019variational, steenkiste19are, dittadi2021on, johnston2023abstract, whittington2023disentanglement, lachapelle23a}. In particular, \textcite{locatello2019challenging} established that learning fully disentangled representations is fundamentally impossible without inductive biases. Prior efforts such as $\beta$-VAE \cite{higgins2017beta} demonstrated that increasing the weight of the KL loss (indicated by $\beta$) promotes disentanglement in VAEs. More recently, \textcite{whittington2023disentanglement} demonstrated that simple biologically inspired constraints such as non-negativity and energy efficiency encourage disentanglement. Here, we hypothesize that another biological inductive bias \cite{sinz2019engineering}, hierarchy in the latent space, will improve representations learned by VAEs.

To evaluate the role of hierarchy, we adopted the DCI framework \cite{eastwood2018framework} which offers a well-rounded evaluation of latent representations. The approach involves training a simple decoder (e.g., lasso regression) that predicts data generative factors $\bm{g}$ from a latent code $\bm{z}$; followed by computing a matrix of relative importances (e.g., based on lasso weights) which is then used to evaluate different aspects of the code quality: \textit{\underline{I}nformativeness}---measures whether $\bm{z}$ contains easily accessible information about $\bm{g}$ (similar to untangling from above). \textit{\underline{D}isentanglement}---measures whether individual latents correspond to individual generative factors. \textit{\underline{C}ompleteness}---measures how many $z_i$ are required to capture any single $g_j$. If a single latent contributes to $g_j$'s prediction, the score will be 1 (complete). If all latent variables equally contribute to $g_j$'s prediction, the score will be 0 (maximally overcomplete). Note that \sayit{completeness} is also referred to as \sayit{compactness} \cite{ridgeway2018learning}. See \cref{sec:dci-details} and \cref{fig:good_g} for more details, ref.~\cite{carbonneau2022measuring} for a review, and ref.~\cite{eastwood2023dcies} for a recent extension of the DCI framework.

Here, we follow the methods outlined by \textcite{eastwood2018framework} with two modifications: (1) we replaced lasso with linear regression to avoid the strong dependence on the lasso coefficient that we observed, and (2) we estimate the matrix of relative importances using a feature permutation-based algorithm (\href{https://scikit-learn.org/stable/modules/generated/sklearn.inspection.permutation_importance.html}{sklearn.inspection.permutation\_importance}), which measures the relative amount of performance drop based on the shuffling of a given feature.

\begin{figure}[ht!]
    \centering
    \includegraphics[width=\linewidth]{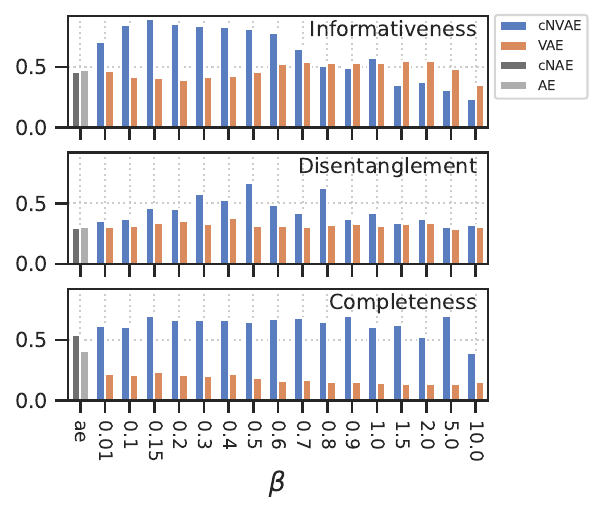}
    \caption[Evaluating learned latent codes using the DCI framework]{Evaluating the learned latent codes using the DCI framework \cite{eastwood2018framework}. Larger values are better for all metrics. Note that \textit{informativeness} is closely related to \textit{untangling} \cite{dicarlo2007untangling, dicarlo2012does}. See \cref{fig:good_mt} for an illustration of the core assumptions behind the DCI framework.}
    \label{fig:5_dci}
\end{figure}

We found that cNVAE outperforms competing models across all metrics for a broad range of $\beta$ values (\cref{fig:5_dci}). The observed pattern of an inverted U shape is consistent with previous work \cite{higgins2017beta}, which suggested there is a sweet spot for $\beta$ that can be found empirically. In our case, cNVAE with $\beta = 0.5$ achieved the best average DCI score. Further, we found that VAEs lacking hierarchical structure learn highly overcomplete codes, such that many latents contribute to predicting a single ground truth variable. In conclusion, the simple inductive bias of hierarchy in the latent space led to a substantial improvement in VAE performance across all components of the DCI metric.

\subsection{Brain-alignment: the cNVAE aligns more closely with MT neurons}

To evaluate the performance of models at predicting the activity of neurons in response to motion stimuli, we used a dataset of $N=141$ previously recorded MT neurons \cite{cui2013diverse, cui2016inferring}. A subset of these neurons ($N=84$) are publicly available on \href{https://crcns.org/data-sets/vc/mt-1}{crcns.org}, and were recently used in \textcite{mineault2021your} that we compare against.

\begin{wrapfigure}[]{r}{0.50\textwidth}
    \begin{center}
        \includegraphics[width=0.50\textwidth]{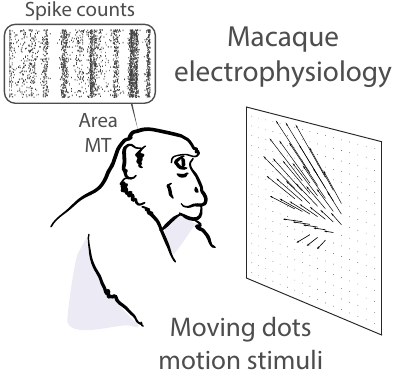}
    \end{center}
    \caption[Electrophysiology experimental setup]{Experimental setup. Macaque monkeys watch random dot kinematograms on a screen, while we record spiking activity from neurons in area MT. The dataset was previously published in \textcite{cui2013diverse, cui2016inferring}.}
    \label{fig:6_ephys}
\end{wrapfigure}

To measure neuronal alignment, we first determined the mapping from each model's latent representations to MT neuron responses (binned spike counts, \cref{fig:6_ephys}). Recall that model representations are defined as the mean of Gaussians for variational models (cNVAE, VAE) and the bottleneck activations for autoencoders (cNAE, AE).

\begin{figure}[ht!]
    \centering
    \includegraphics[width=0.95\linewidth]{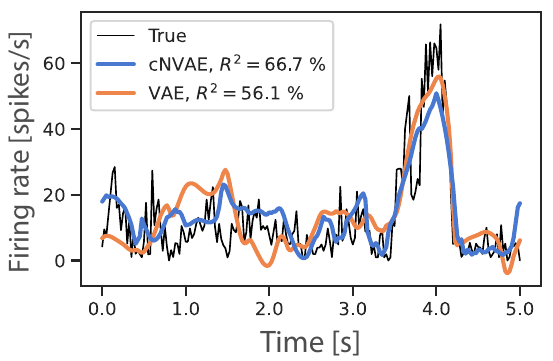}
    \caption[All models explain MT neural variability well]{Both models explain MT neural variability well. The black curve shows the average firing rate of an example MT neuron. The two other curves show model predictions.}
    \label{fig:6_psth_eg}
\end{figure}

\begin{figure}[ht!]
    \centering
    \begin{minipage}{0.57\textwidth}
        \centering
        \includegraphics[width=\linewidth]{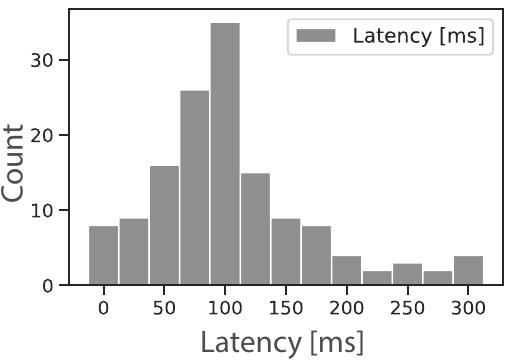}
        \caption[Distribution of best estimated latencies peaks around $100~ms$]{Distribution of best estimated latencies peaks around $100~ms$, which matches known properties of MT neuron responses \cite{cui2013diverse}.}
        \label{fig:6_latencies}
    \end{minipage}
    \hfill
    \begin{minipage}{0.40\textwidth}
        \centering
        \includegraphics[width=\linewidth]{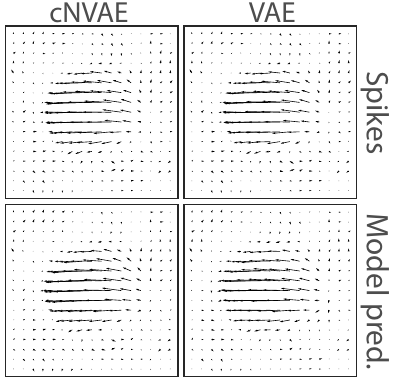}
        \caption[Spike-triggered averages (STA) for an example MT neuron]{Spike-triggered averages (STA) are shown for an example MT neuron (same as \cref{fig:6_psth_eg}). Model-predicted STAs are indistinguishable from true STAs estimated from spike trains.}
        \label{fig:6_sta}
    \end{minipage}
\end{figure}

We learn this linear latent-to-neuron mapping using ridge regression. \Cref{fig:6_psth_eg} shows the average firing rate of an example neuron along with model predictions. Because sensory neurons have a nonzero response latency, for each neuron we determined its response latency that maximized cross-validated performance. The distribution of best-selected latencies (\cref{fig:6_latencies}) peaked around $100~ms$, which is consistent with known MT latencies \cite{cui2013diverse}. We also optimize over 20 orders of magnitude of ridge coefficients to ensure each neuron has its best fit. \Cref{fig:6_sta} shows that the models capture the receptive field properties of MT neurons as measured by the spike-triggered average stimulus.

To evaluate performance, we follow methods established by \textcite{mineault2021your}: whenever repeated trials were available, we report Pearson's $R$ on that held-out data, normalized by maximum explainable variance \cite{sahani2002linear}. When repeats were not available, we performed 5-fold cross-validation and reported the held-out performance using Pearson's $R$ between model prediction and spike trains.

\begin{figure}[ht!]
    \centering
    \includegraphics[width=1.0\linewidth]{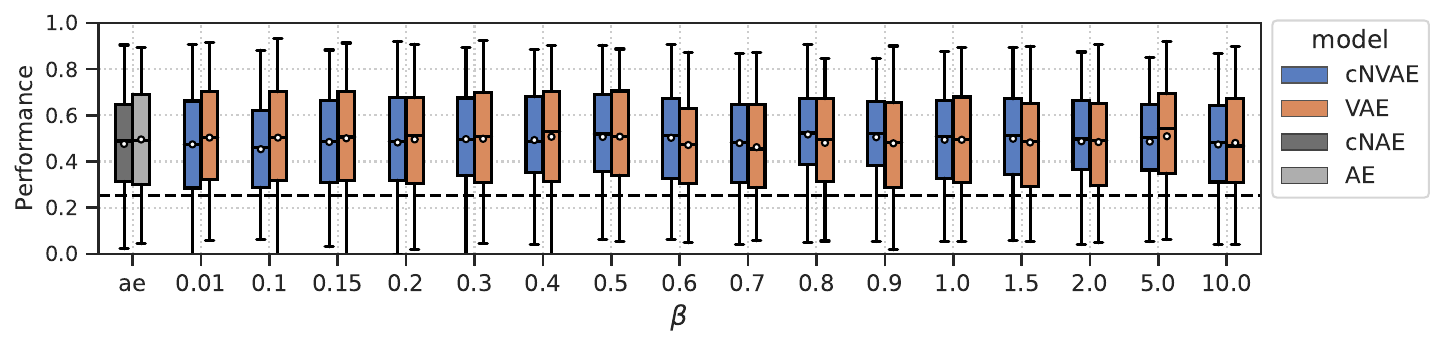}
    \caption[All models pretrained on \fixate{1} perform well]{All models (pretrained on \fixate{1}) perform comparably in predicting MT neuron responses. Dashed line corresponds to the previous state-of-the-art on this data \cite{mineault2012hierarchical}.}
    \label{fig:7_perf}
\end{figure}

\paragraph{\textbf{Evaluating brain alignment.}} We use two measures of brain alignment: the success at predicting the neural response (Pearson’s $R$, \cref{fig:7_perf}, \Cref{tab:mt}); and, the \sayit{alignment} between neurons and individual model latents (\cref{fig:8_align_def}, \cite{higgins2021unsupervised}). These mirror the untangling and completeness metrics described above (more details are provided below).

\paragraph{\textbf{All models predict MT responses well.}} After training a large ensemble of unsupervised models on \fixate{1} and learning the neural mapping, we found that both hierarchical (cNVAE \& cNAE) and non-hierarchical (VAE \& AE) variants performed roughly equally well at predicting neural responses (\cref{fig:7_perf}). We did observe some dependence on the loss function with the variational loss outperforming simple autoencoder reconstruction loss (\Cref{tab:mt}).

\subsection{Hierarchical VAEs are more aligned with MT neurons}

\begin{figure}[ht!]
    \centering
    \includegraphics[width=0.7\linewidth]{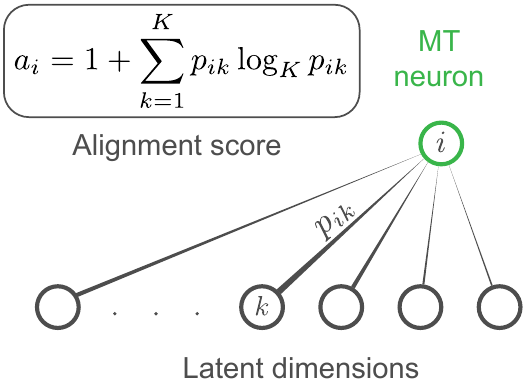}
    \caption[Alignment score measures the sparsity of permutation feature importances]{Alignment score measures the sparsity of permutation feature importances. $p_{ik}$ is some (normalized) measure of the importance of latent $k$ in predicting MT neuron $i$. $a_i = 0$ when all latents are equally important in predicting neuron $i$; and, $a_i = 1$ when a single latent predicts the neuron. See \cref{fig:good_mt} for our motivation behind this.}
    \label{fig:8_align_def}
\end{figure}

We next tested how these factors affect neural alignment, i.e., how closely neurons are related to individual latents in the model. \Cref{fig:8_align_def} demonstrates what we mean by \say{alignment}: a sparse latent-to-neuron relationship means larger alignment, indicative of a similar representational \say{form} \cite{higgins2021unsupervised}. See \cref{fig:good_mt} for the intuition behind this idea.

To formalize this notion, we use feature permutation importance (described above), applied to the ridge regression models. This yields a $420$-dimensional vector per neuron. Each dimension of this vector captures the importance of a given latent variable in predicting the responses of the neuron. We normalize these vectors and interpret them as the probability of importance. We then define alignment score $a_i$ of neuron $i$ as $ a_i = 1 + \sum_{k = 1}^{K} p_{ik} \log_K p_{ik}$, where $p_{ik}$ is interpreted as the importance of $k-$th latent variable in predicting neuron $i$ (\cref{fig:8_align_def}).

\begin{figure}[ht!]
    \centering
    \includegraphics[width=\linewidth]{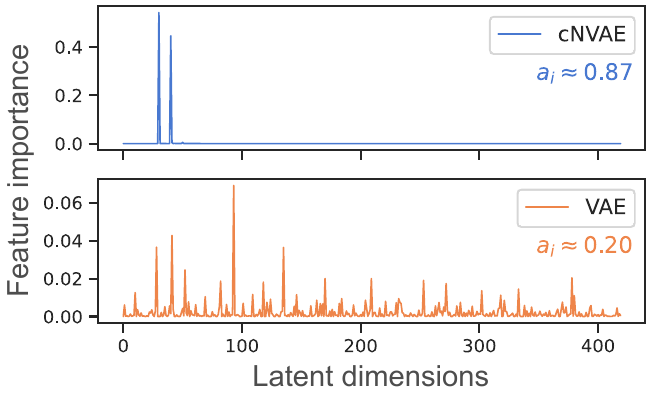}
    \caption[Feature importances are plotted for an example neuron]{Feature importances are plotted for an example neuron (same as in \cref{fig:6_psth_eg} and \cref{fig:6_sta}). Alignment scores (\cref{fig:8_align_def}) are computed from these feature importance vectors and are shown below the legends. cNVAE ($\beta = 0.01$) predicts this neuron's response in a much sparser manner compared to non-hierarchical VAE ($\beta = 5$). See \cref{sec:choosing-beta} for a discussion of our rationale in choosing these specific $\beta$ values.}
    \label{fig:8_align_eg}
\end{figure}

\begin{figure}[ht!]
    \centering
    \includegraphics[width=1.0\linewidth]{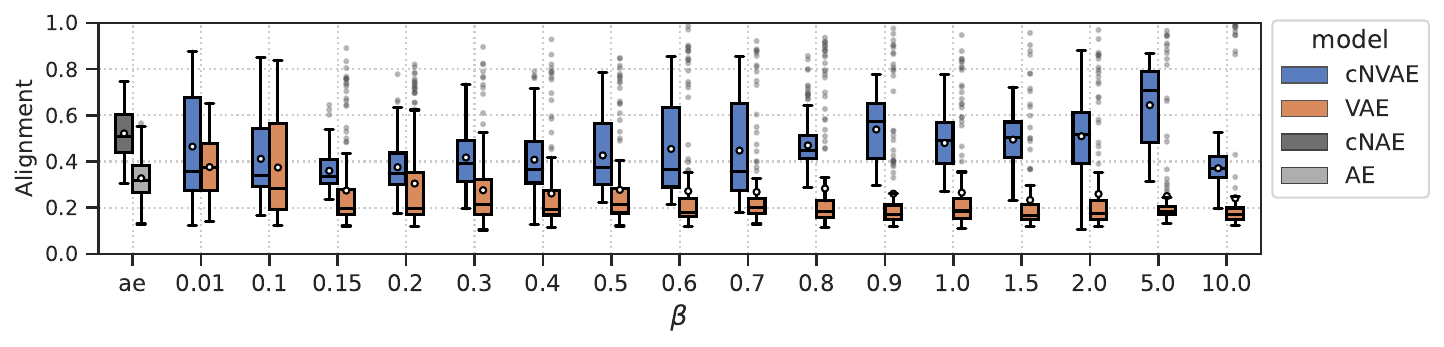}
    \caption[Alignment scores across $\beta$ values and autoencoders (ae) are shown]{Alignment scores across $\beta$ values and autoencoders (ae) are shown. Hierarchical models (cNVAE, cNAE) are more aligned with MT neurons since they enable sparse latent-to-neuron relationships.}
    \label{fig:8_align}
\end{figure}

\begin{figure}[ht!]
    \centering
    \includegraphics[width=1.0\linewidth]{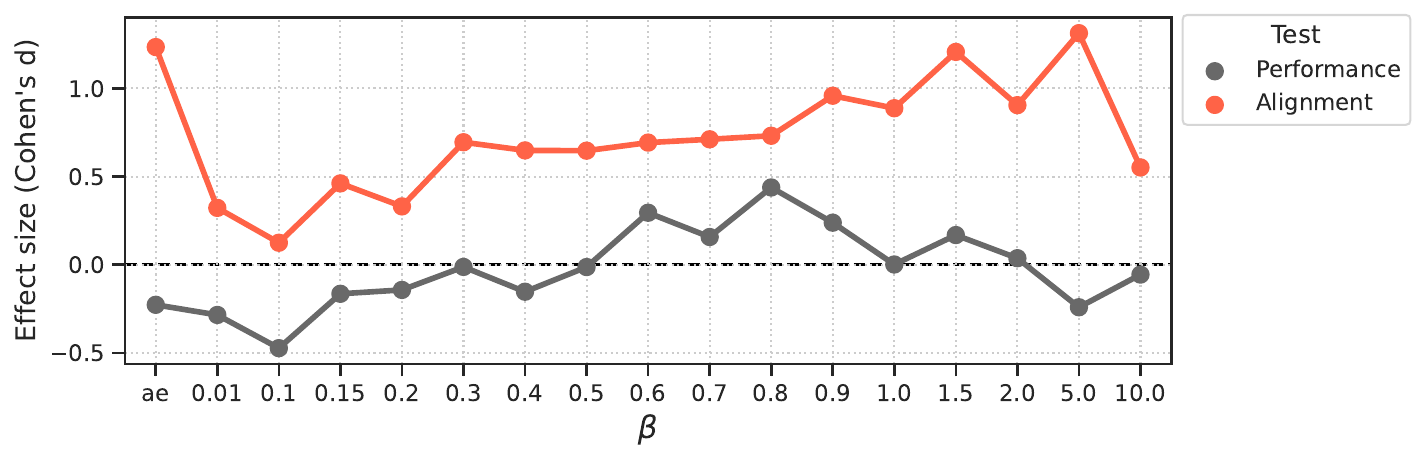}
    \caption[Effect sizes are shown for the two different approaches in model comparisons]{Effect sizes are shown for the two different approaches in model comparisons: using ridge regression \emph{Performance} (related to \cref{fig:7_perf}), versus, \emph{Alignment} scores (related to \cref{fig:8_align}). Alignment demonstrates a substantially larger effect size, leading to a more pronounced differentiation among the models. Positive values indicate a larger value for hierarchical models (i.e., cNVAE $>$ VAE or cNAE $>$ AE). For different $\beta$ values, model \textit{Performance} difference fluctuates between positive and negative values. In contrast, the difference in model \textit{Alignment} scores is consistently positive, indicating a greater brain alignment for hierarchical models.}
    \label{fig:15_effect_sizes}
\end{figure}

\Cref{fig:8_align_eg} shows an example neuron with an extremely sparse feature importance vector only for cNVAE but not VAE (same neuron as in \cref{fig:6_psth_eg}). This suggests that cNVAE-to-MT alignment is much larger than VAE-to-MT alignment. We plotted the distribution of alignment scores across all neurons (\cref{fig:8_align}). For almost all $\beta$ values, the cNVAE exhibited a greater brain alignment than non-hierarchical VAE (cNVAE $>$ VAE, paired $t-$test; see \cref{fig:15_effect_sizes} and \Cref{tab:pvals}). Similarly, for the autoencoders, we found that the hierarchical variant outperformed the non-hierarchical one (cNAE $>$ AE).

Based on these observations, we conclude that higher brain alignment is primarily due to hierarchical latent structure. However, note that hierarchy in the traditional sense did not matter: all these models had approximately the same number of convolutional layers and parameters.

\subsection{Factors leading to brain-alignment}

To test the effect of the training dataset (i.e., category of ROFL) on model performance, we trained cNVAE models using \fixate{0}, \fixate{1}, and \obj{1} categories (\Cref{tab:rofl}), while also exploring a variety of $\beta$ values. We found that \fixate{1} clearly outperformed the other two ROFL categories (\Cref{tab:mt}), suggesting that both global (e.g., self-motion) and local (e.g., object motion) sources of variation are necessary for learning MT-like representations.

The effect of loss function was also visible: some beta values led to more alignment. But this effect was small compared to the effect of hierarchical architecture (\cref{fig:8_align}).

\begin{table}[t]
    \caption[Both cNVAE and VAE perform well in predicting MT neuron responses]{Both cNVAE and VAE perform well in predicting MT neuron responses, surpassing previous state-of-the-art models by more than a twofold improvement. Moreover, the clear gap between \fixate{1} and other categories highlights the importance of pretraining data \cite{simoncelli2001natural}.}
    \label{tab:mt}
    \centering
    \begin{tabular}{ccc}
        \toprule
        Model & \begin{tabular}{c} Pretraining\\dataset\end{tabular} & \begin{tabularx}{92mm}{CCCC} \multicolumn{4}{c}{Performance, $R\,\,\,(\mu \pm se;\,\,N=141)$}\\
        \midrule
        $\beta = 0.5$ & $\beta = 0.8$ & $\beta = 1$ & $\beta = 5$\end{tabularx} \\
        
        \midrule
        {\color{cnvae_color}cNVAE} & \begin{tabular}{c} \fixate{1}\\\fixate{0}\\\obj{1}\end{tabular} & \begin{tabularx}{92mm}{CCCC}$\mathbf{.506 \pm .018}$ & $\mathbf{.517 \pm .017}$ & $.494 \pm .018$ & $.486 \pm .016$\\$.428 \pm .018$ & $.450 \pm .019$ & $.442 \pm .019$ & $.469 \pm .018$\\$.471 \pm .018$ & $.465 \pm .018$ & $.477 \pm .017$ & $.468 \pm .018$\end{tabularx} \\
        
        \midrule
        {\color{vae_color}VAE} & \fixate{1} & \begin{tabularx}{92mm}{CCCC}$\mathbf{.508 \pm .019}$ & $.481 \pm .018$ & $.494 \pm .018$ & $\mathbf{.509 \pm .018}$\end{tabularx} \\

        \midrule
        {\color{cnae_color}cNAE} & \fixate{1} & $.476 \pm .018$ \\

        \midrule
        {\color{ae_color}AE} & \fixate{1} & $.495 \pm .019$ \\

        \midrule
        CPC \cite{oord2019cpc} & AirSim \cite{shah2018airsim} & $.250 \pm .020$ (\textcite{mineault2021your}) \\
        
        \midrule
        DorsalNet & AirSim \cite{shah2018airsim} & $.251 \pm .019$ (\textcite{mineault2021your}) \\
        
        \bottomrule
  \end{tabular}
\end{table}

\section{Discussion}

We introduced a new framework for understanding and evaluating the representation of visual motion learned by artificial and biological neural networks. This framework provides a way to manipulate causes in the world and evaluate whether learned representations untangle and disentangle those causes. In particular, our framework enabled testing the influence of architecture (\cref{fig:2_model}), loss function (\Cref{tab:model}), and training set (\Cref{tab:rofl}) on the learned representations, encompassing 3 out of the 4 core components of a recently proposed neuroconnectionist research programme \cite{doerig2023neuroconnectionist}.

Our framework brings hypothesis-testing to understand [biological] neural processing of vision and provides an interpretive framework to understand neurophysiological data. While the results presented in this Chapter open up many interesting venues for future exploration, it is necessarily simplified, as it focuses on establishing our framework and demonstrating its potential. To this end, we made some simplifying choices that we will discuss below.

\subsection{Study limitations}

\paragraph{\textbf{Using optic flow instead of videos.}} Our simulation generates velocity fields, rather than pixels. This necessarily avoids the complexities of extracting flow patterns from real spatiotemporal movies and thus, does not require the models to capture the intricacies of the early stages of processing that are thought to do exactly this \cite{adelson1985spatiotemporal, nishimoto2011three}. Notably, this allows for direct comparison with recorded neural data in MT and MST using random dot kinematograms \cite{mineault2012hierarchical, cui2013diverse, cui2016inferring}, but would likely fail at explaining V1 responses, which necessarily requires a pixel-computable model as in previous work (e.g., DorsalNet \cite{mineault2021your}).

Similarly, our $>2\times$ improvement in performance in explaining MT neural data over DorsalNet is likely due in part to their network being based on spatiotemporal movies, while the neural data used optic flow patterns as we do. It is possible that a hierarchical VAE trained and tested on video stimuli would align more poorly than the model from \textcite{mineault2021your}. Future work in this space will involve rendering images in simulations and using image-computable models for a fair comparison.

\paragraph{\textbf{Natural optic flow statistics.}} We did not model fully natural optic flow, nor the true complexity of 3-D natural environments and their interaction with eye movements and self-motion as has recently been measured \cite{matthis2022retinal, muller2023retinal}. The simplification of our simulation still demonstrates the importance of including such elements in understanding neural representations and provides a framework for incorporating real eye-tracking and scene data \cite{matthis2022retinal, muller2023retinal} into future work with ROFL.

\paragraph{\textbf{Generalizability of our results to other brain regions.}} We found that hierarchy in the latent space leads to many desirable features of latent representations, including increasing the alignment to biological representations. Here, we tested this on only one dataset from area MT, which leaves the question of whether this is a general principle of brain computation. Addressing this requires testing our approach on more data from other brain areas, such as MST \cite{wild2021primate, Wild2022ElectrophysiologicalDF}. From previous work \cite{mineault2012hierarchical}, we expect that hierarchical computation is even more necessary for MST, which we leave as an open question to address in future work.

\subsection{Interpreting brain-alignment}

We measured the alignment between ANN models and MT neurons using both linear predictive power (\cref{fig:7_perf}), and an alternative measure of alignment that is sensitive to the sparsity of latent-to-neuron relationships (\say{alignment-score}; \cref{fig:8_align_def}). Linear regression has been used extensively to measure similarity, or alignment, between pairs of representations \cite{yamins2014performance, schrimpf2018brain, mineault2021your, conwell2023can}, but often results in a degenerate \cite{schrimpf2018brain, conwell2023can, tuckute2022many}, and unreliable \cite{han2023system}, measure of representational alignment. Our application of linear regression found that it was not a particularly effective signal in differentiating between models: although cNVAE produced the single best model in terms of neuronal prediction, we found that both hierarchical and non-hierarchical VAEs performed similarly in predicting MT neuron responses.

In contrast, the alignment score (\cref{fig:8_align_def}) was much more consistent in distinguishing between models (see \cref{fig:15_effect_sizes}), and revealed that hierarchical models (both cNVAE and cNAE) had significantly sparser latent-to-neuron relationships. The alignment score measures whether a model has learned a similar representational \say{form} to the brain, which would enable the sparsity of latent-to-neuron relationships. This concept is closely related to the \sayit{completeness} score from the DCI framework \cite{eastwood2018framework}. The alignment score shown in \cref{fig:8_align_def} was also employed by \textcite{higgins2021unsupervised}, although they used lasso regression with the magnitude of nonzero coefficients as their feature importance. Finally, we note that this alignment score also has limitations as it is not a proper metric in the mathematical sense \cite{williams2021generalized}. Future work will consider more sophisticated metrics for brain alignment \cite{williams2021generalized, duong2023representational, klabunde2023similarity, canatar2023spectral}.

\subsection{Conclusions}

We used synthetic data to test how causal structure in the world affects the representations learned by autoencoder-based models and evaluated the learned representations based on how they represent ground truth variables and how well they align with biological brains. We found that a single inductive bias, hierarchical latent structure, leads to desirable representations and increased brain alignment.

\section{Code \& Data}
Our code and model checkpoints are available here: \href{https://github.com/hadivafaii/ROFL-cNVAE}{https://github.com/hadivafaii/ROFL-cNVAE}. Our code makes extensive use of the following software packages: PyTorch \cite{paszke2019pytorch}, NumPy \cite{harris2020array}, SciPy \cite{SciPy2020}, scikit-learn \cite{pedregosa2011scikit}, pandas \cite{reback2020pandas}, matplotlib \cite{hunter2007matplotlib}, and seaborn \cite{waskom2021seaborn}.

\section{Additional Methods}

In this section and subsequent ones, I will present additional methods, followed by additional results that support the main ones, and will end with additional discussion points.

\subsection{VAE loss derivation}\label{sec:hier-derive}

Suppose some observed data $\bm{x}$ are sampled from a generative process as follows:

\begin{equation}
    p(\bm{x}) = \int p(\bm{x}, \bm{z}) \, d\bm{z} = \int p(\bm{x} \vert \bm{z}) p(\bm{z}) \, d\bm{z},
\end{equation}

where $\bm{z}$ are latent (or unobserved) variables. In this setting, it is interesting to ask which set of latents $\bm{z}$ are likely, given an observation $\bm{x}$. In other words, we are interested in posterior inference

\begin{equation}
    p(\bm{z} \vert \bm{x} ) \propto p(\bm{x} \vert \bm{z} ) p(\bm{z}).
\end{equation}

The goal of VAEs is to approximate the (unknown) true posterior $p(\bm{z} \vert \bm{x} )$ with a distribution $q(\bm{z} \vert \bm{x}; \theta)$, where $\theta$ are some free parameters to be learned. This goal is achieved by minimizing the Kullback-Leibler divergence between the approximate posterior $q$ and the true posterior $p$:

\begin{equation}
    \text{Goal:\quad minimize}\quad \mathcal{D}_{\mathrm{KL}}\Big[q(\bm{z} \vert \bm{x}; \theta) \,\big\|\, p(\bm{z} \vert \bm{x} )\Big].
\end{equation}

This objective is intractable, but rearranging the terms leads to the following loss that is also the (negative) variational lower bound on $\log p(\bm{x})$:

\begin{equation}\label{eq:vae-loss}
    \mathcal{L}_{\mathrm{VAE}} =
    -\mathbb{E}_q \Big[\log p(\bm{x}\vert\bm{z}; \thdec) \Big] +
    \mathcal{D}_{\mathrm{KL}}\Big[q(\bm{z} \vert \bm{x}; \thenc) \,\big\|\, p(\bm{z} \vert \thdec)\Big],
\end{equation}

where $\thenc$ and $\thdec$ are the parameters for the encoder and decoder neural networks (see \cref{fig:2_model}). The first term in \autoref{eq:vae-loss} is the reconstruction loss, which we chose to be the Euclidean norm of the differences between predicted and reconstructed velocity vectors (i.e., the endpoint error \cite{ilg2017flownet}). We will now focus on the second term in \autoref{eq:vae-loss}, the KL term.

\subsubsection{Prior, approximate posterior, and the KL term in vanilla VAE}

In standard non-hierarchical VAEs, the prior is not parameterized. Instead, it is chosen to be a simple distribution such as a Gaussian with zero mean and unit covariance:

\begin{equation}
    p(\bm{z} \vert \thdec)
    \,\rightarrow\,
    p(\bm{z}) = \mathcal{N}(\bm{0}, \bm{I}).
\end{equation}

The approximate posterior is also a Gaussian with mean $\bm{\mu}(\bm{x}; \thenc)$ and variance $\bm{\sigma}^2(\bm{x}; \thenc)$:

\begin{equation}
    q(\bm{z} \vert \bm{x}; \thenc) = \mathcal{N}(\bm{\mu}(\bm{x}; \thenc), \bm{\sigma}^2(\bm{x}; \thenc))
\end{equation}

As a result, the KL term for a vanilla VAE only depends on encoder parameters $\thenc$.

\subsubsection{Prior, approximate posterior, and the KL term in the cNVAE}
Similar to the NVAE \cite{vahdat2020nvae}, the cNVAE latent space is organized hierarchically such that latent groups are sampled sequentially, starting from \say{top} latents all the way down to the \say{bottom} ones (i.e., $\bm{z}_1$ to $\bm{z}_3$ in \cref{fig:2_model}). In addition to its hierarchical structure, another important difference between cNVAE and vanilla VAE is that the cNVAE prior is learned from data. Note the \say{mid} and \say{bottom} latent groups in \cref{fig:2_model}, indicated as $\bm{z}_2$ and $\bm{z}_3$ respectively: the cNVAE is designed in a way that changing the parameters along the decoder pathway will impact the prior distributions on $\bm{z}_2$ and $\bm{z}_3$ (but not $\bm{z}_1$). Note the \say{$h$} in \cref{fig:2_model}, which is also a learnable parameter. In summary, the \say{top} cNVAE latents (e.g., $\bm{z}_1$ in \cref{fig:2_model}) have a fixed prior distribution similar to vanilla VAEs; whereas, the prior distribution for every other cNVAE latent group is parametrized and learned from data.

More formally, the cNVAE latents are partitioned into disjoint groups, $\bm{z} = \{\bm{z}_1, \bm{z}_2, \dots, \bm{z}_L\}$, where $L$ is the number of groups. Then, the prior is represented by:

\begin{equation}
    p(\bm{z} \vert \thdec) =
    p(\bm{z}_1) \cdot \prod_{\ell = 2}^L p(\bm{z}_\ell \vert \bm{z}_{<\ell}; \thdec),
\end{equation}

where the conditionals are represented by factorial Normal distributions. For the first latent group we have $p(\bm{z}_1) = \mathcal{N}(\bm{0}, \bm{I})$, similar to vanilla VAEs. For every other latent group, we have:

\begin{equation}
    p(\bm{z}_\ell \vert \bm{z}_{<\ell}; \thdec) =
    \mathcal{N}(
    \bm{\mu}(\bm{z}_{<\ell}; \thdec), \bm{\sigma}^2(\bm{z}_{<\ell}; \thdec)
    ),
\end{equation}

where $\bm{\mu}(\bm{z}_{<\ell}; \thdec)$ and $\bm{\sigma}^2(\bm{z}_{<\ell}; \thdec)$ are outputted from the decoder \textit{sampler} layers. Similarly, the approximate posterior in the cNVAE is represented by:

\begin{equation}
    q(\bm{z} \vert \bm{x}; \thenc) =
    \prod_{\ell = 1}^L q(\bm{z}_\ell \vert \bm{z}_{<\ell}, \bm{x}; \thenc).
\end{equation}

We adopt a Gaussian parameterization for each conditional in the approximate posterior:

\begin{equation}
    q(\bm{z}_\ell \vert \bm{z}_{<\ell}, \bm{x}; \thenc) =
    \mathcal{N}(
    \bm{\mu}(\bm{z}_{<\ell}, x; \thenc), \bm{\sigma}^2(\bm{z}_{<\ell}, \bm{x}; \thenc)
    ),
\end{equation}

where $\bm{\mu}(\bm{z}_{<\ell}, x; \thenc)$ and $\bm{\sigma}^2(\bm{z}_{<\ell}, \bm{x}; \thenc)$ are outputted from the encoder \textit{sampler} layers (\cref{fig:2_model}; grey trapezoids). We are now in a position to explicitly write down the KL term from \autoref{eq:vae-loss} for the cNVAE:

\begin{equation}
    \text{KL term } = 
    \mathcal{D}_{\mathrm{KL}}\Big[q(\bm{z}_1 \vert \bm{x}, \thenc) \,\big\|\, p(\bm{z_1}) \Big] +
    \sum_{\ell=2}^L \mathbb{E}_{q(\bm{z}_{<\ell} \vert \bm{x}, \thenc)}\Big[\mathrm{KL}_\ell(\thenc, \thdec)\Big],
\end{equation}

where $\mathrm{KL}_\ell$ refers to the local KL term for group $\ell$ and is given by:

\begin{equation}
    \mathrm{KL}_\ell(\thenc, \thdec) \coloneqq \mathcal{D}_{\mathrm{KL}}\Big[q\left(\bm{z}_\ell \vert \bm{z}_{<\ell}, \bm{x}; \thenc\right) \big\|\, p\left(\bm{z}_\ell \vert \bm{z}_{<\ell}, \thdec\right)\Big],
\end{equation}

and the approximate posterior up to the $(\ell-1)^{th}$ group is defined as:

\begin{equation}
    q(\bm{z}_{<\ell} \vert \bm{x}; \thenc) \coloneqq \prod_{i=1}^{\ell-1} q(\bm{z}_i \vert \bm{z}_{<i},\bm{x}; \thenc).
\end{equation}

This concludes our derivation of the KL terms shown in \Cref{tab:model}.

\subsection{Model hyperparameters \& other details}

The latent space of cNVAE contains a hierarchy of latent groups that are sampled sequentially: \say{top} latents are sampled first, all the way down to \say{bottom} latents that are closest to the stimulus space (\cref{fig:2_model}). \textcite{child2021vdvae} demonstrated that model depth, in this statistical sense, was responsible for the success of hierarchical VAEs. Here, we increased the depth of cNVAE architecture while ensuring that training remained stable. Our final model architecture had a total of 21 hierarchical latent groups: 3 groups operating at the spatial scale of $2 \times 2$, 6 groups at the scale of $4 \times 4$, and 12 groups at the scale of $8 \times 8$ (see \cref{fig:2_model,fig:3_mi}). Each latent group has 20 latents, which yields a $21 \times 20 = 420$ dimensional latent space. See below for more details. We observed that various aspects of model performance such as reconstruction loss and DCI scores increased with depth, corroborating previous work \cite{child2021vdvae}. We set the initial number of channels to $32$ in the first layer of the encoder and doubled the channel size every time we downsample the features spatially. This resulted in a width of $256$ in the final layer. Please refer to our code for more information.

We used the same regular residual cells for both the encoder and the decoder (\say{$r$} in \cref{fig:2_model}). We did not experience memory issues because of the relatively small scale of our experiments. However, rendering the input data at a larger spatial dimension would lead to larger memory usage, which might require using depthwise separable convolutions as in Figure 3b in \textcite{vahdat2020nvae}. Similar to NVAE, we found that the inclusion of squeeze \& excitation \cite{hu2018squeeze} helped. Contrary to the suggestion of NVAE \cite{vahdat2020nvae} and others \cite{sonderby2016ladder}, batch norm \cite{ioffe2015batch} destabilized training in our case. Instead, we found that weight normalization \cite{salimans2016weight} and swish activation \cite{ramachandran2018searching, elfwing2018sigmoid} were both instrumental in training. We used weight normalization without data-dependent initialization. Our observations also deviate from those of \textcite{vahdat2020nvae} with regard to spectral regularization \cite{yoshida2017spectral}. We achieved stable training and good results without the need for spectral regularization. This is perhaps because our dataset is synthetic and has a much smaller dimensionality compared to real-world datasets such as faces considered in the NVAE. Finally, we also found that residual Normal parameterization of the approximate posterior was helpful.

\subsection{cNVAE vs.~NVAE latent dimensionality}\label{sec:latent-dims}
As an illustrative example, here we count the number of latent variables in cNVAE and compare them to an otherwise identical but non-compressed NVAE. The total number of latent variables for a hierarchical VAE is given by:

\begin{equation}
    D = \sum_{\ell = 1}^L d_\ell,
\end{equation}

where $d_\ell$ is the number of latents in level $\ell$, and $L$ is the total number of hierarchical levels. As described above, $d_\ell = d \cdot s_\ell^2$ for NVAE, where $d$ is the number of latents per level and $s_\ell$ is the spatial scale of level $\ell$. In contrast, this number is a constant for the cNVAE: $d_\ell = d$. In this Chapter, we set $d = 20$ and had the following configuration for the cNVAE:

\begin{itemize}
    \item \say{top} latents: 3 groups operating at the spatial scale of $2 \times 2$
    \item \say{mid} latents: 6 groups operating at the spatial scale of $4 \times 4$
    \item \say{bottom} latents: 12 groups operating at the spatial scale of $8 \times 8$
\end{itemize}

This resulted in an overall latent dimensionality of $D = (3 + 6 + 12) \times d = 21 \times 20 = 420$ for the cNVAE. In contrast, an NVAE with a similar number of hierarchical levels would have a dimensionality of $D = (3 \cdot 2^2 + 6 \cdot 4^2 + 12 \cdot 8^2) \times d = 876 \times 20 = 17,520$.

Note that in \cref{fig:4_untangle}, we necessarily had to reduce the NVAE depth to roughly match its latent dimensionality to other models. Specifically, the NVAE in \cref{fig:4_untangle} had $2/3/6$ latent groups at top/mid/bottom with $d = 1$. This configuration resulted in latent dimensionality of $D = (2 \cdot 2^2 + 3 \cdot 4^2 + 6 \cdot 8^2) \times 1 = 440$, which is roughly comparable to other models at $D = 420$. See \Cref{tab:latent-dims} for a comparison between cNVAE and the alternative models used in this study.

\begin{table}[t]
    \caption[Model latent structure and dimensionality]{The hierarchical models (cNVAE, cNAE) process information at various spatial scales. These models have a total of $21$ latent groups, each containing $d = 20$ latent variables. For hierarchical models, we concatenate the latent representations across all levels of the hierarchy, resulting in a single latent vector with dimensionality $D = 420$. In contrast, non-hierarchical models (VAE, AE) do not possess such a multi-scale organization. Instead, their non-hierarchical latent spaces contain a single latent group, operating at a single spatial scale. This single latent group contains the entire set of $d = D = 420$ latent variables. See \Cref{tab:model} and \cref{fig:2_model}.}
    \label{tab:latent-dims}
    \centering
    \begin{tabular}{cccc}
        \toprule
        Model &
        \begin{tabularx}{52mm}{CCCC} \multicolumn{4}{c}{Number of latent groups}
        \\
        \midrule
        $2 \times 2$ & $4 \times 4$ & $8 \times 8$ & total \end{tabularx} &
        \begin{tabular}{c} Number of latent\\
        variables per group\end{tabular} &
        \begin{tabular}{c} Latent space\\dimensionality\end{tabular}
        \\

        \midrule
        {\color{cnvae_color}cNVAE} &
        \begin{tabularx}{52mm}{CCCC} $3$ & $6$ & $12$ & $21$ \end{tabularx} &
        $d = 20$ &
        $D = 21 \times 20 = 420$
        \\
        {\color{vae_color}VAE} &
        \begin{tabularx}{52mm}{CCCC} $1$ & $-$ & $-$ & $1$ \end{tabularx} &
        $d = 420$ &
        $D = 1 \times 420 = 420$
        \\
        {\color{cnae_color}cNAE} &
        \begin{tabularx}{52mm}{CCCC} $3$ & $6$ & $12$ & $21$ \end{tabularx} &
        $d = 20$ &
        $D = 21 \times 20 = 420$
        \\
        {\color{ae_color}AE} &
        \begin{tabularx}{52mm}{CCCC} $1$ & $-$ & $-$ & $1$ \end{tabularx} &
        $d = 420$ &
        $D = 1 \times 420 = 420$
        \\
        \bottomrule
  \end{tabular}
\end{table}

\subsection{Model training details}\label{sec:train-details}

We generated $750,000$ samples for each ROFL category, with a $600,000/75,000/75,000$ split for train/validation/test datasets. Model performance (reconstruction loss, NELBO) was evaluated using the validation set (\cref{fig:loss-v-beta,fig:loss-v-i}). Similarly, we used the validation set to estimate mutual information between latents, $\bm{z}$, and ground truth factors, $\bm{g}$, as shown in \cref{fig:3_mi}. To evaluate the latent codes, we used the validation set to train linear regressors to predict $\bm{g}$ from $\bm{z}$. We then test the performance using the test set. This includes results presented in \cref{fig:4_untangle} and \cref{fig:5_dci}.

\paragraph{\textbf{KL annealing and balancing.}} We annealed the KL term during the first half of the training, which is known to be an effective trick in training VAEs \cite{vahdat2020nvae, sonderby2016ladder, bowman2016generating, fu2019cyclical}. Additionally, during the annealing period, we employed a KL balancing mechanism which ensures an equal amount of information is encoded in each hierarchical latent group \cite{vahdat2020nvae, vahdat2018dvae++, chen2017variational}.

\paragraph{\textbf{Gradient norm clipping.}} During training, we occasionally encountered updates with large gradient norms. Similar to vdvae \cite{child2021vdvae}, we used gradient clipping, with an empirically determined clip value of $250$. However, we did not skip the updates with large gradients, as we found that the model usually gets stuck and does not recover after skipping an update.

\paragraph{\textbf{Training hyperparameters.}} We used batch size of $600$, learning rate of $0.002$, and trained models for $160$ epochs (equivalent to $160k$ steps). Each training session took roughly 24 hours to conclude on a single Quadro RTX 5000 GPU. Similar to NVAE, we used the AdaMax optimizer \cite{kingma2014adam} with a cosine learning rate schedule \cite{loshchilov2017sgdr} but without warm restarts.

\subsection{Choosing \texorpdfstring{$\beta$}{beta} values for different figures}\label{sec:choosing-beta}

It is known that different choices of the $\beta$ value in the VAE loss (\Cref{tab:model}) will result in different model properties \cite{higgins2017beta}. Therefore, while we scanned across many $\beta$ values spanning two orders of magnitude, we displayed results for certain betas to best demonstrate our points.

To ensure a fair model comparison in the results shown in \cref{fig:3_mi,fig:4_untangle,fig:indiv,fig:samples}, we selected $\beta$ values that maximized the overall untangling score for each architecture (cNVAE, $\beta = 0.15$; VAE, $\beta = 1.5$). This can be seen in \cref{fig:5_dci}, which presents DCI scores over the entire $\beta$ spectrum. The top row of \cref{fig:5_dci} displays informativeness (= untangling) scores, revealing an inverted U shape, consistent with previous work \cite{higgins2017beta}. Different architectures peak at different $\beta$ values.

For the neuron shown in \cref{fig:8_align_eg}, we found that different $\beta$ values lead to qualitatively similar outcomes, that is, extreme sparsity in cNVAE feature importances, in sharp contrast to that of VAE. We deliberately chose to show a larger $\beta = 5$ for VAE because previous work by \textcite{higgins2021unsupervised} suggested that increasing $\beta$ values increases neuronal alignment, which we did not observe here. In contrast, even a very small $\beta = 0.01$ for the cNVAE results in a large alignment. This result (paired with other observations) suggests that brain alignment, as we measured it here, emerges due to the architecture rather than from large $\beta$ values alone---although we do observe some dependence on $\beta$, so ultimately a combination of both architecture and loss is important (but mostly architecture).

\subsection{Details on latent code evaluation}\label{sec:approach}
Each ROFL sample $\bm{x} \in \mathbb{R}^{2 \times N \times N}$ is rendered from a low dimensional ground truth vector $\bm{g} \in \mathbb{R}^{K}$, where $N$ is the number of pixels (stimulus size) and $K$ is the true underlying dimensionality of the data (i.e., number of independent ground truth factors). In other words, the data lies on a curved $K$-dimensional manifold embedded in a $2N^2$-dimensional space.

In this Chapter, we worked with relatively small stimuli with $N = 17$. This choice allowed us to train more models and explore a larger combination of hyperparameters. We chose an odd value for $N$ so that there was a pixel at the exact center, which served as the center of gaze or the fixation point (\cref{fig:1_rofl}). The true dimensionality (number of generative factors) of different ROFL categories is given in \Cref{tab:rofl}. For example, it is $K = 11$ for \fixate{1}.

We trained unsupervised models by feeding them raw samples $\bm{x}$ and asking them to produce a reconstruction $\hat{\bm{x}}$. The models accomplish this by mapping their inputs to latent codes $\bm{z}$ during inference (red pathways in \cref{fig:2_model}), and mapping $\bm{z}$ to $\hat{\bm{x}}$ during generation (blue pathways in \cref{fig:2_model}). Note that for all models we have $\bm{z} \in \mathbb{R}^{420}$ (see \Cref{tab:latent-dims}).

In sum, our approach consists of four steps:

\begin{enumerate}
    \item Data generation: $\bm{g} \rightarrow \text{\fbox{ROFL}} \rightarrow \bm{x}$
    \item Unsupervised training: $\bm{x} \rightarrow \text{\fbox{Model}} \rightarrow \hat{\bm{x}}$
    \item Inference: $\bm{x} \rightarrow \text{\fbox{Model}} \rightarrow \bm{z}$
    \item Evaluation: $\bm{z} \xleftrightarrow{\hspace{1mm}} \bm{g}$ ?
\end{enumerate}

Within this framework, we investigated the relationships between data generative factors $\bm{g}$ and model-learned latent codes $\bm{z}$ by using the DCI metrics, which we discuss in detail below.

\subsection{Metrics \& Baselines}
We trained a variety of unsupervised models (\Cref{tab:model}) and evaluated them based on (1) their ability to represent the generative variables of the ROFL stimuli; and (2) the neural alignment of their representation.
This resulted in a total of five different metrics used to evaluate model performance. Below, we describe the different metrics used for each.

\paragraph{\textbf{\#1: untangling \& disentangling of data generative factors.}} To evaluate the latent codes, we employed the DCI framework \cite{eastwood2018framework} and made use of the fact that ROFL provides ground truth factors for each data sample. DCI is comprised of three different metrics, \textit{\underline{D}isentanglement}, \textit{\underline{C}ompleteness}, and \textit{\underline{I}nformativeness}, which we will discuss in detail below.

\paragraph{\textbf{\#2: explaining variability in biological neurons \& brain alignment.}} We used ridge regression to predict macaque MT neuron responses from model latents. To measure the alignment between different latent codes and biological neurons, we used two standard metrics: linear predictive power and brain-alignment score defined in \cref{fig:8_align_def}.

\subsubsection{The DCI framework}\label{sec:dci-details}

\begin{figure}[t!]
    \centering
    \includegraphics[width=1.0\linewidth]{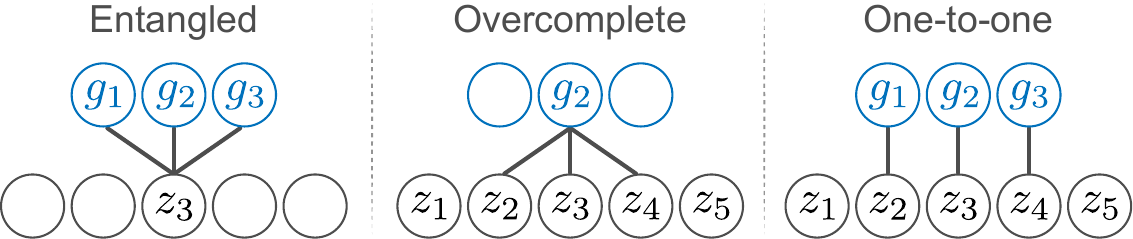}
    \caption[Entangled, overcomplete, and one-to-one codes]{Suppose we train a regression model to predict ground truth factors $g_j$ (blue) from latents $z_i$ (grey). Entanglement (left) refers to a situation in which one latent variable contains information about many generative factors. An overcomplete code (middle) is one in which a single generative factor is predicted from many latent variables. A one-to-one relationship (right) is the ideal outcome.}
    \label{fig:good_g}
\end{figure}

In this section, we provide more details about the DCI framework \cite{eastwood2018framework} and discuss how it was used to evaluate the disentangling performance of our models. The core motivation behind DCI is that a good latent code ideally disentangles as many factors of variation as possible while discarding as little information as possible. In other words, we consider a latent code to be \sayit{good} as long as:

\begin{enumerate}
    \item It can predict the variance of data generative factors, and
    \item It exhibits a one-to-one relationship with the generative factors (\cref{fig:good_g}).
\end{enumerate}

To make this idea concrete, the starting point is to train a regressor $f_j$ (e.g., lasso) that predicts a given data generative factor $g_j$ from a latent code $\bm{z}$. The performance of $\bm{f}$ in predicting $\bm{g}$ (measured using e.g., $R^2$) is referred to as the \textit{Informativeness} score. The notion of \textit{Informativeness}, if measured using a linear regressor, is very similar to \textit{Untangling} \cite{dicarlo2007untangling}.

As the next step, the weights of $\bm{f}$ are used to construct a matrix of relative importances $R$, where $R_{ij}$ denotes the relative importance of $z_i$ in predicting $g_j$. For example, $R$ can be the magnitude of lasso coefficients. The matrix of relative importance is then used to define two pseudo-probability distributions:

\begin{align}
    P_{ij} &= \frac{R_{ij}}{\sum_k R_{ik}} = \text{probability of } z_i \text{ being important for predicting } g_j \\
    \tilde{P}_{ij} &= \frac{R_{ij}}{\sum_k R_{kj}} = \text{probability of } g_j \text{ being predicted by } z_i
\end{align}

These distributions are then used to define the following metrics, which complete the DCI trinity:

\begin{align}
    Disentanglement &:\quad D_i = 1 - H_K(P_{i.}); \quad \text{where} \quad H_K(P_{i.}) = - \sum_j P_{ij} \log_K P_{ij} \\
    Completeness &:\quad C_j = 1 - H_D(\tilde{P}_{.j}); \quad \text{where} \quad H_D(\tilde{P}_{.j}) = - \sum_i \tilde{P}_{ij} \log_D \tilde{P}_{ij} 
\end{align}

Here, $K$ is the number of data generative factors (e.g., $K=11$ for \fixate{1}, \Cref{tab:rofl}), and $D$ is the number of latent variables ($D = 420$ for the models considered in this Chapter, \Cref{tab:latent-dims}). 

Intuitively, if latent variable $z_i$ contributes to predicting a single ground truth factor, then $D_i = 1$. Conversely, $D_i = 0$ when $z_i$ equally contributes to the prediction of all ground truth factors. If the ground truth factor $g_j$ is being predicted from a single latent variable, then $C_j = 1$ (a complete code). On the other hand, if all latents contribute to predicting $g_j$, then $C_j = 0$ (maximally overcomplete). Finally, the overall disentanglement or completeness scores are obtained by averaging across all latents or generative factors.

In the DCI paper, \textcite{eastwood2018framework} used lasso or random forest as their choice of regressors $\bm{f}$. Initially, we experimented with lasso but found that the lasso coefficient had a strong impact on the resulting DCI scores. To mitigate this, we used linear regression instead and estimated the matrix of relative importance using scikit-learns's permutation importance score (\href{https://scikit-learn.org/stable/modules/generated/sklearn.inspection.permutation_importance.html}{sklearn.inspection.permutation\_importance}), which measures how much performance drops based on the shuffling of a given feature.

\subsubsection{Brain alignment score}

\begin{figure}[ht!]
    \centering
    \includegraphics[width=1.0\linewidth]{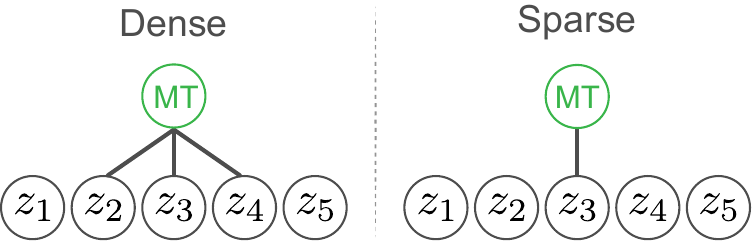}
    \caption[Intuition for why a sparse latent-to-neuron relationship is desirable]{Consider training a regressor to predict MT neuron responses (green) from latent variables $z_i$ (grey). Beyond accurate predictions, we assess the alignment between $\bm{z}$ and MT based on the sparsity of the resulting latent-to-neuron relationships. One possibility is that many latents contribute to predicting the neuron in a dense fashion. This would suggest that the neuron's function is diffusely represented within the latents, hinting that the representational form of the latents is ultimately different from that of MT, perhaps through rotation or affine transformations. Conversely, when few latents effectively predict MT, it suggests a similar representational form, laying the foundation for our rationale in defining our alignment score (\cref{fig:8_align_def}).}
    \label{fig:good_mt}
\end{figure}

As depicted in \cref{fig:8_align_def}, the alignment score emphasizes the degree of correspondence between latent variables and neurons. A sparse relationship between latents and neurons suggests a higher alignment, indicating that the model's representation closely mirrors the neuron's representational \say{form} \cite{higgins2021unsupervised}. This notion of alignment is intimately tied to the \say{completeness} score discussed above. See \cref{fig:good_mt}.

\subsubsection{Baselines}
In \cref{fig:4_untangle} we compared the cNVAE with alternative models introduced in \Cref{tab:model}. We also included an NVAE with matched number of latents, but necessarily fewer hierarchical levels (see \cref{sec:latent-dims} for more details). To demonstrate the difficulty of our untangling task, we included two simple baselines: Raw data and a linear code based on the first $420$ dimensions of data principal components (PCA). Both Raw and PCA baselines performed poorly, only weakly untangling self-motion velocity ($\sim 55\%$) and object velocity in $x$ and $y$ directions ($47\%$). This result demonstrates the importance of nonlinear computation in extracting good representations.

\section{Additional Results}

In this section, we present additional results demonstrating the advantages of hierarchical VAEs versus non-hierarchical ones, which support the main results presented earlier. All models explored in this section were trained on \fixate{1}. For both cNVAE and VAE, we chose $\beta$ values that maximized their informativeness (we discuss this choice in \cref{sec:choosing-beta}). Specifically, we set $\beta = 0.15$ for cNVAE and $\beta = 1.5$ for VAE. These same models trained using the respective $\beta$ values were also used for results presented in \cref{fig:3_mi} and \cref{fig:4_untangle}.

\subsection{Individual cNVAE latents exhibit greater linear correlation with ground truth data-generative factors}

\begin{figure}[ht!]
    \centering
    \includegraphics[width=\linewidth]{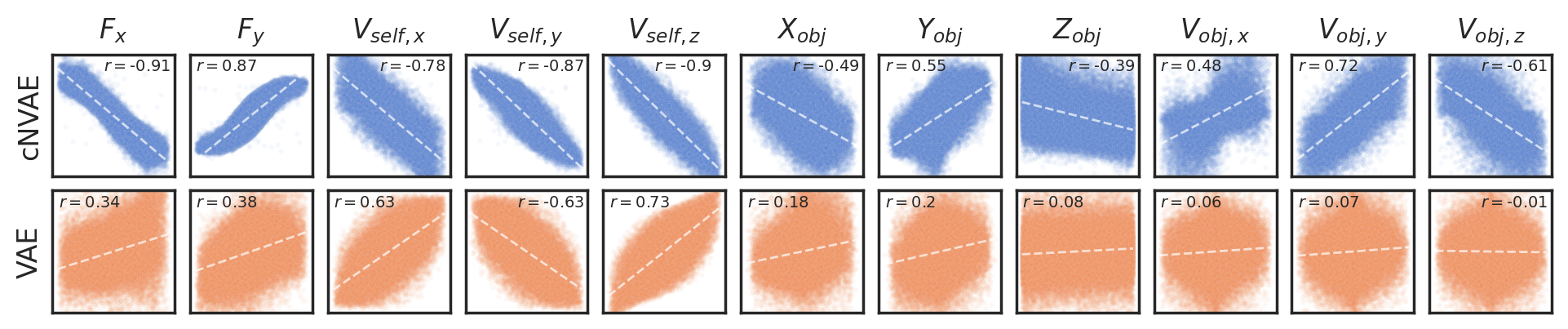}
    \caption[Linear relationship between latent variables and ground truth factors]{For each ground truth factor and model, we identified a single latent variable that displayed the highest (Pearson) correlation with that variable. The ground truth factors exhibited greater linearity within the latent space of cNVAE compared to VAE. Scatter points correspond to data samples from the test set. x-axis, the value of ground truth factors; y-axis, the value of latent variables.}
    \label{fig:indiv}
\end{figure}

In \cref{fig:4_untangle}, we demonstrated that the cNVAE latent representations $\bm{z}$ contained explicit information about ground truth factors $\bm{g}$. We achieved this by training linear regression models that map the $420$-dimensional latent space to each ground truth factor. Here, we investigate the degree that \textit{individual} latents $z_i$ are (Pearson) correlated with individual factors $g_j$ (\cref{fig:indiv}). For some ground truth factors, we find that cNVAE learns latents that are almost perfectly correlated with them. For instance, $r = -0.91$ for $F_x$ and $r=0.87$ for $F_y$, compared to modest correlation values of $r = 0.34$ and $r=0.38$ for VAE. The higher linear correlation is consistently observed for cNVAE over VAE. Especially, we note that the VAE does not capture object-related variables at all, consistent with other results shown earlier (\cref{fig:3_mi,fig:4_untangle}).

For the cNVAE, the indices for the selected highly correlated latents are as follows: $\left[296, 248, 393, 284, 368,  92, 206, 105, 338,  60,  63\right]$, ordered left-to-right in the same manner as in \cref{fig:indiv}. Here, an index of $0$ corresponds to the very top latent operating at the spatial scale of $2 \times 2$, and an index of $419$ corresponds to the very bottom one at scale $8 \times 8$ (see \cref{fig:2_model}). By considering the hierarchical grouping of these latents (see dashed lines in \cref{fig:3_mi}) we see that the majority of the highly correlated latents operate at the spatial scale of $8 \time 8$, with the exception of object-related ones corresponding to $X_{obj}$, $Z_{obj}$, $V_{obj,y}$, and $V_{obj, z}$ which operate at $4 \times 4$.

In conclusion, \cref{fig:indiv} revealed that cNVAE learns individual latents that are highly correlated with individual ground truth factors, which provides complementary insights as to why cNVAE is so successful in untangling (\cref{fig:4_untangle}). However, we note that a high correlation with individual latents is not the best way to capture the overall functional organization, which is better reflected for example in the mutual information heatmap of \cref{fig:3_mi}.

\subsection{Generated samples: non-hierarchical VAE completely neglects objects}

\begin{figure}[ht!]
    \centering
    \includegraphics[width=\linewidth]{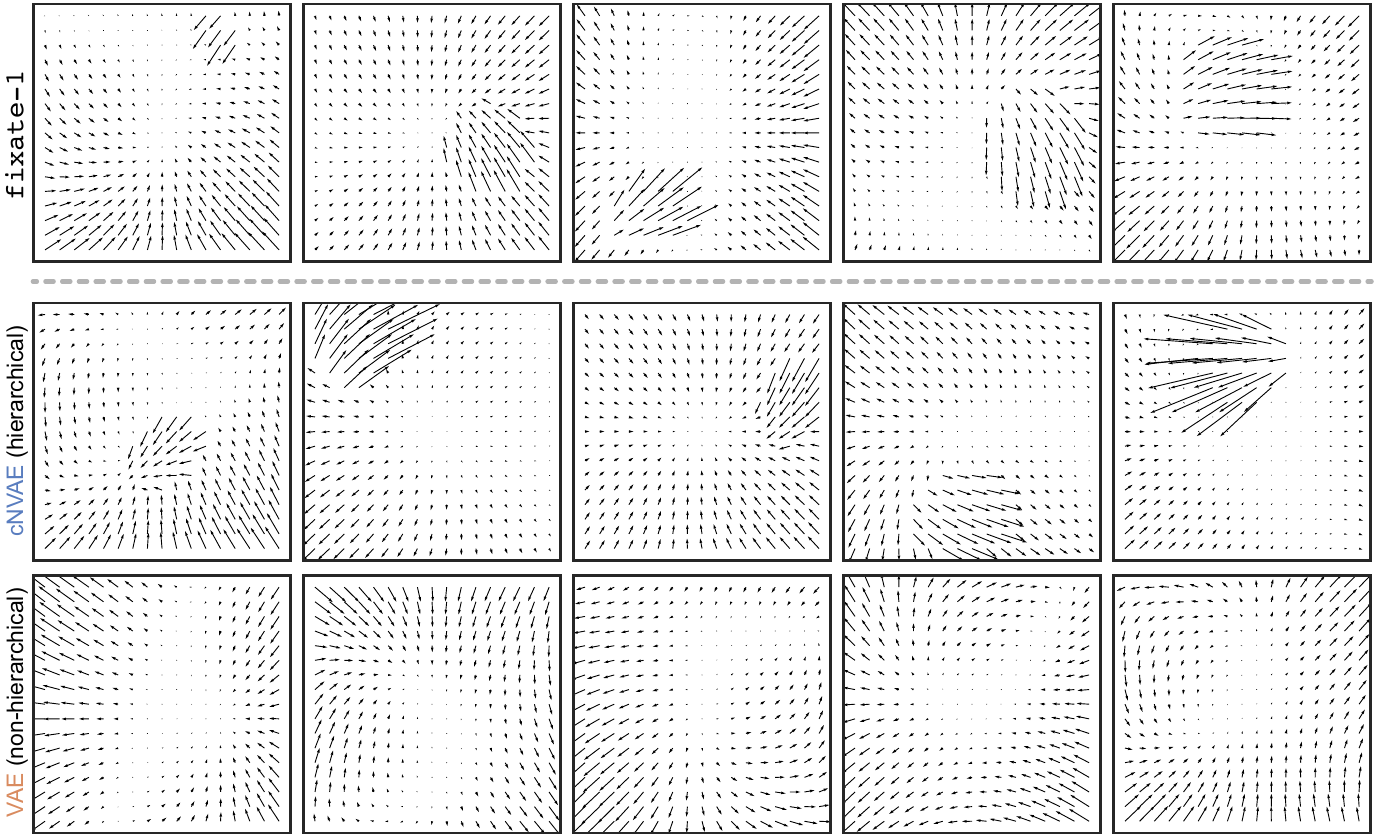}
    \caption[Synthetic samples generated by different models]{The first row shows random samples drawn from ROFL \fixate{1} category. The second and third rows are samples generated from cNVAE and VAE models that were trained on \fixate{1}. Generated samples from cNVAE closely resemble real data; whereas, VAE ignores objects. Samples were generated at a temperature of $t=1$ without cherry-picking.}
    \label{fig:samples}
\end{figure}

If a generative model has truly learned the underlying structure of a dataset, then it should be able to generate fake samples that look like its training data. Here, we examine generated samples from both the cNVAE and VAE models, as illustrated in \cref{fig:samples}. For comparison, we also include five samples randomly drawn from the \fixate{1} dataset that these models were trained on (\cref{fig:samples}, first row). The cNVAE samples appear indistinguishable from true samples; however, VAE fails to capture and generate objects, corroborating our previous results (see \cref{fig:3_mi}, \cref{fig:4_untangle}, and \cref{fig:indiv}). This striking divergence between cNVAE and VAE indicates that a hierarchically organized latent space is necessary for effectively modeling datasets that contain multiple interacting spatial scales.

\subsection{Latent traversal reveals multi-scale understanding in cNVAE}
\begin{figure}[ht!]
    \centering
    \includegraphics[width=0.95\textwidth]{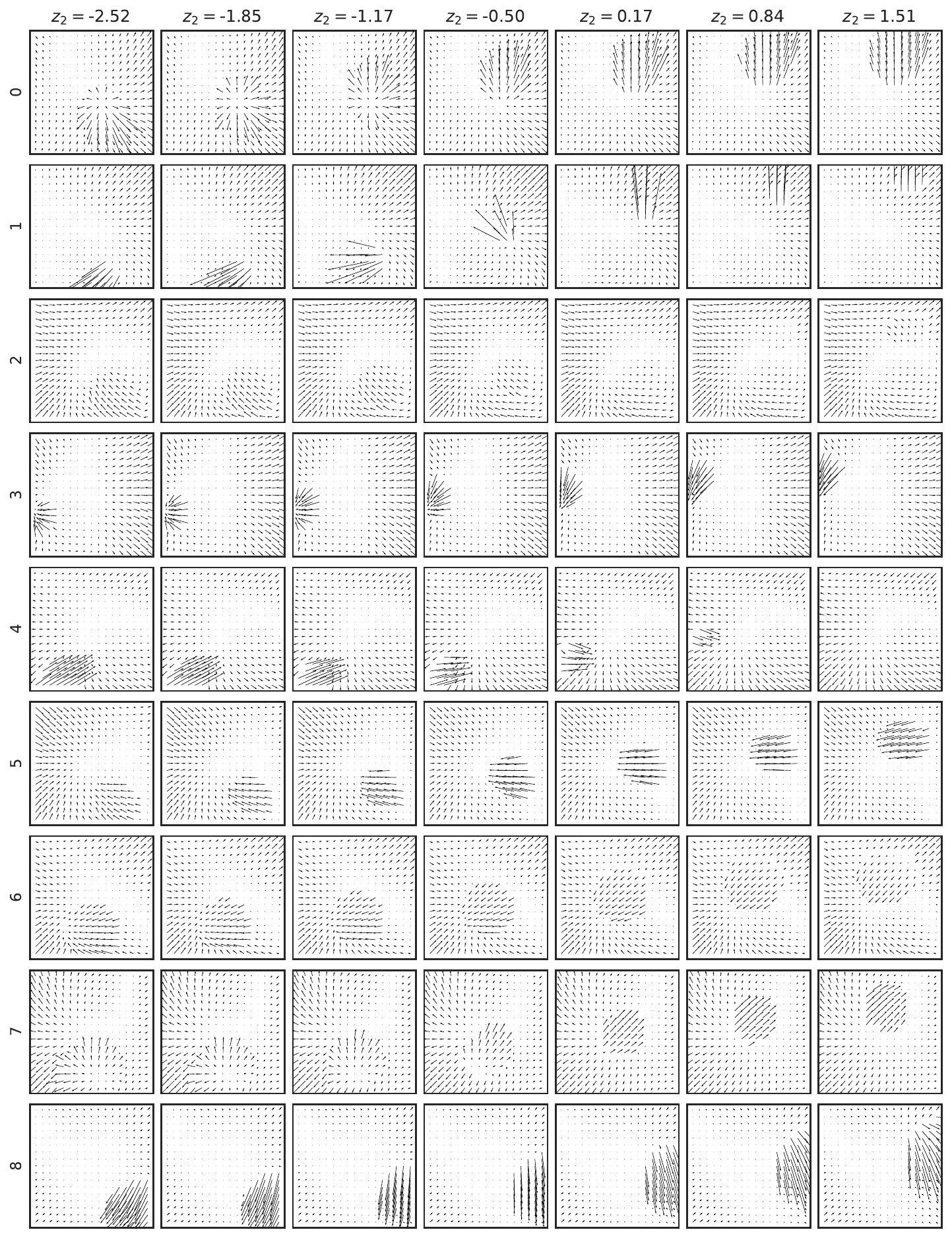}
    \caption[Latent traversal of a latent that captures object y-position]{Latent traversal performed using a latent variable that effectively captures $Y_{obj}$. This variable belongs to a latent group operating at the spatial scale of $2 \times 2$ (see \cref{fig:2_model,fig:3_mi}; \cref{tab:latent-dims}).}
    \label{fig:traverse}
\end{figure}

One desired property of VAE latent representations is that continuous interpolation along a latent dimension results in interpretable effects on the corresponding reconstruction. Here, we provide visual examples of latent traversal in cNVAE, focusing on a latent dimension that effectively captures $Y_{obj}$ (\Cref{fig:traverse}). These examples demonstrate intriguing interactions between object velocity patterns and self-motion within the scene. Notably, in some examples such as rows 0 and 6, relocating the object from the bottom to the top of the scene induces a significant change in its linear velocity along the y-axis. This phenomenon arises due to the influence of self-motion, as seen in the background. These illustrations collectively showcase that flow vectors in the visual scene are shaped both by local variables, such as the object's physical location, and their interactions with global-influencing variables like self-motion. By comprehending data across scales, cNVAE is able to capture such hierarchical and multi-scale properties inherent in naturalistic optic flow patterns. In contrast, non-hierarchical VAEs exhibit a complete failure in this regard.

\subsection{All models achieve good reconstruction performance}

\begin{figure}[ht!]
    \centering
    \includegraphics[width=0.83\linewidth]{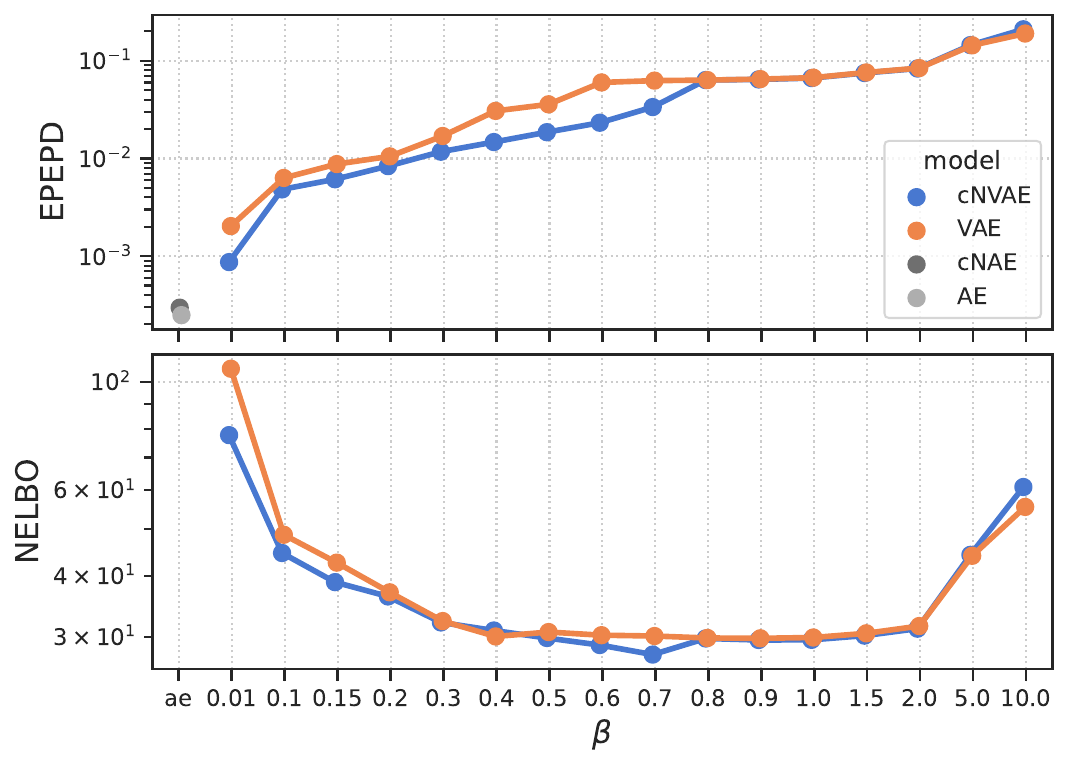}
    \caption[Reconstruction performance of cNVAE vs.~VAE]{Both models demonstrate robust reconstruction performance, with cNVAE exhibiting a slight improvement. EPEPD, endpoint error per dim (lower is better). NELBO, negative evidence lower bound (lower is better).}
    \label{fig:loss-v-beta}
\end{figure}

All models considered in this Chapter are trained based (in part or fully) on reconstruction loss, corresponding to how well they could reproduce a presented stimulus. Here we report the models' reconstruction performance, to ensure that they are all able to achieve a reasonable performance. \Cref{fig:loss-v-beta} illustrates how the reconstruction loss (EPEPD, endpoint error per dim) and negative evidence lower bound (NELBO) depend on different values of $\beta$ and model types. Both hierarchical and non-hierarchical VAE models perform well, with cNVAE exhibiting slightly superior performance for $\beta < 0.8$.

\subsection{Untangling and reconstruction loss are strongly anti-correlated for cNVAE but not VAE}

In this work, we demonstrated that a virtue of hierarchical VAEs is that they make information about ground truth factors easily (linearly) accessible in their latent representations. We were able to quantify this relationship because we used a simulation (ROFL) where ground truth factors were available. However, such a condition is not necessarily the case in real-world datasets that typically lack such direct knowledge \cite{eastwood2018framework}.

\begin{figure}[ht!]
    \centering
    \includegraphics[width=\linewidth]{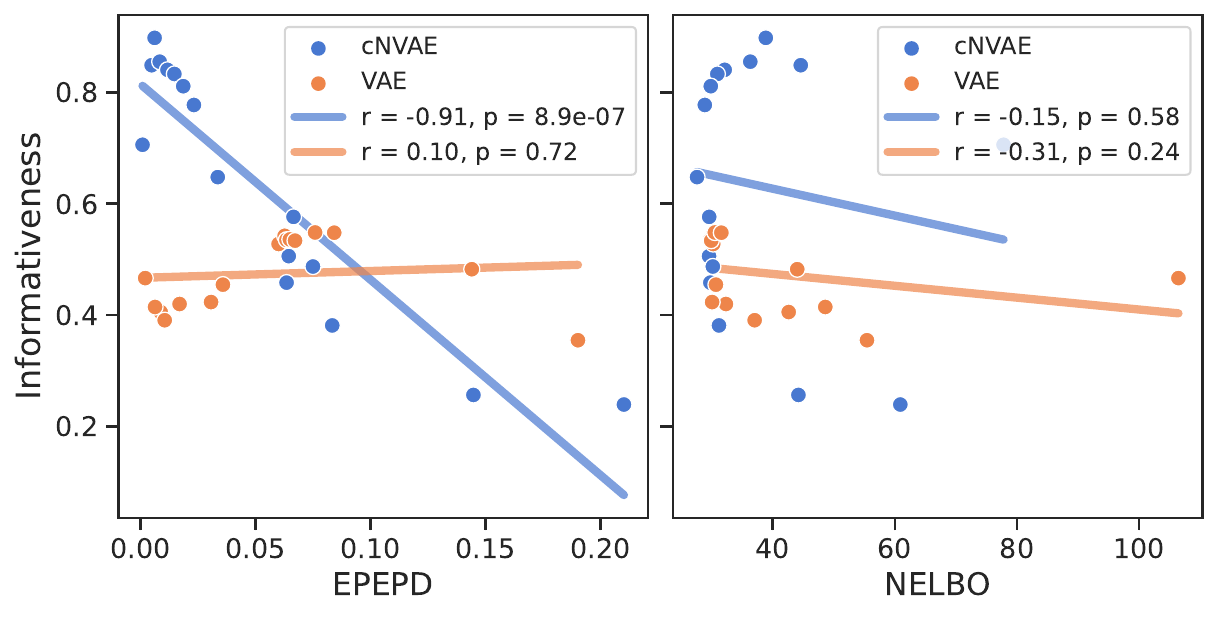}
    \caption[Strong anti-correlation between reconstruction loss and informativeness]{Strong anti-correlation between reconstruction loss and informativeness (i.e., untangling) observed only for cNVAE. Each scatter point corresponds to a model trained with a different $\beta$ value. EPEPD, endpoint error per dim (lower is better). NELBO, negative evidence lower bound (lower is better).}
    \label{fig:loss-v-i}
\end{figure}

To address this, we examined the relationship between informativeness and reconstruction loss, as well as NELBO, depicted in \cref{fig:loss-v-i}, to test whether there is a relationship between them. Strikingly, we observe a significant and strong negative linear relationship between reconstruction loss and informativeness in cNVAE ($r = -0.91, p = \num{8.9e-7}$, $t-$test). Conversely, no such correlation is observed for VAE ($r = 0.10, p = 0.72$, $t-$test). Notably, this relationship is absent for both models when comparing informativeness with NELBO (\cref{fig:loss-v-i}). These findings suggest that reconstruction loss---exclusively in the case of cNVAE---can serve as a viable proxy for informativeness, even in the absence of ground truth factors in real-world datasets.

\subsection{Brain alignment requires more than just linear regression performance}

In neuroscience research, linear regression $R^2$ scores have commonly been employed to measure alignment between biological and artificial neural networks \cite{yamins2014performance, schrimpf2018brain, mineault2021your, conwell2023can}. However, we found that relying solely on regression performance provided a weak signal for model selection, which is consistent with recent work \cite{conwell2023can, tuckute2022many, williams2021generalized, sexton2022reassessing, han2023system}. In contrast, our alignment score defined in \cref{fig:8_align_def} provided a much stronger signal for model discrimination (\cref{fig:7_perf} vs.~\cref{fig:8_align}).

To investigate this further, we conducted statistical tests, comparing pairs of models based on either their performance in explaining the variance of MT neurons or their brain-alignment scores. The results are presented in \Cref{tab:pvals}. We observed that ridge regression performance was not particularly effective in distinguishing between models. We measured the effect sizes between distributions of $N=141$ neurons, measuring how well ridge regression performance (\cref{fig:7_perf}) or alignment scores (\cref{fig:8_align}) discriminated between models. We found that the effect sizes were substantially larger for the alignment scores (\cref{fig:15_effect_sizes}). These findings suggest that while model performance is necessary, it is not sufficient for accurately measuring alignment between artificial and biological representations. In conclusion, refined and more comprehensive approaches are required to properly capture brain alignment \cite{williams2021generalized, duong2023representational, klabunde2023similarity, canatar2023spectral}.

\begin{table}[ht!]
\centering
\begin{tblr}{
    hlines, vlines,
    vline{1}={1.3pt},
    vline{2}={1.3pt},
    vline{5}={1.3pt},
    vline{8}={1.3pt},
    hline{1}={1.3pt},
    hline{3}={1.3pt},
    hline{20}={1.3pt},
    colsep=4pt,
    hspan=minimal,
    vspan=minimal,
    cell{1}{1}={r=2}{c},
    cell{1}{2}={c=3}{c},
    cell{1}{5}={c=3}{c},
    colspec={lllcllc},
}
    $\beta$ & Performance (\Cref{fig:7_perf}) &&& Alignment scores (\cref{fig:8_align}) && \\
    & effect size & $p-$value & significant? & effect size & $p-$value & significant?\\
    ae & -0.23 & 0.01 & \cmark & 1.2 & 8e-29 & \cmark\\
    0.01 & -0.28 & 0.002 & \cmark & 0.32 & 0.0004 & \cmark\\
    0.1 & -0.47 & 2e-07 & \cmark & 0.12 & 0.2 & \xmark\\
    0.15 & -0.16 & 0.07 & \xmark & 0.46 & 4e-07 & \cmark\\
    0.2 & -0.14 & 0.1 & \xmark & 0.33 & 0.0003 & \cmark\\
    0.3 & -0.012 & 0.9 & \xmark & 0.69 & 4e-13 & \cmark\\
    0.4 & -0.15 & 0.09 & \xmark & 0.65 & 7e-12 & \cmark\\
    0.5 & -0.013 & 0.9 & \xmark & 0.65 & 7e-12 & \cmark\\
    0.6 & 0.29 & 0.001 & \cmark & 0.69 & 4e-13 & \cmark\\
    0.7 & 0.16 & 0.08 & \xmark & 0.71 & 1e-13 & \cmark\\
    0.8 & 0.44 & 1e-06 & \cmark & 0.73 & 4e-14 & \cmark\\
    0.9 & 0.24 & 0.008 & \cmark & 0.96 & 1e-20 & \cmark\\
    1.0 & 0.0016 & 1 & \xmark & 0.89 & 1e-18 & \cmark\\
    1.5 & 0.17 & 0.07 & \xmark & 1.2 & 3e-28 & \cmark\\
    2.0 & 0.037 & 0.7 & \xmark & 0.91 & 3e-19 & \cmark\\
    5.0 & -0.24 & 0.008 & \cmark & 1.3 & 7e-31 & \cmark\\
    10.0 & -0.056 & 0.6 & \xmark & 0.55 & 3e-09 & \cmark\\
\end{tblr}
\caption[Hierarchical architectures are significantly more aligned with MT neurons]{Hierarchical architectures are significantly more aligned with MT neurons compared to non-hierarchical ones. Paired $t-$tests were performed, and $p-$values were adjusted for multiple comparisons using the Benjamini-Hochberg correction method \cite{benjamini1995controlling}. Effect sizes were estimated using Cohen's $d$ \cite{cohen1988statistical}, where positive values indicate cNVAE $>$ VAE (or cNAE $>$ AE, first row).}
\label{tab:pvals}
\end{table}

\subsection{Dissecting cross-validated model performance on neural data}

One measure of neural alignment that we consider here used unsupervised latents to predict experimentally recorded neural responses (via linear regression). The neurons in this dataset had long presentations of optic flow stimuli, but only a fraction of the neurons (105 out of 141 neurons) had multiple repeats of a short stimulus sequence. Because these repeats facilitated a much clearer measure of model performance, we used these where available, following previous work \cite{mineault2021your}. However, due to the short stimulus sequence ($\sim\!5$ sec; see \cref{fig:6_ephys}b) used to estimate performance in the neurons with stimulus repeats, here we perform additional tests to verify that our conclusions hold using a more rigorous cross-validation procedure.

Specifically, here we used the long stimulus sequence to both train model parameters (80\% of data) and determine hyperparameters (including regularization and optimal latencies) using a separate validation set (20\%). We then predicted model performance on the repeats. This is in contrast to our approach in the main results, which used the repeat data itself to optimize hyperparameters, but as a result, risked overfitting on the short stimulus sequence.

Indeed, we found that independent determination of hyperparameters (from the validation set) resulted in decreased model performance on the test set for both cNVAE and VAE models (\Cref{tab:repeat}), although with comparable effect sizes. As a result, our conclusions of statistically overlapping performance between cNVAE and VAE, with a slightly better performance of the cNVAE, were robust.

\begin{table}[t]
    \caption[Model performance on the subset of neurons with repeat data ($N=105$)]{Model performance on the subset of neurons for which we had repeat data ($N=105$). Here, we used the repeat data as a held-out test set. We found that model performances remain statistically indistinguishable, providing additional support for our main results.}
    \label{tab:repeat}
    \centering
    \begin{tabular}{cc}
        \toprule
        Model &
        \begin{tabularx}{115mm}{CCCC}
            \multicolumn{4}{c}{Performance, $R\,\,\,(\mu \pm se;\,\,N=105)$} \\
            \midrule
            $\beta = 0.5$ &
            $\beta = 0.8$ &
            $\beta = 1$ &
            $\beta = 5$
        \end{tabularx}
        \\

        \midrule
        {\color{cnvae_color}cNVAE} &
        \begin{tabularx}{115mm}{CCCC}
            $.452 \pm .025$ &
            $.479 \pm .025$ &
            $.486 \pm .024$ &
            $.462 \pm .025$ 
        \end{tabularx}
        \\

        {\color{vae_color}VAE} &
        \begin{tabularx}{115mm}{CCCC}
            $.461 \pm .027$ &
            $.401 \pm .029$ &
            $.466 \pm .028$ &
            $.461 \pm .027$ 
        \end{tabularx}
        \\

        {\color{cnae_color}cNAE} &
        $.460 \pm .026$
        \\

        {\color{ae_color}AE} &
        $.412 \pm .031$
        \\
        
        \bottomrule
  \end{tabular}
\end{table}

\section{Additional Discussion}\label{sec:supp-discussion}

\subsection{Visual-only cues and unsupervised learning are sufficient for untangling optic flow causes}

It is well known that an observer's self-motion through an environment can be estimated using optic flow patterns alone \cite{gibson1950perception}, although it has been proposed that the brain also uses a combination of visual and non-visual cues \cite{DeAngelisAngelaki2012}, such as efference copies of motor commands \cite{von1954relations}, vestibular inputs \cite{macneilage2012vestibular}, and proprioception \cite{cullen2021proprioception}. Indeed, using optic flow alone can be complicated by rotational eye movements and independently moving objects \cite{gibson1954visual, royden1996human, warren1995perceiving}. How can the brain estimate the motion of an object when self-motion is non-trivially affecting the entire flow field \cite{horrocks2023walking, noel2023causal}, including that of the object itself? This is a critical motion processing problem \cite{lee1980optic}, manifested in diverse ecological scenarios such as prey-predator dynamics, crossing the street, or playing soccer. Are retinal visual-only cues sufficient to extract information about different self-motion components, as well as object motion \cite{lappe1999perception}?

Unlike the approach of \textcite{mineault2021your}, our work is based on the premise that retinal-only signals contain sufficient information about their ground truth causes and that these factors can be extracted using unsupervised learning alone. In contrast, \textcite{mineault2021your} trained their DorsalNet by providing dense supervision on self-motion components, arguing that such supervised training is biologically plausible since these signals are approximately available to the observer from vestibular inputs and other such extra-retinal sources.

Our results (\cref{fig:4_untangle}) provide proof-of-concept that it is possible to untangle ground truth causes of retinal optic flow in a completely unsupervised manner, using the raw optic flow patterns alone. Comparing the performance of alternative models revealed two essential ingredients for the success of our approach: (1) an appropriate loss function---inference \cite{helmholtz1867handbuch}; and, (2) a properly designed architecture---hierarchical, like the sensory cortex \cite{mumford1992computational}.

\subsection{ROFL in an active, multimodal setting}

A recent study by \textcite{noel2023causal} examined the role of causal inference in attributing retinal motion to self and object motion during closed-loop navigation. The study reports that in the active self-motion condition, self-motion is estimated using both optic flow and an efference copy of the motor commands from hands controlling a joystick that drove virtual velocity. In contrast, during passive self-motion, there was no efference copy, which led to more uncertain self-motion estimates and greater misattribution between self and object motion.

In our own study, we found that cNVAE was particularly successful in untangling fixation and self-motion-related variables, almost reaching the ceiling value of $R^2 \sim 1.0$. However, it did not reach this maximum value for the object-related variables (\cref{fig:4_untangle}). Based on these observations and the findings of \textcite{noel2023causal}, we speculate that the addition of some extra-retinal signal as weak supervision in an active setting would help cNVAE learn even better representations of its retinal inputs. Training hierarchical VAEs in active settings \cite{ha2018world}, using multimodal signals \cite{DeAngelisAngelaki2012}, while characterizing the interplay between different sensory modalities \cite{fetsch2009dynamic, butler2010bayesian} is an exciting topic for future work.

\subsection{Comparing cNVAE to NVAE, its non-compressed counterpart}

In \cref{fig:4_untangle}, we investigated the performance of the original non-compressed NVAE with an approximately matched latent dimensionality ($440$, slightly larger than $420$ for cNVAE: more details are provided in \cref{sec:latent-dims}). We found that although NVAE did better than non-hierarchical VAE, it still underperformed cNVAE. This is because trying to match the latent dimensionality of NVAE with cNVAE necessitates reducing the number of hierarchical latent groups in the NVAE since most of its convolutional latents capture spatial dimensions rather than more abstract features as cNVAE does. From previous work \cite{child2021vdvae}, we know that the \say{stochastic depth} of hierarchical VAEs is the key reason behind their effectiveness; therefore, it was expected that reduced depth would hurt an NVAE with a matched number of latents.

It is worth noting that cNVAE and NVAE had a balanced performance across all ground truth factors (\cref{fig:4_untangle}), in that they captured both global (self-motion-related) and local (object-related) aspects well. In contrast, non-hierarchical VAE completely ignored object-related variables and focused solely on the global scale determined by fixation point and self-motion. This suggests that hierarchical architecture is crucial for understanding multi-scale datasets such as ROFL.

\subsection{Disentanglement is in the eyes of the beholder}

The conventional definition of disentanglement in machine learning relies on relating the latent representation to predefined generative factors, a choice that is often not uniquely determined, and is often guided by heuristics \cite{higgins2018towards}. For example, in this Chapter, we chose to represent velocities in Cartesian coordinates, rather than polar coordinates and judged the degree of disentanglement based on the relationship of the latents to these factors. Notably, these coordinate systems are related by a nonlinear transform, and latents linearly related to one coordinate system will then be less linearly related to the other.

\begin{figure}[ht!]
    \centering
    \includegraphics[width=\linewidth]{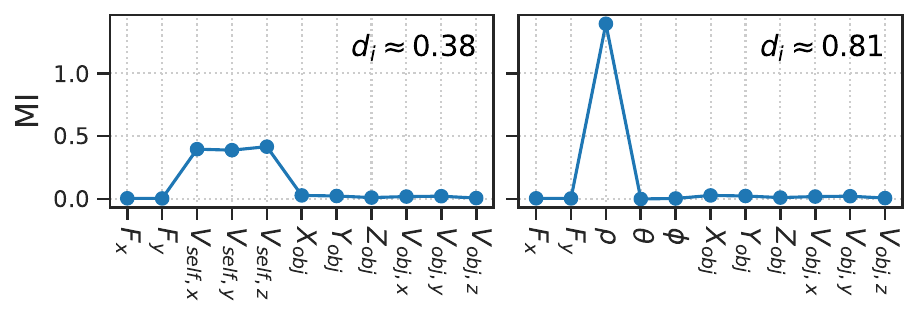}
    \caption[Disentanglement is in the eyes of the beholder]{This latent variable cares about the magnitude of self-motion only. Its disentanglement score depends strongly on what we choose a priori as our independent generative factors.}
    \label{fig:polar2cart}
\end{figure}

Throughout our experiments, we observed one or more cNVAE latent variables that exhibited equally high mutual information with all components of self-motion velocity in Cartesian coordinates (e.g., latent dimensions 154 and 122 in \cref{fig:3_mi}). When we translated the generative factors into polar coordinates and looked at their relationship with these latents, we found that these latents were solely encoding the magnitude of self-motion, irrespective of its direction (\cref{fig:polar2cart}). Thus, these seemingly highly entangled latent variables would have achieved almost perfect disentanglement scores had we employed polar coordinates instead of Cartesian coordinates to represent the ground truth velocities $\left(d_i \approx 0.38 \rightarrow d_i \approx 0.81 \right)$.

Our findings shed light on the fact that disentanglement cannot be considered an inherent property of objective reality; rather, it is significantly influenced by our a priori determination of independent factors of variation within the data. Given the inherent complexity and richness of natural datasets, it becomes challenging to unequivocally determine the true factors of variation a priori, pointing to the need for a more systematic approach.

When it comes to interpreting brain functions, perhaps considering the behavioral relevance of generative factors could guide our choices. For example, an animal directing its own movements does not pick grid coordinates but rather chooses direction and speed, possibly making a polar representation more \say{behaviorally relevant} to representations within its brain.

Another promising approach lies in focusing on the symmetry transformations inherent in the world \cite{higgins2018towards, higgins2022symmetry}. By seeking representations that respect the underlying geometrical and group-theoretical structures of the world \cite{anselmi2016invariance, bronstein2021geometric}, and comparing them in principled ways \cite{williams2021generalized, duong2023representational, klabunde2023similarity}, we may uncover new perspectives that provide valuable insights beyond the conventional and subjective notions of disentanglement.

\section{Appendix: ROFL derivations and details}\label{sec:rofl}

In this section, I will provide more details on ROFL and show how simple rules of projective geometry can give rise to complex motion patterns generated by the ROFL simulation engine.

ROFL aims to generate synthetic optical flow patterns that closely resemble those arising from ecologically relevant types of motion. This entails the consideration of both self-motion and object-motion components.

Specifically, the self-motion component encompasses various subtypes such as translational and rotational motions induced by eye or head movements. Furthermore, it is crucial to have control over the relative prominence of these two types of motion. Depending on the specific configuration, the resulting flow pattern can exhibit an \emph{object-dominated} characteristic when an abundance of objects is incorporated, or conversely, a \emph{self-motion-dominated} nature if fewer or no objects are present.

By carefully manipulating these factors, ROFL enables the generation of optical flow patterns that faithfully capture the dynamics of real-world scenarios involving an interplay between self-motion and object motion.

\subsection{ROFL setup}
Self-motion is meaningful if the observer is moving relative to a static background (i.e., the environment). We consider a solid background wall at a distance $Z=1$ from the observer which extends to infinity in the $X-Y$ plane. Define the following coordinate frames

\begin{enumerate}
   \item \textbf{Fixed coordinates}, represented with capital letters: $\left(X, Y, Z\right)$.
   \item \textbf{Observer's coordinates}, represented with lowercase letters: $\left(x, y, z\right)$.
\end{enumerate}

Both coordinate systems are centered around the observer. The difference is that $\hat{Z}$ is always perpendicular to the background plane, while $\hat{z}$ is defined to be parallel to the fixation point.

\begin{figure}[ht!]
    \centering
    \includegraphics[width=\textwidth]{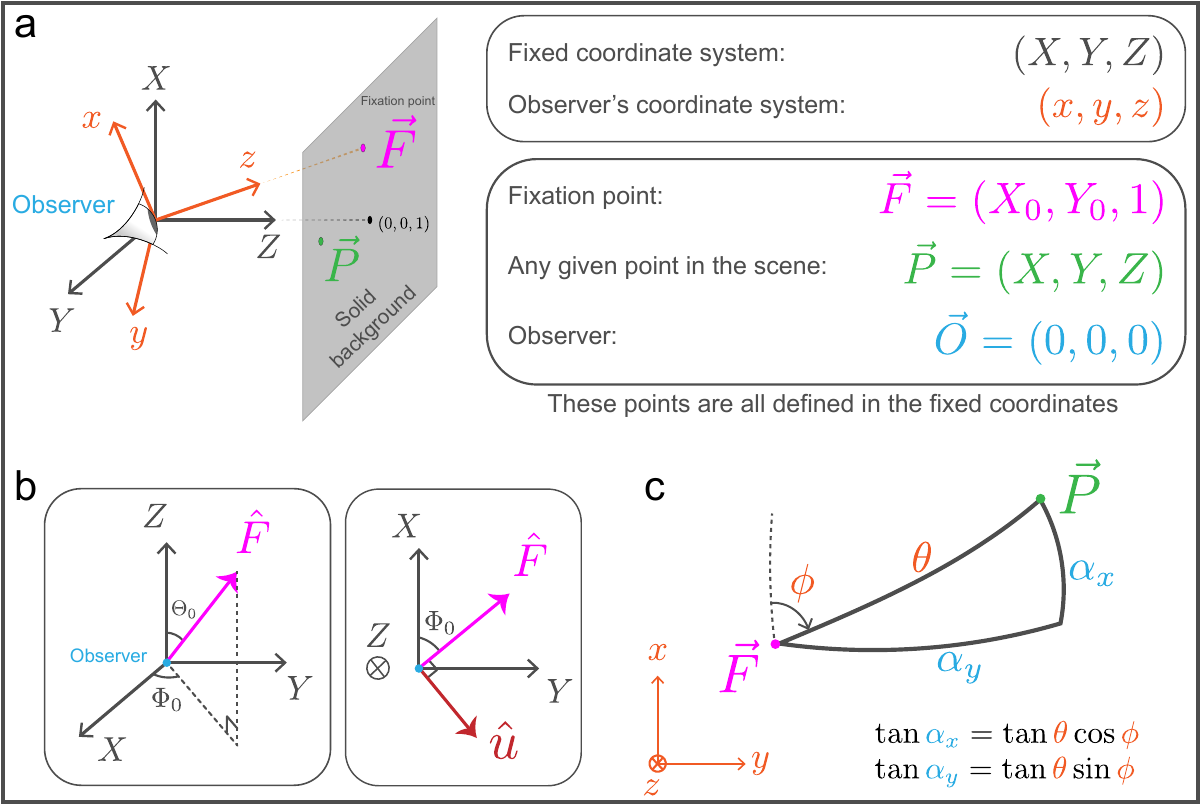}
    \caption[ROFL setup and derivations]{Setup. \textbf{(a)} Background and coordinate systems. \textbf{(b)} The rotation of the fixed coordinates $\left(\hat{X}, \hat{Y}, \hat{Z}\right)$ onto the observer-centric coordinates $\left(\hat{x}, \hat{y}, \hat{z}\right)$ can be accomplished by applying a rotation of angle $\Theta_0$ around the unit vector $\hat{u}$ as defined in \autoref{eq:u}. Note that by definition, $\hat{z}$ is always parallel to the fixation point. See the text for more details. \textbf{(c)} Perspective from the observer’s point of view. The retinotopic coordinates of point $\vec{P}$ are denoted by $\alpha_x$ and $\alpha_y$, measured in radians.}
    \label{fig:setup}
\end{figure}

All points in the space are initially expressed in the fixed coordinates. For example, any given point like $\Vec{P} = \left(X, Y, Z\right)$ has its coordinates defined in the fixed system. The same is true for fixation point, $\Vec{F} = \left(X_0, Y_0, Z\right)$, and the observer's position $\Vec{O} = \left(0, 0, 0\right)$, and so on (\cref{fig:setup}).

\paragraph{\textbf{Relating the two coordinate systems.}} How do we go from the fixed coordinates to the observer's coordinates? We need to find a rotation matrix that maps the $\hat{Z}$ direction to the direction defined by the fixation point. By definition, this will determine $\hat{z}$ and all the other directions in the observer's coordinates.

First, let us express $\vec{F}$ using its polar coordinates

\begin{equation}
    \vec{F} = \left(X_0, Y_0, Z\right) \equiv \left(\norm{\vec{F}}, \Theta_0, \Phi_0\right),
\end{equation}
where $\norm{\vec{F}} = \sqrt{X_0^2 + Y_0^2 + Z^2}$, $\cos\Theta_0 = Z/\norm{\vec{F}}$, and $\tan\Phi_0 = Y_0/X_0$. It turns out what we are looking for is a rotation by an amount $\Theta_0$ about a unit vector $\hat{u}$ that depends only on $\Phi_0$. Formally, this is achieved by

\begin{equation}\label{eq:u}
    \hat{u}(\Phi_0) = \begin{pmatrix}
        -\sin\Phi_0\\
        \cos\Phi_0\\
        0
        \end{pmatrix},
\end{equation}
where the rotation matrix can be written as $R_{\hat{u}}(\Theta_0) = I + \left(\sin\Theta_0\right) L(\hat{u}) + \left(1-\cos\Theta_0\right)L(\hat{u})^2$.

Here, $L(\Vec{\omega})$ is the skew-symmetric matrix representation of a vector $\Vec{\omega}$, that is

\begin{equation}\label{eq:skew}
    L(\Vec{\omega}) \coloneqq \Vec{\omega}\cdot\Vec{L} = \omega_XL_X + \omega_YL_Y + \omega_ZL_Z,
\end{equation}
where $L_i$ are the familiar basis matrices for $\mathfrak{so}(3)$, the Lie algebra of $SO(3)$:

\begin{equation}
L_X = \begin{pmatrix}
    0 & 0 & 0\\
    0 & 0 & -1\\
    0 & 1 & 0\\
\end{pmatrix},
\quad L_Y = \begin{pmatrix}
    0 & 0 & 1\\
    0 & 0 & 0\\
    -1 & 0 & 0\\
\end{pmatrix},
\quad L_Z = \begin{pmatrix}
    0 & -1 & 0\\
    1 & 0 & 0\\
    0 & 0 & 0\\
\end{pmatrix}.
\end{equation}

The rotation matrix in its full form is expressed as

\begin{equation}\label{eq:rot}
R \coloneqq R_{\hat{u}}(\Theta_0) = \begin{pmatrix}
    s^2 + c^2\cos\Theta_0 & -cs\left[1 - \cos\Theta_0\right] & c \sin\Theta_0\\
    -cs\left[1 - \cos\Theta_0\right] & c^2 + s^2\cos\Theta_0 & s \sin\Theta_0\\
    -c \sin\Theta_0 & -s \sin\Theta_0 & \cos\Theta_0\\
\end{pmatrix},
\end{equation}
where we have defined $c \coloneqq \cos\Phi_0$ and $s \coloneqq \sin\Phi_0$.

Any point $\Vec{P} = \left(X, Y, Z\right)$ defined in the fixed coordinate system can be mapped to a point $\Vec{p} = R^T \Vec{P}$ with components $\Vec{p} = \left(x, y, z\right)$ given in the observer's system. For example, it can be shown that $R^T \Vec{F}$ is parallel to $\hat{z}$, which is how we defined $R$ in the first place. See \cref{fig:setup}.

In conclusion, we have a rotation matrix $R$ that is fully determined as a function of the fixation point $\Vec{F}$. Applying $R$ rotates the fixed coordinate system onto the observer's system

\begin{equation}
\begin{aligned}
    \hat{x} &\coloneqq R \hat{X},\\
    \hat{y} &\coloneqq R \hat{Y},\\
    \hat{z} &\coloneqq R \hat{Z}.
\end{aligned}
\end{equation}

This way, the observer's coordinates $\left(x, y, z\right)$ are obtained ensuring that the fixation point is always parallel to $\hat{z}$.

\paragraph{\textbf{Adding motion.}} Consider a point $\Vec{P} = \left(X, Y, Z\right)$. In the previous section, we showed that the rotation matrix in \autoref{eq:rot} can be used to relate $\Vec{P}$ to its observer-centric coordinates

\begin{equation}\label{eq:transform}
    \Vec{P} = R\vec{p}.
\end{equation}

Differentiate both sides to get

\begin{equation}
    \Vec{V} \coloneqq \frac{d\Vec{P}}{dt} = \frac{dR}{dt}\vec{p} + R\frac{d\Vec{p}}{dt},
\end{equation}
where $\Vec{V}$ is the velocity of the point relative to the observer, with components measured in the fixed coordinate system.

\paragraph{\textbf{Derivative of the rotation matrix.}} Here we will derive a closed-form expression for $dR/dt$ in terms of the observer's angular velocity. This is important because it essentially captures the effects of eye movement.

Let us use $\Vec{\omega} = (\omega_X, \omega_Y, \omega_Z)$ to denote the angular velocity of observer's coordinates, with components expressed in fixed coordinates. The time derivative of $R$ is given as follows

\begin{equation}
    \frac{dR}{dt} = L(\Vec{\omega}) R,
\end{equation}
where $L(\Vec{\omega})$ is given in \autoref{eq:skew}. Taken together, equations 3, 4, 8, and 9 yield

\begin{equation}\label{eq:vel}
\boxed{
    \begin{aligned}
        \frac{d\Vec{p}}{dt} &= R^T\Vec{V} - \left[R^T L(\Vec{\omega}) R\right] \Vec{p}\\
        &= R^T\Vec{V} - \left[R^T
        \begin{pmatrix}
            0 & -\omega_Z & \omega_Y\\
            \omega_Z & 0 & -\omega_X\\
            \omega_Y & \omega_X & 0\\
        \end{pmatrix} R\right] \Vec{p}
    \end{aligned}
}
\end{equation}

This equation allows relating the velocity of a moving point expressed in fixed coordinates, $\Vec{V}$, to its observer-centric equivalent, $\Vec{v} \coloneqq d\Vec{p}/dt$. Importantly, \autoref{eq:vel} is a general relationship valid for any combination of velocities and rotations.

\subsection{Motion perceived by the observer}
Before discussing specific applications, we need to solve one remaining piece of the puzzle: what is $\Vec{v}$, and how is it related to the motion perceived by the observer? To address this, we need to map the retinotopic coordinates of the observer (measured in radians), to points in the environment (measured in the observer-centric coordinates).

Denote the retinotopic coordinates using $\left(\alpha_x, \alpha_y\right)$. The goal is to relate each $\left(\alpha_x, \alpha_y\right)$ pair to a point in the world $\Vec{p}$ with coordinates $\left(x, y, z\right)$. It can be shown that

\begin{equation}\label{eq:retina}
\begin{aligned}
    \tan\alpha_x &= x/z = \tan\theta\cos\phi,\\
    \tan\alpha_y &= y/z = \tan\theta\sin\phi,
\end{aligned}
\end{equation}
where $\theta$ and $\phi$ are the polar coordinates of $\Vec{p}$. See \cref{fig:setup}c for an illustration.

Differentiate both sides of \autoref{eq:retina} and rearrange terms to obtain the time derivative of $\left(\alpha_x, \alpha_y\right)$ as a function of $\Vec{v} =\left(\dot{x}, \dot{y}, \dot{z}\right)$ and $\Vec{p}$. That is,

\begin{equation}\label{eq:vel-retina}
\boxed{
\begin{aligned}
    \dot{\alpha_x} &= \frac{\dot{x}z - \dot{z}x}{z^2 + x^2}\\
    \dot{\alpha_y} &= \frac{\dot{y}z - \dot{z}y}{z^2 + y^2}
\end{aligned}
}
\end{equation}

Thus, we obtained a relationship between retinotopic velocity $(\dot{\alpha_x}, \dot{\alpha_y})$ (measured in $radians/s$) and velocity of objects in the world (measured in $m/s$). In conclusion, equations \ref{eq:vel} and \ref{eq:vel-retina} enable us to compute the observer's perceived retinotopic velocity for any type of motion.

\subsection{Example: maintaining fixation}
We are now ready to apply the formalism developed in the preceding section to create optical flow patterns in realistic settings. In this section, we will consider a specific example: self-motion while maintaining fixation on a background point. This is an important and ecologically relevant type of motion, as we often tend to maintain fixation while navigating the environment---think about this next time you are moving around. In fact, one can argue that fixation occurs more naturally during behaviors such as locomotion, as opposed to typical vision experiments where the animals are stationary and passively viewing a screen.

What does it mean to maintain fixation while moving? It means the eyes rotate to cancel out the velocity of the fixation point. Assume the observer is moving with velocity $\Vec{V}_{self}$. Equivalently, we can imagine the observer is stationary while the world is moving with velocity $\Vec{V} = -\Vec{V}_{self}$ (this is the same $\Vec{V}$ that appears in \autoref{eq:vel}).

Any eye movement can be represented mathematically as a rotation of the observer's coordinates with respect to the fixed coordinates. Use $\vec{\omega}$ to denote the angular velocity of the rotation. We can determine $\vec{\omega}$ by requiring that the rotation cancels out the velocity of the fixation point. To this aim, we first compute the normal component of $\Vec{V}$ with respect to the fixation point $\Vec{F}$ as follows

\begin{equation}
    \Vec{V}_{\perp} \coloneqq \Vec{V} - \frac{\Vec{V} \cdot \Vec{F}}{\norm{\Vec{F}}^2} \Vec{F}.
\end{equation}

Use the above equation to compute the angular velocity

\begin{equation}\label{eq:omega}
    \Vec{\omega} = \frac{\Vec{F} \times \Vec{V}_{\perp}}{\norm{\Vec{F}}^2}.
\end{equation}

Plug \autoref{eq:omega} in \autoref{eq:vel} to find $\Vec{v}$, and use \autoref{eq:vel-retina} to find $(\dot{\alpha_x}, \dot{\alpha_y})$. Thus, we calculated the motion perceived by the observer during self-motion while maintaining fixation.

\paragraph{\textbf{The algorithm.}} Start by choosing a fixation point $\Vec{F} = \left(X_0, Y_0, Z\right)$, field-of-view extent ($fov$), and retinal resolution ($res$). For example, $fov = 45^\circ$ and $res = 5.625^\circ$. Use this to create a mesh grid for $\left(\alpha_x, \alpha_y\right)$. The grid covers $\left(-fov, +fov\right)$ degrees in each dimension and has shape $2 * fov/res + 1$, or $17 \times 17$ in our example. Use \autoref{eq:retina} to compute $\left(\theta, \phi\right)$ pair for each point on the $\left(\alpha_x, \alpha_y\right)$ grid.

Because the simulation is scale-invariant, we can always fix the distance between the observer and the background wall. Assume $Z = cte. = 1$, and use equations \ref{eq:rot} and \ref{eq:transform} to find $\Vec{p} = \left(x, y, z\right)$ per grid point (note: the environment points have infinite resolution).

So far, we partitioned the retina into a grid and found the observer-centric coordinates of a point in the real world that falls on each grid point. Now we can sample a random self-motion velocity $\Vec{V}_{self}$ and use the results described in the previous sections to find $(\dot{\alpha_x}, \dot{\alpha_y})$. This provides us with the $2D$ retinal velocity vector on the grid and concludes the algorithm for the simple case of \fixate{0} (i.e., no objects).

Please refer to our code for additional details, including instructions on how to incorporate independently moving objects into the scene.

\subsection{Valid fixation points}
The observer-wall distance is constrained to be a positive value. That is, the observer is always situated on the left side of the wall ($Z > 0$ for all points; \cref{fig:setup}). This constraint, combined with a given $fov$ value and \autoref{eq:rot}, results in a theoretical bound on valid fixation points. To derive this bound, we start by considering the $\hat{Z}$ component of \autoref{eq:transform}

\begin{equation}
    \left(R\Vec{p}\right)_3 = R_{31} x + R_{32} y + R_{33} z = Z > 0.
\end{equation}

Divide both sides by $z$ and use \autoref{eq:retina} to get

\begin{equation}
    R_{31}\tan\alpha + R_{32}\tan\beta + R_{33} > 0.
\end{equation}

Plug the matrix entries from \autoref{eq:rot} and rearrange to get

\begin{equation}
\begin{aligned}
    \sin\Theta_0 (\cos\Phi_0 &\tan\alpha + \sin\Phi_0 \tan\beta ) < \cos\Theta_0,\\
    \quad\quad \Rightarrow \quad X_0 &\tan\alpha + Y_0 \tan\beta < Z.
\end{aligned}
\end{equation}

For a given $fov$ value, we have $- \tan\left( fov \right) \leq \tan\alpha,\tan\beta \leq \tan\left( fov \right)$. Assuming $Z=1$, we find that the $\left(X_0, Y_0\right)$ coordinates of a valid fixation point should satisfy the following inequality

\begin{equation}
    |X_0| + |Y_0| < \frac{1}{\tan\left( fov \right)}.
\end{equation}

In other words, the $L_1$ norm of $\left(X_0, Y_0\right)$ should be less than $1/\tan\left( fov \right)$. For example, when $fov = 45^\circ$, the upper bound is $1$. This result can be used to choose an appropriate $fov$: smaller values allow a larger phase space to explore for fixation points.

\subsection{ROFL conclusions}

We introduced Retinal Optic Flow Learning (ROFL), a novel synthetic data framework to simulate realistic optical flow patterns. The main result is given in \autoref{eq:vel}, which shows the simulation has the potential to generalize to various other types of motion. For instance, the observer might engage in smooth pursuit of a moving object instead of maintaining fixation, a scenario that can be modeled simply by adjusting the choice of $\vec{\omega}$ in \autoref{eq:vel}. This flexibility enables us to generate a diverse repertoire of realistic optical flow patterns, which can be used in training and evaluating unsupervised models, among other applications.

We believe that our framework will prove valuable not only to neuroscientists studying motion perception but also to machine learning researchers interested in disentangled representation learning. By bridging these domains, we aim to foster interdisciplinary collaboration and advance our understanding of both biological and artificial information processing systems.


\renewcommand{\thechapter}{5}

\chapter{Cortical network organization in mice}

The content of this chapter has been published in \textit{Nature Communications} as a research Article entitled: \sayit{Multimodal measures of spontaneous brain activity reveal both common and divergent patterns of cortical functional organization} \cite{vafaii2024multimodal}.

This was joint work between myself, Francesca Mandino, Gabriel Desrosiers-Gr\'egoire, David O'Connor, Marija Markicevic, Xilin Shen, Xinxin Ge, Peter Herman, Fahmeed Hyder, Xenophon Papademetris, Mallar Chakravarty, Michael C.~Crair, and R.~Todd Constable. The work was jointly supervised by Evelyn MR. Lake, and my advisor Luiz Pessoa.

\section{Introduction}
Brains exhibit low-dimensional functional organization across spatiotemporal scales, from synapses to the whole organ. This functional organization varies between individuals and dynamically changes over time, as well as across states such as health, injury, or disease. Understanding the principles that govern brain organization would enable their use as clinical indices. Closing knowledge gaps requires work in humans and model species, across scales, and using complementary sources of image contrast.

Here, we focus on large-scale systems (i.e., \textit{networks}), as a deeper understanding of their characteristics stands to have broad prognostic and diagnostic utility, in part because they can be assessed with noninvasive imaging methods that are applicable in human subjects.

Much of what we know and can access about large-scale systems, especially in humans, comes from the blood-oxygenation-level-dependent (BOLD) contrast obtained with functional magnetic resonance imaging (fMRI). Growing evidence shows that measures of large-scale systems obtained with fMRI-BOLD (or proximal optical measures of hemoglobin) are, to an extent, reflective of neural activity \cite{Lake2020SimultaneousCF, Ma2016RestingstateHA, wright2017functional, vanni2017mesoscale, murphy2018macroscale}. Yet, despite important progress, the relationship between fMRI-BOLD and underlying neural activity is complex and remains incompletely understood \cite{winder2017weak, drew2020ultra, drew2022neurovascular}. This has historically posed several challenges to interpreting network organization obtained using the fMRI-BOLD technique \cite{logothetis2004interpreting, logothetis2008we, Buckner2013OpportunitiesAL, kullmann2020editorial, lu2019origins}.

A powerful tool for investigating the functional organization of large-scale networks is wide-field fluorescence imaging in mouse models bearing genetically encoded calcium (\CA) sensitive indicators \cite{Cardin2020MesoscopicIS, ren2021characterizing}. Critically, \CA imaging affords a large field of view covering much of the mouse cortical mantle and provides image contrast that is a more direct measure of neural activity than BOLD. Applied with fMRI-BOLD (or BOLD-like measures), \CA imaging can reveal the neuronal component captured by the BOLD signal \cite{Matsui2016TransientNC, Ma2016RestingstateHA, murphy2018macroscale, he2018ultra, Lake2020SimultaneousCF, ma2022gaining}. Here, we leverage a novel simultaneous multimodal framework, BOLD-fMRI and \CA imaging \cite{Lake2020SimultaneousCF}, to determine both cross-modal convergent and divergent features of large-scale functional networks in the mouse cortex.

As in previous studies using similar \cite{Matsui2016TransientNC, wright2017functional} or the same experimental approach \cite{Lake2020SimultaneousCF}, we examine \textit{functional connectivity}, a widely employed approach used to characterize large-scale brain networks that are defined based on inter-regional synchrony. Importantly, we consider networks as having \textit{overlapping}, rather than disjoint, functional organization. By overlapping, we mean a network structure in which each brain region, or node, can belong to more than one community. In contrast, in a \textit{disjoint} setting, each node is forced to belong to a single community.

Many complex systems, including biological, technological, and social ones, are inherently overlapping rather than disjoint \cite{palla2005uncovering, barabasi2011network, menche2015uncovering}. In the brain, overlap means that regions participate across multiple communities (to varying degrees), consistent with the notion that functionally flexible regions can contribute to multiple brain processes \cite{cole2013multi, fedorenko2013broad, yeo2015functional, Najafi2016OverlappingCR}. Although evidence for overlap in human brain networks has accrued based on multiple analysis techniques applied to BOLD-fMRI data \cite{Yeo2014EstimatesOS, Najafi2016OverlappingCR, cookson2022evaluating, Faskowitz2020EdgecentricFN, cookson2022evaluating}, it is unclear whether the putative overlapping organization is driven, at least in part, by the nature of BOLD signals. To the best of our knowledge, the potentially overlapping functional organization of cortical networks has not been tested in animal models, where fMRI-BOLD can be obtained together with \CA signals that exhibit greater spatiotemporal resolution and capture neural activity more directly.

Here, we use a highly-sampled and simultaneously recorded wide-field \CA and fMRI-BOLD dataset to resolve whether functional network structure discovered with BOLD are also detected with \CA imaging, while determining their overlapping organization. We use a Bayesian generative algorithm that estimates the membership strength of a given brain region to all communities \cite{Gopalan2013EfficientDO, Najafi2016OverlappingCR, cookson2022evaluating}. Importantly, this approach also allows detection of disjoint organization in a data-driven manner. In addition, region-level properties are quantified including node degree \cite{achard2006resilient, bullmore2009complex} and diversity \cite{Power2013EvidenceFH, Liska2015FunctionalCH, Bertolero2017TheDC}, while a wide range of parameters are explored to test the robustness of our findings (\Cref{table:params}).

Overall, overlapping network organization is robustly detected in simultaneous wide-field \CA and fMRI-BOLD data regardless of the parameters selected. Evidence of rich overlapping organization advances our fundamental understanding of cortical brain organization, helping to further validate the neural origins of clinically accessible fMRI-BOLD network organization.

\begin{table}[ht!]
    \centering
    \begin{tblr}{
        width=0.75\linewidth,
        hlines, vlines,
        colspec={X[1.2,c,m] X[1.9,c,m]},
        colsep=4pt,
        rows={5.5ex},
        hspan=minimal,
    }
    {\large Parameter} & {\large Values and figures}
    \\
    
    Number of communities \break (= networks) & \textbf{3} (\Cref{fig:2_oc3}); \textbf{7} (\Cref{fig:2_ocs7}); \break
    and \textbf{20} (\Cref{fig:2S_oc20})
    \\
    
    Initial parcellation granularity & \textbf{Fine} (542 regions, \Cref{fig:1_roi}b); and \break
    \textbf{Coarse} (152 regions, \Cref{fig:2S_ocs_n128})
    \\
    
    Graph edge density & \textbf{10--25\%} (\Cref{fig:2S_cos-all,fig:6S_deg-all}b)
    \\
    
    fMRI preprocessing pipeline & \Cref{fig:6S_deg-all}a
    \\
    \end{tblr}
\caption[Parameters explored in this Chapter]{Parameters explored in this Chapter. To ensure the robustness of our findings, we explored a range of parameters and found that our results were qualitatively reproduced across all conditions.}
\label{table:params}
\end{table}

\begin{figure}[ht!]
    \centering
    \includegraphics[width=\linewidth]{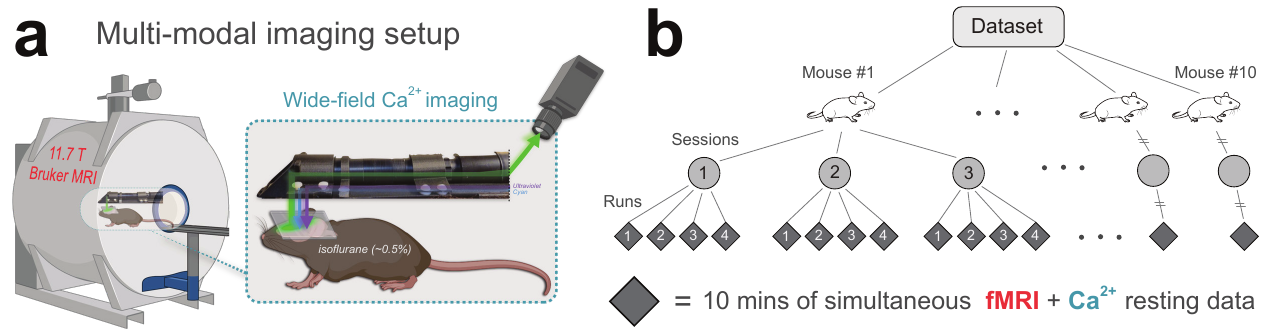}
    \caption[Simultaneous fMRI-BOLD and wide-field \CA imaging]{Experimental setup. \textbf{(a)} Simultaneous fMRI-BOLD and wide-field \CA imaging \cite{Lake2020SimultaneousCF}. \CA data are background-corrected (illustrated by three colored wavelengths; Methods) \textbf{(b)} Hierarchical data structure. $N=10$ mice, scanned across 3 longitudinal sessions, with 4 runs per session, each lasting 10 minutes. Panel a was created with \url{https://BioRender.com}. Mouse illustration in panel b was downloaded from \url{https://scidraw.io/} \cite{mouse}.}
    \label{fig:1_data}
\end{figure}

\section{Results}

Mice (N=10) expressing GCaMP6f in excitatory neurons underwent simultaneous wide-field \CA and BOLD-fMRI, as described previously by \textcite{Lake2020SimultaneousCF} (see Methods, \cref{fig:1_data}a). Animals were lightly anesthetized ($0.50\%$--$0.75\%$ isoﬂurane) and head-fixed. Data were collected in 3 different sessions, with at least one week between the sessions (longitudinal recording). Each session contained 4 runs, each lasting 10 minutes for a total of $1,200$ minutes of data (\cref{fig:1_data}b).

BOLD data (acquisition rate $1~Hz$, Methods) were processed using RABIES (Rodent Automated BOLD Improvement of EPI Sequences) \cite{desrosiers2022rodent, Grandjean2020CommonFN, grandjean2023consensus} and high-pass filtered \cite{smith2013resting} ($0.01$--$0.5~Hz$). Given that \CA and BOLD signals are maximally correlated when \CA is temporally band-passed to match BOLD \cite{Ma2016RestingstateHA, Lake2020SimultaneousCF, he2018ultra}, and the \say{lowpass} nature of the BOLD signal \cite{deyoe1994functional, sauvage2017hemodynamic, mann2021coupling}, we investigated network measures within a slow (BOLD-matched) and fast ($0.5$--$5~Hz$) \CA frequency range (herein, \CAS and \CAF). \CA data were acquired at an effective background-corrected rate of $10~Hz$ and processed as described by us previously \cite{OConnor2022functional}. Critically, we collected both GCaMP-sensitive and GCaMP-insensitive optical measurements for the removal of background fluorescence and hemoglobin signals from the \CA data (\cite{Lake2020SimultaneousCF, OConnor2022functional, allen2017global, valley2020correction}; Methods).

\begin{figure}[ht!]
    \centering
    \includegraphics[width=1.0\linewidth]{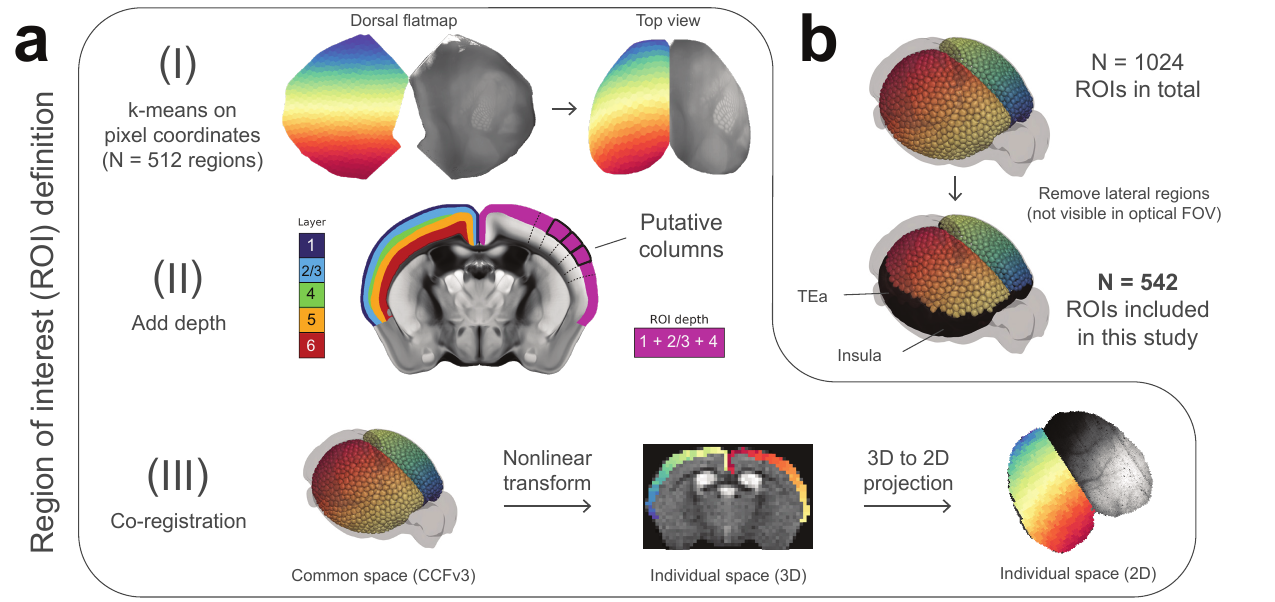}
    \caption[Defining ROIs within the Allen Mouse Brain Common Coordinate Framework]{Defining ROIs within the Allen Mouse Brain Common Coordinate Framework (CCFv3) \cite{Wang2020TheAM}. \textbf{(a)} (I) Division of the mouse dorsal flatmap into $N=1024$ spatially homogeneous ROIs. (II) Add depth by following streamlines normal to the cortical surface. The resulting ROIs are ``column-like". (III) Transform ROIs from common space into 3D and 2D individual spaces (Methods). Dorsal flatmap, layer masks, and columnar streamlines from CCFv3. \textbf{(b)} Analyses were restricted to ROIs that appeared in the \CA imaging FOV after multimodal co-registration (Methods). Lateral areas including the insula and temporal association areas were excluded.}
    \label{fig:1_roi}
\end{figure}

To build functional networks, a common set of regions of interest (ROIs) were defined (\cref{fig:1_roi}a; Methods). To relate 3D BOLD and 2D \CA data, we adopted the CCFv3 space for the mouse brain provided by the Allen Institute for Brain Sciences \cite{Wang2020TheAM}. ROIs covered most of the cortex. Areas not well captured in the wide-field \CA imaging FOV were excluded (\cref{fig:1_roi}b). Correlation matrices were computed for each acquisition run using pairwise Pearson correlation. Matrices were binarized by retaining the top $d \%$ strongest edges (\Cref{table:params}).

\begin{figure}[ht!]
    \centering
    \includegraphics[width=1.0\linewidth]{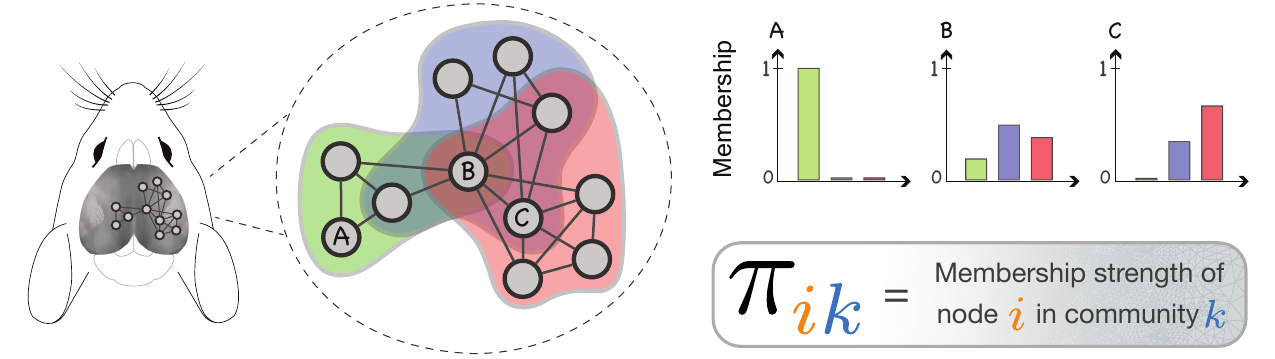}
    \caption[Estimating community structure using a mixed-membership stochastic blockmodel]{We applied a mixed-membership stochastic blockmodel algorithm to estimate overlapping communities \cite{Gopalan2013EfficientDO}. \textit{Membership strength} (values between $0$ and $1$) quantifies the affiliation strength of a node in a network. Here, node $A$ belongs only to the green community, node $B$ belongs to all three communities with varying strengths, and node $C$ belongs to the blue and red communities with varying strengths. Mouse head illustration was downloaded from \url{https://scidraw.io/} \cite{mouse_head}.}
    \label{fig:1_pi}
\end{figure}

We used a mixed-membership stochastic blockmodel algorithm \cite{airoldi2008mixed} that can detect overlapping (or disjoint) communities \cite{Gopalan2013EfficientDO, Najafi2016OverlappingCR, cookson2022evaluating}. The algorithm determines \textit{membership} values for each ROI, with one value per network (\cref{fig:1_pi}). Membership values sum to 1 across networks, which allows these values to be interpreted as probabilities. Overlapping networks, and by extension membership values, were computed at the level of runs and then averaged across sessions to determine an animal-level result. Random-effects group analysis was evaluated based on animal-level estimates and variability. The main results are from 542 ROIs and $d=15\%$.

\subsection{Cortical organization captured by overlapping network solutions}

Hereafter, \textit{network} is used interchangeably with \textit{overlapping community}, as is \textit{node} with \textit{region}. Existing work has shown the decomposition of the mouse cortex into as few as 2--3 networks \cite{Liska2015FunctionalCH, vanni2017mesoscale, Coletta2020NetworkSO}, but 7--10 is more typical \cite{white2011imaging, Sforazzini2014DistributedBA, Zerbi2015MappingTM, Grandjean2020CommonFN}. We explored a range of numbers of networks, from coarse to fine to ultra-fine: $K = 3, 7, \text{ and } 20$.

\subsubsection{The \texorpdfstring{$\bm{K = 3}$}{K = 3} network decomposition}
\begin{figure}[ht!]
    \centering
    \includegraphics[width=1.0\linewidth]{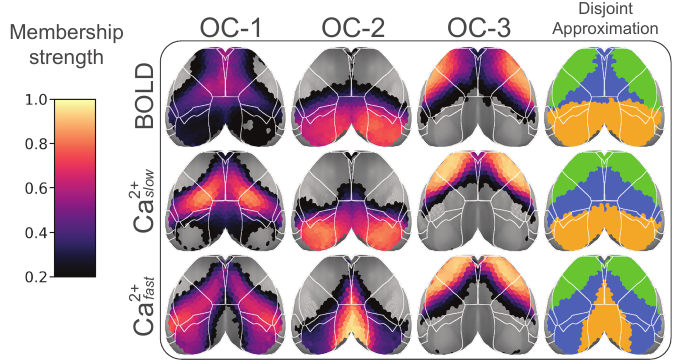}
    \caption[A coarse, 3 networks decomposition of the mouse cortex]{A coarse, 3 networks decomposition of the mouse cortex into overlapping networks. Color scale indicates membership strengths (\cref{fig:1_pi}). The disjoint approximation is obtained by taking each region's maximum membership value. See \cref{fig:2_allen} for a legend showing different brain regions (white overlay lines). OC, overlapping community.}
    \label{fig:2_oc3}
\end{figure}

Our 3-network solution captured previously observed systems, namely, the visual (overlapping community 2, OC-2) and somatomotor (OC-3), as well as a large system (OC-1) that included territories previously classified as the mouse \say{default network} \cite{stafford2014large, Sforazzini2014DistributedBA, Gozzi2016LargescaleFC, Whitesell2021RegionalLA} (\cref{fig:2_oc3}). To facilitate comparisons to standard \textit{disjoint} algorithms, we forced a disjoint version of our solutions by assigning each region to the network with the largest community membership value.

\begin{figure}[ht!]
    \centering
    \includegraphics[width=0.57\linewidth]{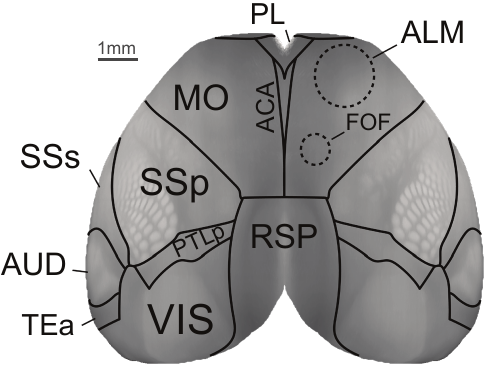}
    \caption[Mouse cortical areas (top view) as defined in the CCFv3 Allen reference atlas]{Mouse cortical areas (top view) as defined in the CCFv3 Allen reference atlas \cite{Wang2020TheAM}. In addition, dashed lines approximately correspond to functionally defined subregions in the secondary motor area \cite{erlich2011cortical, Chen2017AMO}. Abbreviations: ACA, anterior cingulate area; ALM, anterior lateral motor cortex; FOF, frontal orienting field; MO, somatomotor areas; PL, prelimbic area; PTLp, posterior parietal association areas; RSP, retrosplenial area; SSp, primary somatosensory area; SSs, supplemental somatosensory area; VIS, visual areas.}
    \label{fig:2_allen}
\end{figure}

The white overlay lines in \cref{fig:2_oc3} illustrate anatomical boundaries. In \cref{fig:2_allen}, we have visualized anatomically defined cortical regions in the mouse cortex. These regions are defined within the Allen Reference Atlas (ARA) ontology \cite{Wang2020TheAM}.

\subsubsection{The \texorpdfstring{$\bm{K = 7}$}{K = 7} network decomposition}
\begin{figure}[ht!]
    \centering
    \includegraphics[width=1.0\linewidth]{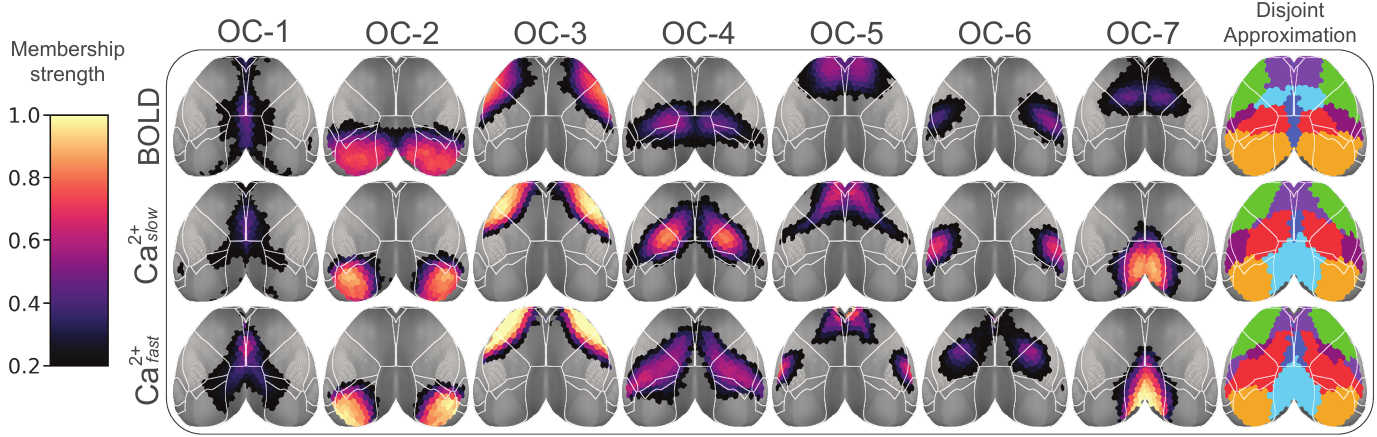}
    \caption[Decomposition with 7 networks]{Decomposition with 7 networks. Color scale indicates membership strengths (\cref{fig:1_pi}). The disjoint approximation is obtained by taking each region's maximum membership value. See \cref{fig:2_allen} for a legend showing different brain regions (white overlay lines). OC, overlapping community.}
    \label{fig:2_ocs7}
\end{figure}

With 7 networks, well-defined visual and somatomotor networks (OC-2 and OC-3, respectively) were again identified \cite{Zerbi2015MappingTM, Grandjean2020CommonFN}, alongside additional systems covering bilateral and well-defined cortical territories (\cref{fig:2_ocs7}). OC-1 encompassed medial areas including the cingulate cortex but also extended more laterally. OC-4 spanned from medial to lateral areas, including the somatosensory cortex. For both \CAF and \CAS, OC-4 also included the frontal orienting field (FOF), a possible homolog of the frontal eye field in primates \cite{Brecht2011MovementCA, erlich2011cortical, Barthas2017SecondaryMC, ebbesen2018more, Sato2019InterhemisphericallyDR}. OC-5 largely overlapped with the anterior lateral motor area, a region involved in motor planning \cite{Chen2017AMO, allen2017global, makino2017transformation, Inagaki2022NeuralAA}; notably, for \CAF this network also included the supplementary somatosensory area. OC-6 overlapped with the barrel field for BOLD and \CAS but captured the upper limb somatosensory cortex for \CAF. Finally, OC-7 was very different for BOLD and both \CA conditions; for BOLD, it was centered around FOF, and for both \CA conditions it was centered around the retrosplenial cortex. We also investigated the 7-network organization in a subset of animals that underwent an additional awake imaging session that measured \CA signal outside the MRI scanner.

\subsubsection{The \texorpdfstring{$\bm{K = 20}$}{K = 20} network decomposition}
The 20-network solution is shown in \cref{fig:2S_oc20}, which revealed finer spatial networks that were again bilateral (like the 3- and 7-network solutions). Notably, even with 20 networks, the FOF did not appear as a separate network for either \CA signal, in contrast to BOLD. In sum, across solutions (3, 7, and 20 networks), recognized functional organization (established brain regions, functional networks, and a high degree of bilateral symmetry) was uncovered using our algorithm across imaging modalities. 

\subsubsection{Inter-modal network organization similarity (\texorpdfstring{$\bm{K = 7}$}{K = 7})}

\begin{figure}[ht!]
    \centering
    \includegraphics[width=1.0\linewidth]{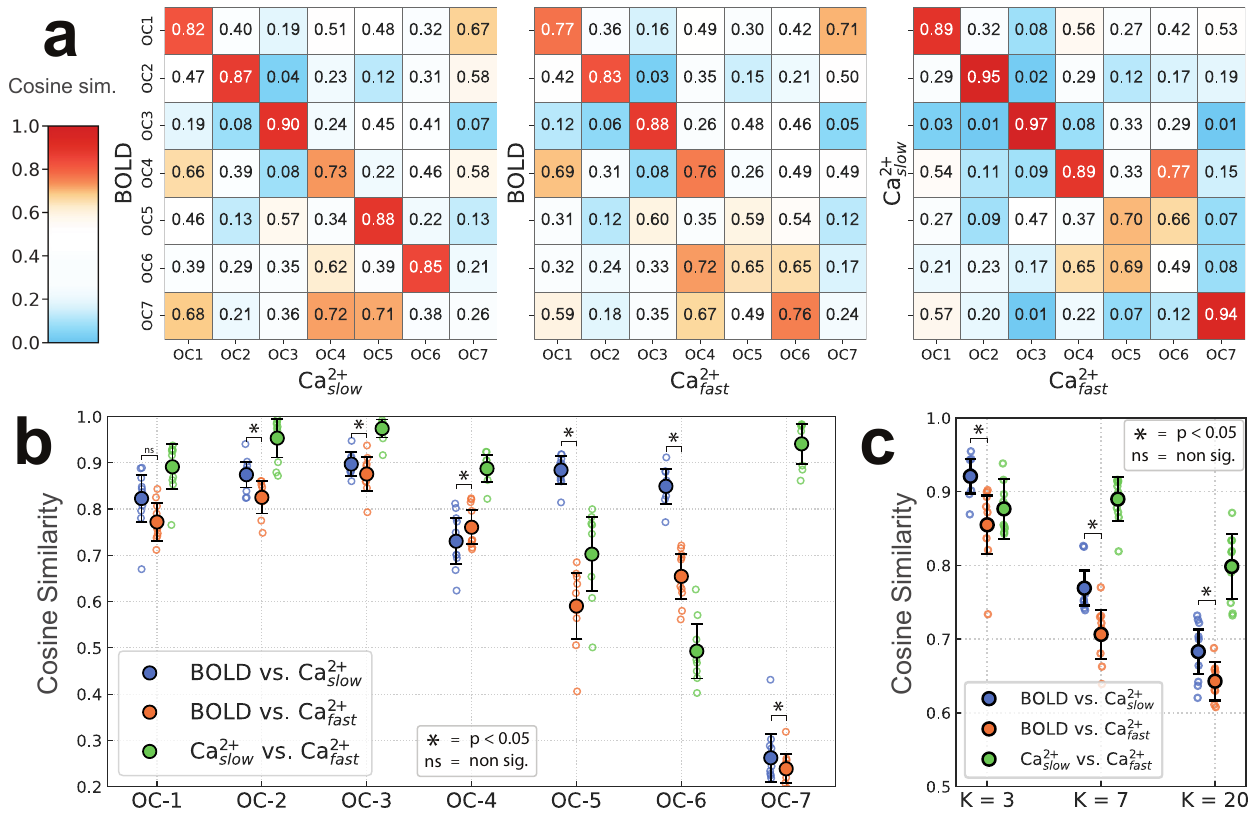}
    \caption[Quantitative comparison of network similarities]{Quantitative comparison of network similarities \textbf{(a)} Network similarity based on cosine similarity (1 = identical, 0.5 = ``orthogonal", 0 = perfectly dissimilar or ``inversely correlated"). Color scale emphasizes similarities and strong dissimilarities. 
    \textbf{(b)} Diagonal elements of matrices in D are plotted. Empty circles correspond to individual animals; the large solid circle is the group average. \textbf{(c)} Overall similarity collapsing across networks as a function of number of networks (3, 7, and 20). \textbf{(b-c)} Comparison of BOLD and \CAS networks ($p<0.05$, Holm-Bonferroni corrected). Error bars are $95\%$ confidence intervals based on hierarchical bootstrap.}
    \label{fig:2_oc7-cos}
\end{figure}

For 7 networks, BOLD, \CAS, and \CAF were quantitatively compared (\cref{fig:2_oc7-cos}a-c) to test the hypothesis that band-pass filtering \CA to match BOLD leads to greater inter-modal agreement. The comparison was based on cosine similarity ($1$ indicates identical organization, $0.5$ indicates \say{orthogonal/unrelated} organization, and $0$ indicates perfectly \say{opposite} organization). The similarity between BOLD and \CAS networks was relatively high ($> 0.73$), except for OC-7 ($0.26$), a network that was evident in both \CA conditions but not captured by BOLD. In comparison, BOLD and \CAF similarity was generally lower but still relatively high for OC-1 to OC-4 ($> 0.77$), though modest for OC-5 and OC-6 ($0.59$ and $0.65$, respectively). Overall, band-pass filtering \CA seemed to have a modest network-dependent impact when comparing network territories across modalities.

To generate a summary metric, we collapsed across networks to generate an overall index of similarity (\cref{fig:2_oc7-cos}c). As expected \cite{Ma2016RestingstateHA, wright2017functional, Lake2020SimultaneousCF}, BOLD and \CAS solutions were more similar than BOLD and \CAF ($p<0.05$, permutation test, Holm–Bonferroni corrected).

\begin{figure}[ht!]
    \centering
    \includegraphics[width=0.81\linewidth]{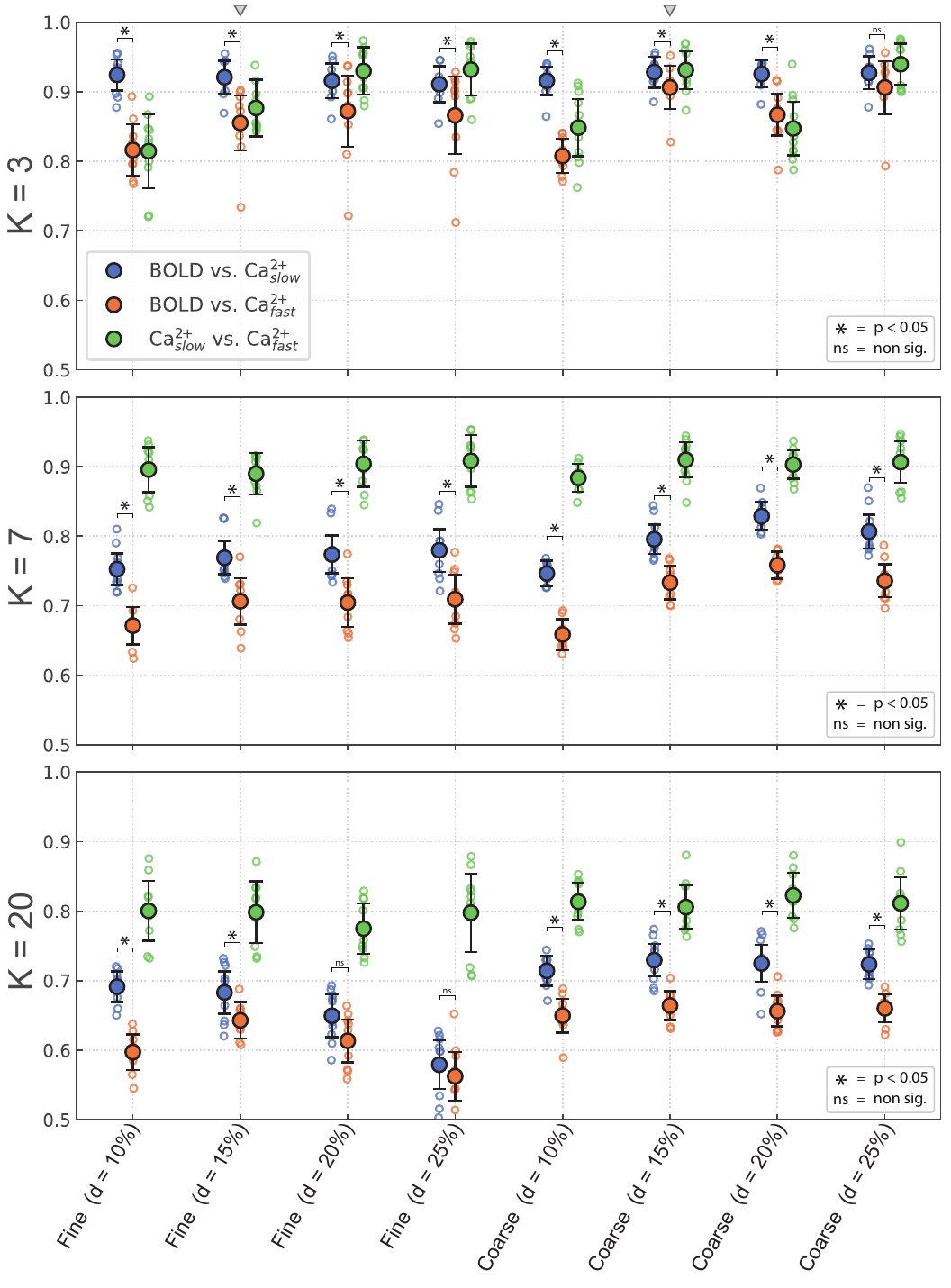}
    \caption[BOLD network organization is more similar to \CAS than it is to \CAF]{BOLD network organization is more similar to \CAS than it is to \CAF. This is a robust finding in this Chapter, as it is reproduced over a large combinatorial space of analysis conditions. The y-axis is cosine similarity and the x-axis corresponds to different conditions. See \cref{fig:2S_ocs_n128}b for a visual comparison of coarse versus fine ROIs. $d$ is graph density after edge-filtering is applied. Small triangles indicate our choices for the main results: $d= 15\%$, fine ROIs. Permutation test, $p<0.05$, Holm-Bonferroni corrected. Error bars show $95\%$ confidence intervals.}
    \label{fig:2S_cos-all}
\end{figure}

Finally, we repeated the analysis across many combinations of parameters to test the robustness of these results. We found that this result was stable across 3, 7, and 20 networks, data processing parameters (\cref{fig:2S_cos-all}), including edge density, and number of ROIs (\cref{fig:2S_ocs_n128}).

\subsubsection{Awake results}
\begin{figure}[ht!]
    \centering
    \includegraphics[width=1.0\linewidth]{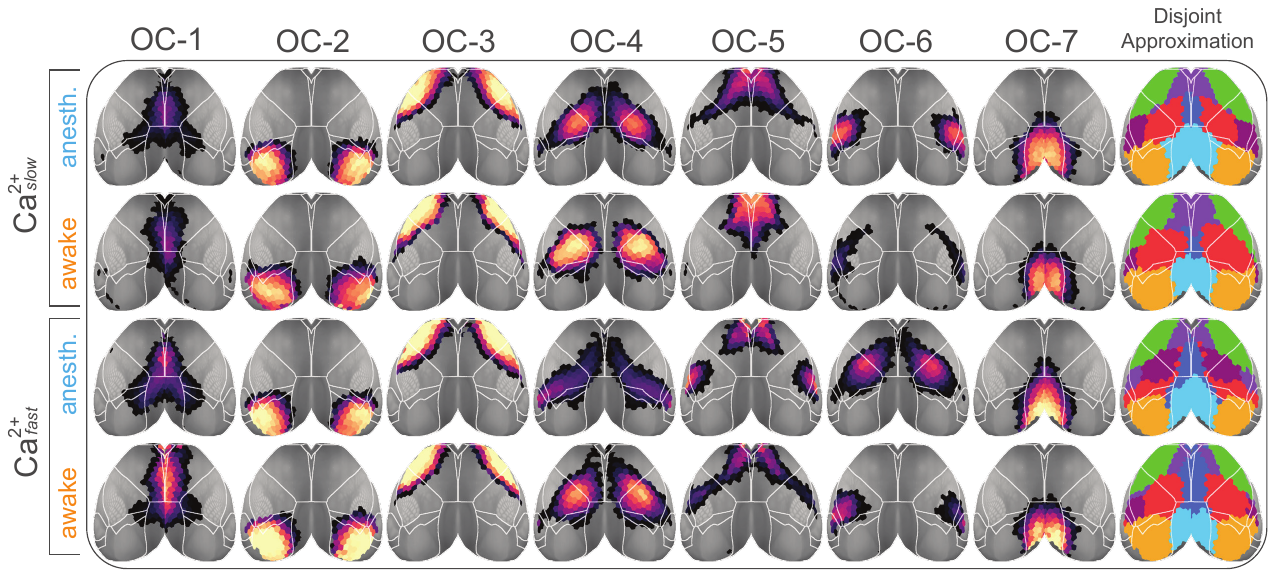}
    \caption[Network structure in the awake state (\CA only)]{Awake results (\CA only). We report exploratory analysis in a group of $N=5$ animals for which we had \CA recordings in both anesthetized and awake states. Along with the awake results, we also plotted group results obtained from the same subset of $N=5$ animals in the anesthetized state. The overall community structure is preserved irrespective of the animal state. This is expected because network structure is a coarse measure of brain organization, and partially because of the low dosage of anesthetics used in our study (Methods). Despite the overall similarities, we still observe some differences, most visibly in \CAF. Further, the group result obtained from half of the anesthetized dataset is almost identical to the full $N=10$ results shown in \cref{fig:2_ocs7}, indicating the robustness and reproducibility of our approach.}
    \label{fig:2S_oc7-awake}
\end{figure}

We repeated the 7 network analysis using a subset of animals for which we had out-of-scanner recordings in the awake state. Notably, the overall organization in awake animals (\cref{fig:2S_oc7-awake}) was qualitatively very similar to that obtained with lightly anesthetized animals.

\subsubsection{Individual results}
All results reported so far were obtained at the group level. Here, we perform the 7-network decomposition at the level of individual animals to test whether our results are reproduced for each mouse separately. We confirmed that the basic organization of the 7-network solution was also observed at the individual level (\cref{fig:2S_oc7-indiv}), indicating the generalizability of our results.

\subsubsection{HRF-filtered \CA networks}
\begin{figure}[ht!]
    \centering
    \includegraphics[width=\linewidth]{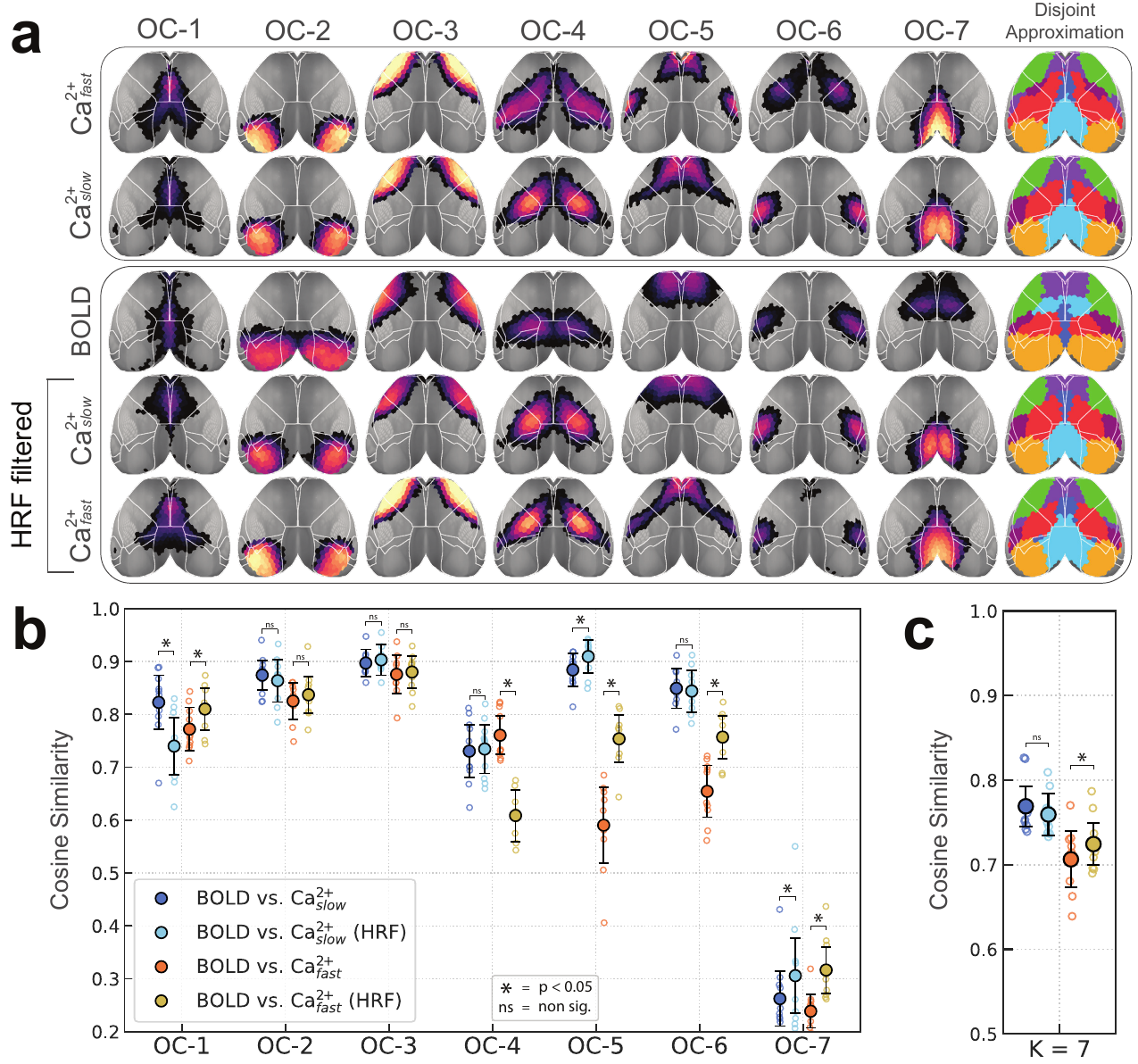}
    \caption[Filtering \CA data with a hemodynamic response function (HRF)]{Filtering \CA data with a hemodynamic response function (HRF) results in modest changes in network structure. We applied the gamma-variate model of \textcite{Ma2016RestingstateHA}, using parameters previously published by us in \textcite{Lake2020SimultaneousCF}. We then inferred the community structure of HRF-filtered \CA data using steps otherwise identical to those in the main results (Methods). \textbf{(a)} The top three rows are reproduced from \cref{fig:2_ocs7} to facilitate visual comparison, with a reordering of rows that puts BOLD adjacent to HRF-\CA results in the bottom two rows. \textbf{(b, c)} Comparison of BOLD vs.~HRF-\CA results relative to BOLD vs.~\CA results. \textbf{(b)} Per-network; and, \textbf{(c)} Overall cosine similarity. Similar to \cref{fig:2_ocs7}b and c, error bars are $95\%$ confidence interval. A paired permutation test was performed to compare conditions ($p<0.05$, Holm-Bonferroni corrected).}
    \label{fig:2S_oc7-hrf}
\end{figure}

To further probe some of the differences between BOLD and \CA results, we repeated our network similarity analysis after applying a gamma-variate hemodynamic filter \cite{Ma2016RestingstateHA} to the calcium signal, which we obtained from our previous work \cite{Lake2020SimultaneousCF} (\cref{fig:2S_oc7-hrf}). By doing so, \CA data would presumably better approximate BOLD data. Filtering \CA signals changed network organization in relatively modest ways for \CAS, although some of the differences were statistically significant (\cref{fig:2S_oc7-hrf}b). Filtering produced more pronounced quantitative changes to \CAF networks, especially OC-4, OC-5, and OC-6.

\subsection{Cortical networks show prominent overlapping organization}

\begin{figure}[ht!]
    \centering
    \includegraphics[width=0.51\linewidth]{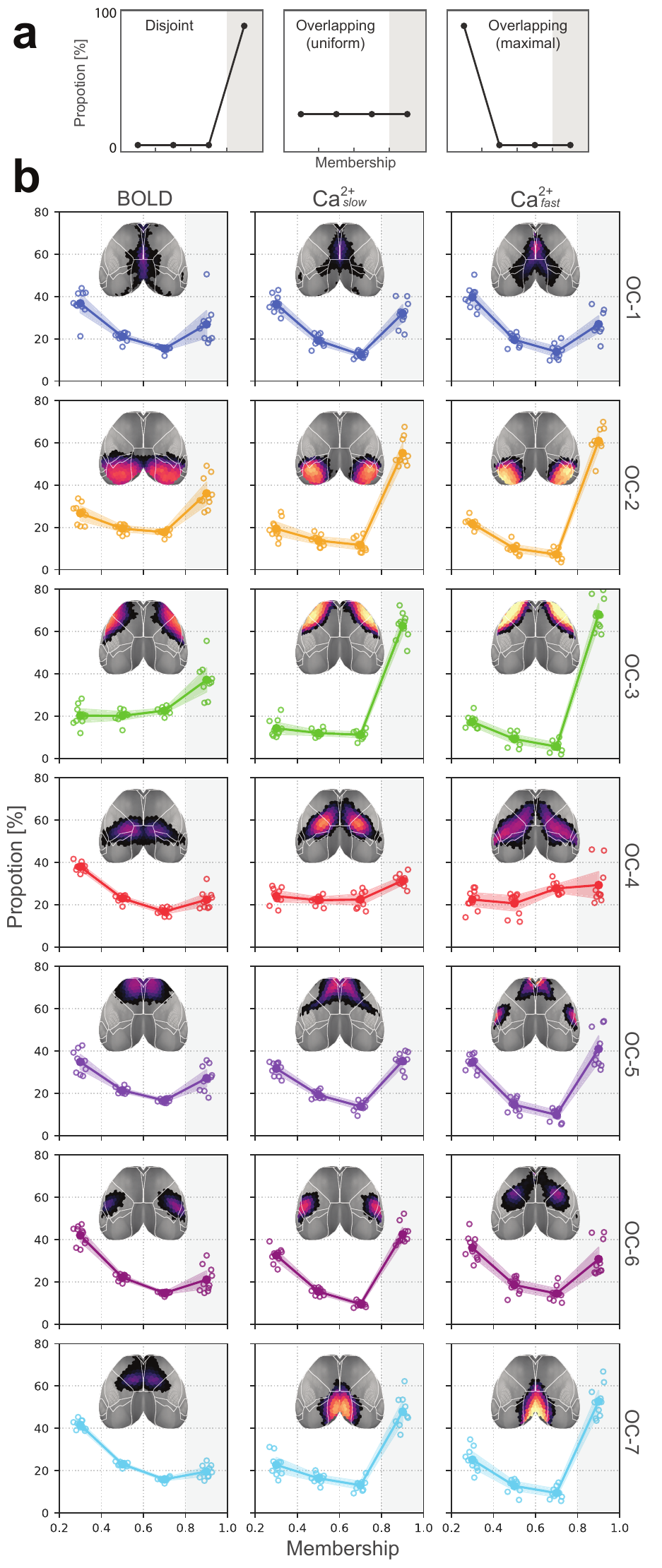}
    \caption[Distribution of membership values]{Distribution of membership values. \textbf{(a)} Three illustrative distributions. Left, disjoint organization; Middle, overlapping with uniform membership values; Right, completely overlapping with no mid-range or strong memberships. \textbf{(b)} Membership distributions computed from multimodal data indicate substantial overlapping organization.}
    \label{fig:3_prop}
\end{figure}

To quantify overlapping organization, we examined the distribution of membership values across networks. Membership values range from 0--1, and sum to 1 across networks. Thus, a disjoint organization would be characterized by all regions having high membership values for a single network (a \say{right-peaked} distribution; \cref{fig:3_prop}a, left). Importantly, this outcome is observed with our algorithm when synthetic, disjoint data are simulated (see \cref{fig:3S_lfr}, also described below). In contrast, a roughly uniform distribution of memberships would correspond to a network whose regions affiliate with multiple networks with varying strengths (\cref{fig:3_prop}a, middle). Finally, extreme overlap would be when regions tend to not affiliate with any network very strongly (\cref{fig:3_prop}a, right).

In simulations, we constructed synthetic disjoint networks with no overlap \cite{Lancichinetti2009BenchmarksFT} and found that the algorithm detected membership values $>0.8$ (\cref{fig:3S_lfr}). In other words, in a completely disjoint network, every node belongs to a single network with $>0.8$ strength and, given that a node's membership strength sums to $1$, the remaining $<0.2$ strength is distributed across the remaining networks. This establishes a floor value ($0.2$) for robust network membership. Thus, to examine membership distributions, we considered the range $(0.2, 1.0]$; membership values $<0.2$ were not considered so as to conservatively characterize network overlap.

Across conditions (BOLD, \CAS, and \CAF), no more than $60\%$ of brain regions within any network were within the \say{disjoint} range ($>0.8$). The least overlapping networks were the visual and somatomotor (OC-2 and OC-3) for all conditions and the retrosplenial network (OC-7) for the two \CA conditions. Networks with the greatest amount of overlap were OC-1 and OC-4 (\cref{fig:3_prop}b), which included the cingulate cortex (OC-1) as well as medial and lateral areas, including somatosensory cortex (OC-4) and the FOF (OC-4, \CA). 

For several of the conditions, we observed a roughly reverse L-shaped distribution (\cref{fig:3_prop}b). In the case of OC-2, OC-3, and OC-7 they were reminiscent of the disjoint pattern of \cref{fig:3_prop}a, except that the \say{base} had a considerably higher level than zero. Thus, these networks have considerable disjoint organization, in some cases with 60\% of the regions affiliating with a single network. OC-1, OC-5, and OC-6 exhibited a more true U shape with relatively more strongly disjoint regions (right side) and weakly affiliated regions (left side), with relatively fewer regions with intermediate membership strengths (middle two values). Nevertheless, it is important to note that in many cases the proportion of regions with intermediate membership values (0.4--0.8 range) was around 40\%.

\subsubsection{LFR analysis}

LFR (Lancichinetti-Fortunato-Radicchi) graphs \cite{Lancichinetti2009BenchmarksFT} are a type of synthetic network used to benchmark community detection algorithms in network science. They are specifically designed to mimic the structure of real-world networks, incorporating features such as power-law distributions for both node degrees and community sizes, which are common in social, biological, and technological networks. LFR graphs can be tailored to exhibit varying levels of community structure, making them highly versatile for testing the effectiveness of clustering algorithms. Here, we used LFR graphs specifically for the purpose of evaluating the robustness of our measures of overlap. To this aim, we generated LFR graphs with varying levels of planted ground truth overlap. We then measured the overlap size exactly in the same way we did for the real graphs.

The following parameters need to be specified to generate a binary and overlapping LFR graph \cite{Lancichinetti2009BenchmarksFT}.  $N$, number of nodes; $k$, average degree; $\mu$, topological mixing parameter; $t_1$, minus exponent for the degree sequence; $t_2$, minus exponent for the community size distribution; $C_{min}$, minimum for the community sizes; $C_{max}$, maximum for the community sizes; $ON$, number of overlapping nodes; $OM$, number of memberships of the overlapping nodes.

\begin{figure}[ht!]
    \centering
    \includegraphics[width=\linewidth]{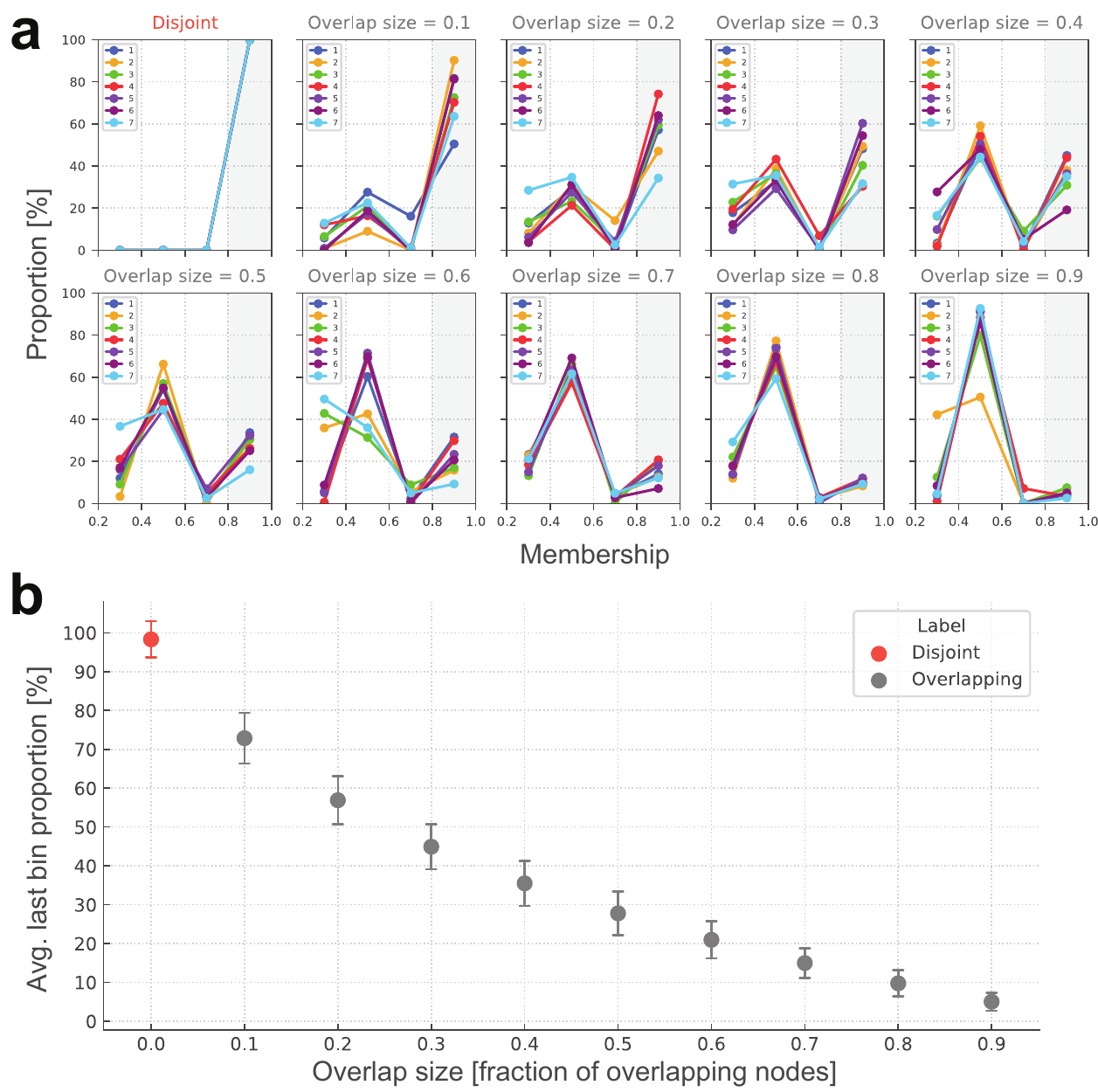}
    \caption[Verifying our analysis procedure using synthetic LFR graphs]{Verifying our analysis procedure using synthetic LFR graphs \cite{Lancichinetti2009BenchmarksFT}. \textbf{(a)} To compute the distribution of node memberships, we divided the interval $(0.2, 1.0]$ into four bins of equal width because it allowed us to distinguish between disjoint and overlapping graphs: for disjoint graphs, the bin corresponding to strong membership values (i.e., last bin) had a proportion of $100\%$, and the proportions were $0\%$ elsewhere. Overlap size, or fraction of overlapping nodes, is a tunable parameter in LFR graphs. Here we show how membership distributions change as a function of overlap size. \textbf{(b)} We averaged the last bin proportions across all communities to get a single statistic per graph. Its mean over thousands of simulated graphs is shown here. The average last bin proportion drops as overlap size increases, thus making it a reliable proxy for network overlap size. Error bars indicate standard deviation. Compare with \cref{fig:3_prop}.}
    \label{fig:3S_lfr}
\end{figure}

To match basic statistics of the real data with LFR graphs we set $N=542$ (\cref{fig:1_roi}b); and, for every run from each data modality, we calculated the average degree $k$ and estimated $t_1$ via an exponential fit to the degree distributions (scipy.stats). We set $t_2 = 0.1$, $C_{min} = 0.05 \times N \approx 27$, $C_{max} = 0.35 \times N \approx 190$. For the fraction of overlapping nodes $ON$, we explored a wide range between $0$ (disjoint) and $0.9$ in incremental steps of $0.1$. This yielded $ON = 0$ up to $ON = 0.9 \times N = 488$. Finally, we used $OM = 2$ and $3$.  This results in a total of $20$ LFR graphs per run, per data modality. We applied the community detection algorithm to LFR graphs in an identical way to the real data but with fewer seeds ($N=10$ compared to $N=500$). The alignment procedure was performed in an identical way as described above. See \cref{fig:3S_lfr}.

\subsection{Membership tiers}
\begin{figure}[ht!]
    \centering
    \includegraphics[width=\linewidth]{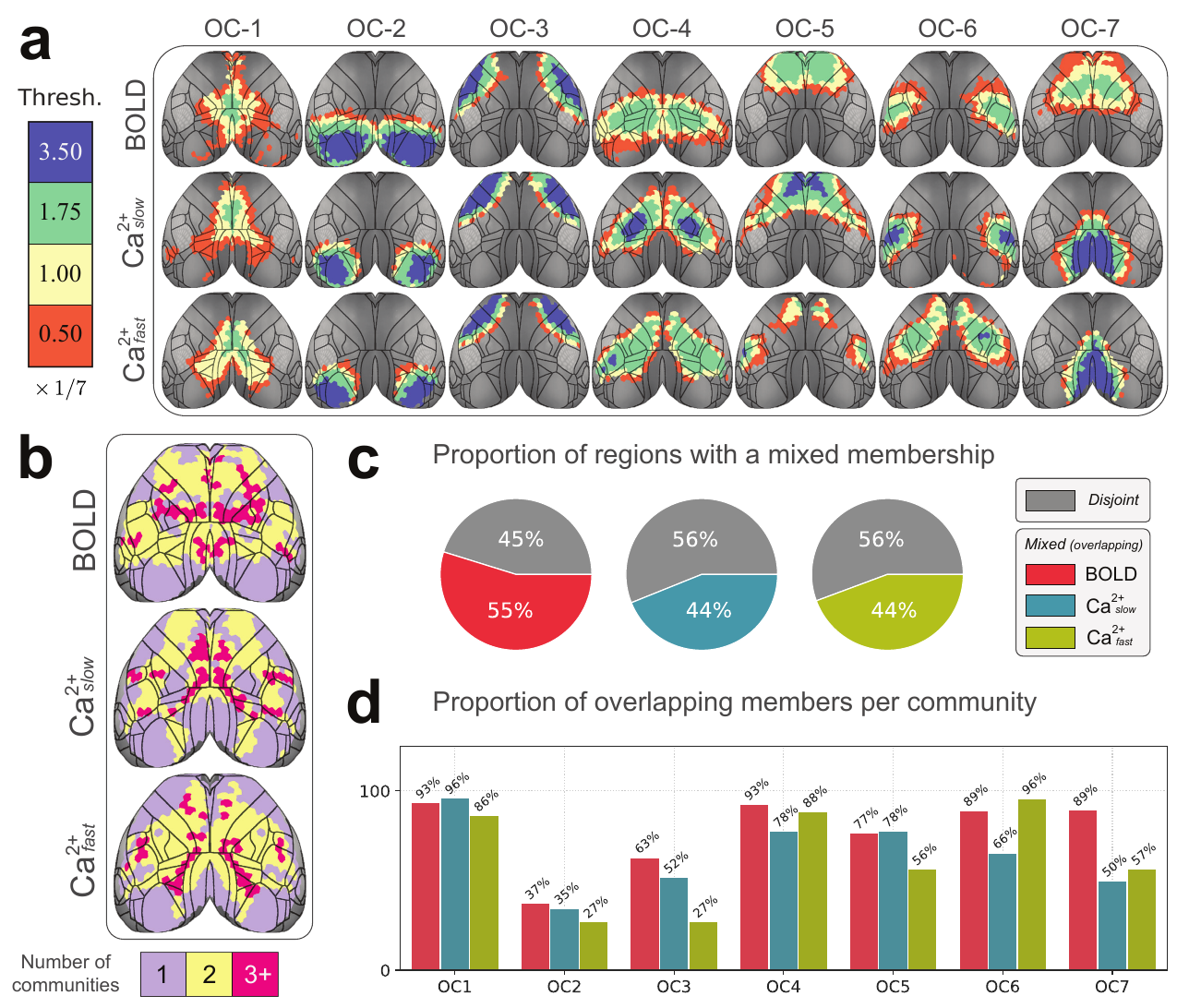}
    \caption[Quantifying overlap extent]{Quantifying overlap extent. \textbf{(a)} Membership values binned by statistical thresholding. Bins were incremented by 1/7 (for the 7 network solution). Blue (membership $> 3.5 \times 1/7$) indicates regions with disjoint-like network affiliation. At the opposite end of the spectrum, orange ($> 0.05 \times 1/7$) indicates regions affiliated with multiple networks. Contour lines correspond to regional divisions in the Allen reference atlas (\cref{fig:2S_oc20}b). \textbf{(b)} Collapsing across networks. Using a statistical threshold of $1/7$ to determine if a region affiliated with a network, we counted the number of networks each region ``belonged to". Most regions belonged to more than one network. \textbf{(c)} We defined a global \textit{overlap score} as the ratio of overlapping regions divided by the total number of regions. \textbf{(d)} Overlap score at the network level shows that regions that affiliate with more than one network are spread across networks and present in both BOLD as well as \CAF and \CAS. Abbreviations: OC, overlapping community.}
    \label{fig:4_tier}
\end{figure}

To visualize network overlap, we divided membership strength into four categories, or membership \emph{tiers}. Because we considered 7 networks, bin thresholds were multiples of $1/7$ (all statistics are FDR-corrected). Based on this representation, we observed that overlapping network organization was arranged in a spatially coherent fashion that showed a nested pattern of membership \emph{tiers} (\cref{fig:4_tier}a). To quantify overlap across networks, if a region had a membership value statistically greater than $1/7$ for a given network, we classified it as \say{belonging to} to that network. We then summed the number of networks to which regions belonged (\cref{fig:4_tier}b). By this definition, approximately $50\%$ of brain regions belonged to more than one network (\cref{fig:4_tier}c). Further, brain regions belonging to more than one network were distributed across networks (\cref{fig:4_tier}d). In even the most disjoint-like cases (OC-2 and OC-3), $> 25\%$ of regions affiliated significantly with more than one network across all conditions.

\subsection{Membership diversity reveals inter-modal differences}

\begin{figure}[ht!]
    \centering
    \includegraphics[width=0.80\textwidth]{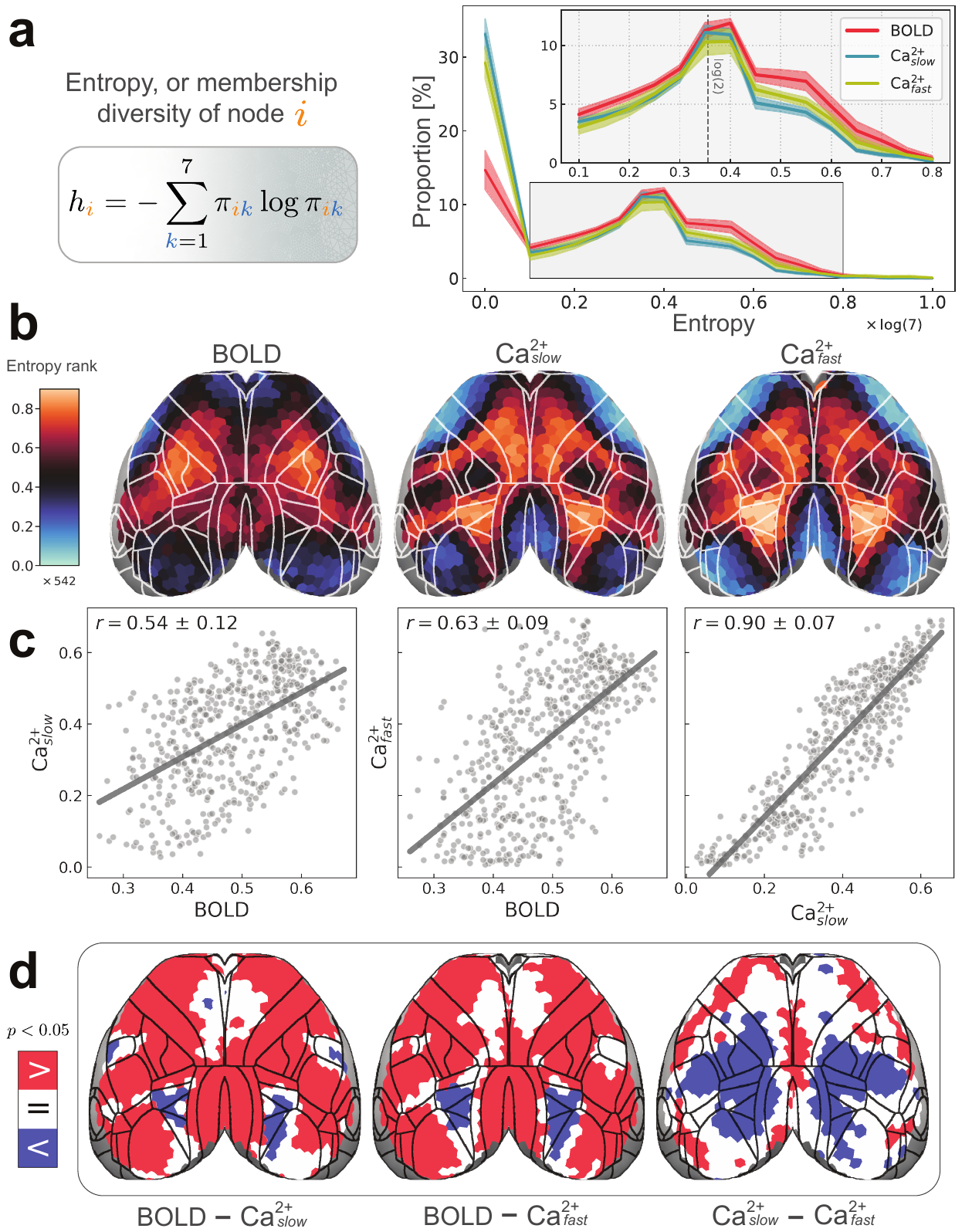}
    \caption[Regional entropy, or membership diversity]{\footnotesize Regional entropy, or membership diversity. \textbf{(a)} Left: equation for Shannon entropy (Methods). Values are normalized $[0, 1]$. $h_i = 0$ if a node $i$ belongs to a single network (is disjoint); $h_i = 1$ if a node belongs to all networks with equal strength (is maximally overlapping). Right: distribution of entropies for all regions. The peak at $h=0$, corresponds to disjoint regions. The second peak at $h \approx \log(2)$ corresponds to regions with membership values of $0.5$ for two networks and $0$ elsewhere. \textbf{(b)} Spatial patterns of regional entropies rank-ordered (total of 542 regions) to facilitate comparisons across conditions (BOLD, \CAS, and \CAF). The non-rank-ordered version is shown in \cref{fig:5S_particip}a. Unimodal areas such as visual and somatomotor areas have low entropy (cool colors), whereas transmodal regions have high entropy (hot colors). \textbf{(c)} Entropy was positively (Pearson) correlated across modalities (variability obtained based on hierarchical bootstrapping; Methods). \textbf{(d)} Differences in entropy between conditions are quantified by subtracting each pair of conditions. Permutation testing revealed BOLD $>$ \CA in most regions, except for some frontal areas where BOLD $=$ \CA, and higher visual areas where BOLD $<$ \CA (left and middle). \CAS exhibited a large territory of regions with entropy $<$ \CAF (right).}
    \label{fig:5_ent}
\end{figure}

The preceding analyses showed clear evidence for overlapping organization alongside disjoint organization in the mouse cortex across imaging modalities and frequency bands. The characteristics of this overlapping organization were further quantified using (normalized) Shannon entropy, a continuous measure of \textit{membership diversity} computed from regional membership values (Methods, \cref{fig:5_ent}a, left). A region that belongs to all networks with equal membership strengths will have maximal diversity. Conversely, a region that belongs to a single network will have minimal diversity. Thus, membership diversity is indicative of a region's multi-functionality and/or involvement in multiple processes. 

\begin{figure}[ht!]
    \centering
    \includegraphics[width=1.0\linewidth]{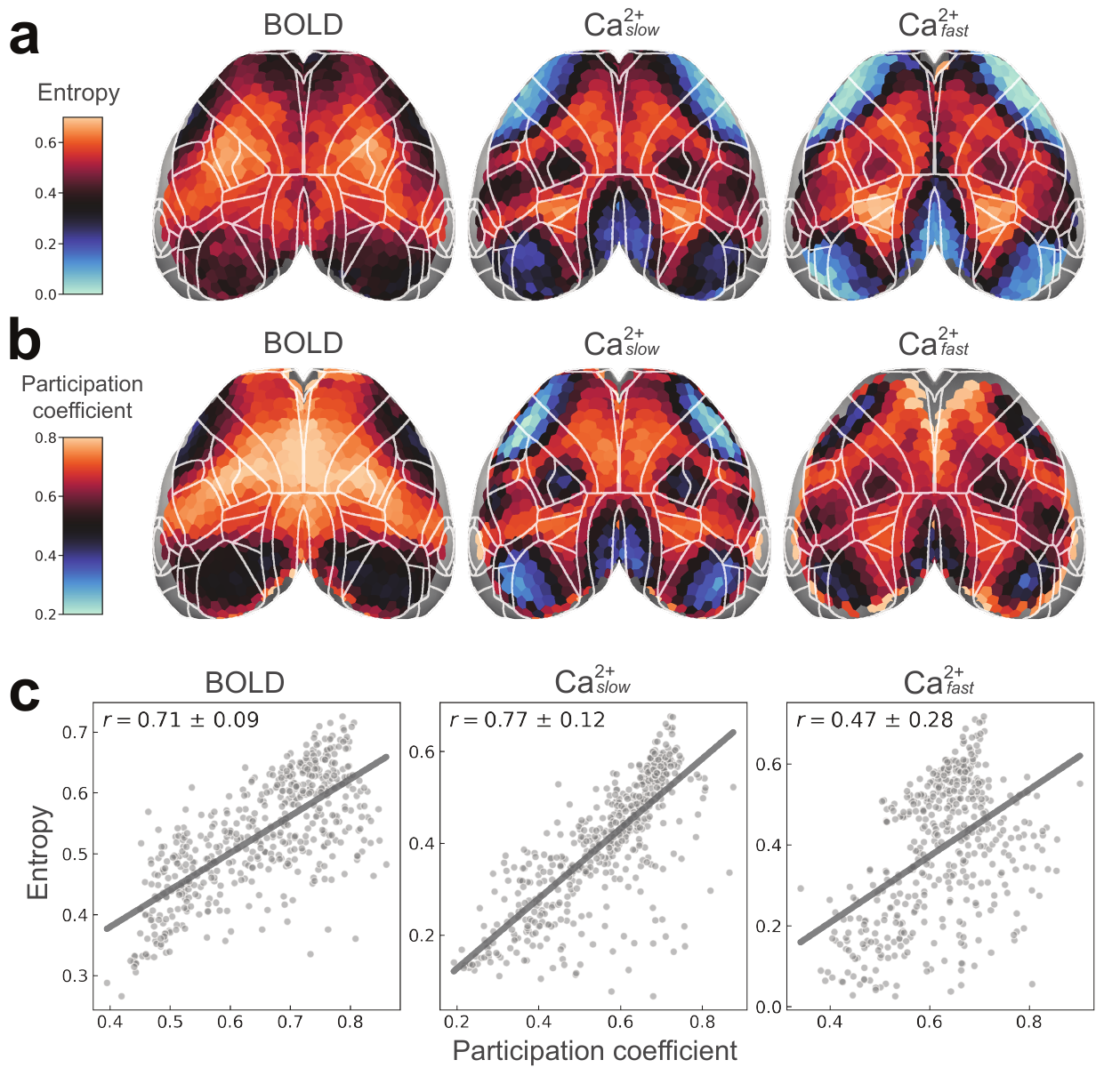}
    \caption[Entropy and participation coefficient exhibit comparable spatial patterns]{Entropy and participation coefficient uncover similar spatial patterns. \textbf{(a)} Here, we show entropy maps with actual values (compare with \cref{fig:5_ent} for the rank-ordered version). Note the positive shift in BOLD values. \textbf{(b)} Participation coefficient is commonly used to quantify how a node's links are distributed across (disjoint) communities \cite{Guimer2005FunctionalCO, van2013network, Power2013EvidenceFH, Liska2015FunctionalCH, Bertolero2017TheDC}. To compute participation coefficients, we used the disjoint approximation obtained from taking the maximum membership of a given node (see the last column in \cref{fig:2_ocs7}). Note that for low-degree nodes with degree $<7$, participation coefficient estimates become unreliable. For example, see frontal regions in \CAF.  \textbf{(c)} The two centrality measures are correlated, suggesting that they capture similar underlying phenomena. The concordance between these measures is most clearly visible for \CAS, but also for BOLD to some extent.}
    \label{fig:5S_particip}
\end{figure}

The distribution of membership diversity values across all regions is shown in \cref{fig:5_ent}a (right). The peak near zero captures a group of regions, approximately $30\%$ for \CAS and \CAF, and $15\%$ for BOLD, that are primarily associated with one network. Beyond this peak, the majority of regions displayed values more or less along a continuum, with a second smaller peak (at approximately 0.35) with regions affiliated with two networks (dashed line in \cref{fig:5_ent}a, right inset). For visualization purposes, we rank-ordered membership diversity values to inspect the overall pattern across conditions (BOLD, \CAS, and \CAF; \cref{fig:5_ent}b; for the non-rank-ordered version, see \cref{fig:5S_particip}a). The resulting patterns revealed modest agreement between BOLD and both \CA conditions, and especially strong agreement between the two \CA frequency bands.

To quantify this agreement, we (Pearson) correlated membership diversity values: between BOLD and \CAS: $r = 0.54 \pm 0.11$; BOLD and \CAF: $r = 0.63 \pm 0.09$; and \CAS and \CAF: $r = 0.90 \pm 0.07$ (\cref{fig:5_ent}c). Contrary to expectations, measures of BOLD membership diversity were \textit{not} more similar to those obtained from \CAS relative to between BOLD and \CAF (\cref{fig:5_ent}c).

We also identified regions that showed significant differences in membership diversity magnitude between conditions by subtracting each pair of measures (\cref{fig:5_ent}d; FDR corrected). Spatially broad differences (BOLD versus both \CA frequency bands) were observed. Diversity was consistently larger for BOLD compared to both \CA conditions, except for two bilateral sectors that showed the opposite pattern (one in higher-order visual areas and one, for \CAS only, in a primary somatosensory area). This final observation was made alongside \CAF exhibiting a large territory of regions with higher membership diversity than \CAS. Overall, differences were observed despite similar proportions across modalities for values above 0.1 (\cref{fig:5_ent}a, right).  

\subsubsection{Comparing entropy to participation coefficient}

Brain regions of high functional diversity are more likely to participate in multiple networks. This region-level property can be characterized using appropriate node centrality measures. In the results above, we took advantage of having access to continuous membership values and defined node entropy centrality (\cref{fig:5_ent}). Another measure, \textit{\say{participation coefficient}} \cite{Guimer2005FunctionalCO, van2013network, Power2013EvidenceFH, Liska2015FunctionalCH, Bertolero2017TheDC}, has also been commonly used to quantify a similar concept: a node's participation coefficient measures how well-distributed its links are among different communities. Large values of participation coefficient indicate higher amounts of link diversity, which could be potentially related to high membership diversity.

To uncover the relationship between the two measures, we visualized their spatial patterns and found that entropy and participation coefficient maps were largely in agreement (\cref{fig:5S_particip}a and b). To quantify this, we performed node-wise (Pearson) correlation between the two measures and found membership diversities and participation coefficients to be in good agreement for BOLD: $r = 0.71 \pm 0.09$, and \CAS: $r = 0.77 \pm 0.12$, and to be more weakly related for \CAF: $r = 0.47 \pm 0.28$.

Entropy and participation coefficient are defined in very different ways. Entropy is computed from membership probability vectors within our overlapping framework; whereas, participation coefficient depends on how links are distributed across communities within a disjoint framework. Despite this, the two measures were highly correlated indicating that they probably capture similar underlying phenomena. 

\subsection{Region degree is substantially different across modalities}

\begin{figure}[ht!]
    \centering
    \includegraphics[width=\textwidth]{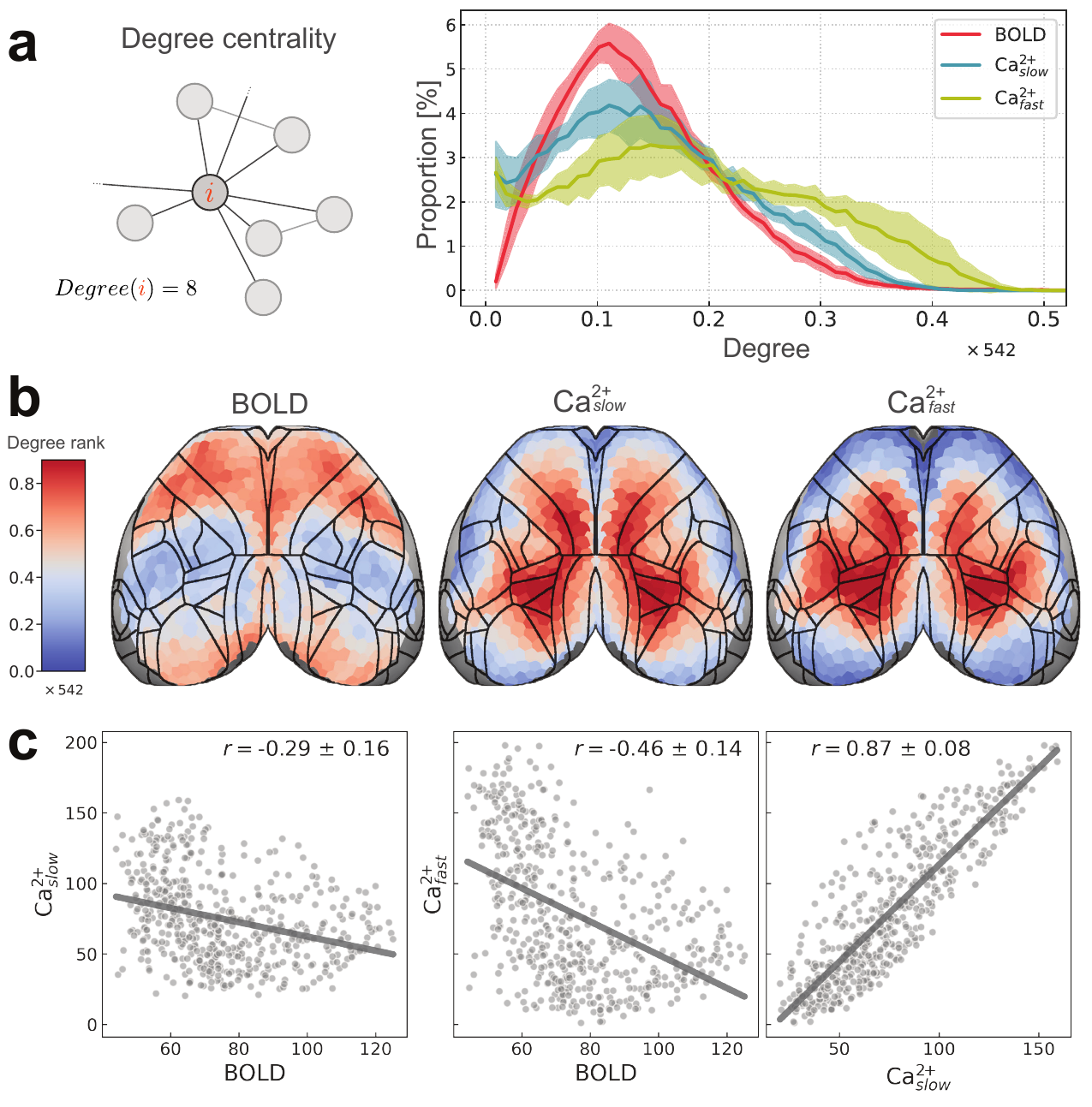}
    \caption[Brain region degree centrality]{Regional degree. \textbf{(a)} Left: schematic of regional degree. Right: distribution of regional degree normalized by the number of brain regions (total of 542). \textbf{(b)} Spatial patterns of regional degrees rank-ordered (total of 542 regions) to facilitate comparisons across conditions (BOLD, \CAS, and \CAF). Actual values are shown in \cref{fig:6S_deg-vals}a. Densely connected regions have high degree (hot colors), whereas sparsely connected regions have low degree (cool colors). BOLD and \CA conditions show opposing spatial patterns. \textbf{(c)} Similarity between conditions is quantified using (Pearson) correlation. BOLD and \CA are negatively correlated whereas \CA conditions are highly positively correlated (variability obtained based on hierarchical bootstrapping; Methods). Weighted (non-thresholded) versions of degree maps are provided in \cref{fig:6S_deg-all}.}
    \label{fig:6_deg}
\end{figure}

\begin{figure}[ht!]
    \centering
    \includegraphics[width=\linewidth]{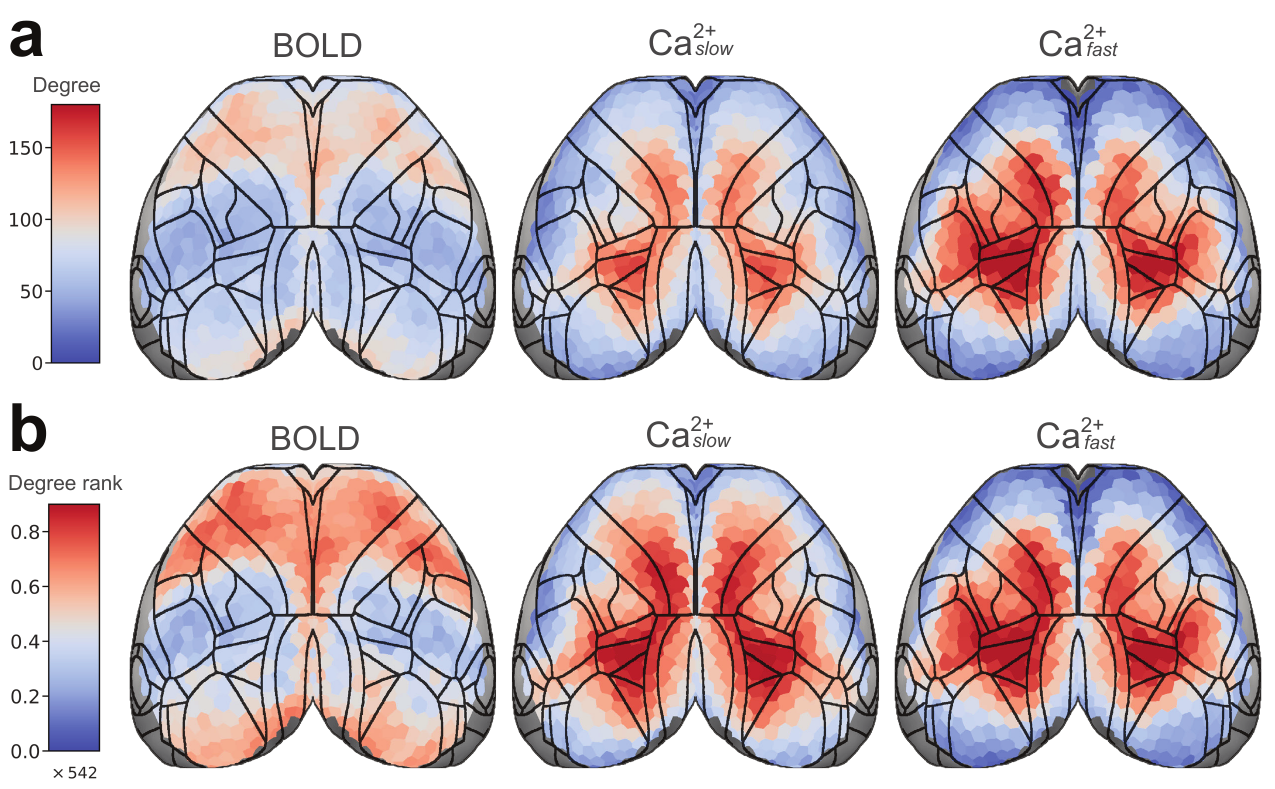}
    \caption[Degree centrality magnitude and ranks]{\textbf{(a)} Both the magnitude and spatial patterns of degree centrality values are different across modalities. \textbf{(b)} Degree ranks are reproduced from \cref{fig:6_deg} to facilitate visual comparison.}
    \label{fig:6S_deg-vals}
\end{figure}

\textit{Degree} is a measure of centrality that quantifies the number of functional connections of a region \cite{achard2006resilient, bullmore2009complex} (\cref{fig:6_deg}a, left). Importantly, degree differs from membership diversity by being independent of community (regions have functional connections both within and between communities). As in the previous section, the distribution of degree across regions was plotted for each condition (BOLD, \CAS, and \CAF) (\cref{fig:6_deg}a, right), and the spatial patterns of degree ranks were displayed on the cortex (\cref{fig:6_deg}b; actual values in \cref{fig:6S_deg-vals}), and (Pearson) correlation was used to measure agreement between conditions (\cref{fig:6_deg}c).

Across conditions, degree distributions showed weak similarity (\cref{fig:6_deg}a, right). As expected, \CAS and \CAF were more similar to one another than to BOLD, and \CAS was more like BOLD than \CAF. BOLD and both \CA measures showed differences at the low-extreme (near zero) as well as across the range: \CA having fewer low-mid degree regions, and more high-degree regions than BOLD. Inter-modal differences were more pronounced when we looked at the spatial distribution of degree (\cref{fig:6_deg}b), and the (Pearson) correlation of degree across conditions (\cref{fig:6_deg}c). Similar to the case of membership diversity, degree showed a consistent spatial pattern across \CA conditions (\cref{fig:6_deg}b), and was highly correlated: $r = 0.87 \pm 0.08$ (\cref{fig:6_deg}c, right). Unlike membership diversity, BOLD and both \CA degree measures showed opposing spatial patterns (\cref{fig:6_deg}b), and were negatively correlated with BOLD: between BOLD and \CAS: $r = -0.29 \pm 0.16$, and BOLD and \CAF: $r = -0.46 \pm 0.14$ (\cref{fig:6_deg}c, left and middle).

\subsection{Differences in data smoothness does not explain degree differences}
\begin{figure}[ht!]
    \centering
    \includegraphics[width=1.0\linewidth]{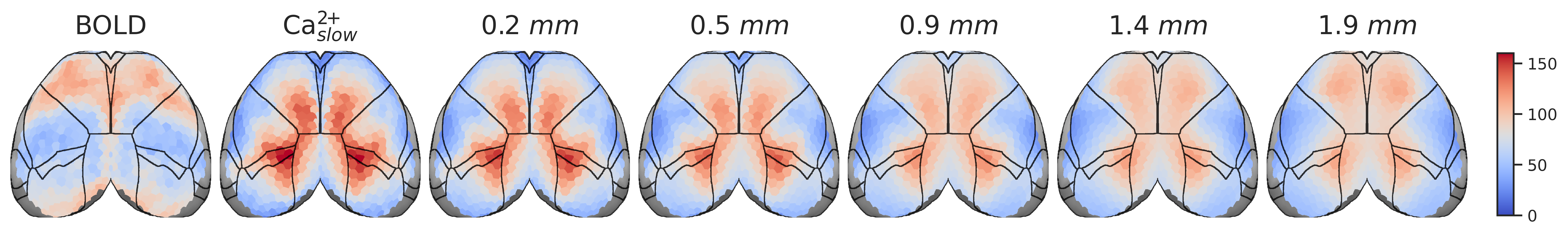}
    \caption[\CA degree maps at different levels of data smoothness]{\CA degree maps at different levels of data smoothness. To obtain these results, we applied a Gaussian filter to raw \CA data followed by otherwise identical steps in our pipeline, including \CAS bandpassing. A considerable amount of data smoothing (full width at half maximum = $1.9~mm$) results in a \CAS degree map that is comparable to that of BOLD in terms of smoothness; however, the spatial patterns remain substantially different across modalities.}
    \label{fig:6S_deg-smooth}
\end{figure}

Different levels of data smoothness can potentially contribute to some of the observed differences across modalities. We evaluated the impact of spatial smoothing on degree maps. To do so, we applied a Gaussian kernel to smoothen \CAS data, followed by computing degree maps using steps that were otherwise identical to those used originally (\cref{fig:6S_deg-smooth}). We explored various degrees of smoothness quantified by the full width at half maximum (FWHM) of the filter (indicated on top of the maps in \cref{fig:6S_deg-smooth}). 

Irrespective of the amount of smoothing of the calcium signals, the resulting degree maps were spatially very similar to that of the original degree maps, and substantially different from the BOLD degree map. It is worth noting that \CAS smoothing did produce degree maps that were substantially smoother than the original one, but which were nevertheless quite different from the BOLD spatial organization.

These results provide strong evidence that the reason the degree map for \CAS is qualitatively different from that of the BOLD degree map is not due to data smoothness, and instead reflects differences in the underlying contrast signals themselves (i.e., BOLD vs. \CA).

\subsection{Variability does not explain inter-modal differences in centrality measures}

\begin{figure}[ht!]
    \centering
    \includegraphics[width=\linewidth]{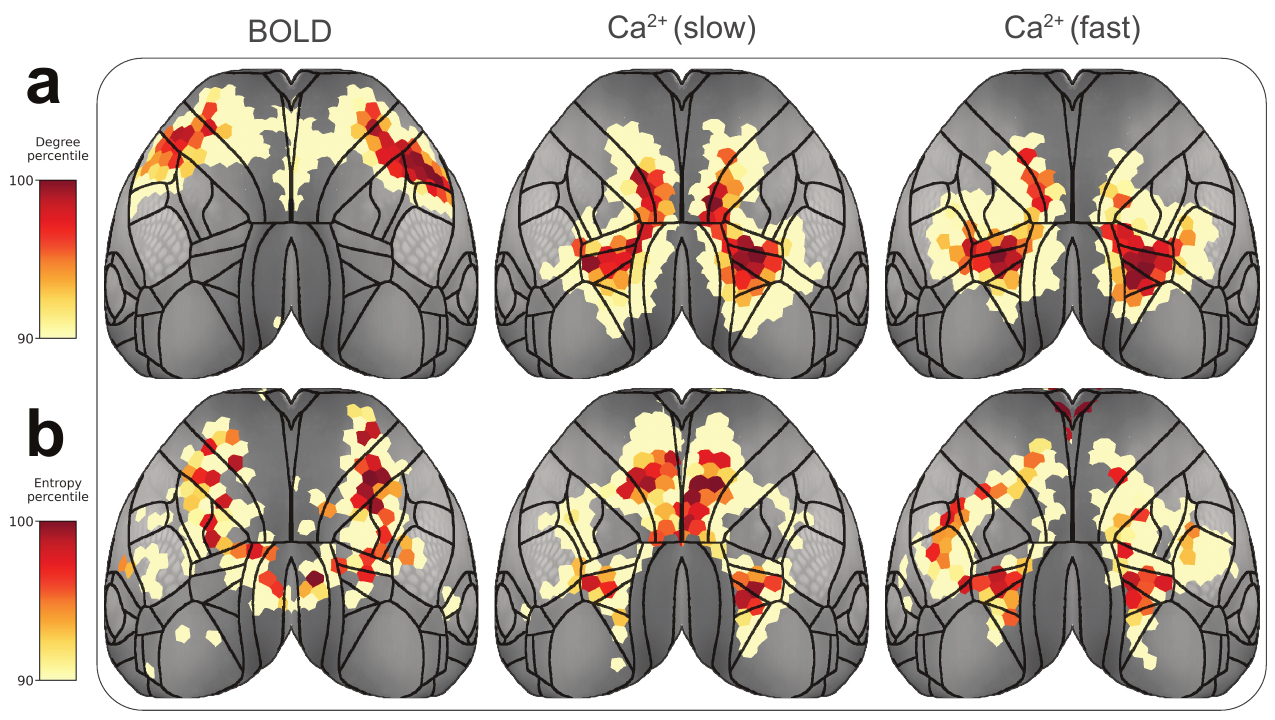}
    \caption[Brain region centrality measures obtained as percentile maps]{Percentile maps are obtained by calculating t-statistics (hierarchical bootstrapping, see Methods) followed by rank-ordering. Values below $80\%$ are not shown. \textbf{(a)} Degree \textbf{(b)} Entropy. Related to \cref{fig:5_ent} and \cref{fig:6_deg}}
    \label{fig:5S-6S_perc}
\end{figure}

To account for differences based solely on the magnitude/variability of degree and entropy values, we computed percentile maps by calculating \textit{t}-statistics followed by rank-ordering \cite{Liska2015FunctionalCH} and found the patterns to be unchanged (\cref{fig:5S-6S_perc}). Further, differences across modalities persisted across edge thresholds and changes in data preprocessing steps (\cref{fig:6S_deg-all}).

\subsection{Different entropy-degree relationships across modalities}

How a given brain region affiliates across multiple networks (as indexed by membership diversity/entropy) is closely linked to its roles as an integrative and/or coordination hub \cite{Bertolero2017TheDC}. Furthermore, membership diversity and degree are measures that, when combined, can further uncover brain organization \cite{Power2013EvidenceFH, Liska2015FunctionalCH}. In particular, regions with low entropy and high degree have few inter-network functional connections (low entropy) and many intra-network functional connections (high degree) and can be thought of as \textit{provincial} hubs \cite{Guimer2005FunctionalCO, van2013network, Liska2015FunctionalCH}. Regions with high entropy and low degree have few functional connections but link many networks, and can be conceptualized as \textit{connector} hubs.

\begin{figure}[ht!]
    \centering
    \includegraphics[width=0.97\textwidth]{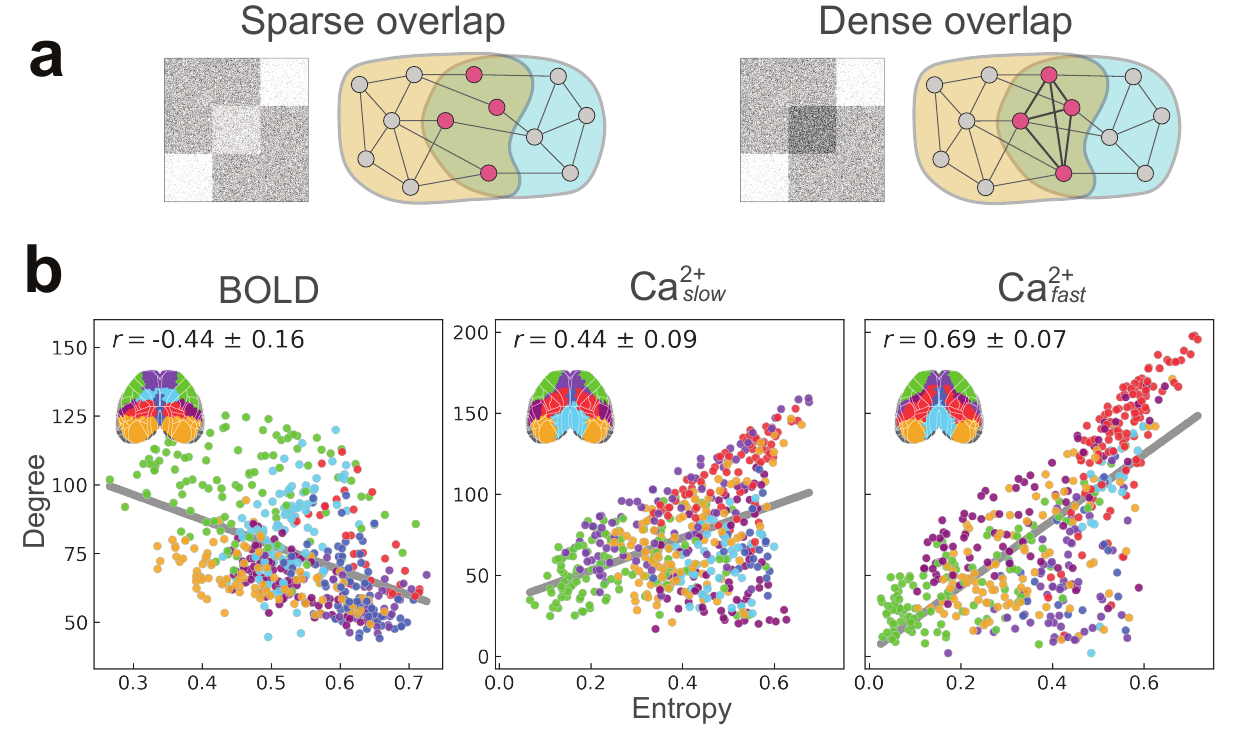}
    \caption[Entropy-degree relationships across modalities]{Entropy-degree relationships across modalities. \textbf{(a)} Illustrated examples of sparse and dense overlapping organization. Figure inspired by Yang and Leskovec \cite{yang2014overlapping}. \textbf{(b)} Entropy versus degree for BOLD (left), \CAS (middle), and \CAF (right). Each point corresponds to a brain region, color-coded by their disjoint network assignment (inset; see last column in \cref{fig:2_ocs7}). See also \cref{fig:5S_particip}.}
    \label{fig:7_carto}
\end{figure}

Inspired by the work of \textcite{yang2014overlapping}, such organization reveals what can be called \say{sparse} network overlap (\cref{fig:7_carto}a, left). Finally, regions with high entropy and high degree interlink many networks via an organization that can be called \say{dense} overlap (\cref{fig:7_carto}a, right). To determine cortical functional organization based on these measures, we visualized region entropy-degree relationships for each condition (BOLD, \CAS, and \CAF) (\cref{fig:7_carto}b) color-coded by disjoint network assignment (\cref{fig:7_carto}b; inset). 

Entropy and degree were inversely (Pearson) correlated for BOLD ($r = -0.44 \pm 0.16$; \cref{fig:7_carto}b, left). This pattern was partly driven by a concentration of regions showing sparse overlap (lower right quadrant) with connector hubs present in most networks, alongside two networks (overlapping with OC-3 and to a lesser extent OC-2; \cref{fig:2_ocs7}), that included regions with a more provincial hub characterization (upper left quadrant). In contrast, entropy and degree were positively correlated for \CAS ($r = 0.44 \pm 0.09$), and \CAF ($r = 0.69 \pm 0.07$) (\cref{fig:7_carto}b, middle and right).

Like BOLD (but to a lesser extent), \CAS results included regions with sparse overlap (lower right quadrant); these overlapped with OC-6, and to a lesser extent OC-1 and OC-7 (\cref{fig:2_ocs7}). However, unlike BOLD, there were regions with high entropy (densely overlapping regions; upper right quadrant), as well as regions with low overall functional connectivity (lower left quadrant). This pattern was more pronounced in the \CAF results where fewer regions exhibited sparse overlap (lower right quadrant; \cref{fig:7_carto}b, right). Together, these results uncovered distinct functional cortical organization observed with BOLD, \CAS, and \CAF, such that \CA signals expressed patterns of denser overlap not captured by BOLD signals.

\subsection{Nearly identical principal functional gradient across modalities}

Large-scale gradients characterize brain regions along continuous axes of variation, complementing parcellation and clustering approaches that emphasize discreteness \cite{Huntenburg2018LargeScaleGI, Margulies2016SituatingTD}. Previous work in mice has explored different measures of structural \cite{Fulcher2019multimodal} and functional \cite{Huntenburg2021GradientsOF} gradients, and the relationships between them \cite{Fulcher2019multimodal, Coletta2020NetworkSO, Huntenburg2021GradientsOF}. 
Here, we tested the extent to which functional connectivity gradients determined with fMRI data are replicated with \CA signals.

\begin{figure}[ht!]
    \centering
    \includegraphics[width=1.0\linewidth]{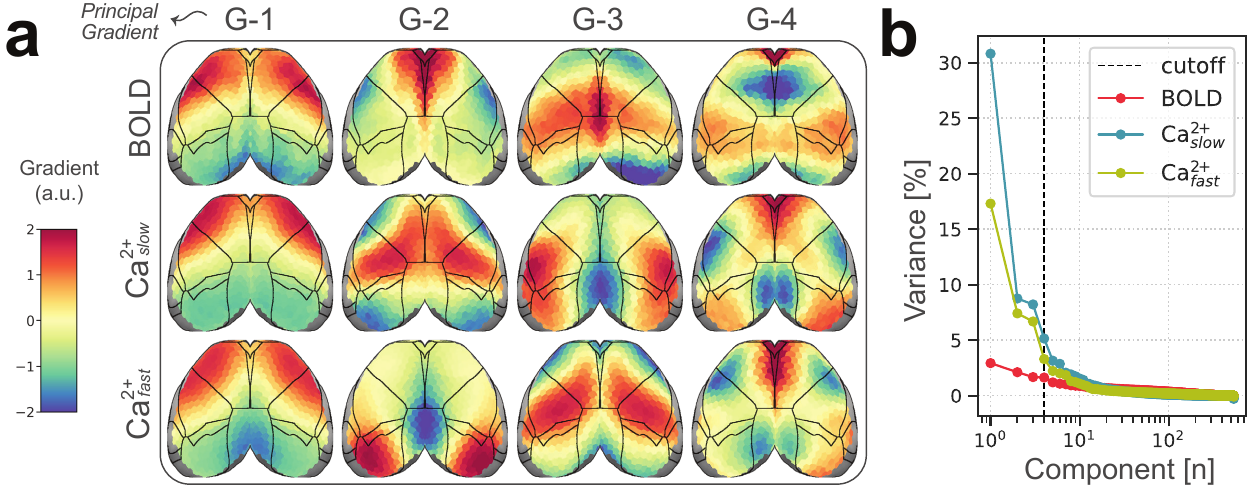}
    \caption[Functional connectivity gradients]{Functional connectivity gradients. \textbf{(a)} Top four gradients are visualized (z-scored; see Methods), and ordered based on the magnitude of the corresponding eigenvalues \cite{coifman2006diffusion}. \textbf{(b)} Portion of variance explained.}
    \label{fig:8_grads}
\end{figure}

We estimated group-level functional connectivity matrices (separately for each modality), which were used to compute gradients (Methods). Here, we focus on the top four gradients (\cref{fig:8_grads}a), as they captured a large portion of the variance (\cref{fig:8_grads}b). The principal gradient (G-1) was organized in terms of primary visual cortex on one extreme, and somatomotor regions on the other end, consistent with previous findings \cite{Coletta2020NetworkSO, Huntenburg2021GradientsOF}.

\begin{figure}[ht!]
    \centering
    \includegraphics[width=1.0\linewidth]{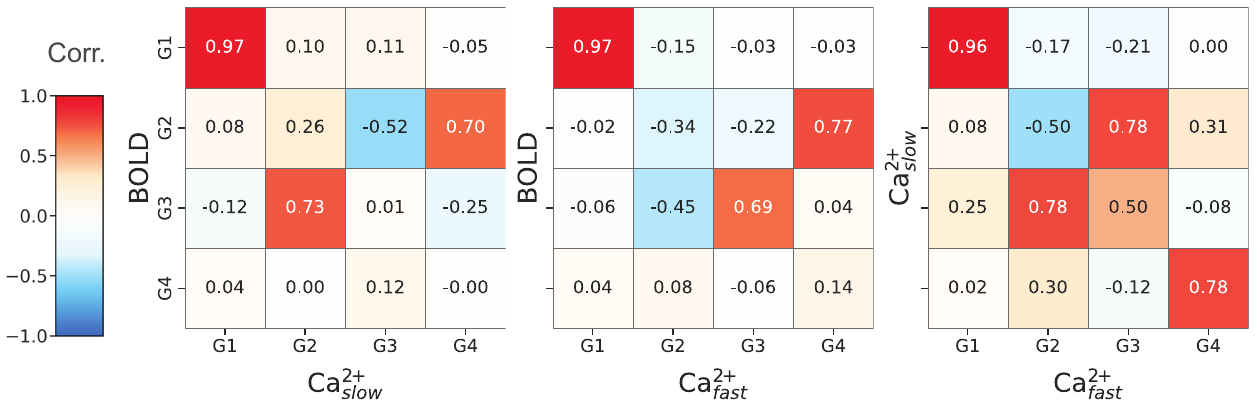}
    \caption[Pearson correlations between pairs of functional connectivity gradients]{Pearson correlations between all pairs of gradients shown in \cref{fig:8_grads}a.}
    \label{fig:8_grads-sim}
\end{figure}

Notably, across conditions (BOLD, \CAS, \CAF), the spatial pattern of G-1 was nearly identical (Pearson $r > 0.96$; \cref{fig:8_grads-sim}) but the amount of variance explained (\cref{fig:8_grads}b) was much lower for BOLD ($<5\%$) relative to \CAF ($\sim 17\%$) and especially \CAS ($\sim 30\%$). Although differences in spatial pattern were observed across conditions for the remaining gradients, strong similarities were also observed (note that we ordered gradients as typically done in the literature based on the magnitude of the corresponding eigenvalues); for example, BOLD G-2 was similar to G-4 for both \CAS ($r = 0.70$) and \CAF ($r = 0.77$), and  BOLD G-3 was similar to \CAS G-2 ($r = 0.73$) and \CAF G-3 ($r = 0.69$). Finally, we note that BOLD G-4 exhibited a very dissimilar spatial pattern to that observed with \CA signals.\\[1mm]

\subsection{Functional gradients of the cortex}

\begin{figure}[ht!]
    \centering
    \includegraphics[width=1.0\linewidth]{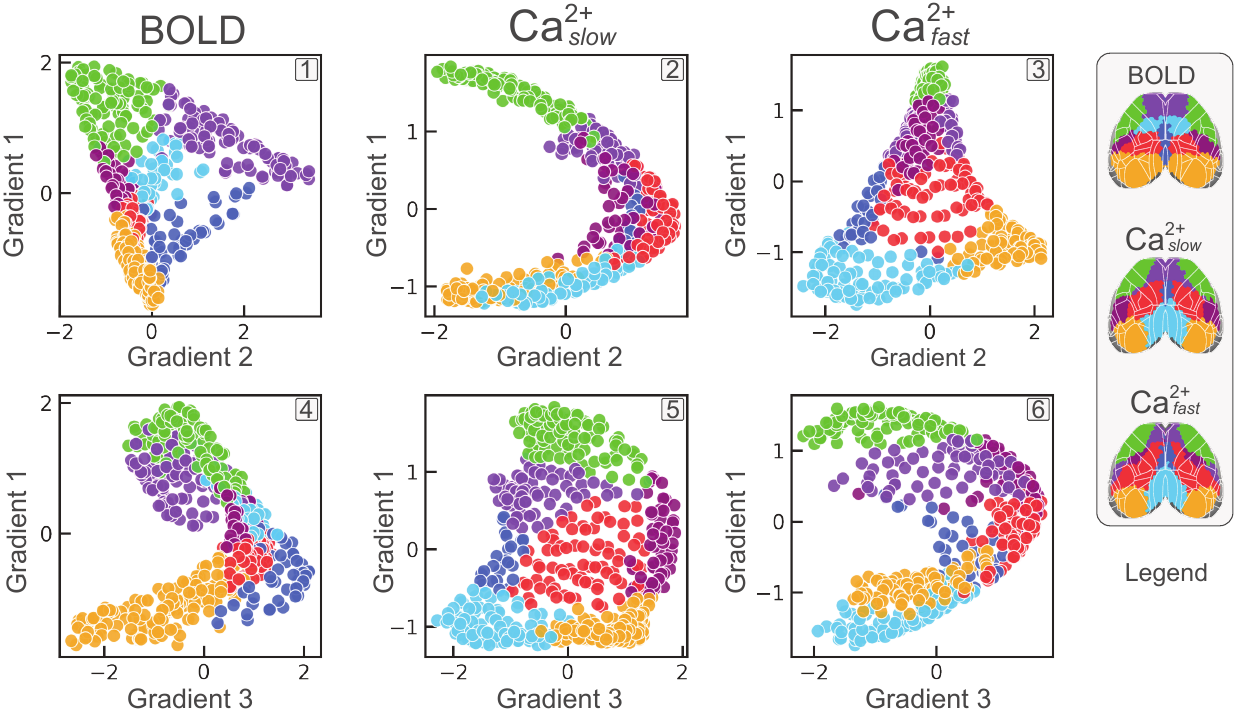}
    \caption[Scatter plots display relationships between pairs of gradient axes]{Scatter plots display relationships between pairs of gradient axes: the $y$ coordinate refers to the G-1 value in column 1 in A, and the $x$ coordinate refers to the G2/G-3 value in columns 2/3 in A. Similar to \cref{fig:7_carto}b, each point is a brain region color-coded according to its disjoint network assignments (legend from last column of \cref{fig:2_ocs7}). Two patterns emerged. A \emph{triangular} shape (panels 1, 3, 5), reminiscent of previous work in mice \cite{Coletta2020NetworkSO} and humans \cite{Margulies2016SituatingTD}, and an \emph{arch} (panels 2, 4, 6) connecting sensory and somatomotor areas on two ends through transmodal areas in the middle.}
    \label{fig:8_grads-pair}
\end{figure}

An individual gradient can be thought of as a continuous representation of an organizational feature \cite{Huntenburg2018LargeScaleGI}, where the position of a brain region along the corresponding axis provides information about its function \cite{Margulies2016SituatingTD}. The overarching organization can be understood by examining how regions are situated in the space spanned by a few top gradients. We visualized this organization (\cref{fig:8_grads-pair}) as two-dimensional maps spanned by the principal gradient (G-1) and G-2/G-3 (specifically, the $y$ coordinate refers to the G-1 value in column 1 of \cref{fig:8_grads}a, and the $x$ coordinate refers to the G2/G-3 value in columns 2/3 of \cref{fig:8_grads}a).

Two patterns emerged that were largely reproduced across modalities. First, a \say{triangular} shape with sensory and somatomotor areas on two extremes, and transmodal areas on the other extreme (panels 1, 3, and 5 of \cref{fig:8_grads-pair}). Notably, two ends of the triangular shape were anchored by visual (orange) and somatomotor (green) areas across all conditions. In contrast, the third anchoring point differed across conditions. For BOLD, it was populated by frontal areas including the anterior lateral motor cortex (purple), whereas for \CA the third anchor was the retrosplenial area (cyan). The second shape was an \say{arch} (panels 2, 4, and 6 of \cref{fig:8_grads-pair}) starting from visual areas on one end of the spectrum (orange), progressing towards areas usually considered part of the default network, and ending in regions within the somatomotor network (green).

These findings align with a theory by \textcite{Mesulam1998FromST}, stating that the lower-dimensional intrinsic cortical organization can be succinctly described as a hierarchy of processing, from unimodal sensory regions to increasingly abstract transmodal regions \cite{Margulies2016SituatingTD}, ultimately converging back to unimodal somatomotor areas. Thus, within this perspective, \sayit{sensation} to \sayit{cognition} to \sayit{action} can be understood as a unified and continuous functional gradient.

\section{Discussion}

Understanding the organization of large-scale brain networks remains a central problem in neuroscience, both in humans \cite{Raichle2001ADM, fox2005human, damoiseaux2006consistent, Yeo2011TheOO, Power2011FunctionalNO, smith2013resting, Faskowitz2020EdgecentricFN} and rodents \cite{hutchison2010functional, white2011imaging, jonckers2011functional, becerra2011robust, nasiriavanaki2014high, stafford2014large, Sforazzini2014DistributedBA, Zerbi2015MappingTM, sierakowiak2015default, Liska2015FunctionalCH, Gozzi2016LargescaleFC, Coletta2020NetworkSO, Grandjean2020CommonFN, bergmann2020individual, mandino2020animal, Markicevic2021EmergingIM, mandino2022triple, GutierrezBarragan2022UniqueSF, grandjean2023consensus}. We used recently developed simultaneous wide-field \CA and fMRI-BOLD acquisition to characterize the functional network architecture of the mouse cortex. The spatial organization of large-scale networks discovered by both modalities showed many similarities, with some temporal frequency dependence: BOLD networks were generally more similar to \CAS than \CAF.

Functional connectivity interrogated using a mixed-membership algorithm, instead of traditional disjoint approaches, confirmed the hypothesis that mouse cortical networks exhibit robust overlap in addition to disjoint organization when either BOLD or \CA signals were considered. Further, despite the considerable agreement, we also uncovered important differences in organizational properties across signal modalities.

Previous multimodal studies comparing cortical functional organization via concurrent GCaMP6 \CA and hemoglobin-sensitive imaging have predominantly employed seed-based analyses \cite{Matsui2016TransientNC, wright2017functional, brier2019separability}. Such work provides information on how one or a limited set of \textit{a priori} regions are functionally related to other areas but does not reveal how all regions are interrelated, which was the goal of the present Chapter. A few studies using optical imaging have gone beyond seed-based analysis; however, the number of identified networks in these studies was limited.
For example, \textcite{vanni2017mesoscale} investigated cortical networks in GCaMP6 mice and reported 3 networks based on slow temporal frequencies ($< 1~Hz$) and two based on faster temporal frequencies ($3~Hz$) (see also \cite{murphy2018macroscale}).

Here, when we decomposed the cortex into 3 networks, we observed visual (OC-2) and somatomotor (OC-3) networks and a network that overlapped with territories possibly linked to the default network (OC-1)  \cite{stafford2014large, Sforazzini2014DistributedBA, Gozzi2016LargescaleFC, Whitesell2021RegionalLA} (\cref{fig:2_oc3}). At this coarse scale, our results agreed with \textcite{vanni2017mesoscale} and other seed-based approaches \cite{stafford2014large, Matsui2016TransientNC, wright2017functional, Grandjean2020CommonFN}.

Importantly, we sought to determine functional organization at finer spatial levels, too. With 7 networks, we still observed visual (OC-2) and somatomotor (OC-3) networks, now together with a finer decomposition of other cortical systems (\cref{fig:2_ocs7}). Overall, our analyses reproduced previous observations at a coarse scale but characterized a more fine-grained decomposition of cortical functional organization.

We quantified the concordance between BOLD and \CA networks using cosine similarity. Overall, collapsing across networks, outcomes from \CAS (BOLD-frequency matched) were more similar than \CAF to BOLD, as expected \cite{deyoe1994functional, Ma2016RestingstateHA, wright2017functional, sauvage2017hemodynamic, he2018ultra, Lake2020SimultaneousCF, mann2021coupling}. However, when networks were characterized separately, three scenarios emerged: ($1$) Low and high frequency \CA signals both manifested networks that were also recovered by BOLD (e.g., OC-1 to 4); ($2$) Low, relative to high, frequency \CA networks were a better match to their BOLD counterparts (e.g., OC-5 and OC-6); and ($3$) Networks that were dissimilar across modalities regardless of \CA temporal frequency (e.g., OC-7).

This observation should be qualified by the finding that \CAS and \CAF results were in close agreement. Overall, linking the functional organization obtained with BOLD to slow \CA signals is not fully supported by our findings. In particular, the proposal that different bands capture distinct neurophysiological properties \cite{mitra2018spontaneous} was not supported for the large-scale system organization uncovered in the present work. 

The pronounced differences involving network OC-7 merit particular attention (\cref{fig:2_ocs7}). In the case of BOLD, this network was centered around parts of the secondary motor cortex known as the frontal orienting field (FOF), considered a homolog of the primate frontal eye field \cite{Brecht2011MovementCA, erlich2011cortical, Barthas2017SecondaryMC, ebbesen2018more, Sato2019InterhemisphericallyDR}. For \CA signals, FOF was consistently detected as part of a large medial network spanning somatosensory, motor, and parietal cortex (OC-4 in \cref{fig:2_ocs7}), without forming an independent network (even in the 20-network solution; \cref{fig:2S_oc20}). 

In contrast, the \CA OC-7 network was centered around the retrosplenial area in a manner that was not captured by any of the seven BOLD networks. We note, however, that the spatially finer 20-network solution for BOLD detected a retrosplenial network, although not to the same extent as identified with \CA signals (panel RSP in \cref{fig:2S_oc20}). 

Overall, we speculate that the differences observed in the case of network OC-7 may reflect particularities of the signal contrasts of the two modalities. To evaluate this question, it will be useful to interrogate \CA indicators that are sensitive to different cell populations such as inhibitory neurons (see below).

We quantified whether networks discovered for each condition (BOLD, \CAS, and \CAF) showed significant overlapping architecture. This was accomplished by examining the distribution of membership values and by quantifying the number of networks each region \say{belonged to}. Notably, our algorithm detects disjoint organization in synthetic data, and the robustness of our findings was tested using a range of parameters (\Cref{table:params}).

Without exception, across conditions, parameter choices, and for all networks, we observed evidence of overlapping organization. On average, slightly over half of brain regions were affiliated with more than one network. Critically, although the extent of network overlap was largest for BOLD, it was also detected in \CA data regardless of temporal frequency. These results lend strong support to the validity of overlapping organization in the human brain discovered with BOLD \cite{Yeo2014EstimatesOS, Najafi2016OverlappingCR, Faskowitz2020EdgecentricFN, cookson2022evaluating}.

Measures of network overlap consistently identified the visual (OC-2) and somatomotor (OC-3) networks as among the most disjoint across all conditions, a finding that is well aligned with their sensory and motor roles and their potential involvement in fewer brain processes. 

Although visual and somatomotor networks (OC-2 and OC-3) exhibited the least amount of overlap among the seven networks, they were still far from being entirely disjoint. This observation is in line with previous proposals that cortical territories should be regarded as essentially multisensory \cite{ghazanfar2006neocortex}, such that mechanisms of multisensory integration extend even into early sensory areas (see \cite{kayser2005integration, iurilli2012sound, ibrahim2016cross, knopfel2019audio}).

Unexpectedly, in the case of \CA data, the retrosplenial area (OC-7) also appeared predominantly as a disjoint network. Given the retrosplenial area's recognized multisensory \cite{keshavarzi2022multisensory, stacho2022mechanistic} and multifunctional \cite{alexander2022rethinking, stacho2022mechanistic} characteristics, further investigation is warranted to understand the underlying factors contributing to the organization detected.

Properties of network overlap, membership diversity (entropy), and degree, were quantified at the region level and compared across conditions (BOLD, \CAS, \CAF). As expected, \CA results showed low diversity in sensorimotor regions relative to areas that have been implicated in multiple processes and have widespread anatomical connections such as the posterior parietal cortex (PTLp; which includes higher-order visual areas) \cite{Lyamzin2018TheMP, hovde2019architecture}.

Despite a positive correlation with \CA results, BOLD membership diversity measures showed some peculiarities. Specifically, in contrast to \CA, the posterior parietal cortex exhibited low diversity, while parts of somatosensory areas exhibited high diversity. This was unexpected given that these regions are not known to be functionally diverse (again, \CA data produced the anticipated outcome). These discrepancies did not disrupt a positive correlation between BOLD and \CA but raised questions about the extent to which the two imaging techniques are capturing the same phenomena. 

The story became more complicated when we examined region degree, a measure of centrality used at times as an indicator of brain region \say{hubness} and/or importance. Degree showed a negative correlation between BOLD and \CA measurements with clearly different spatial patterns. 
\CA degree maps highlighted association areas such as the posterior parietal cortex (PTLp), an area that is anatomically and functionally thought to be important for multimodal integration \cite{Lyamzin2018TheMP}. In contrast, BOLD degree maps produced counter-intuitive results where, for example, lateral somatomotor areas exhibited high degrees though they are likely to be involved in relatively fewer processes. Thus, it appears that while \CA degree maps capture region \say{importance} well, the same cannot be said for BOLD. Indeed, the present finding agrees with, and reinforces, previous suggestions that degree centrality alone should not be used to determine the hubness status of brain areas \cite{Power2013EvidenceFH}, particularly in the case of BOLD data.

The relationship between entropy and degree helped to uncover additional properties of cortical functional organization. For BOLD, membership diversity and degree were inversely correlated, a pattern indicative of \say{sparse overlap} alongside some networks that included \say{provincial} hubs. Notably, this pattern has been observed in human BOLD data \cite{Power2013EvidenceFH, Bertolero2017TheDC}. In contrast, \CA data exhibited a positive correlation (that was more pronounced for \CAF), suggestive of \say{dense overlap} alongside regions showing few connections. Together, these results indicated that BOLD and \CA capture distinct forms of overlapping network organization, with \CA signals particularly able to uncover a \say{diverse club} of regions \cite{Bertolero2017TheDC} that are densely overlapping \cite{yang2014overlapping}, reminiscent of the \say{communication core} organization of structural connections in nonhuman primates \cite{Modha2010NetworkAO, Markov2013CorticalHC}.

Functional gradients characterize continuous axes of variation and/or organization across the cortex \cite{Huntenburg2018LargeScaleGI}. Here, we investigated the top four gradients, which revealed that the principal gradient was nearly identical across modalities, and captured a visual-to-somatomotor axis (\cref{fig:8_grads}a; \cref{fig:8_grads-pair}). Indeed, other studies have identified this gradient as a key organizational feature of the mouse cortex \cite{Huntenburg2021GradientsOF, Coletta2020NetworkSO}. Two other gradients (G-2 and G-3) were also well captured across modalities, although their ordering (based on eigenvalues) was different (but note that the percentage of variance explained was very similar).

Finally, the spatial organization of BOLD G-4 was not reflected in the gradients identified based on \CA signals. Intriguingly, this gradient highlights regions overlapping with the frontal orienting field captured by BOLD OC-7 (\cref{fig:2_ocs7}), a network that was not detected by \CA even in the case of a 20-network decomposition (\cref{fig:2S_oc20}). It is also noteworthy that some of the gradients detected spatial organization that was not reflected in the network organization maps (\cref{fig:2_ocs7}). Overall, our results unveiled three functional gradients that exhibited strong to intermediate correspondence across conditions, as well as one spatial organization seen in BOLD that was not present in \CA signals.

It is possible that some of the discrepancies between BOLD and \CA results stemmed from \CA signals originating from excitatory neurons, while the BOLD signal is cell-type agnostic. Despite excitatory neurons being the most populous cell type in the cortex, it is still unclear to what extent the activity of other cell populations such as inhibitory neurons \cite{moon2021contribution, howarth2021more} and astrocytes \cite{takata2018optogenetic}, or vascular effects \cite{hillman2014coupling, o2016neural, drew2020ultra, drew2022neurovascular}, influence the BOLD signal. This will be explored in our future work utilizing the methods established here.

Another important consideration is our use of anesthesia. Due to the challenges of imaging awake mice \cite{ferris2022applications}, especially head motion, we opted to use low levels of anesthesia. Head motion systematically alters the correlation structure of functional data and was therefore of particular concern in our analyses \cite{satterthwaite2013improved, power2014methods} (Methods). However, the effects of anesthesia on brain activity and neurovascular coupling are complex and may vary by region, anesthetic agent, and dose \cite{Grandjean2014OptimizationOA, brier2019separability, xie2020differential, tsurugizawa2021impact}. Our future studies will evaluate how functional networks, and their properties, differ between awake and anesthetized animals.

Nevertheless, to evaluate the impact of anesthesia on our results, we performed initial, exploratory analyses in a group of $N=5$ animals for which we had \CA recordings in both anesthetized and awake states. These animals, which participated in the sessions reported in the main results, underwent an additional \CA recording session outside of the scanner while awake. The results (\cref{fig:2S_oc7-awake}) clearly show that the large-scale network organization that we reported in the lightly anesthetized state is also observed in awake mice. Relatively minor changes were also observed but require a larger sample for proper statistical comparisons. Nevertheless, the findings with this small group of animals considerably strengthen the generalizability of our findings.

The effects of brain state on functional organization were minimized because the same \say{ground truth} brain activity underlies the results from each modality, given the simultaneous nature of our multimodal data. Thus, moment-to-moment brain-state differences were not driving factors behind our findings. However, we note that the analysis methods we used do not strictly require simultaneous data collection. Further, although the primary findings reported in this Chapter were at the group level, our highly-sampled dataset allowed network organization to be determined at the level of the individual, lending considerable strength to our group-level findings, and underscoring the translational potential of our approach \cite{Elliott2021StrivingTT}. Future work will further explore individualized network properties while exploiting the simultaneous nature of these data.

Processing and analyzing multimodal data entails making several parameter choices that potentially affect outcome measures. In particular, network overlap could be inflated by spatial misalignment. We took great care in co-registering our data and optimizing our parameter set (from the Advanced Normalization Tools package \cite{avants2009advanced}). Further, issues of misalignment were considerably reduced by estimating network measures at the level of runs and combining values subsequently. Thus, modest misalignment after registration did not inflate the overall evaluation of overlap (Methods).

Furthermore, the quantification of membership strength was applied to values that were thresholded based on statistical significance. We also used relatively sparse graphs ($15\%$ density), such that only the strongest correlations were considered; further analyses that quantified the extent of overlap considered only membership values that statistically exceeded $1/7$ (for the 7-network solution). We also probed the effects of parameter changes (\Cref{table:params}) and found that our results were qualitatively robust.

Finally, we emphasize that, while we believe our results provide evidence for considerable network overlap, estimating the extent of overlap quantitatively remains a challenging problem in network science. Important progress in this direction involves developing principled approaches based on statistical inference and generative modeling \cite{peel2022statistical, peixoto2015model}.

In conclusion, we employed novel simultaneous wide-field \CA imaging and BOLD in a highly sampled group of mice expressing GCaMP6f in excitatory neurons to determine the relationship between large-scale networks discovered by the two techniques. Our findings demonstrated that:

\begin{enumerate}
    \item Most BOLD networks were detected via \CA signals.
    \item Considerable overlapping, in addition to disjoint, network organization was recovered from both modalities.
    \item The large-scale functional organization determined by \CA signals at low temporal frequencies ($0.01-0.5~Hz$), relative to high frequencies ($0.5-5~Hz$), was more similar to those recovered with BOLD.
    \item The principal functional connectivity gradient was nearly identical across all modalities.
    \item Key differences were uncovered between the two modalities in the spatial distribution of membership diversity and the relationship between region entropy (i.e., network affiliation diversity) and degree.\\
\end{enumerate}

Together, these findings uncovered a distinct overlapping network phenotype across modalities. This work revealed that the mouse cortex is functionally organized in terms of overlapping large-scale networks that are observed with BOLD, lending support for the neural basis of such a property, which is also observed in human subjects. The robust differences that were uncovered demonstrate that \CA and BOLD also capture some complementary features of brain organization.

Future work exploring these commonalities and differences, using the simultaneous multimodal acquisition used here, promises to help uncover how large-scale networks are supported by underlying brain signals in health and disease.

\section{Supplementary Discussion}

\subsection{Intuition behind the algorithm: SVINET}
Understanding the large-scale organization of brain interaction graphs is an important goal in network neuroscience \cite{bullmore2009complex, Bassett2017NetworkN}. A core axiom of network neuroscience is that the global topology of the graph alone encodes information about the functional architecture of the brain. Network topology, on the other hand, depends on how links are placed locally. In this sense, the existence or non-existence of a link between two nodes is informative and merits an explanation.

The community detection algorithm used here, SVINET \cite{Gopalan2013EfficientDO}, assumes a specific generative process to explain the existence of a link. Namely, SVINET is built under the assumption that pairs of nodes are more likely to be connected if they belong to one or more overlapping communities together \cite{Gopalan2013EfficientDO}. A generative model is then constructed based on this idea, where shared latent community memberships predict links. The algorithm uses variational inference and stochastic gradient ascent \cite{Blei2016VariationalIA, Hoffman2013StochasticVI} to infer latent community memberships from observed networks.

\subsection{Choosing number of communities}
Clustering is in the eyes of the beholder (or the algorithm) \cite{von2012clustering}. Community detection is inherently ill-defined:~algorithms do not find communities, what they do is \textit{\say{define}} communities and later find them according to their definition. Here, we employed a mixed-membership stochastic blockmodel \cite{airoldi2008mixed}, where each community corresponds to a \textit{latent functional role} (see \cite{airoldi2015handbook} for an extensive review). Choosing a specific number of communities is thus equivalent to deciding how many functional roles (or clusters) best describe the observed graphs. In this sense, there is no \textit{\say{true}} number of communities. Therefore, instead of focusing on a single result as a true decomposition, we explored network organization at different levels of granularity. We decided the number of communities empirically, using criteria such as bilateral symmetry. At the most coarse level, our $K=3$ communities recapitulated previous seed-based reports in \CA data and coarse fMRI-ICA findings. Our $K=7$ decomposition was similar to previous fMRI-ICA reports \cite{Zerbi2015MappingTM, Grandjean2020CommonFN}. Overall, the highly bilateral nature of the resulting community structure (observed for up to $K=20$ for \CA) and their similarity to previous reports increased our confidence about our results.

\subsection{Thresholding the graphs}

The algorithm used in this study requires binary graphs as its input \cite{Gopalan2013EfficientDO}, which necessitates choosing an edge-filtering threshold. To mitigate this, we employed an approach known as proportional thresholding which is known to perform better than alternatives such as absolute thresholding \cite{van2010comparing}. In addition, theoretical work has shown that network topology is highly robust against different thresholds \cite{Yan2018WeightTO}.

Here we reported results at a graph density of $d = 15\%$. To ensure the robustness of our findings to this arbitrary choice, we also explored densities going from $d = 10\%$ all the way up to $d = 25\%$ with incremental steps of $5\%$. Empirically, we found that our results were robust to the choice of edge density, as well as other hyperparameters. Overall, our data partially confirmed previous theoretical and simulation work that network topology is robust across a wide range of sparsity levels.

\subsection{Region of interest (ROI) definition scheme}
We started by using brain region masks from Allen Reference Atlas (ARA, \cite{Wang2020TheAM}) as our initial choice of ROIs but observed some mismatch between functional and ARA parcellations. This is probably because ARA regions were delineated using various anatomical and structural criteria; but crucially, function was not one of them. Here, we introduced a new parcellation scheme illustrated in \cref{fig:1_roi}, which increased the robustness of our results. In conclusion, we found that spatially homogeneous ROIs worked well for the purpose of functional network construction, consistent with previous reports in humans \cite{albers2021using}.

Defining appropriate ROIs for our multimodal dataset was challenging because different modalities occupy spaces with different geometries. Namely, fMRI data is defined within a 3D volumetric space, while mesoscopic \CA imaging data exists only on the 2D cortical surface. To address this, we started from the 2D space of cortical flatmap (\cref{fig:1_roi}a; step I), which fits \CA data well. We then added depth to obtain 3D volumetric ROIs (\cref{fig:1_roi}a; step II), suitable for fMRI data. In the depth dimension, we included cortical layers 1 to 4. Crucially, our goal was not to obtain layer-specific results (BOLD resolution was $0.4~mm$ isotropic). Instead, we wanted to consider depths of the cortex that most likely contribute to \CA signal \cite{Yizhar2011OptogeneticsIN, allen2017global, ren2021characterizing, Barson2019SimultaneousMA, Peters2021StriatalAT}, which would render our network-level cross-modality comparisons more meaningful.

\subsection{Mouse cortical areas visible from the top view}
\begin{figure}[t!]
    \centering
    \includegraphics[width=\linewidth]{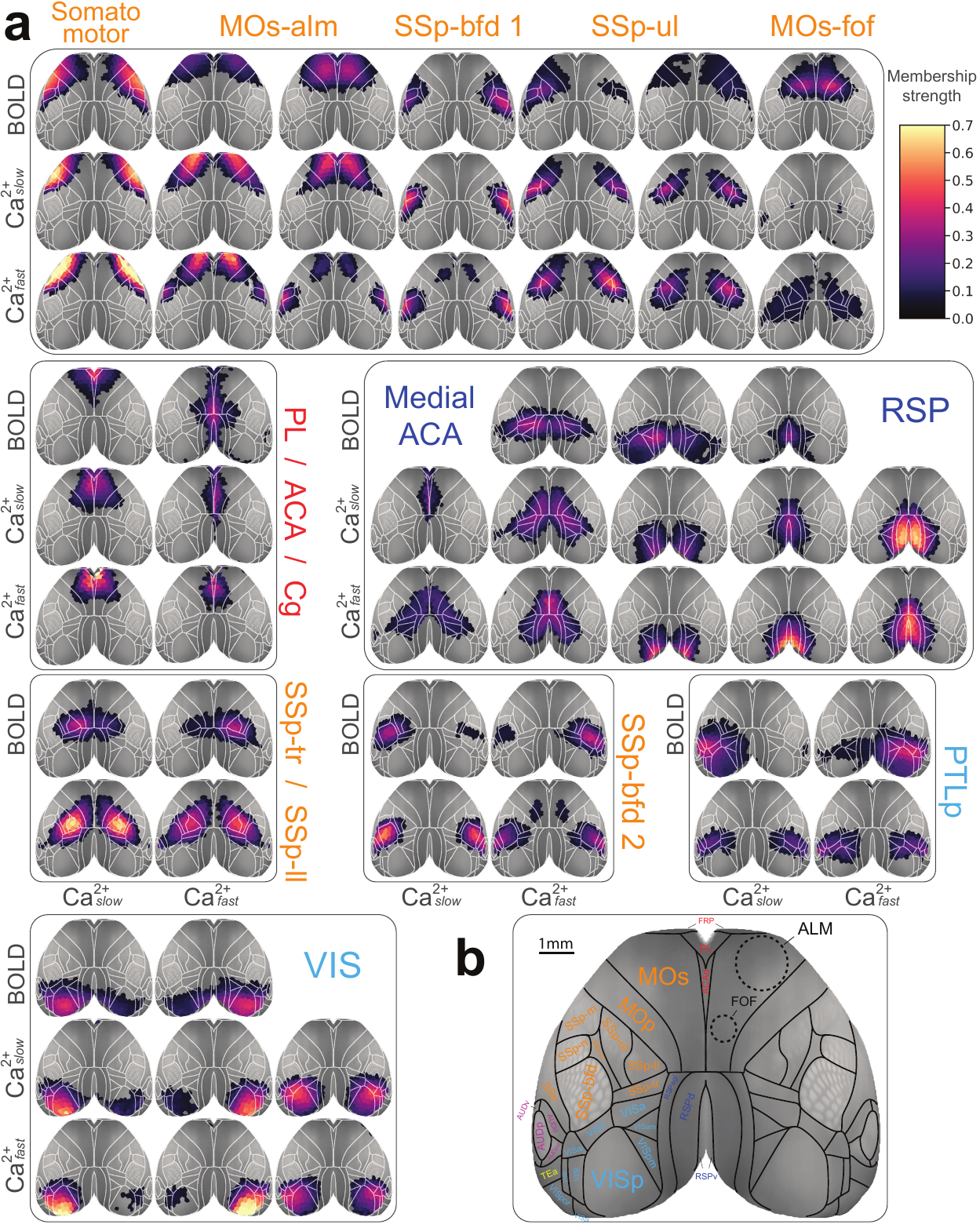}
    \caption[$K=20$ network decomposition]{\footnotesize $K=20$ decomposition. \textbf{(a)} Even at $K=20$, most networks maintain their bilateral symmetry, especially for \CA. A network centered around FOF appears as its own separate network for BOLD, similar to the $K=7$ solution (top-right). In contrast, this network did not appear separately for \CA, even at $K=20$. Instead, FOF partially overlaps with a large medial \CA network that spans parts of SSp-tr/ll. \textbf{(b)} Fine divisions of the cortical regions in Allen reference atlas, along with ALM and FOF (dashed lines). Label colors inspired from Figure 1F in ref.~\cite{harris2019hierarchical}. See \Cref{table:abbrev} for a list of abbreviations. Compare with \cref{fig:2_oc3} for $K=3$ and \cref{fig:2_ocs7} for $K=7$ solutions.}
    \label{fig:2S_oc20}
\end{figure}

The goal of the present study was to examine the cross-modality correspondences between mouse cortical networks obtained from the simultaneous fMRI-BOLD and wide-field \CA imaging. As such, we restricted our analyses to the cortical surface that appears in the \CA imaging field-of-view (see \cref{fig:1_roi}b). This top view includes a good portion of the mouse cortical regions. We provided both a coarse (\cref{fig:2_allen}), and a fine (\cref{fig:2S_oc20}b) delineation of the mouse cortical areas visible in this top view. In these delineations, solid lines correspond to anatomical regions as defined in the CCFv3 Allen reference atlas \cite{Wang2020TheAM}. In addition, dashed lines approximately correspond to functionally defined subregions in the secondary motor area \cite{erlich2011cortical, Chen2017AMO}. The abbreviations for all the brain regions that appear in \cref{fig:2_allen} and \cref{fig:2S_oc20}b are provided in \Cref{table:abbrev}. See the original CCFv3 publication for the full list of brain regions in Allen reference atlas ontology \cite{Wang2020TheAM}.

\section{Code Availability}
Our code is available in the following repository: \url{https://github.com/hadivafaii/Ca-fMRI} \cite{vafaii2023cafmri}. All of our code used for this project is written in Python, making extensive use of Python scientific computing environments, including NumPy v1.20.2 \cite{harris2020array}, SciPy v1.7.0 \cite{SciPy2020}, statsmodels v0.13.5 \cite{seabold2010statsmodels}, scikit-learn v1.2.0 \cite{pedregosa2011scikit}, pandas v1.3.3 \cite{reback2020pandas}, matplotlib v3.7.1 \cite{hunter2007matplotlib}, and seaborn v0.10.1 \cite{waskom2021seaborn}.

\begin{table}[ht!]
    \centering
    \captionsetup{skip=4pt}
    \caption[Brain region abbreviations.]{Brain region abbreviations.}
    \label{table:abbrev}
    \begin{tblr}{
        width=0.7\linewidth,
        hlines, vlines,
        colspec={X[0.4,l,m] X[2.0,l,m]},
        rows={3mm},
        cell{1}{2}={c=1}{c},
        hspan=minimal,
        vspan=minimal,
    }
    \textbf{Abbrev.} & \textbf{Full structure name} \\
    
    ACA & Anterior cingulate area \\

    AUD & Auditory areas \\
    AUDd & Dorsal auditory area \\
    AUDp & Primary auditory area \\
    AUDpo & Posterior auditory area \\
    AUDv & Ventral auditory area \\

    MO & Somatomotor areas \\
    MOs & Secondary motor area \\
    MOp & Primary motor area \\

    PL & Prelimbic area \\

    PTLp & Posterior parietal association areas \\

    RSP & Retrosplenial area \\
    RSPagl & Retrosplenial area, lateral agranular part \\
    RSPd & Retrosplenial area, dorsal part \\
    RSPv & Retrosplenial area, ventral part \\

    SS & Somatosensory areas \\
    SSp & Primary somatosensory area \\
    
    SSp-n & Primary somatosensory area, nose \\
    SSp-bfd & Primary somatosensory area, barrel field \\
    SSp-ll & Primary somatosensory area, lower limb \\
    SSp-m & Primary somatosensory area, mouth \\
    SSp-ul & Primary somatosensory area, upper limb \\
    SSp-tr & Primary somatosensory area, trunk \\
    SSp-un & Primary somatosensory area, unassigned \\

    SSs & Supplemental somatosensory area \\

    TEa & Temporal association areas \\

    VIS & Visual areas \\
    VISal & Anterolateral visual area \\
    VISam & Anteromedial visual area \\
    VISl & Lateral visual area \\
    VISp & Primary visual area \\
    VISpl & Posterolateral visual area \\
    VISpm & posteromedial visual area \\
    VISli & Laterointermediate area \\
    VISpor & Postrhinal area \\
    
    \end{tblr}
\end{table}

\begin{figure}[ht!]
    \centering
    \includegraphics[width=\linewidth]{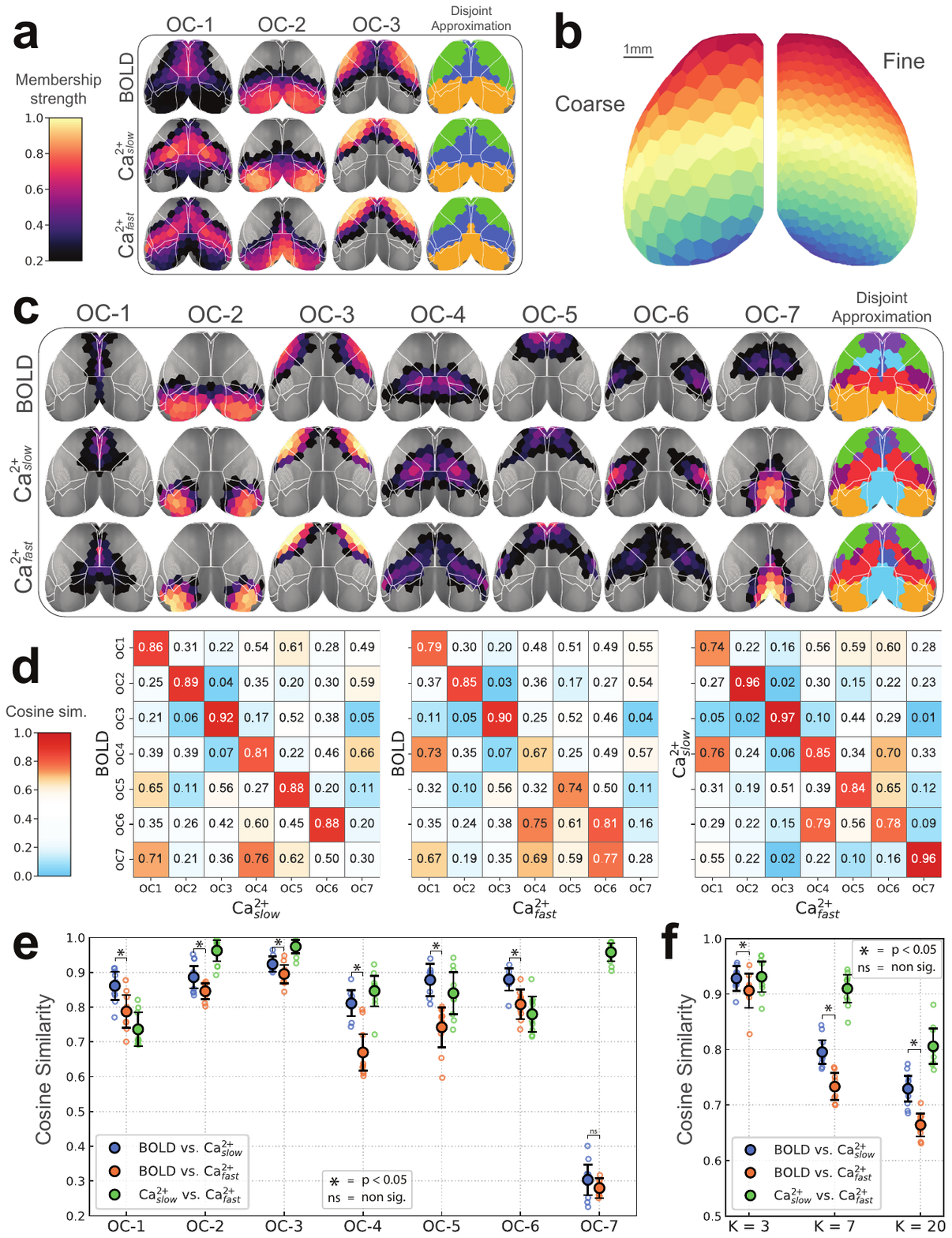}
    \caption[Network structure is robust to the choice of ROI granularity]{Network structure is robust to the choice of ROI granularity. \textbf{(a, c-f)} Similar to \cref{fig:2_ocs7}. \textbf{(b)} Left, results with coarse ROIs are presented here; Right, fine ROIs were used for the main results.}
    \label{fig:2S_ocs_n128}
\end{figure}

\begin{sidewaysfigure}[ht!]
    \centering
    \includegraphics[width=\linewidth]{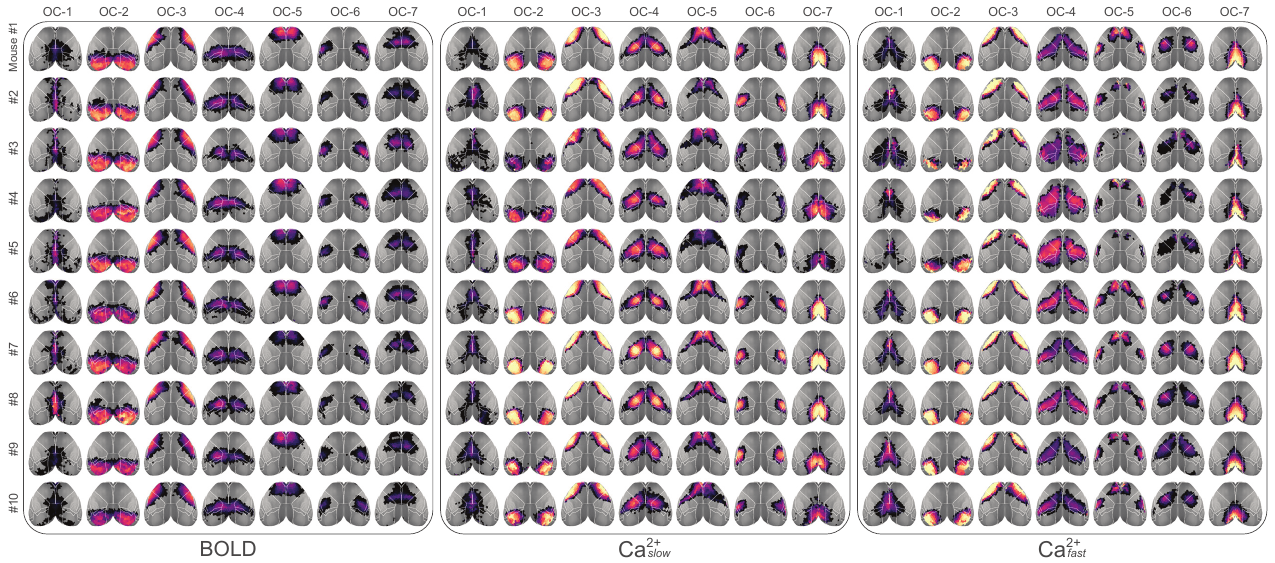}
    \caption[Overlapping networks at the level of individuals]{Overlapping networks at the level of individuals. Each mouse was highly sampled (\cref{fig:1_data}b), which allowed robust estimation of individualized networks. The high similarity of network structures across individuals is anticipated; however, there are still visible differences. Overall, the reproducibility of results at the individual level is notable and lends support to our group results (\cref{fig:2_ocs7}).}
    \label{fig:2S_oc7-indiv}
\end{sidewaysfigure}

\begin{figure}[ht!]
    \centering
    \includegraphics[width=\linewidth]{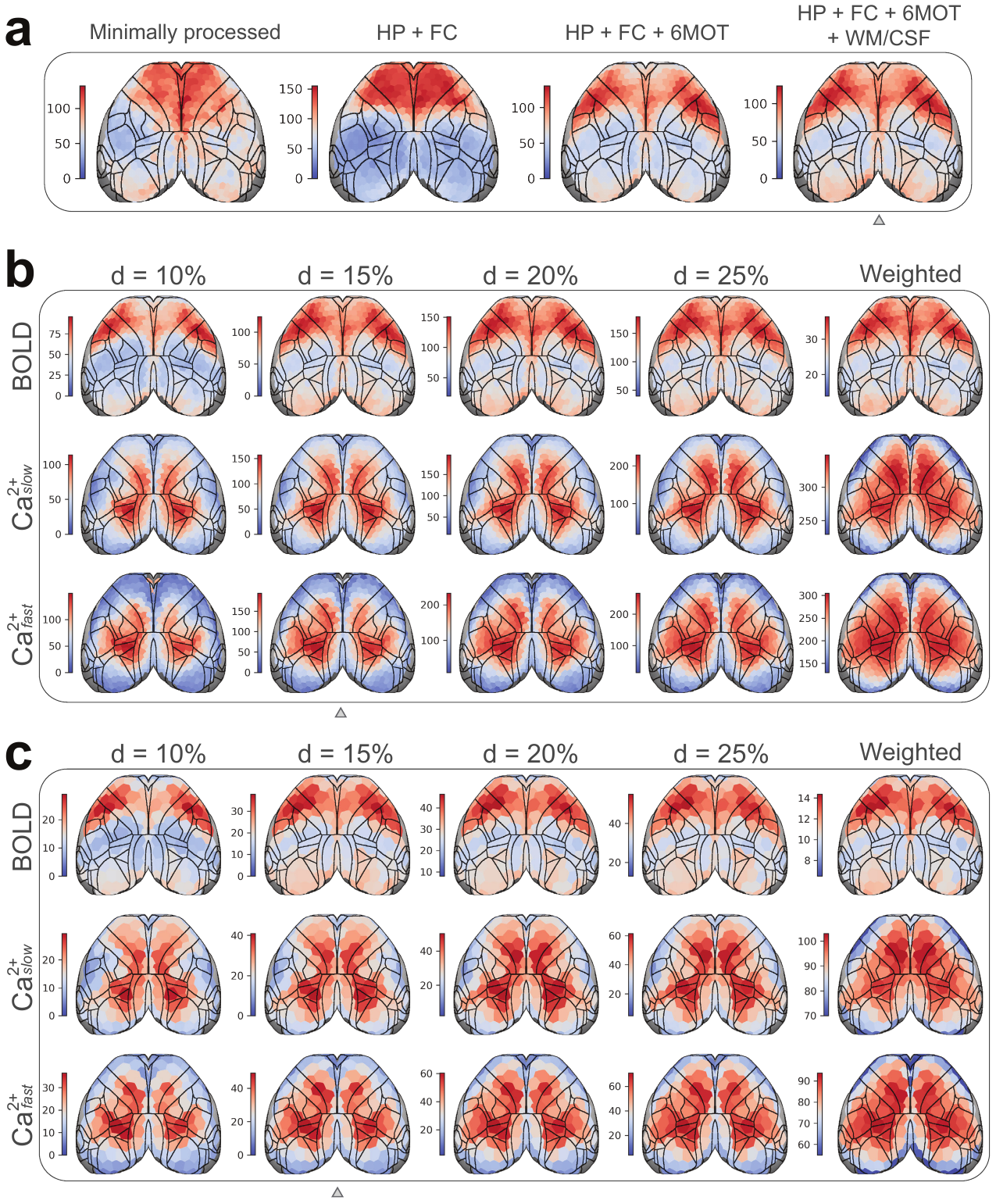}
    \caption[Dependence of degree centrality to preprocessing and analysis choices]{\footnotesize Dependence of degree centrality to preprocessing and analysis choices. \textbf{(a)} Spatial patterns of average node degree are somewhat altered depending on which BOLD preprocessing steps are used. Minimally processed, motion correction (rigid transformations) and detrending; HP, high-pass filtering ($0.01~Hz$); FC, frame censoring; 6MOT, motion regression (6 parameters); WM/CSF, average signal from white matter and ventricles regressed out. \textbf{(b)} Degree patterns are robust to the choice of edge filtering threshold. Difference thresholds result in different scales, but the spatial patterns remain relatively similar. \textbf{(c)} Similar to B but for a coarse parcellation (see \cref{fig:2S_ocs_n128}b). Small triangles indicate the pipeline and parameters used for the main results. Related to \cref{fig:6_deg}.}
    \label{fig:6S_deg-all}
\end{figure}

\clearpage
\section{Materials and Methods}

\subsection{Experimental model and subject details}
All procedures were performed in accordance with the Yale Institutional Animal Care and Use Committee and are in agreement with the National Institute of Health Guide for the Care and Use of Laboratory Animals. All surgeries were performed under anesthesia.

\paragraph{\textbf{Animals.}} Mice were group-housed on a 12-hour light/dark cycle. Food and water were available ad libitum. Cages were individually ventilated. Animals were 6-8 weeks old, 25-30g, at the time of the first imaging session. Animals (\textit{SLC}, \textit{Slc17a7-cre/Camk2$\alpha$-tTA/TITL-GCaMP6f} also known as \textit{Slc17a7-cre/Camk2$\alpha$-tTA/Ai93}) were generated from parent 1 (\textit{Slc17a7-IRES2-Cre-D}) and parent 2 (\textit{Ai93(TITL-GCaMP6f)-D;CaMK2a-tTA}). Both were on a C57BL/6J background. To generate these animals, male CRE mice were selected from the offspring of parents with different genotypes, which is necessary to avoid leaking of CRE expression. Animals were originally purchased from Jackson Labs.

\paragraph{\textbf{Head-plate surgery.}} All mice underwent a minimally invasive surgical procedure enabling permanent optical access to the cortical surface (previously described here: \cite{Lake2020SimultaneousCF}). Briefly, mice were anesthetized with $5\%$ isoflurane (70/30 medical air/O$_2$) and head-fixed in a stereotaxic frame (KOPF). After immobilization, isoflurane was reduced to $2\%$. Paralube was applied to the eyes to prevent dryness, meloxicam (2 mg/kg body weight) was administered subcutaneously, and bupivacaine ($0.1\%$) was injected under the scalp (incision site). Hair was removed (NairTM) from the surgical site and the scalp was washed with betadine followed by ethanol $70\%$ (three times). The scalp was removed along with the soft tissue overlying the skull and the upper portion of the neck muscle. Exposed skull tissue was cleaned and dried. Antibiotic powder (Neo-Predef) was applied to the incision site, and isoflurane was further reduced to $1.5\%$. Skull-thinning of the frontal and parietal skull plates was performed using a hand-held drill (FST, tip diameter: $1.4$ and $0.7~mm$). Superglue (Loctite) was applied to the exposed skull, followed by transparent dental cement (C$\&$B Metabond\textregistered, Parkell) once the glue dried. A custom in-house-built head plate was affixed using dental cement. The head-plate was composed of a double-dovetail plastic frame (acrylonitrile butadiene styrene plastic, TAZ-5 printer, $0.35~mm$ nozzle, Lulzbot) and a hand-cut microscope slide designed to match the size and shape of the mouse skull. Mice were allotted at least 7 days to recover from head-plate implant surgery before undergoing the first of three multimodal imaging sessions.

\subsection{Multimodal image acquisition}
All mice, $N=10$, underwent 3 multimodal imaging sessions with a minimum of 7 days between acquisitions. All animals underwent all imaging sessions. None were excluded prior to the study end-point. Data exclusion (based on motion etc.) is described below. During each acquisition, fMRI-BOLD and wide-field \CA imaging data we acquired simultaneously using a custom apparatus and imaging protocol described previously \cite{Lake2020SimultaneousCF}.

Functional MRI data were acquired on an $11.7~T$ system (Bruker, Billerica, MA), using ParaVision version 6.0.1 software. During each imaging session, 4 functional resting-state runs (10 min each) were acquired. In addition, 3 runs (10 mins each) of unilateral light stimulation data were acquired. These data are not used in this Dissertation. Structural MRI data were acquired to allow both multimodal registration and registration to a common space. Mice were scanned while lightly anesthetized ($0.5-0.75 \%$ isoflurane in 30/70 O$_2$/medical air) and freely breathing. Body temperature was monitored (Neoptix fiber) and maintained with a circulating water bath.

\paragraph{\textbf{Functional MRI.}} We employed a gradient-echo, echo-planar-imaging sequence with a 1.0 sec repetition time (TR) and 9.1 ms echo time (TE). Isotropic data ($0.4~mm \times 0.4~mm \times 0.4~mm$) were acquired along 28 slices providing near whole-brain coverage.

\paragraph{\textbf{Structural MRI.}} We acquired 4 structural images for multimodal data registration and registration to a common space. (1) A multi-spin-multi-echo (MSME) image sharing the same FOV as the fMRI data, with a TR/TE of $2500/20~ms$, 28 slices, two averages, and a resolution of $0.1~mm \times 0.1~mm \times 0.4~mm$. (2) A whole-brain isotropic ($0.2~mm \times 0.2~mm \times 0.2~mm$) 3D MSME image with a TR/TE of $5500/20~ms$, 78 slices, and two averages. (3) A fast-low-angle-shot (FLASH) time-of-flight (TOF) angiogram with a TR/TE of $130/4~ms$, resolution of $0.05~mm \times 0.05~mm \times 0.05~mm$ and FOV of $2.0~cm \times 1.0~cm \times 2.5~cm$ (positioned to capture the cortical surface). (4) A FLASH image of the angiogram FOV, including four averages, with a TR/TE of $61/7~ms$, and resolution of $0.1~mm \times 0.1~mm \times 0.1~mm$. 

\paragraph{\textbf{Wide-field fluorescence \CA imaging.}} Data were recorded using CamWare version 3.17 at an effective rate of $10~Hz$. To enable frame-by-frame background correction, cyan (470/24, \CA\hspace{-0.40em}-sensitive) and violet (395/25, \CA\hspace{-0.40em}-insensitive) illumination (controlled by an LLE 7Ch Controller from Lumencor) were interleaved at a rate of $20~Hz$. The exposure time for each channel (violet and cyan) was $40~ms$ to avoid artifacts caused by the rolling shutter refreshing. Thus, the sequence was: $10~ms$ blank, $40~ms$ violet, $10~ms$ blank, $40~ms$ cyan, and so on. The custom-built optical components used for in-scanner wide-field \CA imaging have been described previously \cite{Lake2020SimultaneousCF}. 

\subsection{Image preprocessing}
\paragraph{\textbf{Multimodal data registration.}} All steps were executed using tools in BioImage Suite (BIS) specifically designed for this purpose \cite{Lake2020SimultaneousCF}. For each animal, and each imaging session, the MR angiogram was masked and used to generate a view that recapitulates what the cortical surface would look like in 2D from above. This treatment of the data highlights the vascular architecture on the surface of the brain (notably the projections of the middle cerebral arteries, MCA) which are also visible in the static wide-field \CA imaging data. Using these and other anatomical landmarks, we generated a linear transform that aligns the MR and wide-field \CA imaging data. The same static wide-field image was used as a reference for correcting motion in the time series. To register the anatomical and functional MRI data, linear transforms were generated and then concatenated before being applied.

Data were registered to a reference space (CCFv3, \cite{Wang2020TheAM}) using isotropic whole-brain MSME images via affine followed by non-linear registration (ANTS, Advanced normalization tools; \cite{avants2009advanced}). The histological volume in CCFv3 was used because of a better contrast match with MRI images. The goodness of fit was quantified using mutual information and a hemispheric symmetry score that captured the bilateral symmetry of major brain structures. A large combination of registration hyperparameters was explored, and the top $10$ fits per animal were selected. The best transformation out of this pool was selected for each animal by visual inspection.

\paragraph{\textbf{RABIES fMRI data preprocessing.}} For fMRI preprocessing, we used RABIES (Rodent automated BOLD improvement of EPI sequences) v0.4.2 \cite{desrosiers2022rodent}. We applied functional inhomogeneity correction N3 (nonparametric nonuniform intensity normalization; \cite{manjon2010adaptive, sled1998nonparametric}), motion correction (ANTS, Advanced normalization tools; \cite{avants2009advanced}), and slice time correction, all in native space. A within-dataset common space was created by nonlinearly registering and averaging the isotropic MSME anatomical images (one from each mouse at each session), which was registered to the Allen CCFv3 reference space using a nonlinear transformation (see above).

For each run, fMRI data were motion-corrected and averaged to create a representative mean image. Each frame in the time series was registered to this reference. To move the fMRI data to the common space, the representative mean image was registered to the isotropic structural MSME image acquired during the same imaging session. This procedure minimizes the effects of distortions caused by susceptibility artifacts \cite{wang2017evaluation}. Then, the three transforms --- (1) representative mean to individual mouse/session isotropic MSME image, (2) individual mouse/session isotropic MSME image to within-dataset common space, and (3) within-dataset common space to out-of-sample common space --– were concatenated and applied to the fMRI data. Functional data ($0.4~mm$ isotropic) were upsampled to match anatomical MR image resolution ($0.2~mm$ isotropic). Registration performance was visually inspected and verified for all sessions. Motion was regressed (6 parameters). Current best practices in both human \cite{smith2013resting} and mouse \cite{desrosiers2022rodent} fMRI preprocessing include the application of a high-pass or band-pass filter to remove physiological and other sources of noise. Specifically, high-pass filtering removes slow ($< 0.01~Hz$) drifts from data \cite{smith2013resting}. Therefore, a high-pass filter (3rd order Butterworth) between $[0.01-0.5]~Hz$ was applied, and $15$ time points ($15s$ of data) were discarded from both the beginning and the end of the time series to avoid filtering-related edge artifacts. Finally, average white matter and ventricle time courses were regressed.

\paragraph{\textbf{Fluorescence \CA imaging data preprocessing.}} All steps in this pipeline are previously described in \textcite{OConnor2022functional}. Briefly, the raw signal was split between GCaMP-sensitive and GCaMP-insensitive imaging frames. Spatial smoothing with a large kernel (16-pixel kernel, median filter) was applied to reduce and/or remove focal artifacts (e.g., dust or dead pixels from broken fibers). Focal artifacts do not move with the subject and can bias motion correction. Motion correction parameters were estimated on these data using normalized mutual information. Rigid image registration was performed between each imaging frame in the time series and the reference frame. Registration parameters were saved, and the large kernel-smoothed images were discarded. Modest spatial smoothing (4-pixel kernel, median filter) was applied to the raw data, and these data were motion-corrected by applying the parameters estimated in the previous step. Data were down-sampled by a factor of two in both spatial dimensions, which yielded a per pixel spatial resolution of $50 \times 50~\mu m^2$ (original was $25 \times 25~\mu m^2$). Photo bleach correction was applied to reduce the exponential decay in fluorescence at the onset of imaging \cite{demchenko2020photobleaching}. The fluorophore-insensitive time series were regressed from the fluorophore-sensitive time series. The first 50 seconds of data were discarded due to the persistent effects of photobleaching in the \CA data. fMRI-BOLD and wide-field calcium imaging have unmatched temporal sampling rates ($1$ versus $10~Hz$) and different sources of noise. Accordingly, a band-pass filter (3rd order Butterworth) was applied that matched the frequencies of BOLD and \CAS ($\left[0.01, 0.5\right]~Hz$), and a complementary band with higher frequencies (\CAF: $\left[0.5, 5.0\right]~Hz$). The application of band-passing also aids with the removal of noise, including slow drifts. Finally, $15$ time points ($1.5s$ of data) were discarded from both the beginning and the end of the time series to avoid filtering-related edge artifacts \cite{power2014methods}.

\paragraph{\textbf{Frame censoring.}} Data were scrubbed for motion using a conservative $0.1~mm$ threshold. High-motion frames were selected based on estimates from the fMRI time series and applied to both fMRI and \CA data. Runs were removed from the data pool if half of the imaging frames exceeded this threshold for a given run. In this dataset, 2 runs (or $\sim1.7\%$ of all runs) were removed for this reason. Additionally, 2 more runs were removed because they did not pass our quality control criteria.

\subsection{Parcellating the cortex into columnar regions of interest (ROI)}
To create regions of interest (ROIs), we employed the Allen CCFv3 (2017) reference space \cite{Wang2020TheAM} and used their anatomical delineations as our initial choice of ROIs. However, this led to poor performance (see supplemental discussion). Here, we introduce a novel spatially homogeneous parcellation of the mouse cortex that can be adopted for both 3D fMRI and 2D wide-field \CA imaging data.

The procedure worked as follows. (1) We generated a cortical flatmap within the CCFv3 space using code published in ref.~\cite{Knox2018HighresolutionDM}(\url{https://github.com/AllenInstitute/mouse_connectivity_models}). (2) We subdivided the left hemisphere into $N$ regions via $k$-means clustering applied to pixel coordinates (for most analyses reported, $N=512$).  The right hemisphere was obtained by simple mirror-reversal to obtain a total of $2N$ regions. (3) Depth was added to the ROIs to obtain column-shaped regions. To do so, a path was generated by following streamlines normal to the surface descending in the direction of white matter (streamline paths were available at $10~\mu m$ resolution in CCFv3; see Figure 3F in \cite{Wang2020TheAM}). Here, we chose ROI depths so that we included potential signals from approximate layers 1 to 4 (layer masks were obtained from CCFv3). Evidence from wide-field \CA imaging suggests that signals originate from superficial layers but can extend into the cortex to some extent \cite{Yizhar2011OptogeneticsIN, allen2017global, ren2021characterizing, Barson2019SimultaneousMA, Peters2021StriatalAT}. (4) Finally, ROIs were downsampled from $10~\mu m$ to $100~\mu m$ resolution. See \cref{fig:1_roi}.

After co-registration, ROIs were transformed from the CCFv3 space into each individual's 3D and 2D anatomical spaces (see above). On average, ROIs had a size of $8 \pm 3$ voxels (3D, fMRI) and $48 \pm 20$ pixels (2D, \CA) in individual spaces (mean $\pm$ standard deviation).

\subsection{Functional network construction}
Time series data were extracted and averaged from all voxels/pixels within an ROI in native space to generate a representative time series per ROI. For each modality, for each run, an adjacency matrix was calculated by applying Pearson correlation to time series data to each ROI pair.
Next, we binarized the adjacency matrices by rank ordering the connection weights and maintaining the top $15\%$; thus, after binarization, the resulting graphs had a fixed density of $d = 15\%$ across runs and modalities. This approach aims to keep the density of links fixed across individuals and runs and better preserves network properties compared to absolute thresholding \cite{van2010comparing}. 
To establish the robustness of our results to threshold values, we also tested values of $10\%$ to $25\%$ in $5\%$ increments. 

\subsection{Finding overlapping communities}
Overlapping network analysis was performed using SVINET, a mixed-membership stochastic blockmodel algorithm \cite{Gopalan2012ScalableIO, Gopalan2013EfficientDO}, which has been previously applied to human fMRI data by \textcite{Najafi2016OverlappingCR} and \textcite{cookson2022evaluating}. SVINET models the observed graph within a latent variable framework by assuming that the existence of links between pairs of nodes can be explained by their latent community memberships. For binary adjacency matrix $A$ and membership matrix $\pi$, the model assumes the conditional probability of a connection as follows:

\begin{equation}
    p(A_{ij} \vert \pi_i, \pi_j) \propto \sum_{k=1}^K \pi_{ik} \pi_{jk},
\end{equation}
where $K$ is the number of communities, and $A_{ij} = 1$ if nodes $i$ and $j$ are connected and $0$ otherwise.  Intuitively, pairs of nodes are more likely to be connected if they belong to the same community or to (possibly several) overlapping communities. More formally, SVINET assumes the following generative process:

\begin{enumerate}
    \item For each node, draw community memberships $\pi_i \sim \mathrm{Dirichlet}(\alpha)$
    \item For each pair of nodes $i$ and $j$:
    \begin{enumerate}
        \item Draw community indicator $z_{i \rightarrow j} \sim \pi_i$
        \item Draw community indicator $z_{i \leftarrow j} \sim \pi_j$
        \item Assign link between $i$ and $j$ if $z_{i \rightarrow j} = z_{i \leftarrow j}$.
    \end{enumerate}
\end{enumerate}
Model parameters $\alpha$ are fit using stochastic gradient ascent \cite{Hoffman2013StochasticVI, Blei2016VariationalIA}. The algorithm was applied to data from each run using 500 different random seeds. Results across seeds were combined to obtain a final consensus for a run.

\subsection{Aligning community results}

Communities were identified in random order due to the stochastic nature of our algorithm. Maximum cosine similarity of the cluster centroids was used to match communities across calculations (runs or random seeds), as follows.

For each run, membership vectors from all random seeds were submitted to \textit{k}-means clustering (sklearn.cluster.KMeans) to determine $K$ clusters (e.g., $K=7$ for analyses with $7$ communities). The similarity between membership vectors from each random seed (source) and the cluster centroids (target) was then established via pairwise cosine similarity, yielding a $K \times K$ similarity matrix from source to target per random seed. A permutation of the rows of this similarity matrix was identified such that diagonals had maximum average similarity, a procedure known as the \textit{Hungarian algorithm} (scipy.optimize.linear\_sum\_assignment). Finally, the identified permutation was applied to seed results, thereby aligning them with the targets. The aligned communities were then averaged.

The outcome of this procedure is a membership matrix $\pi$ as shown in \cref{fig:1_pi}. Importantly, this matching procedure was done for each condition (BOLD, \CAS, \CAF) separately.

\subsection{Group results}
Crucially, all measures discussed below were computed at the run level first before combining at the group level.

\paragraph{\textbf{Membership matrices.}} This is what we have visualized in \cref{fig:2_oc3}, \cref{fig:2_ocs7}, \cref{fig:2S_oc7-awake}, \cref{fig:2S_oc7-hrf}, \cref{fig:2S_oc20}, \cref{fig:2S_ocs_n128}a and c, and \cref{fig:2S_oc7-indiv}.

\paragraph{\textbf{Thresholding membership values.}} To enhance the robustness of our estimates of network overlap, membership values were thresholded to zero if they did not pass a test rejecting the null hypothesis that the value was zero. After thresholding, the surviving membership values were rescaled to sum to 1.
Thresholding was performed for each animal separately by performing a one-sample \textit{t}-test and employing a false discovery rate of 5\%. 
All results shown utilized this step, with the exception of figures that illustrate the spatial patterns of membership values (and do not estimate network overlap).
Note that almost all ($\sim 99\%$) memberships that did not reach significance had values in the range $[0, 0.2]$.

\paragraph{\textbf{Region functional diversity.}} Shannon entropy was applied to membership matrices for each run separately before averaging. That is, given a membership matrix $\pi$ from a run, node entropies were computed to get an entropy estimate per node at the run level:

\begin{equation}
\begin{aligned}
    \text{(normalized) entropy of node } i \text{ at run level} = f(\pi_i)&\\
    \equiv h_i = - \sum_{k=1}^K \pi_{ik} \log \pi_{ik} \,/ \log K,&
\end{aligned}
\end{equation}
where $K=7$ is the number of communities. Entropy values were combined by averaging across runs to get the group-level estimates. This is what's visualized in \cref{fig:5_ent}b. Similarly, group averages were used to calculate the correlations between modalities in \cref{fig:5_ent}c.

\paragraph{\textbf{Computing distributions.}} Similar to the above, distributions were computed for each run separately before combining at the group level. For example, consider $h_i$ (entropy of node $i$) from a run. We computed percentage values using $20$ bins of width $0.05$ that covered the entire range of normalized entropy values $[0, 1]$. We then averaged over run-level histogram values to get group-level estimates shown in \cref{fig:5_ent}a. Other distributions were computed in an identical way. Specifically, $57$ bins of size $5$ were used for \cref{fig:6_deg}a, and $4$ bins of size $0.2$ were used for \cref{fig:3_prop}b.

\subsection{Gradient analysis}
To obtain functionally connectivity (FC) gradients, we closely followed methods outlined by \textcite{Huntenburg2021GradientsOF} and adapted their code (\url{https://github.com/juhuntenburg/mouse_gradients}) for this purpose. Briefly, \emph{per-run} FC matrices were Fisher r-to-z transformed, averaged across all runs, sessions, and animals, and back-transformed to Pearson's r values. The resulting group-averaged FC matrix was decomposed using \emph{diffusion maps} \cite{coifman2006diffusion}, a nonlinear dimensionality reduction method commonly employed by human \cite{Margulies2016SituatingTD} and mouse \cite{Huntenburg2021GradientsOF, Coletta2020NetworkSO} literature for estimating FC gradients. Gradients have arbitrary units and their absolute value is not meaningful. We z-scored each gradient separately, which allowed us to use a shared color bar in \cref{fig:8_grads}a.

\subsection{Statistical analysis}
\paragraph{\textbf{Hierarchical bootstrapping.}} Statistical results were performed at the group level by taking into consideration the hierarchical structure of the data (for each animal, runs within sessions), which can be naturally incorporated into computational bootstrapping to estimate variability respecting the organization of the data \cite{saravanan2020application}. For each iteration (total of $1,000,000$), we sampled (with replacement) $N=10$ animals, $N=3$ sessions, and $N=4$ runs, while guaranteeing sessions were yoked to the animal selected and runs were yoked to the session selected (\cref{fig:1_data}b). In this manner, the multiple runs were always from the same session, which originated from a specific animal. Overall, the procedure allowed us to estimate population-level variability based on the particular sample studied here. To estimate $95\%$ confidence intervals, we used the bias-corrected and accelerated (BCa) method \cite{efron1987better}, which is particularly effective when relatively small sample sizes are considered (SciPy's scipy.stats.bootstrap). Hierarchical bootstrapping was used to estimate variability in \cref{fig:2_oc7-cos}b and c, \cref{fig:3_prop}b, \cref{fig:5_ent}c, \cref{fig:6_deg}c, \cref{fig:7_carto}b, \cref{fig:2S_ocs_n128}e and f, \cref{fig:2S_cos-all}, \cref{fig:2S_oc7-hrf}b and c, \cref{fig:5S_particip}c, and \cref{fig:5S-6S_perc}.

\paragraph{\textbf{One-sample \textit{t}-test.}} We used a one-sample \textit{t}-test to define \textit{belonging} in networks as a function of a threshold $\mu$. The \textit{t}-statistic for membership of node $i$ in network $k$ is given as:

\begin{equation}
    t_{ik} = \frac{\bar{\pi}_{ik} - \mu}{SE_{ik}},
\end{equation}
where $\bar{\pi}_{ik}$ is the group averaged membership of node $i$ in network $k$, and $SE_{ik}$ is the standard error estimated using hierarchical bootstrapping (see above).  We calculated p values using \textit{t}-statistics for all nodes and networks and declared a node a member of a network if its p-value reached significance $p=0.05$.  The results as a function of various $\mu$ are shown in \cref{fig:4_tier}a. We applied Benjamini-Hochberg correction \cite{Benjamini1995ControllingTF} using Python statsmodels' implementation (statsmodels.stats.multitest.multipletests) to correct for multiple comparisons.

\paragraph{\textbf{Permutation test.}} A paired permutation test was used to compare conditions in \cref{fig:2_oc7-cos}c and d, \cref{fig:2S_ocs_n128}e and f, \cref{fig:2S_cos-all}, and \cref{fig:2S_oc7-hrf}b and c; and to perform a node-wise comparison across modalities in \cref{fig:5_ent}d. We used SciPy's implementation with $N=1,000,000$ resamples. Holm–Bonferroni correction \cite{holm1979simple} was applied to correct for multiple comparisons.


\renewcommand{\thechapter}{6}

\chapter{Conclusions and future directions}

In Chapter 2 (\cref{sec:alhazen}), I presented a thought-provoking quote from Alhazen in which he advises \say{the seeker after the truth} on the necessity of conducting thorough and unbiased scientific research, emphasizing the importance of critically examining everything while steering clear from prejudice and leniency.

In this concluding Chapter, as a committed seeker of truth, I will rigorously evaluate the potential limitations of my own research, in line with Alhazen's guiding principles, which is now a standard of scientific conduct. I will also outline possible future research directions that could build upon and extend the findings reported in this Dissertation.

In closing, I will share some reflections on the current state of neuroscience research. Specifically, I will explore the need for \say{interlocking theories} which I believe are crucial for integrating fragmented neuroscience facts into a cohesive body of knowledge.

\section{Vision \& Perception}

In Chapter 4, I drew inspiration from the idea of perception as inference which dates back at least to the 19th century and possibly earlier. This theory states that our perceptions are constructed from the interactions between two components: The first component is an intrinsic one and it determines the brain's prior expectations about the state of the world. This component constitutes knowledge about how the world works, and it contains both innate and learned aspects. The brain's prior expectations are shaped by its ecological niche over evolutionary time scales (genes, brain architecture), as well as learned from experience during an organism's lifetime. The second component is the sensory data that originates from the external world and is picked up by our sensory hardware such as our eyes. These two components interact to \sayit{construct} our perceptions, or our inferences, about the hidden causes of the sensory data.

I applied this theory to answer questions in the domain of motion perception because the causes of retinal motion are well understood (e.g., self and object motion), and they can be systematically manipulated. Specifically, I introduced a synthetic data framework, Retinal Optic Flow Learning (ROFL), and I built a model, the compressed Nouveau VAE (cNVAE), and used these ideas to show that the cNVAE can solve the fundamental problem of \say{optic flow parsing} in motion perception. Additionally, I showed that ROFL and the cNVAE can be used as an interpretive framework for understanding motion processing in primate brains. Finally, comparing the cNVAE to an ensemble of alternative models revealed that its success depends on two critical choices: (1) an appropriate loss function---inference \cite{helmholtz1867handbuch}; and, (2) a properly designed architecture---hierarchical, like the sensory cortex \cite{mumford1992computational}.

\subsection{Limitations and future work}

My work in Chapter 4 was in a \say{passive viewing} paradigm. I used data from experiments in which the animals were passively fixating on a dot on the screen while low-pass randomly moving dots were played on the screen. I introduced a hierarchical VAE model, the cNVAE, which was also built within a passive paradigm and there was no action involved: only passive perception of static stimuli. This has been (and still is) a major limitation of vision science, as it undermines the purpose of seeing. We see because it enables us to act \cite{Cisek2010NeuralMF}. Otherwise, what would be the point of seeing? The majority of experimental efforts have been in the passive paradigm, but things are changing for the better. Concerted efforts by vision scientists are focused on overcoming this limitation, transitioning from passive to active viewing paradigms \cite{yates2023detailed}.

I should note that this action-centric approach to vision is not new. Starting from the 1950s, Gibson spent his entire career advocating for an \say{Ecological Approach to Visual Perception}. His main thesis is that the animals would adapt to perceive things that can eventually lead to action. Actions that are \textit{afforded} by the ecological niche of an animal. For obvious reasons, Gibson's concept of \textit{affordances} has been embraced by the reinforcement learning community. However, the visual cortex is still viewed largely as a passive feature extractor. Nevertheless, this view has led to many discoveries that will likely generalize to active settings.

One lesson we have learned from this passive approach is the significance of nonlinear computations in visual perception. Even in my own results, untangling object and self-motion was made possible by successive applications of nonlinear computations. In stark contrast to this intricate, computational perspective, the ecological approach to vision avoids an in-depth exploration of the underlying mechanics of visual processing. In his book, Gibson almost completely refrains from discussing the processes underlying perception. This is a major limitation of the Gibsonian approach.

Both views on perception emphasize different but equally important aspects of visual perception. The \textit{Helmholtzians} correctly identified that perception is more than just direct viewing. Rather, perception requires processing, and it is a question of how the brain represents the world. The \textit{Gibsonians} correctly identified that the point of perception is to inform subsequent actions, thus, closing the action-perception loops. Therefore, it seems like the most comprehensive approach for understanding vision is a \textit{Gibsholtzian} one---or \textit{Helbsonian}, if you prefer. I wanted to expand upon this idea more, but found a fine piece of scholarly work from 2002 that explores this very issue. Therefore, I will simply cite it for interested readers to check out: \textcite{norman2002two}. See also \textcite{dicarlo2021does, peters2024does} for a more recent discussion of similar ideas.

\begin{figure}[ht!]
    \centering
    \includegraphics[width=\linewidth]{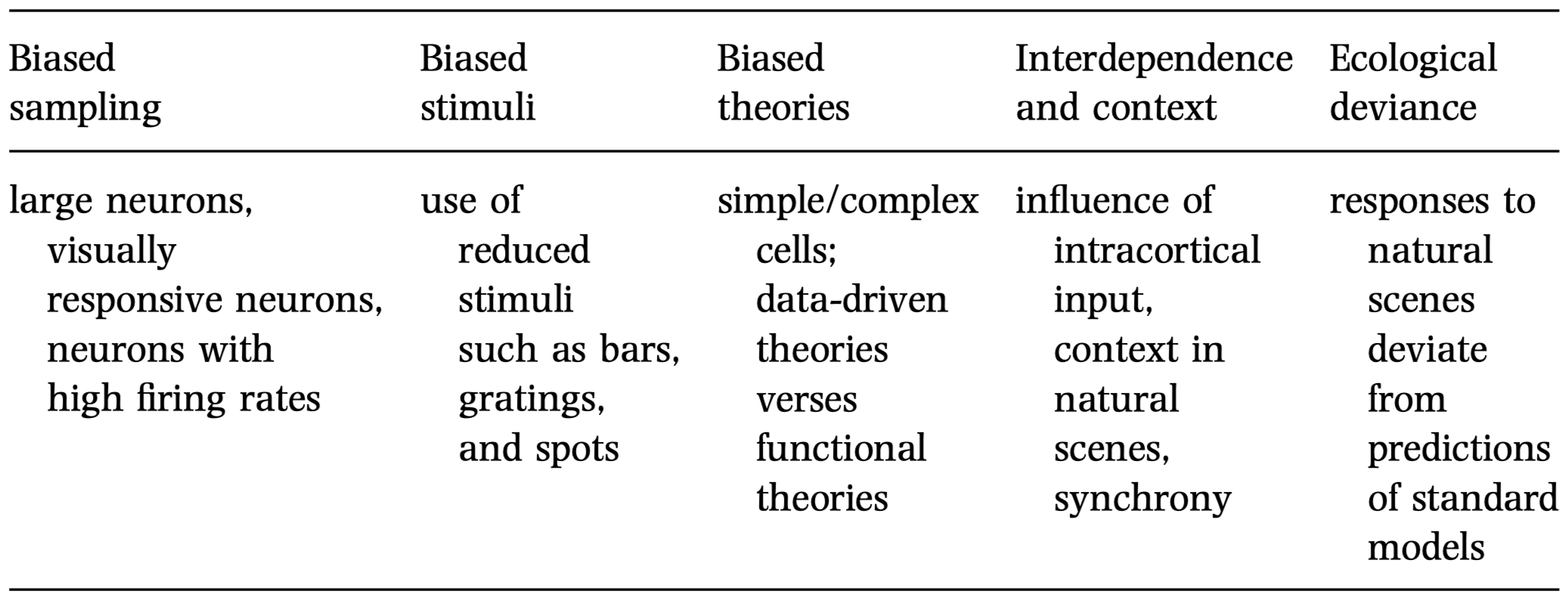}
    \caption[Five problems with the current approaches in vision science]{Five problems with the current view of V1. From \textcite{olshausen2005other}, 2005.}
    \label{fig:biases}
\end{figure}

Another limitation was working with static optic flow rather than image computable videos. Optic flow is not real; photons are. Therefore, using natural videos is preferred, both in collecting experimental data, and training models for in silico experimentation. Using natural movies would result in stronger conclusions, and could overcome artificial biases that may be introduced by using stimuli that don't exist in the ecological niche of the animal. This applies to Hubel \& Wiesel's bar of light stimuli, all the way to the random dot kinematogram I used in Chapter 4. This bias, along with other biases that have long pervaded vision research, are examined by \textcite{olshausen2005other} in a book chapter entitled \sayit{What is the other 85\% of V1 doing?} (\cref{fig:biases}). This incisive work stands as one of the best critical analyses of the field of vision science, and it remains as relevant today as it was in the mid-2000s when it was published.

\section{Intrinsic functional organization of the brain}

In Chapter 5, I used a simultaneously recorded dataset, combining fMRI-BOLD and wide-field \CA imaging, to study functional cortical organization in mice. The results indicate that certain aspects of fMRI--derived network structure are reproduced by \CA\!\!\!--derived network structure. For instance, several communities exhibit similar spatial organization across modalities, and the principal functional gradient is nearly identical for fMRI-BOLD and \CA signals. These observations lend fundamental support for the neural basis of fMRI--derived network structure, which is also observed in human subjects. Despite the similarities, important differences also emerged in a subset of communities and brain region centrality measures, indicating that \CA and fMRI-BOLD signals capture complementary features of brain organization.

Our results provide a foundational reference point for future studies venturing into the intrinsic organization of the mouse brain. However, our work had certain limitations that I will discuss below, along with possible future directions that can address them.

\subsection{The untapped potential of simultaneous multimodal imaging}

Throughout history, advancements in neuroscience have been closely tied to developments in technology. The neuron doctrine did not stand a chance if it wasn't for Golgi's staining method. The neural code would not have evolved so rapidly if it wasn't for Adrian's amplifier invention. The list goes on and on, from Berger's invention of EEG to Hubel's tungsten electrode, all the way to the invention of the MRI machine, optogenetics, neuropixels, and whatever will come next. The future of neuroscience will surely rely on emerging technologies that enable continued elucidation of the brain's mysteries. 

Related to this point, Sydney Brenner, who used electron microscopy technology to map the connectome of the roundworm C.~elegans, has famously said:\\[-3mm]

\quote{Progress in science depends on new techniques, new discoveries and new ideas, probably in that order.}{Sydney Brenner}\\

More recently, multimodal imaging has emerged as an exciting new frontier, which brings together complementary benefits of more than one imaging modality. However, we still lack theoretical concepts and computational tools that would allow us to exploit the full potential of such multimodal datasets. Along these lines, I will discuss several limitations of the work presented in Chapter 5.

\subsection{Future work addressing methodological limitations}

A key limitation of our approach was that we never actually used the simultaneous nature of the dataset. We inferred network structures, as well as all the downstream measures and statistics, independently for each data modality. The results reported in Chapter 5 likely would have worked fine if we had separate fMRI and \CA imaging sessions from separate groups of animals. Despite this limitation, our work lays the foundation for future work that should utilize the simultaneous aspect of the data.

Another issue was modeling the functional brain as a graph, obtained from thresholding covariance matrices. Binarizing covariance matrices as a method of inferring interaction graphs is limited for several reasons. First, this method does not offer a measure of uncertainty. After binarization, links either exist or do not exist, and we have no idea which ones are more likely to be false positives or which links were left undetected as false negatives \cite{peel2022statistical}. Second, it is well-known that applying correlation to short chunks of low-pass time series data can produce spurious outcomes \cite{yule1926we, leonardi2015spurious, vigen2015spurious, lurie2020questions, harris2021nonsense, peel2022statistical}. We used binarized Pearson correlation as our measure of brain interactions because this method involves the least amount of additional assumptions \cite{cliff2023unifying}, and it has been used as the standard approach for a long time \cite{brockwell1991time}. However, we recognize this limitation and recommend that in the future, more sophisticated measures should be used.

Modeling the functional architecture of the brain as a static graph does not take into account its rich dynamics \cite{breakspear2017dynamic}, with moment-to-moment changes in functional connectivity \cite{hutchison2013dynamic, lurie2020questions}. Most of the interesting variance in brain fluctuations gets buried and washed away when we collapse across time to obtain a static \say{snapshot}. This snapshot approach has had some limited success in the past, but the research community has largely moved on from such measures due to its limitations outweighing the benefits \cite{calhoun2014the, lurie2020questions, benisty2023rapid}. In multimodal settings, accounting for brain dynamics is even more important and potentially insightful \cite{Lake2020SimultaneousCF, Barson2019SimultaneousMA, lohani2022spatiotemporally, higley2022spatiotemporal}. Accordingly, future work should go beyond static measures of brain interactions.

\subsection{Conceptual and fundamental limitations}

One of the research questions driving the investigations in Chapter 5 was whether mouse cortical networks are consistent with an overlapping architecture, and if so, to what extent. Our data indicate the presence of a substantially overlapping structure, in both fMRI--derived and \CA\!\!--derived networks. But what if this is just noise? To ensure the observed overlap was real and not driven by noise, we performed a series of rigorous tests and undertook extremely prudent measures to minimize the chance of inflated overlap measures. Even with all the precautions and rigorous control analyses, the overlap did not go away. There is a strong degree of truth to it.

However, our study did not answer one critical question that was looming in the background the whole time. And the question is: \textit{so what}?

What is the point of showing that the brain is an overlapping system and calling it a day? What do we gain from going the extra mile? Does it provide a richer explanation of animal behavior? Does it allow us to distinguish animals in different states, like health and disease? Perhaps, but we don't know yet. In Chapter 5, we also documented a disjoint approximation alongside overlapping results. The disjoint approximation may be just as good when it comes to predicting behavior, which is arguably a primary goal in neuroscience \cite{krakauer2017neuroscience}. We cannot assume, a priori, that there is something inherently more fundamental about overlapping network structure compared to a disjoint one. This must be demonstrated, rather than assumed.

One argument goes that an overlapping network decomposition contains \say{more information} because it provides dense membership strength of brain regions in communities, as opposed to a binary all-or-none type of belonging as is common in disjoint frameworks. While this is factually true, unfortunately, it does not provide a strong case. If losing information is a key consideration, then we could have just printed the entire brain activations in thousands of journal pages.

The argument appears to be that although both disjoint and overlapping accounts are different maps of the same territory \cite{wuppuluri2018map}, one contains just the \textit{right amount} of extra information. But is this truly the case? We don't know. Our work was descriptive. We only showed overlap exists---a trivial and well-known property of correlation-based networks \cite{peixoto2015model, Yeo2014EstimatesOS, Najafi2016OverlappingCR, Faskowitz2020EdgecentricFN}. We did not, however, address the \textit{so what}. We did not demonstrate any functional consequences of network overlap. Neither did we provide a theoretical explanation for possible causes of this overlap.

In retrospect, it appears like this work could have been enhanced with better methodologies (see above), along with better, deeply thought-out, and more consequential research questions.

The \say{overlapping vs.~disjoint} results presented in Chapter 5 don't exactly stand out as the most imaginative or groundbreaking outcome that could have been derived from a simultaneous neuronal and MRI imaging dataset. There are at least a dozen more creative, more impactful, research questions one could have asked instead. However, this should not distract from the fact that our other results---comparing network structures across modalities---simple as they may be, were necessary to lay a foundation upon which future work can build.

To conclude, future directions must go beyond merely descriptive accounts of brain function (\say{the brain is organized in such and such ways. period.}) and embrace a normative perspective, answering why questions (\say{the brain is organized in such and such ways \textit{\underline{because}}\dots}).

\section{Unifying different camps of neuroscience through interlocking theories}

In this final section, I aim to reflect on a critical and unresolved issue in neuroscience: The development of interlocking theories that would enable fragmented camps of neuroscience to contribute towards an integrated corpus of knowledge.

A comprehensive understanding of the brain necessitates a collaborative approach, integrating insights across different areas of neuroscience research. For instance, the work of a vision scientist should inform and be informed by the research of a neuroscientist focusing on the motor system, recognizing that both are probing the same system after all.

Drawing from my background in condensed matter theory, I find a pertinent example in the evolution of this field. Emerging from the early days of quantum mechanics in the 1920s, condensed matter physics (then known as solid-state physics) is unified by a single foundational equation, the Schr\"odinger's equation:

\begin{equation}
    i \hbar \frac{\partial \psi}{\partial t} = \hat{H} \psi.
\end{equation}

This equation's generality lies in identifying the appropriate Hamiltonian, $\hat{H}$, to effectively describe a given experimental phenomenon \cite{coleman2015introduction}. The majority of condensed matter theory research involves developing effective Hamiltonians to explain experimental observations \cite{laughlin2000theory}, sometimes earning Nobel prizes.

The significance of this example for neuroscience is the efficient \say{transfer of knowledge} it enables across different condensed matter research groups. Even when working with diverse experimental setups and Hamiltonians, condensed matter theorists can still learn from each other's work because of the shared foundation in quantum mechanics and Schr\"odinger's equation.

In contrast, the current state of neuroscience often resembles a disconnected tale of \say{hammers and nails}, where specialized tools and approaches are developed and applied in isolation, lacking a unifying theoretical framework. Such an insular approach is not conducive to long-term progress, and the field must evolve to foster broader collaboration and integration through the development of interlocking theories.

With that said, the complexity of living systems presents challenges that almost certainly cannot be addressed by standard approaches such as a Hamiltonian framework, which is more suited to closed, energy-conserving systems. The dynamic exchange of matter, energy, and information, coupled with the adaptive nature of biological information processing, calls for alternative approaches.

A promising direction might be the application of the Fokker-Planck equation to describe neural dynamics, where different time-dependent drift and diffusion terms can capture the varied experimental scenarios in neuroscience, analogous to the varied Hamiltonians in condensed matter physics. The Fokker-Planck equation reads:

\begin{equation}
    \frac{\partial p(\bm{x}, t)}{\partial t}
    =
    -\sum_{i=1}^{N} \frac{\partial}{\partial x_i} \big[ A_i(\bm{x}, t) p(\bm{x}, t) \big]
    +
    \sum_{i,j=1}^{N} \frac{\partial^2}{\partial x_i \partial x_j} \big[ B_{ij}(\bm{x}, t) p(\bm{x}, t) \big].
\end{equation}

Here, $\bm{x}$ is a vector in some state space where each element captures the activity of a given degree of freedom (e.g., neuron, voxel, region of interest, etc.), and $p$ is the probability density. $\bm{A}$ and $\bm{B}$ are generic time-dependent components that correspond to drift and diffusion, respectively.

The stochastic nature of neural activity makes the Fokker-Planck framework a suitable candidate for explaining diverse neural phenomena. A practical strategy would be to initially define a broad hypothesis space for the drift and diffusion terms of the Fokker-Planck equation. This space can then be constrained based on the inherent structure and symmetries of the system, followed by employing machine learning techniques to learn these terms from data. 

It's important to note that the Fokker-Planck equation is just one of several potential models. The key idea here is that if two neuroscientists, each focusing on different aspects of brain function, base their research on a common set of equations, their findings can be interconnected, enabling a meaningful transfer of knowledge.

\subsection{Neuroscience needs concepts, as much as it needs theories}

Devising an appropriate mathematical formalism to theorize about neuroscience data presents its own set of challenges. However, an even greater challenge lies in developing a conceptual framework that is broad enough to encompass the diverse aspects of neuroscience, yet precise enough to be practically useful. 

Overly broad statements like \say{everything is information [or computation]}, \say{everything is emergent}, or \say{everything is free energy minimization} are too vague and unspecific. Claude Shannon, the father of information theory, aptly summarized this challenge in his prophetic piece, \sayit{The Bandwagon}:\\[-2mm]

\hspace{0.020\textwidth}\begin{tabular}{|p{0.84\textwidth}}\say{{Seldom do more than a few of nature's secrets give way at one time. It will be all too easy for our somewhat artificial prosperity to collapse overnight when it is realized that the use of a few exciting words like information, entropy, redundancy, do not solve all our problems.}} --- {Claude Shannon, \textit{The Bandwagon}, 1956 \cite{shannon1956bandwagon}}\end{tabular}\\[2mm]

In conclusion, future neuroscience theories (whatever their nature may be) must be built upon the foundational knowledge we have today. These foundational elements---ranging from distinct cell types to gene expressions to connectivity and neurotransmitters---represent the essential building blocks of nervous systems, shaped by millions of years of evolutionary trial and error.
Whether the objective is to obtain a deeper understanding of brain function, advance AI technologies, or treat neurological disorders, these foundational details about the underlying biology are indispensable. Consequently, our theoretical approaches must be firmly rooted in the actual mechanisms of the system. Without such a foundation, we risk wandering aimlessly, devising unfounded theories that, in Cajal's words, would ultimately \sayit{desert} us \cite{cajal1897advice}.


\renewcommand{\baselinestretch}{1}
\small\normalsize

\printbibliography

@article{harris2021nonsense,
    author = {Kenneth D. Harris},
    title = {Nonsense correlations in neuroscience},
    year = {2021},
    publisher = {Cold Spring Harbor Laboratory},
    journal = {bioRxiv},
    doi = {10.1101/2020.11.29.402719},
}

@article{leonardi2015spurious,
  title={On spurious and real fluctuations of dynamic functional connectivity during rest},
  author={Leonardi, Nora and Van De Ville, Dimitri},
  journal={Neuroimage},
  volume={104},
  pages={430--436},
  year={2015},
  publisher={Elsevier},
  doi={10.1016/j.neuroimage.2014.09.007},
}

@article{lurie2020questions,
  title={Questions and controversies in the study of time-varying functional connectivity in resting fMRI},
  author={Lurie, Daniel J and Kessler, Daniel and Bassett, Danielle S and Betzel, Richard F and Breakspear, Michael and Kheilholz, Shella and Kucyi, Aaron and Li{\'e}geois, Rapha{\"e}l and Lindquist, Martin A and McIntosh, Anthony Randal and others},
  journal={Network neuroscience},
  volume={4},
  number={1},
  pages={30--69},
  year={2020},
  publisher={MIT Press},
  doi={10.1162/netn_a_00116},
}

@article{Cisek2010NeuralMF,
  title={Neural mechanisms for interacting with a world full of action choices.},
  author={Paul Cisek and John F. Kalaska},
  journal={Annual review of neuroscience},
  year={2010},
  volume={33},
  pages={269--98},
  doi={10.1146/annurev.neuro.051508.135409},
}

@book{cajal1897advice,
  title={Advice for a young investigator},
  author={Cajal, Santiago Ramon Y},
  year={1897},
  url={https://mitpress.mit.edu/9780262681506/advice-for-a-young-investigator/},
}

@article{shannon1956bandwagon,
  title={The Bandwagon},
  author={Shannon, Claude E},
  journal={IRE transactions on Information Theory},
  volume={2},
  number={1},
  pages={3},
  year={1956},
  doi={10.1109/TIT.1956.1056774},
}

@article{yates2023detailed,
  title={Detailed characterization of neural selectivity in free viewing primates},
  author={Yates, Jacob L and Coop, Shanna H and Sarch, Gabriel H and Wu, Ruei-Jr and Butts, Daniel A and Rucci, Michele and Mitchell, Jude F},
  journal={Nature Communications},
  volume={14},
  number={1},
  pages={3656},
  year={2023},
  publisher={Nature Publishing Group UK London},
  doi={10.1038/s41467-023-38564-9},
}

@article{norman2002two,
  title={Two visual systems and two theories of perception: An attempt to reconcile the constructivist and ecological approaches},
  author={Norman, Joel},
  journal={Behavioral and brain sciences},
  volume={25},
  number={1},
  pages={73--96},
  year={2002},
  publisher={Cambridge University Press},
  doi={10.1017/S0140525X0200002X},
}

@article{olshausen2005other,
  title={What is the other 85 percent of V1 doing},
  author={Olshausen, Bruno A and Field, David J},
  editor={L. van Hemmen, \& T. Sejnowski (Eds.)},
  publisher={Oxford University Press},
  pages={182--211},
  year={2005},
  doi={10.1093/ACPROF:OSO/9780195148220.001.0001},
  url={https://www.rctn.org/bruno/papers/V1-chapter.pdf},
}

@article{krakauer2017neuroscience,
  title={Neuroscience needs behavior: correcting a reductionist bias},
  author={Krakauer, John W and Ghazanfar, Asif A and Gomez-Marin, Alex and MacIver, Malcolm A and Poeppel, David},
  journal={Neuron},
  volume={93},
  number={3},
  pages={480--490},
  year={2017},
  publisher={Elsevier},
  doi={10.1016/j.neuron.2016.12.041},
}

@book{coleman2015introduction,
  title={Introduction to Many-Body Physics},
  author={Coleman, Piers},
  edition={1},
  publisher={Cambridge University Press},
  year={2015},
  isbn={9781139020916},
  doi={10.1017/CBO9781139020916},
}

@article{laughlin2000theory,
  title={The theory of everything},
  author={Laughlin, Robert B and Pines, David},
  journal={Proceedings of the national academy of sciences},
  volume={97},
  number={1},
  pages={28--31},
  year={2000},
  publisher={National Acad Sciences},
  doi={10.1073/pnas.97.1.28}
}

@article{yule1926we,
  title={Why do we sometimes get nonsense-correlations between Time-Series?--a study in sampling and the nature of time-series},
  author={Yule, G Udny},
  journal={Journal of the Royal Statistical Society},
  volume={89},
  number={1},
  pages={1--63},
  year={1926},
  publisher={JSTOR},
  doi={10.2307/2341482},
}

@book{vigen2015spurious,
  title={Spurious correlations},
  author={Vigen, Tyler},
  year={2015},
  publisher={Hachette UK},
  url = {http://www.tylervigen.com/spurious-correlations},
}

@inproceedings{dicarlo2021does,
    title={How does the brain combine generative models and direct discriminative computations in high-level vision?},
    author={DiCarlo, James J and Haefner, Ralf and Isik, Leyla and Konkle, Talia and Kriegeskorte, Nikolaus and Peters, Benjamin and Rust, Nicole and Stachenfeld, Kim and Tenenbaum, Joshua B and Tsao, Doris and others},
    booktitle = {GAC @ CCN},
    year      = {2021},
    url       = {https://openreview.net/forum?id=zlTiwFtLlR4}
}

@article{peters2024does,
    title={How does the primate brain combine generative and discriminative computations in vision?},
    author={Benjamin Peters and James J. DiCarlo and Todd Gureckis and Ralf Haefner and Leyla Isik and Joshua Tenenbaum and Talia Konkle and Thomas Naselaris and Kimberly Stachenfeld and Zenna Tavares and Doris Tsao and Ilker Yildirim and Nikolaus Kriegeskorte},
    year={2024},
    eprint={2401.06005},
    archivePrefix={arXiv},
    primaryClass={q-bio.NC},
}

@book{wuppuluri2018map,
  title={The map and the territory: Exploring the foundations of science, thought and reality},
  author={Wuppuluri, Shyam and Doria, Francisco Antonio},
  year={2018},
  publisher={Springer},
  doi={10.1007/978-3-319-72478-2},
}

@book{brockwell1991time,
  title={Time series: theory and methods},
  author={Brockwell, Peter J and Davis, Richard A},
  year={1991},
  publisher={Springer science \& business media},
  doi={10.1007/978-1-4419-0320-4},
}

@article{cliff2023unifying,
  title={Unifying pairwise interactions in complex dynamics},
  author={Cliff, Oliver M and Bryant, Annie G and Lizier, Joseph T and Tsuchiya, Naotsugu and Fulcher, Ben D},
  journal={Nature Computational Science},
  volume={3},
  number={10},
  pages={883--893},
  year={2023},
  publisher={Nature Publishing Group US New York},
  doi={10.1038/s43588-023-00519-x},
}

@article{hutchison2013dynamic,
  title={Dynamic functional connectivity: promise, issues, and interpretations},
  author={Hutchison, R Matthew and Womelsdorf, Thilo and Allen, Elena A and Bandettini, Peter A and Calhoun, Vince D and Corbetta, Maurizio and Della Penna, Stefania and Duyn, Jeff H and Glover, Gary H and Gonzalez-Castillo, Javier and others},
  journal={Neuroimage},
  volume={80},
  pages={360--378},
  year={2013},
  publisher={Elsevier},
  doi={10.1016/j.neuroimage.2013.05.079},
}

@article{breakspear2017dynamic,
  title={Dynamic models of large-scale brain activity},
  author={Breakspear, Michael},
  journal={Nature Neuroscience},
  volume={20},
  number={3},
  pages={340--352},
  year={2017},
  publisher={Nature Publishing Group US New York},
  doi={},
}

@article{calhoun2014the,
  title = {The Chronnectome: Time-Varying Connectivity Networks as the Next Frontier in fMRI Data Discovery},
  journal = {Neuron},
  volume = {84},
  number = {2},
  pages = {262-274},
  year = {2014},
  issn = {0896-6273},
  doi = {10.1016/j.neuron.2014.10.015},
  author = {Vince D. Calhoun and Robyn Miller and Godfrey Pearlson and Tulay Adalı},
}

@article{benisty2023rapid,
  title={Rapid fluctuations in functional connectivity of cortical networks encode spontaneous behavior},
  author={Benisty, Hadas and Barson, Daniel and Moberly, Andrew H and Lohani, Sweyta and Tang, Lan and Coifman, Ronald R and Crair, Michael C and Mishne, Gal and Cardin, Jessica A and Higley, Michael J},
  journal={Nature Neuroscience},
  pages={1--11},
  year={2023},
  publisher={Nature Publishing Group US New York},
  doi={10.1038/s41593-023-01498-y},
}

@article{higley2022spatiotemporal,
  title={Spatiotemporal dynamics in large-scale cortical networks},
  author={Higley, Michael J and Cardin, Jessica A},
  journal={Current Opinion in Neurobiology},
  volume={77},
  pages={102627},
  year={2022},
  publisher={Elsevier},
  doi={10.1016/j.conb.2022.102627},
}

@book{bohr1934atomic,
  title={Atomic Theory and the Description of Nature},
  author={Niels Bohr},
  year={1934},
  publisher={Cambridge University Press},
  url={https://www.goodreads.com/book/show/3492109-atomic-theory-and-the-description-of-nature}
}

@book{gibson1947motion,
  title={Motion picture testing and research},
  author={Gibson, James Jerome},
  number={7},
  year={1947},
  publisher={US Government Printing Office},
  url={https://apps.dtic.mil/sti/citations/AD0651783}
}

@book{zurek1990complexity,
  title={Complexity, entropy and the physics of information},
  author={John Wheeler},
  editor={Wojciech H. Zurek},
  year={1990},
  publisher={SFI Press},
  url={https://www.sfipress.org/books/complexity-entropy-and-the-physics-of-information}
}

@Inbook{fuchs2017on,
    author="Fuchs, Christopher A.",
    editor="Durham, Ian T. and Rickles, Dean",
    title="On Participatory Realism",
    bookTitle="Information and Interaction: Eddington, Wheeler, and the Limits of Knowledge",
    year="2017",
    publisher="Springer International Publishing",
    pages="113--134",
    isbn="978-3-319-43760-6",
    doi="10.1007/978-3-319-43760-6_7",
}

@book{mach1914analysis,
  title={The analysis of sensations, and the relation of the physical to the psychical},
  author={Mach, Ernst},
  year={1914},
  publisher={Open Court Publishing Company},
  url={https://archive.org/details/analysisofsensat00mach}
}

@book{knott1911life,
  title={Quote from undated letter from Maxwell to Tait},
  author={Knott, Cargill Gilston},
  year={1911},
  publisher={``Life and scientific work of Peter Guthrie Tait", Cambridge University Press}
}

@article{landauer1961irreversibility,
  title={Irreversibility and heat generation in the computing process},
  author={Landauer, Rolf},
  journal={IBM journal of research and development},
  volume={5},
  number={3},
  pages={183--191},
  year={1961},
  publisher={IBM},
  doi={10.1147/rd.53.0183}
}

@article{bennett2003notes,
  title={Notes on Landauer's principle, reversible computation, and Maxwell's Demon},
  author={Bennett, Charles H},
  journal={Studies In History and Philosophy of Science Part B: Studies In History and Philosophy of Modern Physics},
  volume={34},
  number={3},
  pages={501--510},
  year={2003},
  publisher={Elsevier},
  doi={10.1016/S1355-2198(03)00039-X}
}

@article{maruyama2009colloquium,
  title={Colloquium: The physics of Maxwell’s demon and information},
  author={Maruyama, Koji and Nori, Franco and Vedral, Vlatko},
  journal={Reviews of Modern Physics},
  volume={81},
  number={1},
  pages={1},
  year={2009},
  publisher={APS},
  doi={10.1103/RevModPhys.81.1},
}

@article{mandal2012work,
  title={Work and information processing in a solvable model of Maxwell’s demon},
  author={Mandal, Dibyendu and Jarzynski, Christopher},
  journal={Proceedings of the National Academy of Sciences},
  volume={109},
  number={29},
  pages={11641--11645},
  year={2012},
  publisher={National Acad Sciences},
  doi={10.1073/pnas.12042631},
}

@article{parrondo2015thermodynamics,
  title={Thermodynamics of information},
  author={Parrondo, Juan MR and Horowitz, Jordan M and Sagawa, Takahiro},
  journal={Nature physics},
  volume={11},
  number={2},
  pages={131--139},
  year={2015},
  publisher={Nature Publishing Group UK London},
  doi={10.1038/nphys3230}
}

@book{baker2012oxford,
  title={The Oxford handbook of the history of psychology: Global perspectives},
  author={Baker, David B},
  year={2012},
  publisher={Oxford University Press},
  doi={10.1093/OXFORDHB/9780195366556.001.0001}
}

@book{steffens2007ibn,
  title={Ibn Al-Haytham: First Scientist},
  author={Bradley Steffens},
  year={2007},
  publisher={Morgan Reynolds Pub},
  url={https://www.goodreads.com/book/show/720632.Ibn_Al_Haytham},
}

@book{adamson2016philosophy,
  title={Philosophy in the Islamic World: A history of philosophy without any gaps, Volume 3},
  author={Adamson, Peter},
  year={2016},
  publisher={Oxford University Press},
  url={https://www.goodreads.com/book/show/38254683-a-history-of-philosophy-without-any-gaps-volume-3},
}

@article{lindberg1967alhazen,
  title={Alhazen's theory of vision and its reception in the West},
  author={Lindberg, David C},
  journal={Isis},
  volume={58},
  number={3},
  pages={321--341},
  year={1967},
  doi={10.1086/350266},
}

@online{sep22locke,
  author={William Uzgalis},
  title={John Locke},
  note={In: \textit{The Stanford Encyclopedia of Philosophy}},
  url={https://plato.stanford.edu/archives/fall2022/entries/locke/},
  urldate={2022-07-07},
  year={2022},
}

@book{locke1690an,
  title={An essay concerning human understanding},
  author={John Locke},
  editor={Peter H. Nidditch},
  year={1975},
  publisher={Oxford University Press},
  doi={10.1093/actrade/9780198243861.book.1},
}

@article{warren1984helmholtz,
  title={Helmholtz and His Continuing Influence},
  author={Richard M. Warren},
  journal={Music Perception: An Interdisciplinary Journal},
  year={1984},
  pages={253--275},
doi={10.2307/40285260}
}

@article{warren1968helmholtz,
  title={Helmholtz on perception: Its physiology and development},
  author={Warren, Richard M and Warren, Roslyn P},
  year={1968},
  publisher={John Wiley and Sons}
}

@article{navarro2009optical,
  title={The optical design of the human eye: a critical review},
  author={Navarro, Rafael},
  journal={Journal of Optometry},
  volume={2},
  number={1},
  pages={3--18},
  year={2009},
  publisher={Elsevier},
  doi={10.3921/joptom.2009.3},
}

@book{ditzinger2021illusions,
  title={Illusions of Seeing: Exploring the World of Visual Perception},
  author={Ditzinger, Thomas},
  year={2021},
  publisher={Springer},
  doi={10.1007/978-3-030-63635-7},
}

@article{boring1942sensation,
  title={Sensation and perception in the history of experimental psychology.},
  author={Boring, Edwin Garrigues},
  year={1942},
  publisher={Appleton-Century},
  url={https://archive.org/details/sensationpercept0000bori},
}

@book{cropper2001great,
  title={Great physicists: The life and times of leading physicists from Galileo to Hawking},
  author={Cropper, William H},
  year={2001},
  publisher={Oxford University Press},
  url={https://www.goodreads.com/book/show/321329.Great_Physicists}
}

@book{helmholtz2009sensations,
  title={On the Sensations of Tone as a Physiological Basis for the Theory of Music},
  author={Helmholtz, Hermann LF},
  year={2009},
  publisher={Cambridge University Press},
  doi={10.1017/CBO9780511701801}
}

@article{fukushima1980neocognitron,
  title={Neocognitron: A self-organizing neural network model for a mechanism of pattern recognition unaffected by shift in position},
  author={Fukushima, Kunihiko},
  journal={Biological cybernetics},
  volume={36},
  number={4},
  pages={193--202},
  year={1980},
  publisher={Springer},
  doi={10.1007/BF00344251},
}

@article{lecun1995convolutional,
  title={Convolutional networks for images, speech, and time series},
  author={LeCun, Yann and Bengio, Yoshua and others},
  journal={The handbook of brain theory and Neural Networks},
  volume={3361},
  number={10},
  pages={1995},
  year={1995},
  publisher={Citeseer}
}

@article{lecun1998gradient,
  title={Gradient-based learning applied to document recognition},
  author={LeCun, Yann and Bottou, L{\'e}on and Bengio, Yoshua and Haffner, Patrick},
  journal={Proceedings of the IEEE},
  volume={86},
  number={11},
  pages={2278--2324},
  year={1998},
  publisher={Ieee},
  doi={10.1109/5.726791},
}

@article{bock2013cajal,
  title={Cajal, Golgi, Nansen, Sch{\"a}fer and the neuron doctrine},
  author={Bock, Ortwin},
  journal={Endeavour},
  volume={37},
  number={4},
  pages={228--234},
  year={2013},
  publisher={Elsevier},
  doi={10.1016/j.endeavour.2013.06.006}
}

@article{jones1994neuron,
  title={The neuron doctrine 1891},
  author={Jones, Edward G},
  journal={Journal of the History of the Neurosciences},
  volume={3},
  number={1},
  pages={3--20},
  year={1994},
  publisher={Taylor \& Francis},
  doi={10.1080/09647049409525584},
}

@article{Burke2006SirCS,
  title={Sir Charles Sherrington's the integrative action of the nervous system: a centenary appreciation},
  author={Robert E Burke},
  journal={Brain : a journal of neurology},
  year={2006},
  volume={130 Pt 4},
  pages={887-94},
  doi={10.1093/brain/awm022},
}

@book{sherrington1906integrative,
  title={The integrative action of the nervous system},
  author={Sherrington, Charles},
  year={1906},
  publisher={Yale University Press},
  url={https://liberationchiropractic.com/wp-content/uploads/research/1906Sherrington-IntegrativeAction.pdf}
}

@article{shepherd1997centenary,
  title={Centenary of the synapse: from Sherrington to the molecular biology of the synapse and beyond},
  author={Shepherd, Gordon M and Erulkar, Solomon D},
  journal={Trends in Neurosciences},
  volume={20},
  number={9},
  pages={385--392},
  year={1997},
  publisher={Elsevier},
  doi={10.1016/S0166-2236(97)01059-X},
}

@article{Swazey1968SherringtonsCO,
  title={Sherrington's concept of integrative action},
  author={Judith P. Swazey},
  journal={Journal of the History of Biology},
  year={1968},
  volume={1},
  pages={57--89},
  doi={10.1007/BF00149776}
}

@article{Grant2006The1A,
  title={The 1932 and 1944 Nobel Prizes in Physiology or Medicine: Rewards for Ground-Breaking Studies in Neurophysiology},
  author={Gunnar Grant},
  journal={Journal of the History of the Neurosciences},
  year={2006},
  volume={15},
  pages={341--357},
  doi={10.1080/09647040600638981},
}

@article{Schuetze1983TheDO,
  title={The discovery of the action potential},
  author={Stephen M. Schuetze},
  journal={Trends in Neurosciences},
  year={1983},
  volume={6},
  pages={164--168},
  doi={10.1016/0166-2236(83)90078-4}
}

@inproceedings{Finger1999MindsBT,
  title={Minds behind the brain : a history of the pioneers and their discoveries},
  author={Stanley Finger},
  publisher={Oxford University Press},
  year={1999},
  url={10.1093/acprof:oso/9780195181821.001.0001}
}

@article{Adrian1926TheIP,
  title={The impulses produced by sensory nerve-endings: Part II. The response of a Single End-Organ},
  author={Edgar Douglas Adrian and Yngve Zotterman},
  journal={The Journal of Physiology},
  year={1926},
  pages={151--71},
  doi={10.1113/jphysiol.1926.sp002281}
}

@article{lettvin1959frog,
  title={What the frog's eye tells the frog's brain},
  author={Lettvin, Jerome Y and Maturana, Humberto R and McCulloch, Warren S and Pitts, Walter H},
  journal={Proceedings of the IRE},
  volume={47},
  number={11},
  pages={1940--1951},
  year={1959},
  publisher={IEEE},
  doi={10.1109/JRPROC.1959.287207},
}

@article{Hubel1959ReceptiveFO,
  title={Receptive fields of single neurones in the cat's striate cortex},
  author={David H. Hubel and Torsten N. Wiesel},
  journal={The Journal of Physiology},
  year={1959},
  volume={148},
  doi={10.1113/jphysiol.1959.sp006308}
}

@article{Hubel1968ReceptiveFA,
  title={Receptive fields and functional architecture of monkey striate cortex},
  author={David H. Hubel and Torsten N. Wiesel},
  journal={The Journal of Physiology},
  year={1968},
  volume={195},
  doi={10.1113/jphysiol.1968.sp008455},
}

@article{Scoville1957LOSSOR,
  title={LOSS OF RECENT MEMORY AFTER BILATERAL HIPPOCAMPAL LESIONS},
  author={William Beecher Scoville and B. Milner},
  journal={Journal of Neurology, Neurosurgery \& Psychiatry},
  year={1957},
  volume={20},
  pages={11--21},
  doi={10.1136/jnnp.20.1.11}
}

@article{Behrens2018WhatIA,
  title={What Is a Cognitive Map? Organizing Knowledge for Flexible Behavior},
  author={Timothy Edward John Behrens and Timothy H. Muller and James C. R. Whittington and Shirley Mark and Alon Baram and Kimberly L. Stachenfeld and Zeb Kurth-Nelson},
  journal={Neuron},
  year={2018},
  volume={100},
  pages={490--509},
  doi={10.1016/j.neuron.2018.10.002},
}

@article{Strange2014FunctionalOO,
  title={Functional organization of the hippocampal longitudinal axis},
  author={Bryan Andrew Strange and Menno P. Witter and Ed S. Lein and Edvard I. Moser},
  journal={Nature Reviews Neuroscience},
  year={2014},
  volume={15},
  pages={655--669},
  doi={10.1038/nrn3785}
}

@book{OKeefe1978TheHA,
  title={The Hippocampus as a Cognitive Map},
  author={John O’Keefe and Lynn Nadel},
  year={1978},
  publisher={Oxford University Press},
  url={https://www.cmor-faculty.rice.edu/~cox/neuro/HCMComplete.pdf}
}

@article{passingham79the,
    title = {The hippocampus as a cognitive map},
    journal = {Neuroscience},
    volume = {4},
    number = {6},
    pages = {863},
    year = {1979},
    issn = {0306-4522},
    doi={10.1016/0306-4522(79)90015-0},
    author = {R.E. Passingham}
}

@article{Hafting2005MicrostructureOA,
  title={Microstructure of a spatial map in the entorhinal cortex},
  author={Torkel Hafting and Marianne Fyhn and Sturla Molden and May-Britt Moser and Edvard I. Moser},
  journal={Nature},
  year={2005},
  volume={436},
  pages={801--806},
  doi={10.1038/nature03721}
}

@article{Danjo2018SpatialRO,
  title={Spatial representations of self and other in the hippocampus},
  author={Teruko Danjo and Taro Toyoizumi and Shigeyoshi Fujisawa},
  journal={Science},
  year={2018},
  volume={359},
  pages={213--218},
  doi={10.1126/science.aao3898}
}

@article{Omer2018SocialPI,
  title={Social place-cells in the bat hippocampus},
  author={David B. Omer and Shir Maimon and Liora Las and Nachum Ulanovsky},
  journal={Science},
  year={2018},
  volume={359},
  pages={218--224},
  doi={10.1126/science.aao3474},
}

@article{Taube1990HeaddirectionCR,
  title={Head-direction cells recorded from the postsubiculum in freely moving rats. I. Description and quantitative analysis},
  author={JS Taube and R.U. Muller and JB Ranck},
  booktitle={Journal of Neuroscience},
  year={1990},
  doi={10.1523/JNEUROSCI.10-02-00420.1990}
}

@article{Hydal2019ObjectvectorCI,
  title={Object-vector coding in the medial entorhinal cortex},
  author={{\O}yvind Arne H{\o}ydal and Emilie Ranheim Skyt{\o}en and Sebastian O Andersson and May-Britt Moser and Edvard I. Moser},
  journal={Nature},
  year={2019},
  volume={568},
  pages={400--404},
  doi={10.1038/s41586-019-1077-7},
}

@article{Gauthier2018ADP,
  title={A Dedicated Population for Reward Coding in the Hippocampus},
  author={Jeffrey L. Gauthier and David W. Tank},
  journal={Neuron},
  year={2018},
  volume={99},
  pages={179--193},
  doi={10.1016/j.neuron.2018.06.008},
}

@article{Lever2009BoundaryVC,
  title={Boundary Vector Cells in the Subiculum of the Hippocampal Formation},
  author={Colin Lever and Stephen Burton and Ali Jeewajee and John O’Keefe and Neil Burgess},
  journal={The Journal of Neuroscience},
  year={2009},
  volume={29},
  pages={9771--9777},
  doi={10.1523/JNEUROSCI.1319-09.2009}
}

@article{Sarel2017VectorialRO,
  title={Vectorial representation of spatial goals in the hippocampus of bats},
  author={Ayelet Sarel and Arseny Finkelstein and Liora Las and Nachum Ulanovsky},
  journal={Science},
  year={2017},
  volume={355},
  pages={176--180},
  doi={10.1126/science.aak9589}
}

@article{Garcia2020AnatomyAF,
  title={Anatomy and Function of the Primate Entorhinal Cortex},
  author={Aaron D. Garcia and Elizabeth A. Buffalo},
  journal={Annual review of vision science},
  year={2020},
  doi={10.1146/annurev-vision-030320-041115}
}

@article{Seelig2015NeuralDF,
  title={Neural dynamics for landmark orientation and angular path integration},
  author={Johannes D. Seelig and Vivek Jayaraman},
  journal={Nature},
  year={2015},
  volume={521},
  pages={186--191},
  doi={10.1038/nature14446}
}

@article{Kandel2014APA,
  title={A Place and a Grid in the Sun},
  author={Eric R. Kandel},
  journal={Cell},
  year={2014},
  volume={159},
  pages={1239--1242},
  doi={10.1016/j.cell.2014.11.033}
}

@article{Douglas2004NeuronalCO,
  title={Neuronal circuits of the neocortex.},
  author={Rodney J. Douglas and Kevan A. C. Martin},
  journal={Annual Review of Neuroscience},
  year={2004},
  volume={27},
  pages={419--51},
  doi={10.1146/annurev.neuro.27.070203.144152}
}

@article{Mountcastle97Columnar,
    author = {Mountcastle, V B},
    title = "{The columnar organization of the neocortex.}",
    journal = {Brain},
    volume = {120},
    number = {4},
    pages = {701-722},
    year = {1997},
    month = {04},
    issn = {0006-8950},
    doi = {10.1093/brain/120.4.701},
}

@article{hegde2007reappraising,
  title={Reappraising the functional implications of the primate visual anatomical hierarchy},
  author={Hegde, Jay and Felleman, Daniel J},
  journal={The Neuroscientist},
  volume={13},
  number={5},
  pages={416--421},
  year={2007},
  publisher={Sage Publications},
  doi={10.1177/1073858407305201},
}

@article{Zeki1974FunctionalOO,
  title={Functional organization of a visual area in the posterior bank of the superior temporal sulcus of the rhesus monkey},
  author={Semir Zeki},
  journal={The Journal of Physiology},
  year={1974},
  volume={236},
  doi={10.1113/jphysiol.1974.sp010452}
}

@article{Maunsell1983FunctionalPO,
  title={Functional properties of neurons in middle temporal visual area of the macaque monkey. I. Selectivity for stimulus direction, speed, and orientation.},
  author={John H. R. Maunsell and David C. van Essen},
  journal={Journal of Neurophysiology},
  year={1983},
  volume={49 5},
  pages={1127--47},
  doi={10.1152/JN.1983.49.5.1127},
}

@article{Albright1984DirectionAO,
  title={Direction and orientation selectivity of neurons in visual area MT of the macaque.},
  author={Thomas D. Albright},
  journal={Journal of Neurophysiology},
  year={1984},
  volume={52 6},
  pages={1106--30},
  doi={10.1152/JN.1984.52.6.1106}
}

@article{Tanaka1986AnalysisOL,
  title={Analysis of local and wide-field movements in the superior temporal visual areas of the macaque monkey},
  author={K. Tanaka and Kazuo Hikosaka and H. Saito and Masao Yukie and Yoshiro Fukada and Eiichi Iwai},
  booktitle={Journal of Neuroscience},
  year={1986},
  doi={10.1523/JNEUROSCI.06-01-00134.1986}
}

@article{Wild2021PrimateEC,
  title={Primate Extrastriate Cortical Area MST: A Gateway between Sensation and Cognition.},
  author={Benedict Wild and Stefan Treue},
  journal={Journal of Neurophysiology},
  year={2021},
  doi={10.1152/jn.00384.2020}
}

@article{Born2005StructureAF,
  title={Structure and function of visual area MT.},
  author={Richard T. Born and David C. Bradley},
  journal={Annual Review of Neuroscience},
  year={2005},
  volume={28},
  pages={157--89},
  doi={10.1146/annurev.neuro.26.041002.131052}
}

@article{Duffy1991SensitivityOM,
  title={Sensitivity of MST neurons to optic flow stimuli. I. A continuum of response selectivity to large-field stimuli.},
  author={Charles Duffy and Robert H. Wurtz},
  journal={Journal of Neurophysiology},
  year={1991},
  volume={65 6},
  pages={1329--45},
  doi={10.1152/JN.1991.65.6.1329},
}

@article{Graziano1994TuningOM,
  title={Tuning of MST neurons to spiral motions},
  author={Mart{\'i}n Graziano and R. A. Andersen and R J Snowden},
  booktitle={Journal of Neuroscience},
  year={1994},
  doi={10.1523/JNEUROSCI.14-01-00054.1994}
}

@article{Page2003HeadingRI,
  title={Heading representation in MST: sensory interactions and population encoding.},
  author={William Page and Charles Duffy},
  journal={Journal of Neurophysiology},
  year={2003},
  volume={89 4},
  pages={1994--2013},
  doi={10.1152/JN.00493.2002}
}

@ARTICLE{alonso2009receptive,
AUTHOR = {Alonso, J.  and Chen, Y. },
TITLE   = {{R}eceptive field},
YEAR    = {2009},
JOURNAL = {Scholarpedia},
VOLUME  = {4},
NUMBER  = {1},
PAGES   = {5393},
DOI     = {10.4249/scholarpedia.5393},
NOTE    = {revision \#136681}
}

@book{Hubel1988EyeBA,
  title={Eye, brain, and vision},
  author={David H. Hubel},
  year={1988},
  note={Scientific American Library Series},
  url={https://www.goodreads.com/book/show/2629579}
}

@book{Squire2008TheHO,
  title={The History of Neuroscience in Autobiography Volume 6},
  author={Larry R. Squire},
  publisher={Oxford University Press},
  year={2009},
  doi={10.1093/acprof:oso/9780195380101.001.0001}
}

@article{gross1972visual,
  title={Visual properties of neurons in inferotemporal cortex of the macaque},
  author={Gross, Charles G and Rocha-Miranda, CE de and Bender, DB},
  journal={Journal of Neurophysiology},
  volume={35},
  number={1},
  pages={96--111},
  year={1972},
  doi={10.1152/jn.1972.35.1.96}
}

@article{bruce1981visual,
  title={Visual properties of neurons in a polysensory area in superior temporal sulcus of the macaque},
  author={Bruce, Charles and Desimone, Robert and Gross, Charles G},
  journal={Journal of Neurophysiology},
  volume={46},
  number={2},
  pages={369--384},
  year={1981},
  doi={10.1152/jn.1981.46.2.369},
}

@article{desimone1984stimulus,
  title={Stimulus-selective properties of inferior temporal neurons in the macaque},
  author={Desimone, Robert and Albright, Thomas D and Gross, Charles G and Bruce, Charles},
  journal={Journal of Neuroscience},
  volume={4},
  number={8},
  pages={2051--2062},
  year={1984},
  publisher={Soc Neuroscience},
  doi={10.1523/JNEUROSCI.04-08-02051.1984}
}

@article{kanwisher1997fusiform,
  title={The fusiform face area: a module in human extrastriate cortex specialized for face perception},
  author={Kanwisher, Nancy and McDermott, Josh and Chun, Marvin M},
  journal={Journal of Neuroscience},
  volume={17},
  number={11},
  pages={4302--4311},
  year={1997},
  publisher={Soc Neuroscience},
  doi={10.1523/JNEUROSCI.17-11-04302.1997},
}

@article{tsao2003faces,
  title={Faces and objects in macaque Cerebral Cortex},
  author={Tsao, Doris Y and Freiwald, Winrich A and Knutsen, Tamara A and Mandeville, Joseph B and Tootell, Roger BH},
  journal={Nature Neuroscience},
  volume={6},
  number={9},
  pages={989--995},
  year={2003},
  publisher={Nature Publishing Group US New York},
  doi={10.1038/nn1111},
}

@article{tsao2006cortical,
  title={A cortical region consisting entirely of face-selective cells},
  author={Tsao, Doris Y and Freiwald, Winrich A and Tootell, Roger BH and Livingstone, Margaret S},
  journal={Science},
  volume={311},
  number={5761},
  pages={670--674},
  year={2006},
  publisher={American Association for the Advancement of Science},
  doi={10.1126/science.1119983},
}

@article{wu2006complete,
  title={Complete functional characterization of sensory neurons by system identification},
  author={Wu, Michael C-K and David, Stephen V and Gallant, Jack L},
  journal={Annu. Rev. Neurosci.},
  volume={29},
  pages={477--505},
  year={2006},
  publisher={Annual Reviews},
  doi={10.1146/annurev.neuro.29.051605.113024},
}

@article{berger1929elektroenkephalogramm,
  title={{\"U}ber das elektroenkephalogramm des menschen},
  author={Berger, Hans},
  journal={Archiv f{\"u}r psychiatrie und nervenkrankheiten},
  volume={87},
  number={1},
  pages={527--570},
  year={1929},
  doi={10.1007/BF01797193},
}

@article{buzsaki2004neuronal,
  title={Neuronal oscillations in cortical networks},
  author={Buzsaki, Gyorgy and Draguhn, Andreas},
  journal={Science},
  volume={304},
  number={5679},
  pages={1926--1929},
  year={2004},
  publisher={American Association for the Advancement of Science},
  doi={10.1126/science.10997}
}

@article{raichle2006dark,
  title={The brain's dark energy},
  author={Raichle, Marcus E},
  journal={Science},
  volume={314},
  number={5803},
  pages={1249--1250},
  year={2006},
  publisher={American Association for the Advancement of Science},
  doi={10.1126/science.113440},
}

@article{raichle2002appraising,
  title={Appraising the brain's energy budget},
  author={Raichle, Marcus E and Gusnard, Debra A},
  journal={Proceedings of the National Academy of Sciences},
  volume={99},
  number={16},
  pages={10237--10239},
  year={2002},
  publisher={National Acad Sciences},
  doi={10.1073/pnas.172399499},
}

@book{llinas2002vortex,
  title={I of the vortex: From neurons to self},
  author={Llin{\'a}s, Rodolfo R},
  year={2002},
  publisher={MIT press},
  doi={10.7551/mitpress/3626.001.0001}
}

@article{bullock2005neuron,
  title={The Neuron Doctrine, Redux},
  author={Bullock, Theodore H and Bennett, Michael VL and Johnston, Daniel and Josephson, Robert and Marder, Eve and Fields, R Douglas},
  journal={Science},
  volume={310},
  number={5749},
  pages={791--793},
  year={2005},
  publisher={American Association for the Advancement of Science},
  doi={10.1126/science.1114394}
}

@article{Yuste2015FromTN,
  title={From the neuron doctrine to Neural Networks},
  author={Rafael Yuste},
  journal={Nature Reviews Neuroscience},
  year={2015},
  volume={16},
  pages={487--497},
  doi={10.1038/nrn3962}
}

@article{marder2012neuromodulation,
  title={Neuromodulation of neuronal circuits: back to the future},
  author={Marder, Eve},
  journal={Neuron},
  volume={76},
  number={1},
  pages={1--11},
  year={2012},
  publisher={Elsevier},
  doi={10.1016/j.neuron.2012.09.010},
}

@article{lee2012neuromodulation,
  title={Neuromodulation of brain states},
  author={Lee, Seung-Hee and Dan, Yang},
  journal={Neuron},
  volume={76},
  number={1},
  pages={209--222},
  year={2012},
  publisher={Elsevier},
  doi={10.1016/j.neuron.2012.09.012},
}

@article{mccormick2020neuromodulation,
  title={Neuromodulation of brain state and behavior},
  author={McCormick, David A and Nestvogel, Dennis B and He, Biyu J},
  journal={Annual Review of Neuroscience},
  volume={43},
  pages={391--415},
  year={2020},
  publisher={Annual Reviews},
  doi={10.1146/annurev-neuro-100219-105424},
}

@article{shine2023neuromodulatory,
  title={Neuromodulatory control of complex adaptive dynamics in the brain},
  author={Shine, James M},
  journal={Interface Focus},
  volume={13},
  number={3},
  pages={20220079},
  year={2023},
  publisher={The Royal Society},
  doi={},
}

@article{shine2019neuromodulatory,
  title={Neuromodulatory influences on integration and segregation in the brain},
  author={Shine, James M},
  journal={Trends in Cognitive Sciences},
  volume={23},
  number={7},
  pages={572--583},
  year={2019},
  publisher={Elsevier},
  doi={10.1016/j.tics.2019.04.002},
}

@article{london2005dendritic,
  title={Dendritic computation},
  author={London, Michael and H{\"a}usser, Michael},
  journal={Annu. Rev. Neurosci.},
  volume={28},
  pages={503--532},
  year={2005},
  publisher={Annual Reviews},
  doi={10.1146/annurev.neuro.28.061604.135703},
}

@article{mu2019glia,
  title={Glia accumulate evidence that actions are futile and suppress unsuccessful behavior},
  author={Mu, Yu and Bennett, Davis V and Rubinov, Mikail and Narayan, Sujatha and Yang, Chao-Tsung and Tanimoto, Masashi and Mensh, Brett D and Looger, Loren L and Ahrens, Misha B},
  journal={Cell},
  volume={178},
  number={1},
  pages={27--43},
  year={2019},
  publisher={Elsevier},
  doi={10.1016/j.cell.2019.05.050},
}

@article{saxena2019towards,
  title={Towards the neural population doctrine},
  author={Saxena, Shreya and Cunningham, John P},
  journal={Current Opinion in Neurobiology},
  volume={55},
  pages={103--111},
  year={2019},
  publisher={Elsevier},
  doi={10.1016/j.conb.2019.02.002},
}

@article{ebitz2021population,
  title={The population doctrine in cognitive neuroscience},
  author={Ebitz, R Becket and Hayden, Benjamin Y},
  journal={Neuron},
  volume={109},
  number={19},
  pages={3055--3068},
  year={2021},
  publisher={Elsevier},
  doi={10.1016/j.neuron.2021.07.011},
}

@article{cull2007mathematical,
  title={The mathematical biophysics of Nicolas Rashevsky},
  author={Cull, Paul},
  journal={BioSystems},
  volume={88},
  number={3},
  pages={178--184},
  year={2007},
  publisher={Elsevier},
  doi={10.1016/j.biosystems.2006.11.003},
}

@book{rosen1991life,
  title={Life itself: a comprehensive inquiry into the nature, origin, and fabrication of life},
  author={Rosen, Robert},
  year={1991},
  publisher={Columbia University Press},
  url={https://cup.columbia.edu/book/life-itself/9780231075640},
}

@book{Poincar1906ScienceAH,
  title={Science and Hypothesis},
  author={Henri Poincar{\'e}},
  year={1902},
  url={https://www.gutenberg.org/files/37157/37157-pdf.pdf}
}

@article{watts1998collective,
  title={Collective dynamics of ‘small-world’networks},
  author={Watts, Duncan J and Strogatz, Steven H},
  journal={nature},
  volume={393},
  number={6684},
  pages={440--442},
  year={1998},
  publisher={Nature Publishing Group},
  doi={10.1038/30918},
}

@article{sejnowski2014putting,
  title={Putting big data to good use in neuroscience},
  author={Sejnowski, Terrence J and Churchland, Patricia S and Movshon, J Anthony},
  journal={Nature Neuroscience},
  volume={17},
  number={11},
  pages={1440--1441},
  year={2014},
  publisher={Nature Publishing Group US New York},
  doi={10.1038/nn.3839},
}

@article{newman2006modularity,
  title={Modularity and community structure in networks},
  author={Newman, Mark EJ},
  journal={Proceedings of the National Academy of Sciences},
  volume={103},
  number={23},
  pages={8577--8582},
  year={2006},
  publisher={National Acad Sciences},
  doi={10.1073/pnas.0601602103},
}

@article{fosdick2018configuring,
  title={Configuring random graph models with fixed degree sequences},
  author={Fosdick, Bailey K and Larremore, Daniel B and Nishimura, Joel and Ugander, Johan},
  journal={Siam Review},
  volume={60},
  number={2},
  pages={315--355},
  year={2018},
  publisher={SIAM},
  doi={10.1137/16M108717},
}

@article{bassett2018nature,
  title={On the nature and use of models in network neuroscience},
  author={Bassett, Danielle S and Zurn, Perry and Gold, Joshua I},
  journal={Nature Reviews Neuroscience},
  volume={19},
  number={9},
  pages={566--578},
  year={2018},
  publisher={Nature Publishing Group UK London},
  doi={10.1038/s41583-018-0038-8},
}

@article{vavsa2022null,
  title={Null models in network neuroscience},
  author={V{\'a}{\v{s}}a, Franti{\v{s}}ek and Mi{\v{s}}i{\'c}, Bratislav},
  journal={Nature Reviews Neuroscience},
  volume={23},
  number={8},
  pages={493--504},
  year={2022},
  publisher={Nature Publishing Group UK London},
  doi={10.1038/s41583-022-00601-9},
}

@article{blondel2008fast,
  title={Fast unfolding of communities in large networks},
  author={Blondel, Vincent D and Guillaume, Jean-Loup and Lambiotte, Renaud and Lefebvre, Etienne},
  journal={Journal of statistical mechanics: theory and experiment},
  volume={2008},
  number={10},
  pages={P10008},
  year={2008},
  publisher={IOP Publishing},
  doi={10.1088/1742-5468/2008/10/P10008},
}

@article{traag2019louvain,
  title={From Louvain to Leiden: guaranteeing well-connected communities},
  author={Traag, Vincent A and Waltman, Ludo and Van Eck, Nees Jan},
  journal={Scientific reports},
  volume={9},
  number={1},
  pages={5233},
  year={2019},
  publisher={Nature Publishing Group UK London},
  doi={10.1038/s41598-019-41695-z},
}

@article{rosvall2008maps,
  title={Maps of random walks on complex networks reveal community structure},
  author={Rosvall, Martin and Bergstrom, Carl T},
  journal={Proceedings of the National Academy of Sciences},
  volume={105},
  number={4},
  pages={1118--1123},
  year={2008},
  publisher={National Acad Sciences},
  doi={10.1073/pnas.070685110},
}

@article{good2010performance,
  title={Performance of modularity maximization in practical contexts},
  author={Good, Benjamin H and De Montjoye, Yves-Alexandre and Clauset, Aaron},
  journal={Physical review E},
  volume={81},
  number={4},
  pages={046106},
  year={2010},
  publisher={APS},
  doi={10.1103/PhysRevE.81.046106},
}

@book{peixoto2023descriptive,
  title={Descriptive vs. inferential community detection in networks: Pitfalls, myths and half-truths},
  author={Peixoto, Tiago P},
  year={2023},
  publisher={Cambridge University Press},
  doi={10.1017/9781009118897}
}

@article{holland1983stochastic,
  title={Stochastic blockmodels: First steps},
  author={Holland, Paul W and Laskey, Kathryn Blackmond and Leinhardt, Samuel},
  journal={Social networks},
  volume={5},
  number={2},
  pages={109--137},
  year={1983},
  publisher={Elsevier},
  doi={10.1016/0378-8733(83)90021-7},
}

@article{fortunato202220,
  title={20 years of network community detection},
  author={Fortunato, Santo and Newman, Mark EJ},
  journal={Nature Physics},
  volume={18},
  number={8},
  pages={848--850},
  year={2022},
  publisher={Nature Publishing Group UK London},
  doi={10.1038/s41567-022-01716-7},
}

@article{barabasi2023neuroscience,
  title={Neuroscience needs network science},
  author={Barab{\'a}si, D{\'a}niel L and Bianconi, Ginestra and Bullmore, Ed and Burgess, Mark and Chung, SueYeon and Eliassi-Rad, Tina and George, Dileep and Kov{\'a}cs, Istv{\'a}n A and Makse, Hern{\'a}n and Nichols, Thomas E and others},
  journal={Journal of Neuroscience},
  volume={43},
  number={34},
  pages={5989--5995},
  year={2023},
  publisher={Soc Neuroscience},
  doi={10.1523/JNEUROSCI.1014-23.2023},
}

@article{weimann1991neurons,
  title={Neurons that form multiple pattern generators: identification and multiple activity patterns of gastric/pyloric neurons in the crab stomatogastric system},
  author={Weimann, James M and Meyrand, Pierre and Marder, E},
  journal={Journal of Neurophysiology},
  volume={65},
  number={1},
  pages={111--122},
  year={1991},
  doi={10.1152/jn.1991.65.1.111},
}

@article{bargmann2013connectome,
  title={From the connectome to brain function},
  author={Bargmann, Cornelia I and Marder, Eve},
  journal={Nature methods},
  volume={10},
  number={6},
  pages={483--490},
  year={2013},
  publisher={Nature Publishing Group US New York},
  doi={10.1038/nmeth.2451}
}

@article{marder2007understanding,
  title={Understanding circuit dynamics using the stomatogastric nervous system of lobsters and crabs},
  author={Marder, Eve and Bucher, Dirk},
  journal={Annu. Rev. Physiol.},
  volume={69},
  pages={291--316},
  year={2007},
  publisher={Annual Reviews},
  doi={10.1146/annurev.physiol.69.031905.161516},
}

@article{shine2018principles,
  title={Principles of dynamic network reconfiguration across diverse brain states},
  author={Shine, James M and Poldrack, Russell A},
  journal={NeuroImage},
  volume={180},
  pages={396--405},
  year={2018},
  publisher={Elsevier},
  doi={10.1016/j.neuroimage.2017.08.010},
}

@article{ogawa1990oxygenation,
  title={Oxygenation-sensitive contrast in magnetic resonance image of rodent brain at high magnetic fields},
  author={Ogawa, Seiji and Lee, Tso-Ming and Nayak, Asha S and Glynn, Paul},
  journal={Magnetic resonance in medicine},
  volume={14},
  number={1},
  pages={68--78},
  year={1990},
  publisher={Wiley Online Library},
  doi={10.1002/mrm.1910140108},
}

@article{kwong1992dynamic,
  title={Dynamic magnetic resonance imaging of human brain activity during primary sensory stimulation.},
  author={Kwong, Kenneth K and Belliveau, John W and Chesler, David A and Goldberg, Inna E and Weisskoff, Robert M and Poncelet, Brigitte P and Kennedy, David N and Hoppel, Bernice E and Cohen, Mark S and Turner, Robert},
  journal={Proceedings of the National Academy of Sciences},
  volume={89},
  number={12},
  pages={5675--5679},
  year={1992},
  publisher={National Acad Sciences},
  doi={10.1073/pnas.89.12.5675},
}

@article{biswal1995functional,
  title={Functional connectivity in the motor cortex of resting human brain using echo-planar MRI},
  author={Biswal, Bharat and Zerrin Yetkin, F and Haughton, Victor M and Hyde, James S},
  journal={Magnetic resonance in medicine},
  volume={34},
  number={4},
  pages={537--541},
  year={1995},
  publisher={Wiley Online Library},
  doi={10.1002/mrm.1910340409},
}

@article{raichle2006brain,
  title={Brain work and brain imaging},
  author={Raichle, Marcus E and Mintun, Mark A},
  journal={Annu. Rev. Neurosci.},
  volume={29},
  pages={449--476},
  year={2006},
  publisher={Annual Reviews},
  doi={10.1146/annurev.neuro.29.051605.112819},
}

@article{fox2007spontaneous,
  title={Spontaneous fluctuations in brain activity observed with functional magnetic resonance imaging},
  author={Fox, Michael D and Raichle, Marcus E},
  journal={Nature reviews neuroscience},
  volume={8},
  number={9},
  pages={700--711},
  year={2007},
  publisher={Nature Publishing Group UK London},
  doi={10.1038/nrn2201},
}

@article{liu2022decoding,
  title={Decoding cognition from spontaneous neural activity},
  author={Liu, Yunzhe and Nour, Matthew M and Schuck, Nicolas W and Behrens, Timothy EJ and Dolan, Raymond J},
  journal={Nature Reviews Neuroscience},
  volume={23},
  number={4},
  pages={204--214},
  year={2022},
  publisher={Nature Publishing Group UK London},
  doi={10.1038/s41583-022-00570-z},
}

@article{uddin2020bring,
  title={Bring the noise: reconceptualizing spontaneous neural activity},
  author={Uddin, Lucina Q},
  journal={Trends in Cognitive Sciences},
  volume={24},
  number={9},
  pages={734--746},
  year={2020},
  publisher={Elsevier},
  doi={10.1016/j.tics.2020.06.003},
}

@article{beckmann2005investigations,
  title={Investigations into resting-state connectivity using independent component analysis},
  author={Beckmann, Christian F and DeLuca, Marilena and Devlin, Joseph T and Smith, Stephen M},
  journal={Philosophical Transactions of the Royal Society B: Biological Sciences},
  volume={360},
  number={1457},
  pages={1001--1013},
  year={2005},
  publisher={The Royal Society London},
  doi={10.1098/rstb.2005.1634},
}

@article{vincent2007intrinsic,
  title={Intrinsic functional architecture in the anaesthetized monkey brain},
  author={Vincent, Justin L and Patel, Gaurav H and Fox, Michael D and Snyder, Abraham Z and Baker, Justin T and Van Essen, David C and Zempel, John M and Snyder, Lawrence H and Corbetta, Maurizio and Raichle, Marcus E},
  journal={Nature},
  volume={447},
  number={7140},
  pages={83--86},
  year={2007},
  publisher={Nature Publishing Group UK London},
  doi={10.1038/nature05758},
}

@article{mittner2016neural,
  title={A neural model of mind wandering},
  author={Mittner, Matthias and Hawkins, Guy E and Boekel, Wouter and Forstmann, Birte U},
  journal={Trends in Cognitive Sciences},
  volume={20},
  number={8},
  pages={570--578},
  year={2016},
  publisher={Elsevier},
  doi={10.1016/j.tics.2016.06.004},
}

@article{greicius2007resting,
  title={Resting-state functional connectivity in major depression: abnormally increased contributions from subgenual cingulate cortex and thalamus},
  author={Greicius, Michael D and Flores, Benjamin H and Menon, Vinod and Glover, Gary H and Solvason, Hugh B and Kenna, Heather and Reiss, Allan L and Schatzberg, Alan F},
  journal={Biological psychiatry},
  volume={62},
  number={5},
  pages={429--437},
  year={2007},
  publisher={Elsevier},
  doi={10.1016/j.biopsych.2006.09.020},
}

@article{huang2015there,
  title={Is there a nonadditive interaction between spontaneous and evoked activity? Phase-dependence and its relation to the temporal structure of scale-free brain activity},
  author={Huang, Zirui and Zhang, Jianfeng and Longtin, Andr{\'e} and Dumont, Gr{\'e}gory and Duncan, Niall W and Pokorny, Johanna and Qin, Pengmin and Dai, Rui and Ferri, Francesca and Weng, Xuchu and others},
  journal={Cerebral Cortex},
  pages={bhv288},
  year={2015},
  publisher={Oxford University Press},
  doi={10.1093/cercor/bhv288},
}

@article{northoff2016self,
  title={Is the self a higher-order or fundamental function of the brain? The ``basis model of self-specificity" and its encoding by the brain's spontaneous activity},
  author={Northoff, Georg},
  journal={Cognitive Neuroscience},
  volume={7},
  number={1-4},
  pages={203--222},
  year={2016},
  publisher={Taylor \& Francis},
  doi={10.1080/17588928.2015.1111868},
}

@inproceedings{hoyer02interpreting,
 author = {Hoyer, Patrik and Hyv\"{a}rinen, Aapo},
 booktitle = {Advances in Neural Information Processing Systems},
 publisher = {MIT Press},
 title = {Interpreting Neural Response Variability as Monte Carlo Sampling of the Posterior},
 url = {https://proceedings.neurips.cc/paper_files/paper/2002/hash/a486cd07e4ac3d270571622f4f316ec5-Abstract.html},
 volume = {15},
 year = {2002}
}

@article{fiser2010statistically,
  title={Statistically optimal perception and learning: from behavior to neural representations},
  author={Fiser, J{\'o}zsef and Berkes, Pietro and Orb{\'a}n, Gerg{\H{o}} and Lengyel, M{\'a}t{\'e}},
  journal={Trends in cognitive sciences},
  volume={14},
  number={3},
  pages={119--130},
  year={2010},
  publisher={Elsevier},
  doi={10.1016/j.tics.2010.01.003},
}

@article{orban2016neural,
  title={Neural variability and sampling-based probabilistic representations in the visual cortex},
  author={Orb{\'a}n, Gerg{\H{o}} and Berkes, Pietro and Fiser, J{\'o}zsef and Lengyel, M{\'a}t{\'e}},
  journal={Neuron},
  volume={92},
  number={2},
  pages={530--543},
  year={2016},
  publisher={Elsevier},
  doi={10.1016/j.neuron.2016.09.038},
}

@article{pitkow2016probability,
  title={Probability by Time},
  author={Pitkow, Xaq},
  journal={Neuron},
  volume={92},
  number={2},
  pages={275--277},
  year={2016},
  publisher={Elsevier},
  doi={10.1016/j.neuron.2016.10.007},
}

@article{echeveste2020cortical,
  title={Cortical-like dynamics in recurrent circuits optimized for sampling-based probabilistic inference},
  author={Echeveste, Rodrigo and Aitchison, Laurence and Hennequin, Guillaume and Lengyel, M{\'a}t{\'e}},
  journal={Nature neuroscience},
  volume={23},
  number={9},
  pages={1138--1149},
  year={2020},
  publisher={Nature Publishing Group US New York},
  doi={10.1038/s41593-020-0671-1},
}

@inproceedings{shrinivasan2023taking,
    title={Taking the neural sampling code very seriously: A data-driven approach for evaluating generative models of the visual system},
    author={Suhas Shrinivasan and Konstantin-Klemens Lurz and Kelli Restivo and George Denfield and Andreas S. Tolias and Edgar Y. Walker and Fabian H. Sinz},
    booktitle={Thirty-seventh Conference on Neural Information Processing Systems},
    year={2023},
    url={https://openreview.net/forum?id=Pl416tPkNv}
}

@article{lange2023bayesian,
  title={Bayesian encoding and decoding as distinct perspectives on neural coding},
  author={Lange, Richard D and Shivkumar, Sabyasachi and Chattoraj, Ankani and Haefner, Ralf M},
  journal={Nature Neuroscience},
  pages={1--10},
  year={2023},
  publisher={Nature Publishing Group US New York},
  doi={10.1038/s41593-023-01458-6},
}

@article{bolt2022parsimonious,
  title={A parsimonious description of global functional brain organization in three spatiotemporal patterns},
  author={Bolt, Taylor and Nomi, Jason S and Bzdok, Danilo and Salas, Jorge A and Chang, Catie and Thomas Yeo, BT and Uddin, Lucina Q and Keilholz, Shella D},
  journal={Nature Neuroscience},
  volume={25},
  number={8},
  pages={1093--1103},
  year={2022},
  publisher={Nature Publishing Group US New York},
  doi={10.1038/s41593-022-01118-1},
}

@article{gonzalez2022traveling,
  title={Traveling and standing waves in the brain},
  author={Gonzalez-Castillo, Javier},
  journal={Nature Neuroscience},
  volume={25},
  number={8},
  pages={980--981},
  year={2022},
  publisher={Nature Publishing Group US New York},
  doi={10.1038/s41593-022-01119-0},
}

@article{shadlen1998variable,
  title={The variable discharge of cortical neurons: implications for connectivity, computation, and information coding},
  author={Shadlen, Michael N and Newsome, William T},
  journal={Journal of Neuroscience},
  volume={18},
  number={10},
  pages={3870--3896},
  year={1998},
  publisher={Soc Neuroscience},
  doi={10.1523/JNEUROSCI.18-10-03870.1998},
}

@article{tolhurst1983statistical,
  title={The statistical reliability of signals in single neurons in cat and monkey visual cortex},
  author={Tolhurst, David J and Movshon, J Anthony and Dean, Andrew F},
  journal={Vision Research},
  volume={23},
  number={8},
  pages={775--785},
  year={1983},
  publisher={Elsevier},
  doi={10.1016/0042-6989(83)90200-6},
}

@article{tomko1974neuronal,
  title={Neuronal variability: non-stationary responses to identical visual stimuli},
  author={Tomko, George J and Crapper, Donald R},
  journal={Brain research},
  volume={79},
  number={3},
  pages={405--418},
  year={1974},
  publisher={Elsevier},
  doi={10.1016/0006-8993(74)90438-7},
}

@article{arieli1996dynamics,
  title={Dynamics of ongoing activity: explanation of the large variability in evoked cortical responses},
  author={Arieli, Amos and Sterkin, Alexander and Grinvald, Amiram and Aertsen, AD},
  journal={Science},
  volume={273},
  number={5283},
  pages={1868--1871},
  year={1996},
  publisher={American Association for the Advancement of Science},
  doi={10.1126/science.273.5283.1868},
}

@article{faisal2008noise,
  title={Noise in the nervous system},
  author={Faisal, A Aldo and Selen, Luc PJ and Wolpert, Daniel M},
  journal={Nature reviews neuroscience},
  volume={9},
  number={4},
  pages={292--303},
  year={2008},
  publisher={Nature Publishing Group UK London},
  doi={10.1038/nrn2258},
}

@article{werner1963variability,
  title={The variability of central neural activity in a sensory system, and its implications for the central reflection of sensory events},
  author={Werner, Gerhard and Mountcastle, Vernon B},
  journal={Journal of Neurophysiology},
  volume={26},
  number={6},
  pages={958--977},
  year={1963},
  doi={10.1152/jn.1963.26.6.958},
}

@article{he2013spontaneous,
  title={Spontaneous and task-evoked brain activity negatively interact},
  author={He, Biyu J},
  journal={Journal of Neuroscience},
  volume={33},
  number={11},
  pages={4672--4682},
  year={2013},
  publisher={Soc Neuroscience},
  doi={10.1523/JNEUROSCI.2922-12.2013},
}

@article{ponce2015task,
  title={Task-driven activity reduces the cortical activity space of the brain: experiment and whole-brain modeling},
  author={Ponce-Alvarez, Adri{\'a}n and He, Biyu J and Hagmann, Patric and Deco, Gustavo},
  journal={PLoS Computational Biology},
  volume={11},
  number={8},
  pages={e1004445},
  year={2015},
  publisher={Public Library of Science San Francisco, CA USA},
  doi={10.1371/journal.pcbi.1004445},
}

@article{zhang2022brain,
  title={Brain-wide ongoing activity is responsible for significant cross-trial BOLD variability},
  author={Zhang, Qingqing and Cramer, Samuel R and Ma, Zilu and Turner, Kevin L and Gheres, Kyle W and Liu, Yikang and Drew, Patrick J and Zhang, Nanyin},
  journal={Cerebral Cortex},
  volume={32},
  number={23},
  pages={5311--5329},
  year={2022},
  publisher={Oxford University Press},
  doi={10.1093/cercor/bhac016},
}

@article{churchland2010stimulus,
  title={Stimulus onset quenches neural variability: a widespread cortical phenomenon},
  author={Churchland, Mark M and Yu, Byron M and Cunningham, John P and Sugrue, Leo P and Cohen, Marlene R and Corrado, Greg S and Newsome, William T and Clark, Andrew M and Hosseini, Paymon and Scott, Benjamin B and others},
  journal={Nature Neuroscience},
  volume={13},
  number={3},
  pages={369--378},
  year={2010},
  publisher={Nature Publishing Group US New York},
  doi={10.1038/nn.2501},
}

@article{fairhall2019whither,
  title={Whither variability?},
  author={Fairhall, Adrienne L},
  journal={Nature Neuroscience},
  volume={22},
  number={3},
  pages={329--330},
  year={2019},
  publisher={Nature Publishing Group US New York},
  doi={10.1038/s41593-019-0344-0},
}

@article{nogueira2018neuronal,
  title={Neuronal variability as a proxy for network state},
  author={Nogueira, Ramon and Lawrie, Sof{\'\i}a and Moreno-Bote, Rub{\'e}n},
  journal={Trends in Neurosciences},
  volume={41},
  number={4},
  pages={170--173},
  year={2018},
  publisher={Elsevier},
  doi={10.1016/j.tins.2018.02.003},
}

@article{kenet2003spontaneously,
  title={Spontaneously emerging cortical representations of visual attributes},
  author={Kenet, Tal and Bibitchkov, Dmitri and Tsodyks, Misha and Grinvald, Amiram and Arieli, Amos},
  journal={Nature},
  volume={425},
  number={6961},
  pages={954--956},
  year={2003},
  publisher={Nature Publishing Group UK London},
  doi={10.1038/nature02078},
}

@article{ringach2003states,
  title={States of mind},
  author={Ringach, Dario L},
  journal={Nature},
  volume={425},
  number={6961},
  pages={912--913},
  year={2003},
  publisher={Nature Publishing Group UK London},
  doi={10.1038/425912a},
}

@article{berkes2011spontaneous,
  title={Spontaneous cortical activity reveals hallmarks of an optimal internal model of the environment},
  author={Berkes, Pietro and Orb{\'a}n, Gerg{\H{o}} and Lengyel, M{\'a}t{\'e} and Fiser, J{\'o}zsef},
  journal={Science},
  volume={331},
  number={6013},
  pages={83--87},
  year={2011},
  publisher={American Association for the Advancement of Science},
  doi={10.1126/science.1195870},
}

@article{carrillo2015endogenous,
  title={Endogenous sequential cortical activity evoked by visual stimuli},
  author={Carrillo-Reid, Luis and Miller, Jae-eun Kang and Hamm, Jordan P and Jackson, Jesse and Yuste, Rafael},
  journal={Journal of Neuroscience},
  volume={35},
  number={23},
  pages={8813--8828},
  year={2015},
  publisher={Soc Neuroscience},
  doi={10.1523/JNEUROSCI.5214-14.2015},
}

@article{fox2010clinical,
  title={Clinical applications of resting state functional connectivity},
  author={Fox, Michael D and Greicius, Michael},
  journal={Frontiers in systems neuroscience},
  volume={4},
  pages={1443},
  year={2010},
  publisher={Frontiers},
  doi={10.3389/fnsys.2010.00019},
}

@article{broyd2009default,
  title={Default-mode brain dysfunction in mental disorders: a systematic review},
  author={Broyd, Samantha J and Demanuele, Charmaine and Debener, Stefan and Helps, Suzannah K and James, Christopher J and Sonuga-Barke, Edmund JS},
  journal={Neuroscience \& biobehavioral reviews},
  volume={33},
  number={3},
  pages={279--296},
  year={2009},
  publisher={Elsevier},
  doi={10.1016/j.neubiorev.2008.09.002},
}

@article{lau2019resting,
  title={Resting-state abnormalities in autism spectrum disorders: a meta-analysis},
  author={Lau, Way KW and Leung, Mei-Kei and Lau, Benson WM},
  journal={Scientific reports},
  volume={9},
  number={1},
  pages={3892},
  year={2019},
  publisher={Nature Publishing Group UK London},
  doi={10.1038/s41598-019-40427-7},
}

@article{northoff2016abnormalities,
  title={How do abnormalities in the brain’s spontaneous activity translate into symptoms in schizophrenia? From an overview of resting state activity findings to a proposed spatiotemporal psychopathology},
  author={Northoff, Georg and Duncan, Niall W},
  journal={Progress in neurobiology},
  volume={145},
  pages={26--45},
  year={2016},
  publisher={Elsevier},
  doi={10.1016/j.pneurobio.2016.08.003},
}

@article{watanabe2017brain,
  title={Brain network dynamics in high-functioning individuals with autism},
  author={Watanabe, Takamitsu and Rees, Geraint},
  journal={Nature communications},
  volume={8},
  number={1},
  pages={16048},
  year={2017},
  publisher={Nature Publishing Group UK London},
  doi={10.1038/ncomms16048},
}

@article{perovnik2023functional,
  title={Functional brain networks in the evaluation of patients with neurodegenerative disorders},
  author={Perovnik, Matej and Rus, Toma{\v{z}} and Schindlbeck, Katharina A and Eidelberg, David},
  journal={Nature Reviews Neurology},
  volume={19},
  number={2},
  pages={73--90},
  year={2023},
  publisher={Nature Publishing Group UK London},
  doi={10.1038/s41582-022-00753-3},
}

@article{sorg2007selective,
  title={Selective changes of resting-state networks in individuals at risk for Alzheimer's disease},
  author={Sorg, Christian and Riedl, Valentin and Muhlau, Mark and Calhoun, Vince D and Eichele, Tom and L{\"a}er, Leonhard and Drzezga, Alexander and F{\"o}rstl, Hans and Kurz, Alexander and Zimmer, Claus and others},
  journal={Proceedings of the National Academy of Sciences},
  volume={104},
  number={47},
  pages={18760--18765},
  year={2007},
  publisher={National Acad Sciences},
  doi={10.1073/pnas.0708803104},
}

@article{zhu2021altered,
  title={Altered spontaneous brain activity in subjects with different cognitive states of biologically defined Alzheimer's Disease: a surface-based functional brain imaging study},
  author={Zhu, Zili and Zeng, Qingze and Kong, Linghan and Luo, Xiao and Li, Kaicheng and Xu, Xiaopei and Zhang, Minming and Huang, Peiyu and Yang, Yunjun},
  journal={Frontiers in Aging Neuroscience},
  volume={13},
  pages={683783},
  year={2021},
  publisher={Frontiers},
  doi={10.3389/fnagi.2021.683783},
}

@book{Northoff2018TheSB,
  title={The Spontaneous Brain: From the Mind–Body to the World–Brain Problem},
  author={Georg Northoff},
  year={2018},
  publisher={The MIT Press},
  url={https://mitpress.mit.edu/9780262038072/the-spontaneous-brain/}
}

@article{he2008multimodal,
  title={Multimodal functional neuroimaging: integrating functional MRI and EEG/MEG},
  author={He, Bin and Liu, Zhongming},
  journal={IEEE reviews in biomedical engineering},
  volume={1},
  pages={23--40},
  year={2008},
  publisher={IEEE},
  doi={10.1109/RBME.2008.2008233},
}

@article{liu2015multimodal,
  title={Multimodal neuroimaging computing: a review of the applications in neuropsychiatric disorders},
  author={Liu, Sidong and Cai, Weidong and Liu, Siqi and Zhang, Fan and Fulham, Michael and Feng, Dagan and Pujol, Sonia and Kikinis, Ron},
  journal={Brain informatics},
  volume={2},
  number={3},
  pages={167--180},
  year={2015},
  publisher={SpringerOpen},
  doi={10.1007/s40708-015-0019-x},
}

@article{tulay2019multimodal,
  title={Multimodal neuroimaging: basic concepts and classification of neuropsychiatric diseases},
  author={Tulay, Emine Elif and Metin, Bar{\i}{\c{s}} and Tarhan, Nevzat and Ar{\i}kan, Mehmet Kemal},
  journal={Clinical EEG and neuroscience},
  volume={50},
  number={1},
  pages={20--33},
  year={2019},
  publisher={Sage Publications Sage CA: Los Angeles, CA},
  doi={10.1177/1550059418782093},
}

@article{cichy2020eeg,
  title={A M/EEG-fMRI fusion primer: resolving human brain responses in space and time},
  author={Cichy, Radoslaw M and Oliva, Aude},
  journal={Neuron},
  volume={107},
  number={5},
  pages={772--781},
  year={2020},
  publisher={Elsevier},
  doi={10.1016/j.neuron.2020.07.001},
}

@article{laufs2012personalized,
  title={A personalized history of EEG--fMRI integration},
  author={Laufs, Helmut},
  journal={Neuroimage},
  volume={62},
  number={2},
  pages={1056--1067},
  year={2012},
  publisher={Elsevier},
  doi={10.1016/j.neuroimage.2012.01.039}
}

@article{uludaug2014general,
  title={General overview on the merits of multimodal neuroimaging data fusion},
  author={Uluda{\u{g}}, K{\^a}mil and Roebroeck, Alard},
  journal={Neuroimage},
  volume={102},
  pages={3--10},
  year={2014},
  publisher={Elsevier},
  doi={10.1016/j.neuroimage.2014.05.018},
}

@article{lohani2022spatiotemporally,
  title={Spatiotemporally heterogeneous coordination of cholinergic and neocortical activity},
  author={Lohani, Sweyta and Moberly, Andrew H and Benisty, Hadas and Landa, Boris and Jing, Miao and Li, Yulong and Higley, Michael J and Cardin, Jessica A},
  journal={Nature Neuroscience},
  volume={25},
  number={12},
  pages={1706--1713},
  year={2022},
  publisher={Nature Publishing Group US New York},
  di={10.1038/s41593-022-01202-6},
}

@article{bassett2006small,
  title={Small-world brain networks},
  author={Bassett, Danielle Smith and Bullmore, ED},
  journal={The neuroscientist},
  volume={12},
  number={6},
  pages={512--523},
  year={2006},
  publisher={Sage Publications Sage CA: Thousand Oaks, CA},
  doi={10.1177/1073858406293},
}

@article{bassett2017small,
  title={Small-world brain networks revisited},
  author={Bassett, Danielle S and Bullmore, Edward T},
  journal={The Neuroscientist},
  volume={23},
  number={5},
  pages={499--516},
  year={2017},
  publisher={Sage Publications Sage CA: Los Angeles, CA},
  doi={10.1177/1073858416667720},
}

@article{humphries2006brainstem,
  title={The brainstem reticular formation is a small-world, not scale-free, network},
  author={Humphries, Mark D and Gurney, Kevin and Prescott, Tony J},
  journal={Proceedings of the Royal Society B: Biological Sciences},
  volume={273},
  number={1585},
  pages={503--511},
  year={2006},
  publisher={The Royal Society London},
  doi={10.1098/rspb.2005.3354},
}

@article{newsome1988selective,
  title={A selective impairment of motion perception following lesions of the middle temporal visual area (MT)},
  author={Newsome, William T and Pare, Edmond B},
  journal={Journal of Neuroscience},
  volume={8},
  number={6},
  pages={2201--2211},
  year={1988},
  publisher={Soc Neuroscience},
  doi={10.1523/JNEUROSCI.08-06-02201.1988},
}

@article{conway2007specialized,
  title={Specialized color modules in macaque extrastriate cortex},
  author={Conway, Bevil R and Moeller, Sebastian and Tsao, Doris Y},
  journal={Neuron},
  volume={56},
  number={3},
  pages={560--573},
  year={2007},
  publisher={Elsevier},
  doi={10.1016/j.neuron.2007.10.008},
}

@article{conway2009color,
  title={Color-tuned neurons are spatially clustered according to color preference within alert macaque posterior inferior temporal cortex},
  author={Conway, Bevil R and Tsao, Doris Y},
  journal={Proceedings of the National Academy of Sciences},
  volume={106},
  number={42},
  pages={18034--18039},
  year={2009},
  publisher={National Acad Sciences},
  doi={10.1073/pnas.0810943106},
}

@article{mayer2022gut,
  title={The gut--brain axis},
  author={Mayer, Emeran A and Nance, Karina and Chen, Shelley},
  journal={Annual Review of Medicine},
  volume={73},
  pages={439--453},
  year={2022},
  publisher={Annual Reviews},
  doi={10.1146/annurev-med-042320-014032},
}

@article{serre2007feedforward,
  title={A feedforward architecture accounts for rapid categorization},
  author={Serre, Thomas and Oliva, Aude and Poggio, Tomaso},
  journal={Proceedings of the National Academy of Sciences},
  volume={104},
  number={15},
  pages={6424--6429},
  year={2007},
  publisher={National Acad Sciences},
  doi={10.1073/pnas.0700622104},
}

@article{hesse2020macaque,
  title={The macaque face patch system: a turtle’s underbelly for the brain},
  author={Hesse, Janis K and Tsao, Doris Y},
  journal={Nature Reviews Neuroscience},
  volume={21},
  number={12},
  pages={695--716},
  year={2020},
  publisher={Nature Publishing Group UK London},
  doi={10.1038/s41583-020-00393-w},
}

@article{grossberg1987competitive,
  title={Competitive learning: From interactive activation to adaptive resonance},
  author={Grossberg, Stephen},
  journal={Cognitive science},
  volume={11},
  number={1},
  pages={23--63},
  year={1987},
  publisher={Elsevier},
  doi={10.1016/S0364-0213(87)80025-3},
}

@article{lillicrap2020backpropagation,
  title={Backpropagation and the brain},
  author={Lillicrap, Timothy P and Santoro, Adam and Marris, Luke and Akerman, Colin J and Hinton, Geoffrey},
  journal={Nature Reviews Neuroscience},
  volume={21},
  number={6},
  pages={335--346},
  year={2020},
  publisher={Nature Publishing Group UK London},
  doi={10.1038/s41583-020-0277-3},
}

@article{mehrer2021ecologically,
  title={An ecologically motivated image dataset for deep learning yields better models of human vision},
  author={Mehrer, Johannes and Spoerer, Courtney J and Jones, Emer C and Kriegeskorte, Nikolaus and Kietzmann, Tim C},
  journal={Proceedings of the National Academy of Sciences},
  volume={118},
  number={8},
  pages={e2011417118},
  year={2021},
  publisher={National Acad Sciences},
  doi={10.1073/pnas.2011417118},
}

@inproceedings{bakhtiari2021functional,
 author = {Bakhtiari, Shahab and Mineault, Patrick and Lillicrap, Timothy and Pack, Christopher and Richards, Blake},
 booktitle = {Advances in Neural Information Processing Systems},
 editor = {M. Ranzato and A. Beygelzimer and Y. Dauphin and P.S. Liang and J. Wortman Vaughan},
 pages = {25164--25178},
 publisher = {Curran Associates, Inc.},
 title = {The functional specialization of visual cortex emerges from training parallel pathways with self-supervised predictive learning},
 url = {https://papers.nips.cc/paper/2021/hash/d384dec9f5f7a64a36b5c8f03b8a6d92-Abstract.html},
 volume = {34},
 year = {2021}
}

@book{cahan2018helmholtz,
  title={Helmholtz: A life in science},
  author={Cahan, David},
  year={2018},
  publisher={University of Chicago Press},
  url={https://press.uchicago.edu/ucp/books/book/chicago/H/bo28246106.html},
}

@article{carandini2005we,
  title={Do we know what the early visual system does?},
  author={Carandini, Matteo and Demb, Jonathan B and Mante, Valerio and Tolhurst, David J and Dan, Yang and Olshausen, Bruno A and Gallant, Jack L and Rust, Nicole C},
  journal={Journal of Neuroscience},
  volume={25},
  number={46},
  pages={10577--10597},
  year={2005},
  publisher={Soc Neuroscience},
  doi={10.1523/JNEUROSCI.3726-05.2005},
}

@article{mcfarland2013inferring,
  title={Inferring nonlinear neuronal computation based on physiologically plausible inputs},
  author={McFarland, James M and Cui, Yuwei and Butts, Daniel A},
  journal={PLoS Computational Biology},
  volume={9},
  number={7},
  pages={e1003143},
  year={2013},
  publisher={Public Library of Science San Francisco, USA},
  doi={10.1371/journal.pcbi.1003143},
}

@article{marder2006variability,
  title={Variability, compensation and homeostasis in neuron and network function},
  author={Marder, Eve and Goaillard, Jean-Marc},
  journal={Nature Reviews Neuroscience},
  volume={7},
  number={7},
  pages={563--574},
  year={2006},
  publisher={Nature Publishing Group UK London},
  doi={10.1038/nrn1949},
}

@article{sender2021distribution,
  title={The distribution of cellular turnover in the human body},
  author={Sender, Ron and Milo, Ron},
  journal={Nature medicine},
  volume={27},
  number={1},
  pages={45--48},
  year={2021},
  publisher={Nature Publishing Group US New York},
  doi={10.1038/s41591-020-01182-9},
}

@article{chevalier2020effect,
  title={Effect of gut microbiota on depressive-like behaviors in mice is mediated by the endocannabinoid system},
  author={Chevalier, Gr{\'e}goire and Siopi, Eleni and Guenin-Mac{\'e}, Laure and Pascal, Maud and Laval, Thomas and Rifflet, Aline and Boneca, Ivo Gomperts and Demangel, Caroline and Colsch, Benoit and Pruvost, Alain and others},
  journal={Nature Communications},
  volume={11},
  number={1},
  pages={6363},
  year={2020},
  publisher={Nature Publishing Group UK London},
  doi={10.1038/s41467-020-19931-2},
}

@article{driscoll2022representational,
  title={Representational drift: Emerging theories for continual learning and experimental future directions},
  author={Driscoll, Laura N and Duncker, Lea and Harvey, Christopher D},
  journal={Current Opinion in Neurobiology},
  volume={76},
  pages={102609},
  year={2022},
  publisher={Elsevier},
  doi={10.1016/j.conb.2022.102609},
}

@article{deitch2021representational,
  title={Representational drift in the mouse visual cortex},
  author={Deitch, Daniel and Rubin, Alon and Ziv, Yaniv},
  journal={Current Biology},
  volume={31},
  number={19},
  pages={4327--4339},
  year={2021},
  publisher={Elsevier},
  doi={10.1016/j.cub.2021.07.062},
}

@article{schoonover2021representational,
  title={Representational drift in primary olfactory cortex},
  author={Schoonover, Carl E and Ohashi, Sarah N and Axel, Richard and Fink, Andrew JP},
  journal={Nature},
  volume={594},
  number={7864},
  pages={541--546},
  year={2021},
  publisher={Nature Publishing Group UK London},
  doi={10.1038/s41586-021-03628-7},
}

@article{rule2019causes,
  title={Causes and consequences of representational drift},
  author={Rule, Michael E and O’Leary, Timothy and Harvey, Christopher D},
  journal={Current Opinion in Neurobiology},
  volume={58},
  pages={141--147},
  year={2019},
  publisher={Elsevier},
  doi={10.1016/j.conb.2019.08.005},
}

@article{grienberger2012imaging,
  title={Imaging calcium in neurons},
  author={Grienberger, Christine and Konnerth, Arthur},
  journal={Neuron},
  volume={73},
  number={5},
  pages={862--885},
  year={2012},
  publisher={Elsevier},
  doi={10.1016/j.neuron.2012.02.011},
}

@article{tian2009imaging,
  title={Imaging neural activity in worms, flies and mice with improved GCaMP calcium indicators},
  author={Tian, Lin and Hires, S Andrew and Mao, Tianyi and Huber, Daniel and Chiappe, M Eugenia and Chalasani, Sreekanth H and Petreanu, Leopoldo and Akerboom, Jasper and McKinney, Sean A and Schreiter, Eric R and others},
  journal={Nature methods},
  volume={6},
  number={12},
  pages={875--881},
  year={2009},
  publisher={Nature Publishing Group US New York},
  doi={10.1038/nmeth.1398},
}

@article{hall2014relationship,
  title={The relationship between MEG and fMRI},
  author={Hall, Emma L and Robson, Si{\^a}n E and Morris, Peter G and Brookes, Matthew J},
  journal={Neuroimage},
  volume={102},
  pages={80--91},
  year={2014},
  publisher={Elsevier},
  doi={10.1016/j.neuroimage.2013.11.005}
}

@article{ritter2006simultaneous,
  title={simultaneous EEG--fMRI},
  author={Ritter, Petra and Villringer, Arno},
  journal={Neuroscience \& Biobehavioral Reviews},
  volume={30},
  number={6},
  pages={823--838},
  year={2006},
  publisher={Elsevier},
  doi={10.1016/j.neubiorev.2006.06.008},
}

@book{barabasi2016network,
  title={Network Science},
  author={Barab{\'a}si, Albert-L{\'a}szl{\'o}},
  isbn={9781107076266},
  lccn={2015938594},
  url={http://networksciencebook.com/},
  year={2016},
  publisher={Cambridge University Press}
}

@book{newman2018networks,
  title={Networks},
  author={Newman, Mark},
  year={2018},
  publisher={Oxford university press},
  url={https://global.oup.com/academic/product/networks-9780198805090},
}

@article{milgram1967small,
  title={The small world problem},
  author={Milgram, Stanley},
  journal={Psychology Today},
  volume={2},
  number={1},
  pages={60--67},
  year={1967}
}

@book{sporns2010networks,
  title={Networks of the Brain},
  author={Sporns, Olaf},
  year={2010},
  publisher={MIT Press},
  url={https://mitpress.mit.edu/9780262528986/networks-of-the-brain}
}

@book{fornito2016fundamentals,
  title={Fundamentals of Brain Network Analysis},
  author={Fornito, Alex and Zalesky, Andrew and Bullmore, Edward},
  year={2016},
  publisher={Academic Press},
  doi={10.1016/C2012-0-06036-X}
}

@article{pessoa2014understanding,
  title={Understanding brain networks and brain organization},
  author={Pessoa, Luiz},
  journal={Physics of life reviews},
  volume={11},
  number={3},
  pages={400--435},
  year={2014},
  publisher={Elsevier},
  doi={10.1016/j.plrev.2014.03.005}
}

@article{stein2005neuronal,
  title={Neuronal variability: noise or part of the signal?},
  author={Stein, Richard B and Gossen, E Roderich and Jones, Kelvin E},
  journal={Nature Reviews Neuroscience},
  volume={6},
  number={5},
  pages={389--397},
  year={2005},
  publisher={Nature Publishing Group UK London},
  doi={10.1038/nrn1668}
}

@book{Kandel2021,
  title={Principles of Neural Science},
  author={Kandel, ER and Koester, JD and Mack, SH and Siegelbaum, SA},
  editor={Kandel, Eric R. and others},
  publisher={McGraw Hill},
  year={2021},
  edition={6},
  url={https://neurology.mhmedical.com/Book.aspx?bookid=3024},
}

@article{kandel2009introduction,
  title={An introduction to the work of David Hubel and Torsten Wiesel},
  author={Kandel, Eric R},
  journal={The Journal of physiology},
  volume={587},
  number={Pt 12},
  pages={2733},
  year={2009},
  publisher={Wiley-Blackwell}
}

@article{waters2020sources,
  title={Sources of widefield fluorescence from the brain},
  author={Waters, Jack},
  journal={Elife},
  volume={9},
  pages={e59841},
  year={2020},
  publisher={eLife Sciences Publications, Ltd},
  doi={10.7554/eLife.59841},
}

@article{drew2019vascular,
  title={Vascular and neural basis of the BOLD signal},
  author={Drew, Patrick J},
  journal={Current Opinion in Neurobiology},
  volume={58},
  pages={61--69},
  year={2019},
  publisher={Elsevier},
  doi={10.1016/j.conb.2019.06.004},
}

@article{arthurs2002well,
  title={How well do we understand the neural origins of the fMRI BOLD signal?},
  author={Arthurs, Owen J and Boniface, Simon},
  journal={Trends in Neurosciences},
  volume={25},
  number={1},
  pages={27--31},
  year={2002},
  publisher={Elsevier},
  doi={10.1016/S0166-2236(00)01995-0},
}

@article{matsui2019neuronal,
  title={Neuronal origin of the temporal dynamics of spontaneous BOLD activity correlation},
  author={Matsui, Teppei and Murakami, Tomonari and Ohki, Kenichi},
  journal={Cerebral Cortex},
  volume={29},
  number={4},
  pages={1496--1508},
  year={2019},
  publisher={Oxford University Press},
  doi={10.1093/cercor/bhy045},
}

@article{sinz2019engineering,
  title={Engineering a less artificial intelligence},
  author={Sinz, Fabian H and Pitkow, Xaq and Reimer, Jacob and Bethge, Matthias and Tolias, Andreas S},
  journal={Neuron},
  volume={103},
  number={6},
  pages={967--979},
  year={2019},
  publisher={Elsevier},
  doi={10.1016/j.neuron.2019.08.034},
}

@article{zhuang2021unsupervised,
  title={Unsupervised neural network models of the ventral visual stream},
  author={Zhuang, Chengxu and Yan, Siming and Nayebi, Aran and Schrimpf, Martin and Frank, Michael C and DiCarlo, James J and Yamins, Daniel LK},
  journal={Proceedings of the National Academy of Sciences},
  volume={118},
  number={3},
  pages={e2014196118},
  year={2021},
  publisher={National Acad Sciences},
  doi={10.1073/pnas.20141961},
}

@inproceedings{McIntosh16Deep,
 author = {McIntosh, Lane and Maheswaranathan, Niru and Nayebi, Aran and Ganguli, Surya and Baccus, Stephen},
 booktitle = {Advances in Neural Information Processing Systems},
 editor = {D. Lee and M. Sugiyama and U. Luxburg and I. Guyon and R. Garnett},
 publisher = {Curran Associates, Inc.},
 title = {Deep Learning Models of the Retinal Response to Natural Scenes},
 url = {https://proceedings.neurips.cc/paper/2016/hash/a1d33d0dfec820b41b54430b50e96b5c-Abstract.html},
 volume = {29},
 year = {2016}
}

@inproceedings{Klindt17neural,
 author = {Klindt, David and Ecker, Alexander S and Euler, Thomas and Bethge, Matthias},
 booktitle = {Advances in Neural Information Processing Systems},
 editor = {I. Guyon and U. Von Luxburg and S. Bengio and H. Wallach and R. Fergus and S. Vishwanathan and R. Garnett},
 publisher = {Curran Associates, Inc.},
 title = {Neural system identification for large populations separating \textquotedblleft what\textquotedblright  and \textquotedblleft where\textquotedblright },
 url = {https://proceedings.neurips.cc/paper/2017/hash/8c249675aea6c3cbd91661bbae767ff1-Abstract.html},
 volume = {30},
 year = {2017}
}

@inproceedings{sinz18stimulus,
 author = {Sinz, Fabian and Ecker, Alexander S and Fahey, Paul and Walker, Edgar and Cobos, Erick and Froudarakis, Emmanouil and Yatsenko, Dimitri and Pitkow, Zachary and Reimer, Jacob and Tolias, Andreas},
 booktitle = {Advances in Neural Information Processing Systems},
 editor = {S. Bengio and H. Wallach and H. Larochelle and K. Grauman and N. Cesa-Bianchi and R. Garnett},
 pages = {},
 publisher = {Curran Associates, Inc.},
 title = {Stimulus domain transfer in recurrent models for large scale cortical population prediction on video},
 url = {https://proceedings.neurips.cc/paper/2018/hash/9d684c589d67031a627ad33d59db65e5-Abstract.html},
 volume = {31},
 year = {2018}
}

@inproceedings{ecker2018rotation,
    title={A rotation-equivariant convolutional neural network model of primary visual cortex},
    author={Alexander S. Ecker and Fabian H. Sinz and Emmanouil Froudarakis and Paul G. Fahey and Santiago A. Cadena and Edgar Y. Walker and Erick Cobos and Jacob Reimer and Andreas S. Tolias and Matthias Bethge},
    booktitle={International Conference on Learning Representations},
    year={2019},
    url={https://openreview.net/forum?id=H1fU8iAqKX},
}

@article {wang23towards,
	author = {Eric Y. Wang and Paul G. Fahey and Kayla Ponder and Zhuokun Ding and Andersen Chang and Taliah Muhammad and Saumil Patel and Zhiwei Ding and Dat Tran and Jiakun Fu and Stelios Papadopoulos and Katrin Franke and Alexander S. Ecker and Jacob Reimer and Xaq Pitkow and Fabian H. Sinz and Andreas S. Tolias},
	title = {Towards a Foundation Model of the Mouse Visual Cortex},
	year = {2023},
	doi = {10.1101/2023.03.21.533548},
	publisher = {Cold Spring Harbor Laboratory},
	journal = {bioRxiv},
}

@article{cadena2019deep,
  title={Deep convolutional models improve predictions of macaque V1 responses to natural images},
  author={Cadena, Santiago A and Denfield, George H and Walker, Edgar Y and Gatys, Leon A and Tolias, Andreas S and Bethge, Matthias and Ecker, Alexander S},
  journal={PLoS Computational Biology},
  volume={15},
  number={4},
  pages={e1006897},
  year={2019},
  publisher={Public Library of Science San Francisco, CA USA},
  doi={10.1371/journal.pcbi.1006897},
}

@article{nayebi2023mouse,
  title={Mouse visual cortex as a limited resource system that self-learns an ecologically-general representation},
  author={Nayebi, Aran and Kong, Nathan CL and Zhuang, Chengxu and Gardner, Justin L and Norcia, Anthony M and Yamins, Daniel LK},
  journal={PLoS Computational Biology},
  volume={19},
  number={10},
  pages={e1011506},
  year={2023},
  publisher={Public Library of Science San Francisco, CA USA},
  doi={10.1371/journal.pcbi.1011506},
}

@article{konkle2022self,
  title={A self-supervised domain-general learning framework for human ventral stream representation},
  author={Konkle, Talia and Alvarez, George A},
  journal={Nature communications},
  volume={13},
  number={1},
  pages={491},
  year={2022},
  publisher={Nature Publishing Group UK London},
  doi={10.1038/s41467-022-28091-4},
}

@article {kubilius18cornet,
    author = {Jonas Kubilius and Martin Schrimpf and Aran Nayebi and Daniel Bear and Daniel L. K. Yamins and James J. DiCarlo},
    title = {CORnet: Modeling the Neural Mechanisms of Core Object Recognition},
    year = {2018},
    doi = {10.1101/408385},
    publisher = {Cold Spring Harbor Laboratory},
    journal = {bioRxiv},
}

@article{rockland1979laminar,
  title={Laminar origins and terminations of cortical connections of the occipital lobe in the rhesus monkey},
  author={Rockland, Kathleen S and Pandya, Deepak N},
  journal={Brain research},
  volume={179},
  number={1},
  pages={3--20},
  year={1979},
  publisher={Elsevier},
  doi={10.1016/0006-8993(79)90485-2},
}

@article{markov2014anatomy,
  title={Anatomy of hierarchy: feedforward and feedback pathways in macaque visual cortex},
  author={Markov, Nikola T and Vezoli, Julien and Chameau, Pascal and Falchier, Arnaud and Quilodran, Ren{\'e} and Huissoud, Cyril and Lamy, Camille and Misery, Pierre and Giroud, Pascale and Ullman, Shimon and others},
  journal={Journal of Comparative Neurology},
  volume={522},
  number={1},
  pages={225--259},
  year={2014},
  publisher={Wiley Online Library},
  doi={10.1002/cne.23458},
}

@article{zeki1978functional,
  title={Functional specialisation in the visual cortex of the rhesus monkey},
  author={Zeki, Semir M},
  journal={Nature},
  volume={274},
  number={5670},
  pages={423--428},
  year={1978},
  publisher={Nature Publishing Group UK London},
  doi={10.1038/274423a0},
}

@article{david2004natural,
  title={Natural stimulus statistics alter the receptive field structure of v1 neurons},
  author={David, Stephen V and Vinje, William E and Gallant, Jack L},
  journal={Journal of Neuroscience},
  volume={24},
  number={31},
  pages={6991--7006},
  year={2004},
  publisher={Soc Neuroscience},
  doi={10.1523/JNEUROSCI.1422-04.2004}
}

@article{butts2019data,
  title={Data-driven approaches to understanding visual neuron activity},
  author={Butts, Daniel A},
  journal={Annual review of vision science},
  volume={5},
  pages={451--477},
  year={2019},
  publisher={Annual Reviews},
  doi={10.1146/annurev-vision-091718-014731}
}

@inbook{VezoliEvolvingHier,
  title={The Evolving Concept of Cortical Hierarchy},
  author={Vezoli, Julien and Hou, Yujie and Kennedy, Henry},
  booktitle={The Cerebral Cortex and Thalamus},
  chapter={37},
  editor={W. Martin Usrey and S. Murray Sherman},
  publisher={Oxford University Press},
  year={2023},
  pages={393--404},
  url={https://global.oup.com/academic/product/the-cerebral-cortex-and-thalamus-9780197676158}
}

@article{kourtzi2002object,
  title={Object-selective responses in the human motion area MT/MST},
  author={Kourtzi, Zoe and B{\"u}lthoff, Heinrich H and Erb, Michael and Grodd, Wolfgang},
  journal={Nature Neuroscience},
  volume={5},
  number={1},
  pages={17--18},
  year={2002},
  publisher={Nature Publishing Group US New York},
  doi={10.1038/nn780}
}

@article{kravitz2011new,
  title={A new neural framework for visuospatial processing},
  author={Kravitz, Dwight J and Saleem, Kadharbatcha S and Baker, Chris I and Mishkin, Mortimer},
  journal={Nature Reviews Neuroscience},
  volume={12},
  number={4},
  pages={217--230},
  year={2011},
  publisher={Nature Publishing Group UK London},
  doi={10.1038/nrn3008},
}

@article{ayzenberg2022does,
  title={Does the brain's ventral visual pathway compute object shape?},
  author={Ayzenberg, Vladislav and Behrmann, Marlene},
  journal={Trends in Cognitive Sciences},
  volume={26},
  number={12},
  pages={1119--1132},
  year={2022},
  publisher={Elsevier},
  doi={10.1016/j.tics.2022.09.019}
}

@article{goodale2023shape,
  title={Shape perception does not require dorsal stream processing},
  author={Goodale, Melvyn A and Milner, A David},
  journal={Trends in Cognitive Sciences},
  year={2023},
  publisher={Elsevier},
  doi={10.1016/j.tics.2022.12.007},
}

@article{goodale2023patients,
  title={Patients with dorsal-stream lesions can perceive global shape},
  author={Goodale, Melvyn A and Milner, A David},
  journal={Trends in Cognitive Sciences},
  year={2023},
  publisher={Elsevier},
  doi={10.1016/j.tics.2023.03.007},
}

@article{ayzenberg2023temporal,
  title={Temporal asymmetries and interactions between dorsal and ventral visual pathways during object recognition},
  author={Ayzenberg, Vladislav and Simmons, Claire and Behrmann, Marlene},
  journal={Cerebral Cortex Communications},
  volume={4},
  number={1},
  pages={tgad003},
  year={2023},
  publisher={Oxford University Press},
  doi={10.1093/texcom/tgad003}
}

@article{rossetti2017rise,
  title={Rise and fall of the two visual systems theory},
  author={Rossetti, Yves and Pisella, Laure and McIntosh, Robert D},
  journal={Annals of physical and rehabilitation medicine},
  volume={60},
  number={3},
  pages={130--140},
  year={2017},
  publisher={Elsevier},
  doi={10.1016/j.rehab.2017.02.002},
}

@article{sheth2016two,
  title={Two visual pathways in primates based on sampling of space: exploitation and exploration of visual information},
  author={Sheth, Bhavin R and Young, Ryan},
  journal={Frontiers in integrative neuroscience},
  volume={10},
  pages={37},
  year={2016},
  publisher={Frontiers Media SA},
  doi={10.3389/fnint.2016.00037},
}

@book{dayan2001theoretical,
    author = {Dayan, Peter and Abbott, Laurence F.},
    title = {Theoretical Neuroscience},
    publisher = {The MIT Press},
    year = {2001},
    url = {https://mitpress.mit.edu/9780262041997/theoretical-neuroscience/},
}

@article{shannon1948mathematical,
  title={A mathematical theory of communication},
  author={Shannon, Claude Elwood},
  journal={The Bell system technical journal},
  volume={27},
  number={3},
  pages={379--423},
  year={1948},
  publisher={Nokia Bell Labs},
  doi={10.1002/j.1538-7305.1948.tb01338.x},
}

@article{attneave1954some,
  title={Some informational aspects of visual perception.},
  author={Attneave, Fred},
  journal={Psychological review},
  volume={61},
  number={3},
  pages={183},
  year={1954},
  publisher={American Psychological Association},
  doi={10.1037/h0054663},
}

@article{barlow1961possible,
  title={Possible principles underlying the transformation of sensory messages},
  author={Barlow, Horace B.},
  journal={Sensory communication},
  volume={1},
  number={01},
  pages={217--233},
  year={1961},
  url={https://www.cnbc.cmu.edu/~tai/microns_papers/Barlow-SensoryCommunication-1961.pdf},
}

@article{bolanos2022efficient,
    author = {Federico Bola{\~n}os and Javier G. Orlandi and Ryo Aoki and Akshay V. Jagadeesh and Justin L. Gardner and Andrea Benucci},
    title = {Efficient coding of natural images in the mouse visual cortex},
    year = {2022},
    doi = {10.1101/2022.09.14.507893},
    publisher = {Cold Spring Harbor Laboratory},
    journal = {bioRxiv},
}

@article{atick1992could,
  title={Could information theory provide an ecological theory of sensory processing?},
  author={Atick, Joseph J},
  journal={Network: Computation in neural systems},
  volume={3},
  number={2},
  pages={213--251},
  year={1992},
  publisher={Taylor \& Francis},
  doi={10.1088/0954-898X_3_2_009},
}

@article{lewicki2002efficient,
  title={Efficient coding of natural sounds},
  author={Lewicki, Michael S},
  journal={Nature Neuroscience},
  volume={5},
  number={4},
  pages={356--363},
  year={2002},
  publisher={Nature Publishing Group US New York},
  doi={10.1038/nn831},
}

@article{simoncelli2001natural,
  title={Natural image statistics and neural representation.},
  author={Eero P. Simoncelli and Bruno A. Olshausen},
  journal={Annual Review of Neuroscience},
  year={2001},
  volume={24},
  pages={1193-216},
  doi={10.1146/annurev.neuro.24.1.1193}
}

@article{olshausen1997sparse,
  title={Sparse coding with an overcomplete basis set: A strategy employed by V1?},
  author={Olshausen, Bruno A and Field, David J},
  journal={Vision Research},
  volume={37},
  number={23},
  pages={3311--3325},
  year={1997},
  publisher={Elsevier},
  doi={10.1016/S0042-6989(97)00169-7},
}

@article{olshausen1996emergence,
  title={Emergence of simple-cell receptive field properties by learning a sparse code for natural images},
  author={Olshausen, Bruno A and Field, David J},
  journal={Nature},
  volume={381},
  number={6583},
  pages={607--609},
  year={1996},
  publisher={Nature Publishing Group UK London},
  doi={10.1038/381607a0},
}

@article{pitkow2012decorrelation,
  title={Decorrelation and efficient coding by retinal ganglion cells},
  author={Pitkow, Xaq and Meister, Markus},
  journal={Nature Neuroscience},
  volume={15},
  number={4},
  pages={628--635},
  year={2012},
  publisher={Nature Publishing Group US New York},
  doi={10.1038/nn.3064},
}

@article{polania2019efficient,
  title={Efficient coding of subjective value},
  author={Polania, Rafael and Woodford, Michael and Ruff, Christian C},
  journal={Nature Neuroscience},
  volume={22},
  number={1},
  pages={134--142},
  year={2019},
  publisher={Nature Publishing Group US New York},
  doi={10.1038/s41593-018-0292-0}
}

@article{geisler2008visual,
  title={Visual perception and the statistical properties of natural scenes},
  author={Geisler, Wilson S},
  journal={Annu. Rev. Psychol.},
  volume={59},
  pages={167--192},
  year={2008},
  publisher={Annual Reviews},
  doi={10.1146/annurev.psych.58.110405.085632},
}

@article{barlow2001redundancy,
  title={Redundancy reduction revisited},
  author={Barlow, Horace},
  journal={Network: computation in neural systems},
  volume={12},
  number={3},
  pages={241},
  year={2001},
  publisher={IOP Publishing},
  doi={10.1088/0954-898X/12/3/301},
  url={https://www.researchgate.net/profile/Horace-Barlow/publication/11783583_Redundancy_reduction_revisited/links/0deec51c9c73ad49a1000000/Redundancy-reduction-revisited.pdf},
}

@book{li2014understanding,
  title={Understanding vision: theory, models, and data},
  author={Li, Zhaoping},
  year={2014},
  publisher={Oxford University Press (UK)},
  url={https://global.oup.com/academic/product/understanding-vision-9780198829362},
}

@article{mlynarski2021efficient,
  title={Efficient and adaptive sensory codes},
  author={M{\l}ynarski, Wiktor F and Hermundstad, Ann M},
  journal={Nature Neuroscience},
  volume={24},
  number={7},
  pages={998--1009},
  year={2021},
  publisher={Nature Publishing Group US New York},
  doi={10.1038/s41593-021-00846-0},
}

@article{lesica2007adaptation,
  title={Adaptation to stimulus contrast and correlations during natural visual stimulation},
  author={Lesica, Nicholas A and Jin, Jianzhong and Weng, Chong and Yeh, Chun-I and Butts, Daniel A and Stanley, Garrett B and Alonso, Jose-Manuel},
  journal={Neuron},
  volume={55},
  number={3},
  pages={479--491},
  year={2007},
  publisher={Elsevier},
  doi={10.1016/j.neuron.2007.07.013},
}

@article{gutnisky2008adaptive,
  title={Adaptive coding of visual information in neural populations},
  author={Gutnisky, Diego A and Dragoi, Valentin},
  journal={Nature},
  volume={452},
  number={7184},
  pages={220--224},
  year={2008},
  publisher={Nature Publishing Group UK London},
  doi={10.1038/nature06563},
}

@article{fairhall2001efficiency,
  title={Efficiency and ambiguity in an adaptive neural code},
  author={Fairhall, Adrienne L and Lewen, Geoffrey D and Bialek, William and de Ruyter van Steveninck, Robert R},
  journal={Nature},
  volume={412},
  number={6849},
  pages={787--792},
  year={2001},
  publisher={Nature Publishing Group UK London},
  doi={10.1038/35090500},
}

@article{ollerenshaw2014adaptive,
  title={The adaptive trade-off between detection and discrimination in cortical representations and behavior},
  author={Ollerenshaw, Douglas R and Zheng, He JV and Millard, Daniel C and Wang, Qi and Stanley, Garrett B},
  journal={Neuron},
  volume={81},
  number={5},
  pages={1152--1164},
  year={2014},
  publisher={Elsevier},
  doi={10.1016/j.neuron.2014.01.025},
}

@article{wolfe2010sparse,
  title={Sparse and powerful cortical spikes},
  author={Wolfe, Jason and Houweling, Arthur R and Brecht, Michael},
  journal={Current Opinion in Neurobiology},
  volume={20},
  number={3},
  pages={306--312},
  year={2010},
  publisher={Elsevier},
  doi={10.1016/j.conb.2010.03.006},
}

@article{rozell2008sparse,
  title={Sparse coding via thresholding and local competition in neural circuits},
  author={Rozell, Christopher J and Johnson, Don H and Baraniuk, Richard G and Olshausen, Bruno A},
  journal={Neural Computation},
  volume={20},
  number={10},
  pages={2526--2563},
  year={2008},
  publisher={MIT Press},
  doi={10.1162/neco.2008.03-07-486},
}

@article{fang2022learning,
  title={Learning and Inference in Sparse Coding Models With Langevin Dynamics},
  author={Fang, Michael Y-S and Mudigonda, Mayur and Zarcone, Ryan and Khosrowshahi, Amir and Olshausen, Bruno A},
  journal={Neural Computation},
  volume={34},
  number={8},
  pages={1676--1700},
  year={2022},
  publisher={MIT Press},
  doi={10.1162/neco_a_01505},
}

@article{rentzeperis2023beyond,
  title={Beyond $\ell_1$ sparse coding in V1},
  author={Rentzeperis, Ilias and Calatroni, Luca and Perrinet, Laurent U and Prandi, Dario},
  journal={PLoS Computational Biology},
  volume={19},
  number={9},
  pages={e1011459},
  year={2023},
  publisher={Public Library of Science San Francisco, CA USA},
  doi={10.1371/journal.pcbi.1011459},
}

@article{olshausen2004sparse,
  title={Sparse coding of sensory inputs},
  author={Olshausen, Bruno A and Field, David J},
  journal={Current Opinion in Neurobiology},
  volume={14},
  number={4},
  pages={481--487},
  year={2004},
  publisher={Elsevier},
  doi={10.1016/j.conb.2004.07.007},
}

@book{northoff2013unlocking,
  title={Unlocking the brain: volume 1: coding},
  author={Northoff, Georg},
  year={2013},
  publisher={Oxford University Press},
  doi={10.1093/acprof:oso/9780199826988.001.0001},
}

@misc{mineault2011sparse,
  title={The Far-Reaching Influence of Sparse Coding in V1},
  author={Mineault, Patrick},
  year={2011},
  month={6},
  day={15},
  url={https://xcorr.net/2011/06/15/the-far-reaching-influence-of-sparse-coding-in-v1/},
  note={Accessed: 2023-11-28},
}

@article{laughlin1998metabolic,
  title={The metabolic cost of neural information},
  author={Laughlin, Simon B and de Ruyter van Steveninck, Rob R and Anderson, John C},
  journal={Nature Neuroscience},
  volume={1},
  number={1},
  pages={36--41},
  year={1998},
  publisher={Nature Publishing Group},
  doi={10.1038/236}
}

@article{hasenstaub2010metabolic,
  title={Metabolic cost as a unifying principle governing neuronal biophysics},
  author={Hasenstaub, Andrea and Otte, Stephani and Callaway, Edward and Sejnowski, Terrence J},
  journal={Proceedings of the National Academy of Sciences},
  volume={107},
  number={27},
  pages={12329--12334},
  year={2010},
  publisher={National Acad Sciences},
  doi={10.1073/pnas.132309911},
}

@article{vinje2000sparse,
  title={Sparse coding and decorrelation in primary visual cortex during natural vision},
  author={Vinje, William E and Gallant, Jack L},
  journal={Science},
  volume={287},
  number={5456},
  pages={1273--1276},
  year={2000},
  publisher={American Association for the Advancement of Science},
  doi={10.1126/science.287.5456.1273}
}

@article{vinje2002natural,
  title={Natural stimulation of the nonclassical receptive field increases information transmission efficiency in V1},
  author={Vinje, William E and Gallant, Jack L},
  journal={Journal of Neuroscience},
  volume={22},
  number={7},
  pages={2904--2915},
  year={2002},
  publisher={Soc Neuroscience},
  doi={10.1523/JNEUROSCI.22-07-02904.2002}
}

@article{young1992sparse,
  title={Sparse population coding of faces in the inferotemporal cortex},
  author={Young, Malcolm P and Yamane, Shigeru},
  journal={Science},
  volume={256},
  number={5061},
  pages={1327--1331},
  year={1992},
  publisher={American Association for the Advancement of Science},
  doi={10.1126/science.1598577}
}

@article{quiroga2008sparse,
  title={Sparse but not 'Grandmother-cell' coding in the medial temporal lobe},
  author={Quiroga, R Quian and Kreiman, Gabriel and Koch, Christof and Fried, Itzhak},
  journal={Trends in Cognitive Sciences},
  volume={12},
  number={3},
  pages={87--91},
  year={2008},
  publisher={Elsevier},
  doi={10.1016/j.tics.2007.12.003},
}

@article{haider2010synaptic,
  title={Synaptic and network mechanisms of sparse and reliable visual cortical activity during nonclassical receptive field stimulation},
  author={Haider, Bilal and Krause, Matthew R and Duque, Alvaro and Yu, Yuguo and Touryan, Jonathan and Mazer, James A and McCormick, David A},
  journal={Neuron},
  volume={65},
  number={1},
  pages={107--121},
  year={2010},
  publisher={Elsevier},
  doi={10.1016/j.neuron.2009.12.005},
}

@article{carlson2011sparse,
  title={A sparse object coding scheme in area V4},
  author={Carlson, Eric T and Rasquinha, Russell J and Zhang, Kechen and Connor, Charles E},
  journal={Current Biology},
  volume={21},
  number={4},
  pages={288--293},
  year={2011},
  publisher={Elsevier},
  doi={10.1016/j.cub.2011.01.013}
}

@article{willmore2011sparse,
  title={Sparse coding in striate and extrastriate visual cortex},
  author={Willmore, Ben DB and Mazer, James A and Gallant, Jack L},
  journal={Journal of Neurophysiology},
  volume={105},
  number={6},
  pages={2907--2919},
  year={2011},
  publisher={American Physiological Society Bethesda, MD},
  doi={10.1152/jn.00594.2010}
}

@article{sachdev2012surround,
  title={Surround suppression and sparse coding in visual and barrel cortices},
  author={Sachdev, Robert NS and Krause, Matthew R and Mazer, James A},
  journal={Frontiers in neural circuits},
  volume={6},
  pages={43},
  year={2012},
  publisher={Frontiers Media SA},
  doi={10.3389/fncir.2012.00043},
}

@article{beyeler2019neural,
  title={Neural correlates of sparse coding and dimensionality reduction},
  author={Beyeler, Michael and Rounds, Emily L and Carlson, Kristofor D and Dutt, Nikil and Krichmar, Jeffrey L},
  journal={PLoS Computational Biology},
  volume={15},
  number={6},
  pages={e1006908},
  year={2019},
  publisher={Public Library of Science San Francisco, CA USA},
  doi={10.1371/journal.pcbi.1006908},
}

@article{zaslaver2015hierarchical,
  title={Hierarchical sparse coding in the sensory system of Caenorhabditis elegans},
  author={Zaslaver, Alon and Liani, Idan and Shtangel, Oshrat and Ginzburg, Shira and Yee, Lisa and Sternberg, Paul W},
  journal={Proceedings of the National Academy of Sciences},
  volume={112},
  number={4},
  pages={1185--1189},
  year={2015},
  publisher={National Acad Sciences},
  doi={10.1073/pnas.1423656112},
}

@article{froudarakis2014population,
  title={Population code in mouse V1 facilitates readout of natural scenes through increased sparseness},
  author={Froudarakis, Emmanouil and Berens, Philipp and Ecker, Alexander S and Cotton, R James and Sinz, Fabian H and Yatsenko, Dimitri and Saggau, Peter and Bethge, Matthias and Tolias, Andreas S},
  journal={Nature Neuroscience},
  volume={17},
  number={6},
  pages={851--857},
  year={2014},
  publisher={Nature Publishing Group US New York},
  doi={10.1038/nn.3707},
}

@article{ecker2010decorrelated,
  title={Decorrelated neuronal firing in cortical microcircuits},
  author={Ecker, Alexander S and Berens, Philipp and Keliris, Georgios A and Bethge, Matthias and Logothetis, Nikos K and Tolias, Andreas S},
  journal={Science},
  volume={327},
  number={5965},
  pages={584--587},
  year={2010},
  publisher={American Association for the Advancement of Science},
  doi={10.1126/science.1179867}
}

@article{renart2010asynchronous,
  title={The asynchronous state in cortical circuits},
  author={Renart, Alfonso and De La Rocha, Jaime and Bartho, Peter and Hollender, Liad and Parga, N{\'e}stor and Reyes, Alex and Harris, Kenneth D},
  journal={Science},
  volume={327},
  number={5965},
  pages={587--590},
  year={2010},
  publisher={American Association for the Advancement of Science},
  doi={10.1126/science.1179850}
}

\end{document}